\documentclass[]{tandf2}

\usepackage{graphicx,epsfig}
\usepackage{xcolor}
\usepackage{amsmath}
\usepackage{amssymb}
\usepackage{amsthm}
\usepackage{gensymb}
\usepackage[hypertexnames=false]{hyperref} % option secures correct links after putting counters to zero
\hypersetup{colorlinks=true, linkcolor=blue, citecolor=blue, urlcolor=blue}
\usepackage{textcomp}
\usepackage{longtable}
\usepackage{siunitx}
\usepackage{comment}
\usepackage{rotating} % for creating rotated table
\usepackage{array, booktabs} % for generatingthe table of figures at the end
\usepackage{pdflscape}

\usepackage{subfigure}
\usepackage[numbers,sort&compress]{natbib}% Citation support using natbib.sty
\bibpunct[, ]{[}{]}{,}{n}{,}{,}% Citation support using natbib.sty
\renewcommand\bibfont{\fontsize{10}{12}\selectfont}% Bibliography support using natbib.sty
\makeatletter
\def\NAT@def@citea{\def\@citea{\NAT@separator}}% Suppress spaces between citations using natbib.sty
\makeatother

\allowdisplaybreaks

\newcommand{\ket}[1]{\left| #1 \right>} % for Dirac bras
\begin{document}

\articletype{REVIEW ARTICLE}
\title{Cavity QED with Quantum Gases: New Paradigms in Many-Body Physics}

\author{
\name{Farokh Mivehvar\textsuperscript{a}
and Francesco Piazza\textsuperscript{b}
and Tobias Donner\textsuperscript{c}$^{\ast}$\thanks{$^\ast$ Corresponding author. Email: donner@phys.ethz.ch}
and Helmut Ritsch\textsuperscript{a}
}
\affil{\textsuperscript{a}Institut f\"ur Theoretische Physik, Universit{\"a}t Innsbruck, A-6020~Innsbruck, Austria\\ \textsuperscript{b}Max-Planck-Institut f\"{u}r Physik komplexer Systeme, D-01187 Dresden, Germany\\ 
\textsuperscript{c}Institute for Quantum Electronics, ETH Zurich, CH-8093 Zurich, Switzerland}
}

\maketitle

\begin{abstract}

We review the recent developments and the current status in the field of quantum-gas cavity QED. Since the first experimental demonstration of atomic self-ordering in a system composed of a Bose-Einstein condensate coupled to a quantized electromagnetic mode of a high-$Q$ optical cavity, the field has rapidly evolved over the past decade. The composite quantum-gas--cavity systems offer the opportunity to implement, simulate, and experimentally test fundamental solid-state Hamiltonians, as well as to realize non-equilibrium many-body phenomena beyond conventional condensed-matter scenarios. This hinges on the unique possibility to design and control in open quantum environments photon-induced tunable-range interaction potentials for the atoms using tailored pump lasers and dynamic cavity fields. Notable examples range from Hubbard-like models with long-range interactions exhibiting a lattice-supersolid phase, over emergent magnetic orderings and quasicrystalline symmetries, to the appearance of dynamic gauge potentials and non-equilibrium topological phases. Experiments have managed to load spin-polarized as well as spinful quantum gases into various cavity geometries and engineer versatile tunable-range atomic interactions. This led to the experimental observation of spontaneous discrete and continuous symmetry breaking with the appearance of soft-modes as well as supersolidity, density and spin self-ordering, dynamic spin-orbit coupling, and non-equilibrium dynamical self-ordered phases among others. In addition, quantum-gas--cavity setups offer new platforms for quantum-enhanced measurements.  In this review, starting from an introduction to basic models, we \emph{pedagogically} summarize a broad range of theoretical developments and put them in perspective with the current and near future state-of-art experiments.

\end{abstract}

\begin{keywords}
cavity quantum electrodynamics (QED), ultracold quantum gases, Bose-Einstein condensate (BEC), Fermi gases, strong matter-field coupling, Dicke superradiance, self-organization, cavity-enhanced metrology
\end{keywords}
\newpage

\tableofcontents
\clearpage
\section{Introduction}
\label{sec:intro}

\subsection{Historical remarks}
Already since the historical conception that matter is built up of  indivisible elementary particles, {\sl a-tomos}, in ancient Greek philosophy, a deeper understanding of the detailed structure, the formation, and the physical properties of solid matter has been at the heart of physics and, in particular, material research. Much later with the advent of the periodic table of elements it became finally clear that matter in all of its uncountable forms and wide ranging properties is composed of only about a hundred atomic elements, which are held together by electromagnetic forces and governed by the laws of quantum mechanics. While the corresponding Schr\"{o}dinger equation can, at least in principle, readily be written down even in quite general circumstances, its full solution can be found solely in very few, exceptionally simple cases. 

Tremendous progress in experimental techniques has made it possible to nowadays analyze the detailed composition, the structure, and essential properties of a wide class of solids down to the atomic level. Simplified but still very powerful theoretical models and methods as well as a plethora of numerical techniques have been developed in parallel to understand and explain the mechanical, optical, electrical, and thermal properties of materials. Many fundamental properties like crystal and electronic band structures, conductivity, or optical properties nowadays can be numerically predicted and simulated to be at least qualitatively understood. In this context a finite set of generic Hamiltonians, such as lattice spin models in various dimensions~\cite{Micheli2006A} or generalized Hubbard models~\cite{Zwerger2003Bose}, have proven to be a viable basis for successful theoretical modeling of many observed phenomena. Despite these enormous efforts, several important phenomena, such as high-$T_c$ superconductivity, quantum magnetism, or quantum phase transitions, still remain elusive due to the complexity of corresponding materials and the hopelessly large Hilbert space needed to study realistic-size quantum many-body problems in full detail. 

As already pointed out by Feynman decades ago~\cite{Lloyd1996Universal}, a possible alternative route to investigate the physics of more complex and realistic Hamiltonians is to implement them experimentally by help of other quantum systems. These systems should be chosen such that they can be much better controlled, observed in real time in a sufficiently precise way, and well understood~\cite{Cirac2012Goals}. While this suggestion of the quantum simulation had been considered as a sort of a \textit{Gedankenexperiment} for decades, breathtaking recent advances of laser technology and the laser manipulation of atomic gases have made it feasible to implement a wide class of analog quantum simulations in atomic-molecular-optical (AMO) laboratories~\cite{Georgescu2014Quantum}. Generic model Hamiltonians developed to describe key properties of solids can now be tested in extremely well-controlled, tunable environments magnified by two to three orders of magnitude compared to original solid materials. Ultracold quantum gases in tailored optical potentials have so become a proven workhorse to explore complex multiparticle quantum dynamics in synthetic solid-state analogs~\cite{Lewenstein2007Ultracold, Bloch2008Many}.  
    
A prominent and fundamentally important class of solids exhibits a regular periodic crystalline structure with a corresponding periodic potential for high energy electrons. The resulting single-particle Bloch energy bands decisively determine electric, magnetic, optical, and thermal properties of such crystalline solids. A finite set of laser beams can be utilized to create readily analogous, defect-free, spatially periodic light intensity distributions mimicking a large class of crystalline potentials at more than a hundred-fold increased scale without any undesired perturbation. Atoms cooled down to almost absolute zero kinetic temperature can then be used to play the role of electrons in crystals and simulate their orbital quantum motion and interactions in these optical lattice structures at a directly observable scale~\cite{Georgescu2014Quantum}.
 
By proper choice of laser-light frequencies and atomic species, virtually back-action-free optical potentials spanning over macroscopic distances can be implemented and tuned in space and in real time. Employing more atomic internal states allows also to incorporate spin degrees of freedom and simulate quantum magnetism. Furthermore, geometric phases can be imprinted on charge-neutral atoms via tailored atom-field coupling, in order to mimic minimal coupling and gauge potentials~\cite{Dalibard2011Artificial, Goldman2014Light}. These vast possibilities have led to a huge number of experimental implementations~\cite{Lewenstein2007Ultracold,Bloch2008Many}, a development which in turn triggered widespread interest in physics communities beyond the AMO. This includes, in particular, the solid-state community seeing a unique opportunity to adapt and implement their paradigmatic model Hamiltonians, and test their predictions in a controlled way using quantum-gas setups, which nowadays can analyze quantum phases with single-site resolution \cite{Bakr2009Quantum, Sherson2010Single, Mazurenko2017Cold, Gross2017Quantum}.

Already in the early stage of the field using then available tools for ultracold atoms, a seminal quantum-gas experiment exhibited a reversible quantum phase transition between a superfluid and a Mott-insulator phase exactly as predicted by the Bose-Hubbard Hamiltonian~\cite{Greiner2002Quantum}. A huge number of further successful demonstrations of intriguing condensed-matter phenomena such as superfluidity, quantum magnetism, Abelian and non-Abelian gauge potentials, just to name a few, has followed since~\cite{Hofstetter2018Quantum}.   
 
Beyond these early, yet very impressive, successful demonstrations, however, several fundamental open challenges for quantum simulation of solid-state materials with quantum gases have remained. Generally, the underlying lattice structure is externally prescribed and is independent of particle number and their state. This hinders studies of self-ordering and crystal formation. Likewise, it is challenging to represent the effect of phonons (i.e., lattice vibrations). More generally,  implementation of long- or infinite-range atom-atom interactions proves to be rather demanding. This includes notably pairing interactions in momentum space as in various condensed-matter models for superconductivity. Furthermore, fundamental questions concerning crystal growth and melting, or lattice-defect formation are not so straightforward to implement. The externally prescribed order generally inhibits structural phase transitions or the appearance of new long-range order through dynamical symmetry breaking. As a practical limitation on the other hand, the observation of any kind of time evolution is generally very slow as measurements are typically destructive in quantum-gas experiments which impedes repeated observations within a single measurement sample.  

The challenge to implement these effects in quantum-gas setups mostly stems from the fact that optical potentials are typically required to operate in regimes where electromagnetic fields generating the optical potentials are far detuned from any optical atomic transition, so that most photons propagate almost freely through the atomic gas. Consequently, the back-action of the trapped particles on lattice fields, which is at the heart of phonon-induced interaction potentials, is extremely small and negligible in many cases, and sizeable effects only appear in very large setups or closer to resonance when dissipation starts to be an issue~\cite{Asboth2008Optomechanical}. To mimic phonon-like long-range coupling in free-space configurations, a diverse set of alternative approaches has been exploited. Among others these encompass the creation of hetero-nuclear molecules, the use of atomic species with strong magnetic dipole moments or the admixture of Rydberg states giving rise to induced electric dipole moments~\cite{Lahaye2009Physics, Baranov2012Condensed} . In all these cases the original lattice remains static, stable, and perfect. As we will see below by help of multi-mode cavity fields and atomic self-ordering this limitation can be largely overcome.

\subsection{Quantum-gas cavity QED}
\begin{figure}[t!]
  \begin{center}
    \includegraphics[width=0.80\columnwidth]{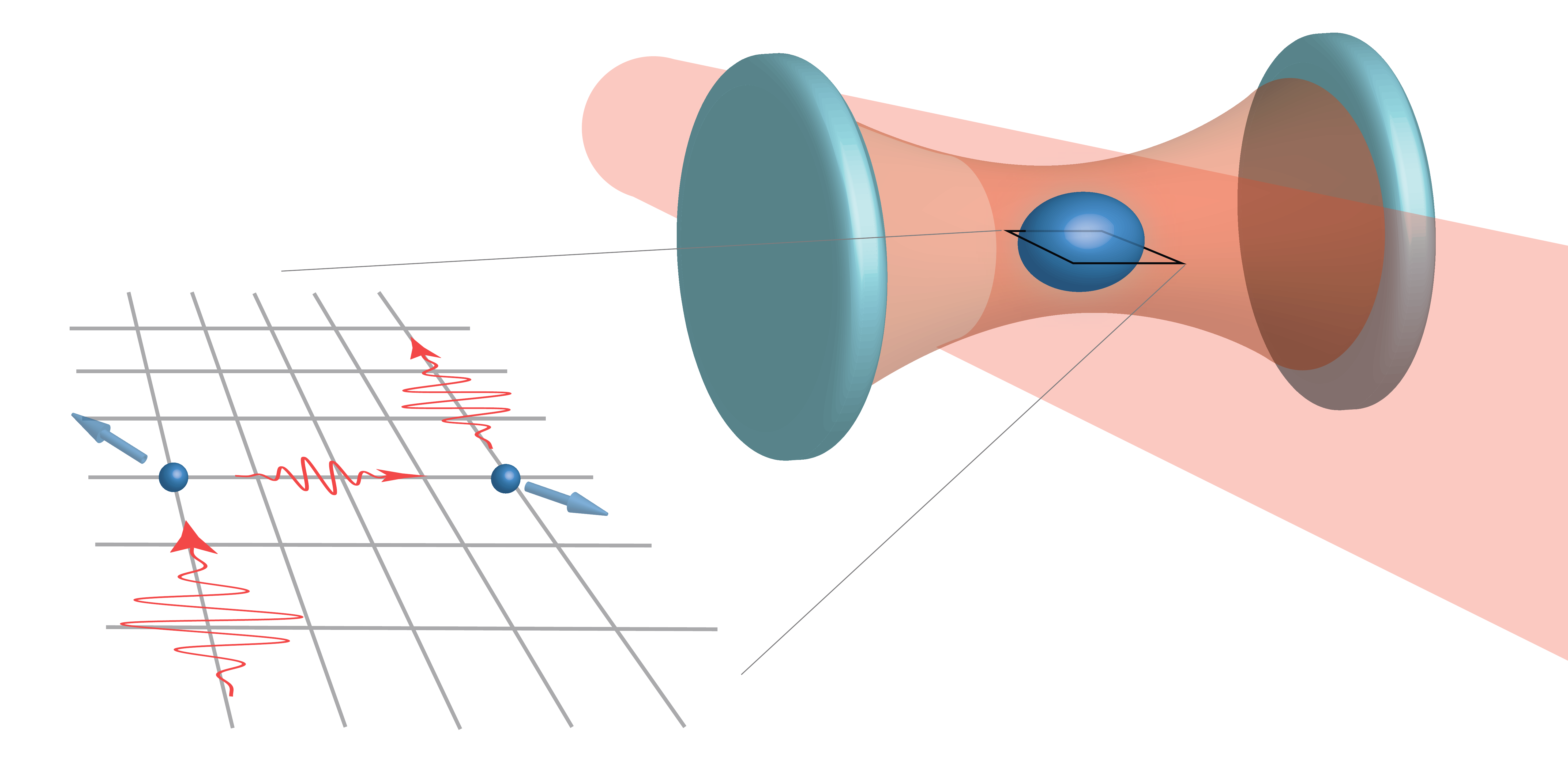}
   \caption{Basic scheme of cavity QED with a quantum gas to realize a self-ordering phase transition. A quantum gas (blue sphere) is loaded inside a standing-wave cavity and driven by a transverse, coherent pump-laser field. The schematic drawing on the left side illustrates the fundamental photon scattering processes inducing long-range interactions between the atoms, where the grid indicates the standing-wave mode structures of the cavity and the pump fields: A photon (red wiggly lines) from the pump beam is scattered off a first atom into the cavity mode and then back into the pump off a second atom. Both atoms acquire a momentum kick (blue arrows) due to the photon recoil. These atoms effectively interact with each other via the exchange of the cavity photon.}
    \label{Fig_intro_scheme_a} 
  \end{center}
\end{figure}

In this review we concentrate on an alternative approach to implement versatile, tailored interatomic interactions and explore novel many-body phenomena using the framework of quantum gases by the help of optical cavities (also referred to as resonators) composed of high-quality mirrors to enhance atom-photon interaction; see Figure~\ref{Fig_intro_scheme_a}. Any conservative optical force on atoms microscopically originates from the coherent photon redistribution among various electromagnetic fields acting on the atoms. In principle, any photon scattered by one atom subsequently interacts with the other atoms of the ensemble and can be re-scattered, resulting in non-local photon-induced interactions among the atoms. Hence, non-local interatomic interactions are already naturally built into optical potentials. 

Sequential photon scattering processes and strong atom-photon coupling in principle always lead to atomic back-action on the field, non-local interatomic interactions, and nonlinear effects~\cite{Asboth2008Optomechanical,Ostermann2016Spontaneous}. However, these sequential photon scattering processes are extremely unlikely and usually negligible for typical free-space configurations operated in far-detuned regimes to reduce dissipation and heating. Luckily the minute back-action of the atoms on the trapping electromagnetic fields can be dramatically enhanced by the help of optical resonators. This hinges on the fact that confining the scattered fields within a high-$Q$ cavity allows each photon to multiply pass through the same atomic ensemble. Hence resonators significantly enhance the tiny probability of a scattered photon to be re-scattered by the same or any other atom in the ensemble resulting in photon-induced interatomic interactions~\cite{Mottl2012Roton,Konya2014Damping,Guo2021An}---reminiscent of phonon-induced interactions in natural crystals~\cite{Altland2010Condensed}. 

Mode functions of a cavity form a basis in its volume. While a single cavity mode covering the whole quantum gas mediates infinite-range interactions among the atoms, by employing a large number of degenerate cavity modes as in confocal or concentric cavities the range of interatomic interactions can be vastly tuned~\cite{Domokos2002Dissipative,Salzburger2002Enhanced,Gopalakrishnan2009Emergent,Vaidya2018Tunable}. Also by help of a manifold of pump frequencies (as in frequency combs) and modes, range and shape of interatomic interactions can be controlled~\cite{Torggler2017Quantum,Torggler2020Self}. Therefore, tailored sequential photon scattering in optical resonators enables to create a large class of sizable tunable-range interatomic interactions.

Let us note here that for spatially coherent light fields, as they are generally employed to create optical potentials, all possible scattering amplitudes for a photon by different atoms interfere. Depending on relative phases, they may collectively add up or cancel destructively all together. The collective enhancement is in close analogy with the well-known Bragg enhancement of coherent wave scattering by periodic structures, which here is further amplified in the presence of an optical resonator.

Some intriguing aspects of the collective motional coupling of a quantum gas to a cavity mode have been pointed out already two decades ago~\cite{Horak2001Dissipative}. It then took several years until the first generation of pioneering experiments with cold atomic gases as well as Bose-Einstein condensates (BECs) in driven optical cavities appeared~\cite{Black2003Observation,Kruse2003Cold,Gupta2007Cavity,Colombe2007Strong,Brennecke2008Cavity}. As it was shown in several theoretical approaches and first experiments, the collectively enhanced atom-field interaction allowed to implement strong coupling of the center-of-mass mode of the atomic ensemble to a single cavity mode. This corresponds to the implementation of a zero-temperature optomechanical Hamiltonian in the so-called strong coupling limit, where single cavity photons create sizable forces and highly nonlinear dynamics~\cite{Nagorny2003Collective,Kruse2003Cold,Slama2007Superradiant,Brennecke2008Cavity, Purdy2010Tunable}.

The potential of cavity-enhanced forces and interatomic interactions was, however, fully highlighted experimentally a bit later in a seminal demonstration of superradiant self-ordering of a transversely-driven homogeneous BEC to an ordered crystalline phase at zero temperature~\cite{Baumann2010Dicke,Nagy2008Self,Nagy2010Dicke}. In detail, quantum fluctuations in the homogeneous atomic density stimulate collective scattering of photons from a transversely applied, standing-wave laser into the cavity mode. In turn, photons scattered into the cavity act back non-linearly on the atoms and amplify the BEC density fluctuations~\cite{Horak2001Dissipative,Nagy2008Self,Nagy2011Critical,Mottl2012Roton}. Above a critical pump power, the system undergoes a reversible zero-temperature quantum-phase transition into a perfectly ordered, crystalline state.  Here, in theory quantum fluctuations of the radiation field play a decisive role~\cite{Ostermann2020Unraveling} and atom-field entanglement appears to be vital in this zero temperature transition~\cite{Maschler2007Entanglement}.

Mathematically, the underlying dynamics near the self-ordering transition can be well described by the Dicke Hamiltonian. Hepp and Lieb predicted more than half a century ago that the Dicke Hamiltonian exhibits a zero-temperature phase transition involving a $\mathbf{Z}_2$ parity-symmetry breaking to a so-called superradiant state in the limit of very strong coupling~\cite{Hepp1973On,Wang1973Phase,Bastidas2012Nonequilibrium}. Truncating the self-ordering Hamiltonian close to threshold to low atomic momenta yields exactly the Dicke Hamiltonian with tunable parameters, allowing thus to emulate the corresponding phase transition~\cite{Nagy2010Dicke}. As the original Hepp and Lieb prediction based on a very simplified toy model created a long-standing and controversial debate on its validity, this first and very clear experimental demonstration of the existence of a superradiant phase transition at exactly the predicted parameter values created huge attention even far beyond the cold-atom community. From a different viewpoint, the experiment can be seen as one of the successful implementations of a quantum simulation of a multi-particle Hamiltonian.

Further calculations and a closer look on the detailed underlying dynamics including local atomic interactions revealed even more remarkable properties of the realized self-ordered state. In particular, the self-ordered state has been shown to simultaneously possess spatial order associated with the spontaneously broken discrete $\mathbf{Z}_2$ symmetry as well as  phase coherence (i.e., off-diagonal long-range order) of the atomic state. As the combination of these two properties is usually attributed to lattice supersolids, the self-ordered state constitutes a prominent candidate for such an intriguing state of matter.  

\begin{figure}[t!]
  \begin{center}
    \includegraphics[width=1\columnwidth]{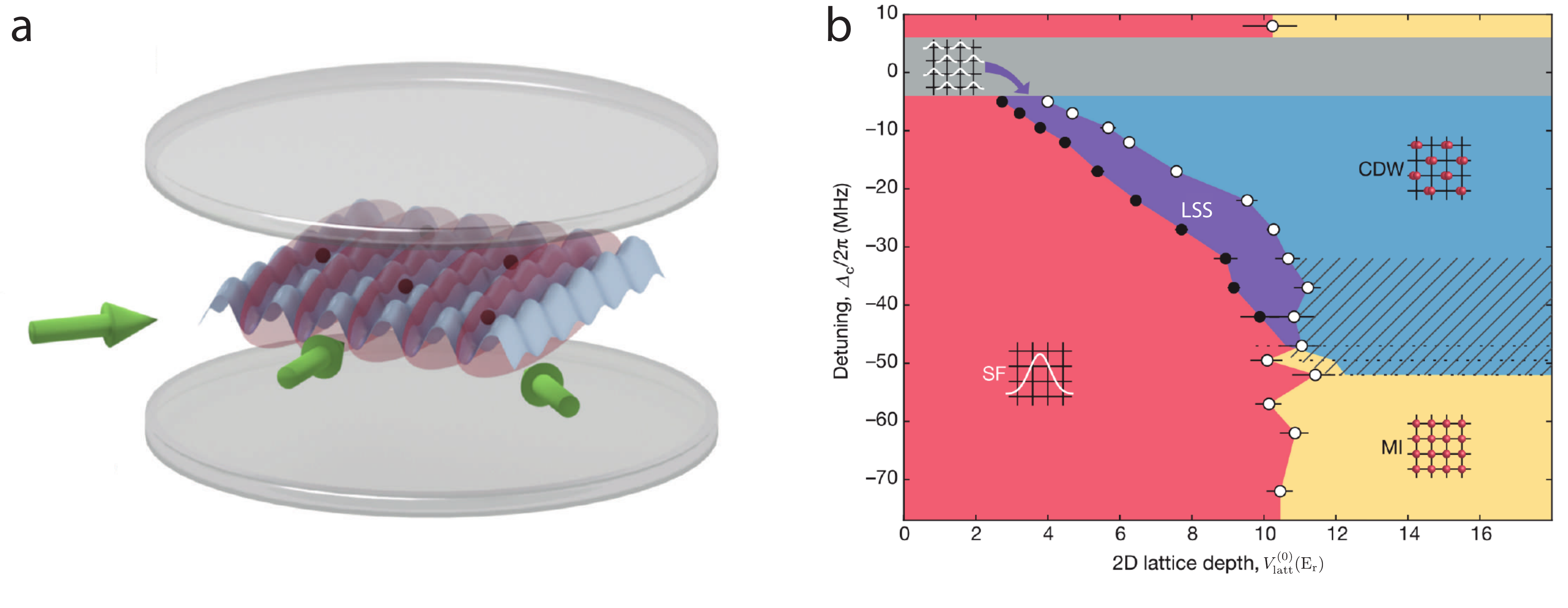}
   \caption{Cavity QED with lattice quantum gases. (a) Schematic sketch of a typical setup of cavity QED with quantum gases trapped in a static lattice. The system is described by an extended Hubbard-like Hamiltonian with cavity-induced global, all-to-all atomic interactions. (b) Phase diagram of an extended Bose-Hubbard model with cavity-induced global atomic interactions obtained in the ETH Zurich lattice-setup experiment. In addition to the common superfluid (SF) and Mott-insulator (MI) states, the system also exhibits a novel lattice supersolid (LSS) and a charge-density-wave (CDW) state.  Figure (a) reprinted from Ref.~\cite{Torggler2019A}, and Figure (b) adapted and reprinted with permission from Ref.~\cite{Landig2016Quantum} published at 2016 by the Nature Publishing Group.} 
    \label{Fig_intro_scheme_b} 
  \end{center}
\end{figure}

After initial theoretical works and these experimental realizations, a plethora of proposals for generalizations and novel directions to pursue was put forward. A very prominent example was the introduction of cavity-mediated long-range interatomic interactions in Hubbard models for bosons~\cite{Mekhov2012Quantum} as well as for fermions~\cite{Jiang2011Exciton}. The phase diagram of the Bose-Hubbard model with cavity-induced global atom-atom interactions was predicted using various approximate theoretical approaches~\cite{Maschler2008Ultracold, Fernandez-Vidal2010Quantum,Li2013Lattice,Bakhtiari2015Nonequilibrium,Niederle2016Ultracold} and subsequently successfully measured experimentally~\cite{Klinder2015Dynamical, Landig2016Quantum}, as shown in Figure~\ref{Fig_intro_scheme_b}. Here, strong evidence of an intermediate lattice-supersolid phase between the homogeneous superfluid and the density-ordered superradiant phase was found in agreement with theoretical models. Subsequent employing of two crossed linear cavities enabled the observation of a supersolid state breaking a continuous symmetry~\cite{Leonard2017Supersolid}.  Nontrivial topological insulator states have also been theoretically predicted to appear in fermionic quantum-gas cavity QED~\cite{Mivehvar2017Superradiant,Chanda2020Self-organized}. With the first experimental demonstrations of interacting ultracold fermions coupled to cavities, experimental confirmation could be envisaged soon~\cite{Roux2020Strongly,Roux2021Cavity}.

Cavity-induced interatomic interactions can be tuned from attractive to repulsive simply by changing the laser frequencies with respect to cavity resonances~\cite{Piazza2015Self,Zupancic2019P}. Already in simple 1D configurations repulsive potentials allow to implement intriguing scenarios such as a nonequilibrium Su-Schrieffer-Heeger model with periodically modulated tunneling amplitude, where symmetry-protected edge states and other topologically nontrivial effects are expected to appear~\cite{Mivehvar2017Superradiant}. Particularly, in 2D configurations interesting phenomena have been predicted due to the repulsive cavity-induced interactions, including controllable dynamical instabilities in conjunction with limit cycles---also referred to as time crystals in this context---as well as strong indications of chaotic dynamics~\cite{Griesser2011Nonlinear,Piazza2015Self,Cosme2018Dynamical,Kessler2019Emergent}.

\begin{figure}[t!]
  \begin{center}
   \includegraphics[width=\columnwidth]{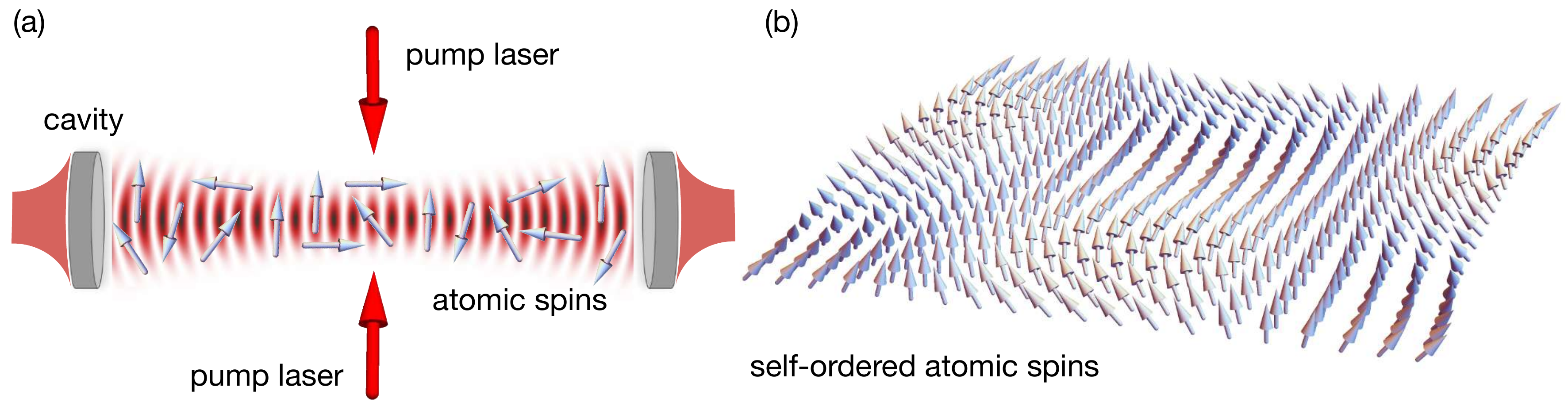}
   \caption{Cavity-QED implementation of quantum magnetism. (a) Schematic sketch of the setup. Spinor ultracold atoms (light-blue arrows) are driven transversely by coherent pump lasers (red arrows) and strongly coupled to a cavity mode. By properly designing the spatial forms and frequencies of the lasers and the cavity field, a variety of long-range spin-spin interactions can be induced between atoms by cavity photons. (b) A typical cavity-induced magnetic self-ordering. Here, the cavity mediates Dzyaloshinskii-Moriya-type interactions between the atoms, resulting in chiral spin-spiral order. Figure reprinted with permission from Ref.~\cite{Mivehvar2019Cavity} \textcopyright~2019 by the American Physical Society.}
    \label{fig:intro-scheme-spin-texture} 
  \end{center}
\end{figure}

As atomic species generally possess isotopes with variable nuclear spin, they come with a large variation of hyperfine level structures. These hyperfine internal states can be exploited to simulate spin degrees of freedom. Again, the direct spin-spin interaction between atoms is quite weak at optical wavelength distances. Nevertheless, it has been suggested that using specifically designed cavity-enhanced optical Raman transitions between hyperfine states of a multi-component quantum gas allows to implement much stronger interatomic spin interactions and simulate a wide class of long-range quantum spin Hamiltonians, including Heisenberg and Dzyaloshinskii-Moriya interactions, across the self-ordering phase transition~\cite{Mivehvar2017Disorder, Mivehvar2019Cavity}; see Figure~\ref{fig:intro-scheme-spin-texture}.  The first fundamental steps towards this goal had already been implemented in experiments~\cite{Landini2018Formation,Kroeze2018Spinor}, opening a new route towards quantum magnetism in atomic quantum gases. The use of cavity-enhanced Raman transitions between hyperfine states also enabled the first realization of a dynamical synthetic spin-orbit coupling~\cite{Kroeze2019Dynamical}, paving the way for the implementation of dynamical artificial gauge potentials.

Interestingly, cavity-induced interactions can also be tuned to couple various species of atoms or even molecules in a single or even distinct traps~\cite{Griesser2012Cooperative}. A recent proposal even suggested to study phenomenon of many-body localization~\cite{Sierant2019Many} as well as special aspects of ergodicity in quantum-gas--cavity-QED systems despite the presence of virtually infinite-range interactions~\cite{Kubala2020Ergodicity}.

The inevitable presence of dissipation via the leaking cavity field has two important consequences. First, it renders the system to be a driven-dissipative one. This not only leads to quantitative modifications of the physics such as shifted critical points or modified scaling of the critical fluctuations \cite{Brennecke2013Real,Landig2015Measuring,Nagy2011Critical,Nagy2015Nonequilibrium,Piazza2013Bose}, but also allows to explore dissipative phase transitions to states that would otherwise not be present. This includes, for example, dissipation-stabilized phases and gauge potentials, and non-stationary states~\cite{Soriente2018Dissipation,Dogra2019Dissipation,Soriente2021Distinctive,Colella2021Open}. 

Second, as a generic, advantageous consequence of the openness of optical resonators, real-time information on dynamics of cavity-QED systems can be obtained simply by non-destructive observation and analysis of cavity output fields~\cite{Mekhov2007Light,Sandner2015Self,Landig2015Measuring}, thus complementing or replacing destructive absorption-imaging measurements of atomic density distributions. Besides a precise and non-destructive, real-time access to atomic quantum currents as recently studied~\cite{Laflamme2017Continuous,Uchino2018Universal}, this in principle paves the way towards atom-based novel and improved implementations of quantum-enhanced sensing~\cite{Peden2009Nondestructive,Leroux2010Implementation,Gietka2019Supersolid,Motazedifard2019Force,Gietka2020Magnetometry}. Moreover, in some cases one can even expect the appearance of genuine consequences of quantum measurement back-action on the system, which can be controlled on a new level of precision~\cite{Elliott2015Multipartite,Ivanov2020Feedback}. Here again genuine quantum phenomena such as spontaneous antiferromagnetic ordering can be induced via controlled measurement projections~\cite{Mazzucchi2016Quantum}.

\subsection{The present review article}

\subsubsection{Scope of the review article}

The experimental merger of quantum gases with cavity QED opened a wide field of possibilities ranging from cavity optomechanics at zero temperature to the real-time exploration of quantum phase transitions~\cite{Ritsch2013Cold}. The central physical mechanism is the quantum back-action between the dynamic cavity fields and the atomic degrees of freedom. Combining the excellent control of quantum-gas experiments over external and internal atomic degrees of freedom with the strong coupling to quantized light fields in high-finesse optical resonators provides completely new opportunities specifically for the quantum simulation of many-body systems~\cite{Cirac2012Goals, Gopalakrishnan2012Exploring, Arguello-Luengo2019Analogue} as well as for quantum-enhanced optimization~\cite{Torggler2017Quantum,Marsh2020Enhancing}, for cavity optomechanics in the bad~\cite{Gupta2007Cavity, Brennecke2008Cavity, Purdy2010Tunable, Brooks2012Non, Spethmann2016Cavity} and in the good~\cite{Wolke2012Cavity, Kessler2014Optomechanical, Kessler2016In,Klinder2016Bose} cavity limits, and for quantum sensing~\cite{Gietka2019Supersolid}.

The field has rapidly expanded in the last decade. In order for the material to fit into a single manuscript with a reasonable length, we restrict this review to scenarios where the back-action between the quantum gas and the cavity field leads to the emergence of correlated many-body phases, some of which go beyond the scope of quantum simulation in that they offer novel paradigms in many-body physics. In particular, the systems of interest in the present review are chiefly \emph{transversely driven quantum gases} inside single- and many-mode cavities. Therefore, we do not include and discuss explicitly thermal gases inside cavities~\cite{Domokos2002Collective, Black2003Observation, Kruse2003Cold, Maschler2005Cold}, cavity optomechanical systems with quantum gases~\cite{Brennecke2008Cavity}, and Dicke-type cavity models with atomic mediums without quantized motion~\cite{Dimer2007Proposed, Gopalakrishnan2011Frustration, Zhiqiang2016Nonequilibrium, Masson2017Cavity, Norcia2018Cavity, Davis2019Photon}. Such scenarios have been extensively considered in the previous review articles~\cite{Domokos2003Mechanical, Ritsch2013Cold, Kirton2018Introduction} and we refer interested readers to these excellent reviews. Despite these restrictions in choosing the systems of interest, the scope of the current review is quite vast and encompasses plethora of intriguing physical phenomena.

\subsubsection{Structure of the review article}

 In Section~\ref{sec:theory} we introduce basic physical notions and models for interacting quantum gases coupled to optical cavities, followed by a short review of typical approximate theoretical approaches mainly focusing on mean-field treatments and collective elementary excitations for both bosons and fermions. Basic principles and fundamental experiments on discrete symmetry breaking in superradiant quantum phase transitions leading to the spontaneous self-ordering and crystallization are reviewed in Chapter~\ref{sec:SR_discrete}, followed by foundational examples for continuous symmetry breaking and the formation of nonequilibrium supersolids in Chapter~\ref{sec:supersolid}. Section~\ref{sec:multimode} is devoted to  quantum gases strongly coupled to many modes of optical cavities featuring tunable-range cavity-mediated interatomic interactions as demonstrated recently, while in Chapter~\ref{sec:spinor-selfordering} we review the dynamics of multilevel quantum gases in cavities introducing the notion of cavity-induced long-range spin-spin interactions as a versatile platform for simulating quantum magnetism. Optical atomic lattice models including cavity-mediated interatomic interactions are introduced and discussed in detail in Section~\ref{sec:extended-BH}, while dynamical synthetic lattice gauge potentials and nonequilibrium topological states are reviewed in Chapter~\ref{sec:synthetic-gauge-fields-TI}. The great versatility of cavity-generated dynamical optical potentials and self-ordering is further highlighted in Chapter~\ref{sec:qc} discussing quasicrystal formation and emergent symmetries in geometries involving a few cavities. In Chapter~\ref{sec:dynamics} we review further prospects and consequences of the openness of optical resonators in the atomic self-ordering, which feature dissipation forces, strongly nonlinear dynamics, and dynamical instabilities, followed by a short discussion on measurement back-action and cavity-enhanced sensing on Section~\ref{sec:quantmeasure}. Finally in Section~\ref{sec:conclusions}, we conclude by adding some remarks on future prospects and challenges.                 
    
We would like to note here that we have taken a pedagogical approach in composing this review article in the expense of a longer manuscript. In the view of this, chapters are, however, almost independent from each other. In particular, after reading Chapter~\ref{sec:theory}, readers based on their interest and need can select and jump to any chapter without a great discontinuity.

\section{Fundamentals: Ultracold atoms dispersively coupled to a single cavity mode}
\label{sec:theory}

In this chapter, we introduce the framework for studying the many-body physics of cavity QED with quantum gases. We start our description with the most fundamental system: Ultracold atoms  transversely driven by a laser field and dispersively interacting with a single quantized electromagnetic radiation field of a standing-wave linear cavity. Most of the basic concepts and methods can be captured in this simple model. Moreover, this system constitutes the first experimental setup where collective phenomena were observed. In the subsequent chapters, we shall generalize this simple model to different scenarios and regimes, including additional atomic and photonic degrees of freedom---most notably spinor gases as well as multiple cavity modes. For the convenience, we summarize the most important and commonly-used symbols throughout the paper in Appendix~\ref{sec_app:symbols}. We note that some of the symbols we use in this review can be different from those used in some literature.

\subsection{The basic model}
\label{sec:basic-model}

\begin{figure}[t!]
  \begin{center}
    \includegraphics[width=0.65\columnwidth]{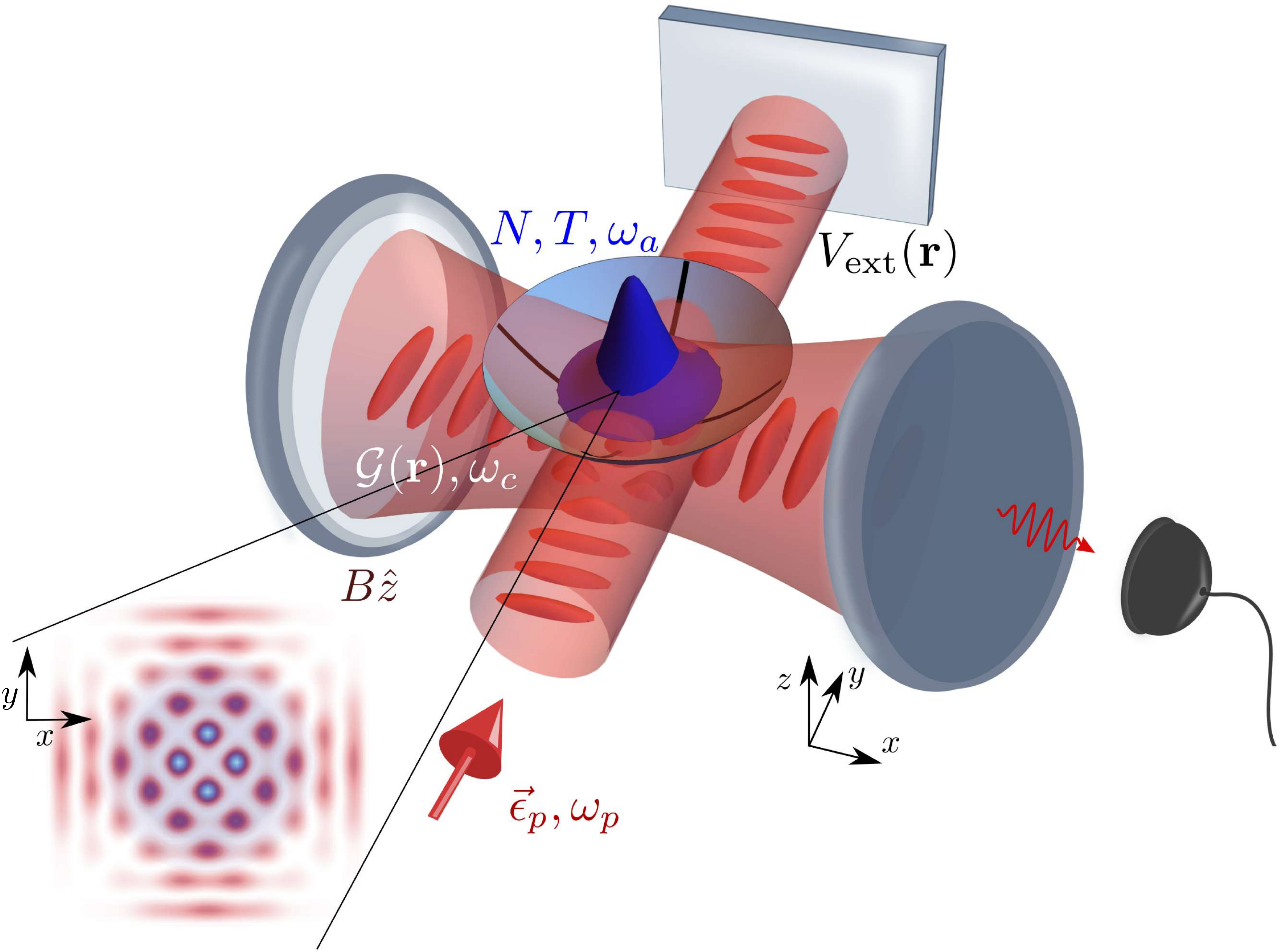}
    \caption{Schematic sketch of a basic quantum-gas--cavity-QED setup. A cloud of $N$ ultracold atoms (blue) at temperature $T$ is confined by an external potential $V_\mathrm{ext}(\mathbf{r})$, centered with respect to the cavity mode (light red). The resonance frequency $\omega_c$ of the cavity is far detuned from the atomic transition frequency $\omega_a$. The coupling between the atomic cloud and the cavity field is captured by the spatially dependent vacuum Rabi rate $\mathcal{G}(\mathbf{r})$. The atomic cloud is driven by a transverse pump laser (red arrows) with a polarization $\boldsymbol{\epsilon}_p$ at a frequency $\omega_p$ close to the cavity resonance frequency. The transverse pump laser is retroreflected by a planar mirror and its coupling to the atoms is described by the spatially dependent Rabi rate $\Omega(\mathbf{r})$. The cavity field leaking out of the resonator at a rate $\kappa$ is detected either via a single-photon counter or a heterodyne system. The quantization axis for the atoms is fixed by an applied magnetic field $\mathbf{B}$.}
    \label{fig:coupling_scheme}
  \end{center}
\end{figure}

Consider ultracold two-level atoms with internal states $\{\ket{g},\ket{e}\}$ and the transition frequency $\omega_a$ inside a single-mode standing-wave linear cavity with the resonance frequency $\omega_c$. The wave length $\lambda_c$ and the wave number $k_c$ of the cavity mode are related to the resonance frequency via $\omega_c=ck_c=2\pi c/\lambda_c$, with $c$ being the speed of light. The atoms are driven by a standing-wave classical coherent laser field in the transverse direction with the position-dependent Rabi frequency $\Omega(\mathbf{r})=\Omega_0\cos(k_cy)$ and are strongly coupled to the quantized mode of the cavity with the strength $\mathcal{G}(\mathbf{r})=\mathcal{G}_0\cos(k_cx)$; see Figure~\ref{fig:coupling_scheme}. $\Omega_0$ and $\mathcal{G}_0$ are the maximum Rabi frequency of the pump field and the maximum single atom-photon cavity-QED coupling strength, respectively. Here we do not take into account the transverse profile of the cavity mode, which is a Gaussian in the simplest case; the transverse profile of cavity modes will be discussed in more detail in Section~\ref{subsubsec:optical_resonators} and will be included in multimode cavities in Section~\ref{sec:multimode}. Note that the Rabi rate $\Omega(\mathbf{r})$ can have a different spatial form (e.g., via using a running-wave pump laser) and the pump laser can also impinge the atomic cloud with an angle different than 90\textdegree\ with respect to the cavity axis; these scenarios will be considered in the following chapters.

The system in the rotating frame of the pump laser is described by the many-body Hamiltonian~\cite{Maschler2008Ultracold},
\begin{align}
\hat{H}&=
-\hbar\Delta_c\hat{a}^\dag\hat{a}
+\sum_{\tau=g,e}\int \hat\psi^\dag_\tau(\mathbf{r}) 
\left[-\frac{\hbar^2}{2M}\boldsymbol{\nabla}^2+V_{\rm ext}(\mathbf{r}) \right] 
\hat{\psi}_\tau(\mathbf{r})d\mathbf{r}
-\hbar\Delta_a \int \hat\psi^\dag_e(\mathbf{r}) \hat\psi_e(\mathbf{r})d\mathbf{r}
\nonumber\\
&+\hbar\int \left\{\hat\psi^\dag_e(\mathbf{r}) 
\left[\Omega(\mathbf{r})+\mathcal{G}(\mathbf{r})\hat{a} \right] 
\hat{\psi}_g(\mathbf{r})+\rm{H.c.}\right\}d\mathbf{r}+\hat{H}_{\rm int},
\end{align}
where $\Delta_a\equiv\omega_p-\omega_a$ is the relative frequency between the pump-laser frequency $\omega_p$ and the atomic transition frequency $\omega_a$, and $\Delta_c\equiv\omega_p-\omega_c$ the relative frequency between the pump-laser and cavity frequencies. Note that although in general $\Delta_c\neq0$, in practice one has $|\Delta_c|/\omega_p\ll1$ implying that $k_p\simeq k_c$, which has been made use of above. Here, $\hat\psi_\tau(\mathbf{r})=\hat\psi_\tau(\mathbf{r},t)$ and $\hat{a}=\hat{a}(t)$ are the slowly-varying bosonic or fermionic atomic, and bosonic photonic annihilation field operators, respectively. $V_{\rm ext}(\mathbf{r})$ is an external trapping potential acting identically on both internal atomic states, and $\hat{H}_{\rm int}$ accounts for two-body interactions between the atoms. For dilute quantum gases at low temperatures, the two-body interaction Hamiltonian is well represented by local contact interactions~\cite{Dalfovo1999Theory},
\begin{align} \label{eq:H_int-2level}
\hat{H}_{\rm int}=\sum_\tau\frac{1}{2}g_{\tau\tau}\int 
\hat\psi_\tau^\dag(\mathbf{r})\psi_\tau^\dag(\mathbf{r})
\hat\psi_\tau(\mathbf{r})\hat\psi_\tau(\mathbf{r})d\mathbf{r}
+g_{eg}\int 
\hat\psi_e^\dag(\mathbf{r})\psi_g^\dag(\mathbf{r})
\hat\psi_g(\mathbf{r})\hat\psi_e(\mathbf{r})d\mathbf{r},
\end{align}
where the two-body contact-interaction strengths $g_{gg}$, $g_{ee}$, and $g_{eg}=g_{ge}$ are related to the respective s-wave scattering lengths $a^{s}_{\tau\tau'}$ by $g_{\tau\tau'}=4\pi a^{s}_{\tau\tau'}\hbar^2/m$ ($\tau,\tau'=g,e$). The strength of the two-body s-wave interactions $g_{\tau\tau'}$ can in principle be tuned using Feshbach resonance techniques. In many experiments in cavity QED, the contact interactions are negligible compared to cavity-mediated interactions (see Section~\ref{subsubsec:cavity-induced-int}) and can be omitted. Note that the s-wave interactions proportional to $g_{gg}$ and $g_{ee}$, i.e., the first term in Equation~\eqref{eq:H_int-2level}, are identically zero for fermionic atoms owing to the Pauli exclusion principle~\cite{Giorgini2008Theory}.

The dynamics of the system is described by the Heisenberg equations of motion for the atomic and photonic field operators, 
\begin{align} \label{eq:Heisberg-eq-ega}
i\hbar\frac{\partial \hat\psi_e(\mathbf{r})}{\partial t}&=
\left[-\frac{\hbar^2}{2M}\boldsymbol{\nabla}^2+V_{\rm ext}(\mathbf{r})-\hbar\Delta_a
+g_{ee}\hat\psi_e^\dag(\mathbf{r})\hat\psi_e(\mathbf{r})
+g_{eg}\hat\psi_g^\dag(\mathbf{r})\hat\psi_g(\mathbf{r}) \right]\hat\psi_e(\mathbf{r})\nonumber\\
&+\hbar\left[\Omega(\mathbf{r})+\mathcal{G}(\mathbf{r})\hat{a} \right]\hat{\psi}_g(\mathbf{r}),\nonumber\\
i\hbar\frac{\partial \hat\psi_g(\mathbf{r})}{\partial t}&=
\left[-\frac{\hbar^2}{2M}\boldsymbol{\nabla}^2+V_{\rm ext}(\mathbf{r})
+g_{gg}\hat\psi_g^\dag(\mathbf{r})\hat\psi_g(\mathbf{r})
+g_{ge}\hat\psi_e^\dag(\mathbf{r})\hat\psi_e(\mathbf{r}) \right]\hat\psi_g(\mathbf{r})\nonumber\\
&+\hbar\left[\Omega(\mathbf{r})+\mathcal{G}(\mathbf{r})\hat{a}^\dag\right]\hat{\psi}_e(\mathbf{r}),\nonumber\\
i\hbar\frac{\partial \hat{a}}{\partial t}&=-\hbar\Delta_c\hat{a}
+\hbar\int \mathcal{G}(\mathbf{r}) \hat\psi^\dag_g(\mathbf{r})\hat{\psi}_e(\mathbf{r})d\mathbf{r},
\end{align}
where we have assumed, without loss of generality, that $\{\Omega_0,\mathcal{G}_0\}\in\mathbb{R}$. The pump and cavity frequencies are assumed to be far detuned from the atomic transition frequency such that $1/|\Delta_a|$ is the fastest time scale in the system. In this ``dispersive'' regime, the atomic excited state is not populated significantly---minimizing heating of the atomic cloud due to spontaneous emission---and its dynamics can be eliminated adiabatically: 
\begin{align} \label{eq:psi-e-ss}
\hat\psi_{e,\rm ss}(\mathbf{r})\simeq\frac{1}{\Delta_a}
\left[\Omega(\mathbf{r})+\mathcal{G}(\mathbf{r})\hat{a} \right]\hat{\psi}_g(\mathbf{r})\,.
\end{align}
Kinetic energy, potential, and contact interactions have been omitted as they are negligible compared to $\hbar\Delta_a$. Substituting the steady-state, excited-state field operator $\hat\psi_{e,\rm ss}(\mathbf{r})$ into Equation~\eqref{eq:Heisberg-eq-ega}, and ignoring the term $\hat\psi_e^\dag(\mathbf{r})\hat\psi_e(\mathbf{r})\propto1/\Delta_a^2$, yields a closed set of two coupled equations for the atomic ground-state and photonic field operators,
\begin{subequations} 
\label{eq:Heisberg-eq-ga}
\begin{align}
\label{eq:Heisberg-eq-g} 
i\hbar\frac{\partial \hat\psi}{\partial t}&=
\left[-\frac{\hbar^2\boldsymbol{\nabla}^2}{2M}+V_{\rm ext}(\mathbf{r})
+\hbar V(\mathbf{r})
+\hbar U(\mathbf{r})\hat{a}^\dag\hat{a}
+\hbar \eta(\mathbf{r})(\hat{a}^\dag+\hat{a})
+g_0\hat{n}(\mathbf{r})\right]\hat\psi,
\end{align}
\begin{align}
i\hbar\frac{\partial \hat{a}}{\partial t}&=
-\hbar\left(\Delta_c- \int U(\mathbf{r})\hat{n}(\mathbf{r})d\mathbf{r} \right)\hat{a}
+\hbar \int \eta(\mathbf{r})\hat{n}(\mathbf{r})d\mathbf{r},
\end{align}
\end{subequations}
where $\hat{\psi}(\mathbf{r},t)\equiv\hat{\psi}_g(\mathbf{r},t)$ (the explicit position and time dependence of the atomic field operator has been suppressed above to save space) and $g_0\equiv g_{gg}$ for the sake of simplicity of the notation, and $\hat{n}(\mathbf{r})=\hat\psi^\dag(\mathbf{r})\hat{\psi}(\mathbf{r})$. Here we have introduced the definition of the transverse pump lattice $V(\mathbf{r})\equiv V_{0}\cos^2(k_cy)$ with the lattice depth $V_0=\Omega_0^2/\Delta_a$, and the intra-cavity lattice per photon $U(\mathbf{r})\equiv U_{0}  \cos^2(k_cx)$ with the single-photon lattice depth $U_0=\mathcal{G}_0^2/\Delta_a$. Finally, the  checkerboard lattice formed via the interference of the transverse pump laser with the cavity field is given by $\eta(\mathbf{r})\equiv \eta_{0} \cos(k_cx)\cos(k_cy)$ with the two-photon Rabi frequency $\eta_0=\Omega_0\mathcal{G}_0/\Delta_a$. We stress again that the non-linear contact interaction term $g_0\hat{n}(\mathbf{r})$ is absent for fermionic atoms due to the Pauli exclusion principle.

The dynamics of the atomic field operator $\hat{\psi}(\mathbf{r},t)$ depend on the photonic operator $\hat{a}(t)$, via the \emph{dynamical} quantum potential $\hbar\hat{V}_{\rm SR}(\mathbf{r})\equiv\hbar U(\mathbf{r})\hat{a}^\dag\hat{a}+\hbar \eta(\mathbf{r})(\hat{a}^\dag+\hat{a})$, and vice versa. That is, in addition to the classical optical potential $\hbar V(\mathbf{r})$ due to the absorption and emission of pump laser photons, the atoms experience two dynamical quantum optical potentials: $\hbar U(\mathbf{r})\hat{a}^\dag\hat{a}$ due to the absorption and emission of cavity photons, and $\hbar \eta(\mathbf{r})(\hat{a}^\dag+\hat{a})$ owing to the redistribution of photons between the pump laser and the cavity field. Note that the atomic field acts back on the photonic field by shifting the cavity resonance frequency by $-\int U(\mathbf{r}) \hat{n}(\mathbf{r})d\mathbf{r}$, leading to the dispersively-shifted cavity detuning $\delta_c\equiv\Delta_c- \int U(\mathbf{r})\hat{n}(\mathbf{r})d\mathbf{r}$. It also serves as an indirect effective pump for the cavity field by Bragg scattering photons from the pump laser into the cavity indicated by the term $\int \eta(\mathbf{r})\hat{n}(\mathbf{r})d\mathbf{r}$.

\begin{figure}[t!]
  \begin{center}
    \includegraphics[width=1\columnwidth]{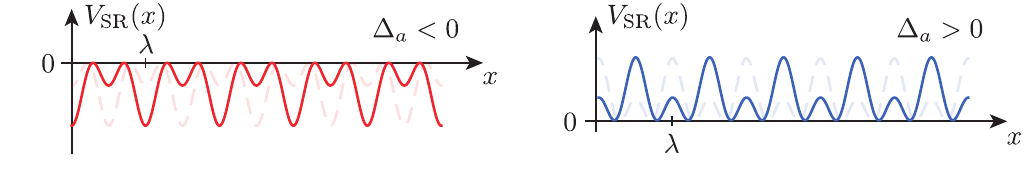}
    \caption{Potential $V_{\rm SR}(x)$ along the cavity axis given by the sum of the cavity standing-wave lattice and the interference lattice between pump and cavity fields. In the case of a pump laser field red detuned with respect to the atomic resonance, $\Delta_a<0$, the potential is heteropolar and attractive, while it is homopolar and repulsive for a blue-detuned pump field, $\Delta_a>0$. The dashed, light-color curves are the respective $\mathbf{Z}_2$-symmetric potentials.}
    \label{fig:potentials}
  \end{center}
\end{figure}

When the pump field is red detuned with respect to the atomic transition, $\Delta_a<0$, the classical and dynamical potentials are attractive; see Figure~\ref{fig:potentials}. That is, the atoms are high-field seekers. If the frequency of the pump field is also red detuned with respect to the dispersively-shifted resonance frequency of the cavity, $\delta_c<0$, the minima of the interference potential $\hbar\eta(\mathbf{r})(\hat{a}^\dag+\hat{a})$ are located at the positions of the atoms scattering photons. This gives rise to an atomic density modulation with maxima at positions where constructive photon scattering from the pump into the cavity is supported. This leads to a  positive runaway feedback mechanism and hence ``self-ordering'' as will be discussed in detail in Section~\ref{subsec:self-ordering_PT}. If in contrast, the dispersively-shifted cavity detuning is blue, $\delta_c>0$, the phase of the intra-cavity field is shifted by $\pi$, such that the emerging checkerboard potential seen by the atoms is inverted and the atoms are pushed away from their initial positions and constructive photon scattering is suppressed. The case of atomic blue detuning $\Delta_a>0$ leads to repulsive potentials as shown in Figure~\ref{fig:potentials} and is discussed in Section~\ref{subsubsec:blue_detuned_SR}.

The effective Hamiltonian of the system which reproduces the coupled, nonlinear dynamics of the system can be read out from Equation~\eqref{eq:Heisberg-eq-ga} as  
\begin{align} 
\label{eq:H_eff_1comp}
\hat{H}_{\rm eff}&=\int \hat\psi^\dag
\left[-\frac{\hbar^2\boldsymbol{\nabla}^2}{2M}+V_{\rm ext}(\mathbf{r})
+\hbar V(\mathbf{r})
+\hbar U(\mathbf{r})\hat{a}^\dag\hat{a}
+\hbar \eta(\mathbf{r})(\hat{a}^\dag+\hat{a})
+\frac{g_0}{2}\hat{n}(\mathbf{r}) \right] 
\hat{\psi}d\mathbf{r}\nonumber\\
&-\hbar\Delta_c\hat{a}^\dag\hat{a}.
\end{align}
The corresponding effective single-particle atomic Hamiltonian density plus the free cavity-field Hamiltonian is then given by
\begin{align} 
\label{eq:H_eff_density_1comp}
\hat{\mathcal H}_{1,\rm eff}=
-\frac{\hbar^2}{2M}\boldsymbol{\nabla}^2+V_{\rm ext}(\mathbf{r})
+\hbar V(\mathbf{r})
+\hbar U(\mathbf{r})\hat{a}^\dag\hat{a}
+\hbar \eta(\mathbf{r})(\hat{a}^\dag+\hat{a})
-\hbar\Delta_c\hat{a}^\dag\hat{a}\,.
\end{align}
This effective model has been the basis for the description of many groundbreaking experiments as well as theoretical works in the field. The generalization of this simple model to more sophisticated and other novel scenarios has been the core of the state-of-the-art research, both in theory and experiment. These will be the subject of the following sections.   

The effective Hamiltonian, Equation~\eqref{eq:H_eff_1comp}, is invariant under the common $U(1)$ gauge symmetry $\hat{\psi}(\mathbf{r})\to\hat{\psi}'(\mathbf{r})=e^{i\chi}\hat{\psi}(\mathbf{r})$, with $\chi$ being a position-independent phase shift, ensuring the mass conservation in the system. In addition, in the absence of the external (harmonic) trap $V_{\rm ext}(\mathbf{r})=0$, the effective Hamiltonian $\hat{H}_{\rm eff}$ possesses a $\mathbf{Z}_2$ parity symmetry: $\hat{a}\to\hat{a}'=-\hat{a}$ and $\mathbf{r}\to \mathbf{r}'=\mathbf{r}+(\lambda_c/2)\hat{e}_x$, where $\hat{e}_x$ is the unit vector along the $x$ direction. That is, the effective Hamiltonian is invariant under the combined phase rotation of the cavity field $\hat{a}$ by $\pi$ and the $\lambda_c/2$ spatial translation along the cavity axis. Note that the effective Hamiltonian $\hat{H}_{\rm eff}$ in the absence of the external (harmonic) trap is only $\lambda_c$-periodic in both $x$ and $y$ directions. 

Since the pump laser and the cavity field are far detuned from the atomic excited state and the atomic excited state is not populated significantly, the spontaneous emission from the atomic excited state is suppressed strongly. However, photon losses out of the cavity constitute another dissipation channel in this intrinsically open system. The photon losses can be accounted for in the master equation for the atom-field density operator $\hat{\rho}$ of the system,
\begin{align}
\label{eq:master-eq}
\frac{d}{dt}\hat{\rho}=\frac{1}{i\hbar}[\hat{H}_{\rm eff},\hat{\rho}]+\mathcal{L}\hat{\rho},
\end{align}
via the Liouvillean terms
\begin{align}
\label{eq:Liouville-op}
\mathcal{L}\hat{\rho}=\kappa\left(2\hat{a}\hat{\rho}\hat{a}^\dag-\hat{a}^\dag\hat{a}\hat{\rho}-\hat{\rho}\hat{a}^\dag\hat{a}\right).
\end{align}
Here, $2\kappa$ is the photon decay rate through the cavity mirrors and determines the band width of the cavity.

Equivalently, the photon losses introduce a damping term and fluctuations in the Heisenberg equation of motion for the cavity electromagnetic-field operator \cite{Carmichael2013Statistical},
\begin{align}
\label{eq:Heisberg-eq-a-loss}
i\hbar\frac{\partial \hat{a}}{\partial t}&=
-\hbar\left(\Delta_c-\int U(\mathbf{r}) 
\hat{n}(\mathbf{r})d\mathbf{r}+i\kappa \right)\hat{a}
+\hbar \int \eta(\mathbf{r}) \hat{n}(\mathbf{r})d\mathbf{r}+\hat{\xi},
\end{align}
where $\hat{\xi}(t)$ is a Gaussian noise operator with zero mean value and the correlation function $\langle \hat{\xi}(t) \hat{\xi}^\dag(t')\rangle=2\kappa\delta(t-t')$ for a zero-temperature bath outside the cavity, a condition which is well satisfied in the optical frequency domain. The noise operator $\hat{\xi}(t)$ induces a Langevin-type stochastic behavior into the field dynamics. The photon losses have important consequences. As discussed in Section~\ref{sec:dynamics}, the most noticeable effect of the photon losses appears in the dynamics, where they can induce new instabilities and dynamical phases via a detuning-dependent phase shift of the cavity field. Moreover, in the presence of loss even the stationary state of the system is not in general in thermal equilibrium. Though in many cases it can be approximately described as such that, the loss-induced noise generates effectively a finite temperature.

\subsection{Cavity-induced long-range interactions between atoms}
\label{subsubsec:cavity-induced-int}

When the cavity-field dynamics evolve on a much faster time scale compared to the center-of-mass motion of the atoms, it follows the latter adiabatically. Consequently, the cavity field can be slaved to the atomic external degrees of freedom in the sense of the Born-Oppenheimer approximation. By setting $\partial \hat{a}/\partial t=0$ in Equation~\eqref{eq:Heisberg-eq-a-loss} (and neglecting the loss-induced noise term $\hat{\xi}$), one obtains the steady-state cavity field operator,
\begin{align}
\label{eq:ss-a}
\hat{a}_{\rm ss}=
\frac{\int \eta(\mathbf{r})\hat{n}(\mathbf{r})d\mathbf{r}}
{\tilde{\Delta}_c-\int U(\mathbf{r})\hat{n}(\mathbf{r})d\mathbf{r}}
\simeq
\frac{1}{\tilde{\Delta}_c} \int \eta(\mathbf{r})  
\hat{n}(\mathbf{r})d\mathbf{r}+\mathcal{O}(\frac{1}{\Delta_a^2}),
\end{align}
by assuming $|\tilde{\Delta}_c|\equiv|\Delta_c+i\kappa|\gg|\int U(\mathbf{r})\hat{n}(\mathbf{r})d\mathbf{r}|$, up to first order in $1/\Delta_a$.

Substituting the steady-state photonic field operator $\hat{a}_{\rm ss}$ in the Heisenberg equation of motion for the atomic field operator $\partial\hat\psi(\mathbf{r},t)/\partial t$, Equation~\eqref{eq:Heisberg-eq-g}, yields an atom-only description of the system with the Hamiltonian,
\begin{align}
\label{eq:H-eff-at-only}
\hat{H}_{\text{eff-at}}&=\int \hat\psi^\dag(\mathbf{r}) 
\left[-\frac{\hbar^2}{2M}\boldsymbol{\nabla}^2+V_{\rm ext}(\mathbf{r})+\hbar V(\mathbf{r})
+\frac{1}{2}g_0\hat{n}(\mathbf{r}) \right] \hat{\psi}(\mathbf{r})d\mathbf{r}\nonumber\\
&+\iint \mathcal{D}(\mathbf{r},\mathbf{r}')  \hat{n}(\mathbf{r}) \hat{n}(\mathbf{r}') d\mathbf{r}'d\mathbf{r}
+\mathcal{O}(\frac{1}{\Delta_a^3}),
\end{align} 
up to $1/\Delta_a^{2}$, where
\begin{align}
\mathcal{D}(\mathbf{r},\mathbf{r'})=
\frac{2\hbar\Delta_c}{|\tilde\Delta_c|^2}
\eta(\mathbf{r}')\eta(\mathbf{r})
=\frac{2\hbar\Delta_c\Omega_0^2\mathcal{G}_0^2}{\Delta_a^2|\tilde\Delta_c|^2}
\cos(k_cx')\cos(k_cy')\cos(k_cx)\cos(k_cy).
\label{eq:cavity-interaction-strength}
\end{align}
Note that in general one has to first symmetrize the atomic Heisenberg equation, Equation~\eqref{eq:Heisberg-eq-g}, due to the freedom in ordering of the photonic $\{\hat{a},\hat{a}^\dag\}$ and atomic $\hat\psi(\mathbf{r})$ field operators as they commute. That is, the term proportional to $(\hat{a}^\dag+\hat{a})\hat\psi(\mathbf{r})$ in Equation~\eqref{eq:Heisberg-eq-g} is symmetrized as $[(\hat{a}^\dag+\hat{a})\hat\psi(\mathbf{r})+\hat\psi(\mathbf{r})(\hat{a}^\dag+\hat{a})]/2$ before substituting the steady-state photonic field operator $\hat{a}_{\rm ss}$. 
The potential term $\hbar U(\mathbf{r})\hat{a}^\dag\hat{a}$ in the atomic Heisenberg equation~\eqref{eq:Heisberg-eq-g} is omitted as it yields interactions which scale as $\Delta_a^{-3}$. 

The penultimate term in the effective atom-only Hamiltonian $\hat{H}_{\text{eff-at}}$, Equation~\eqref{eq:H-eff-at-only}, describes a cavity-mediated infinite-ranged, or ``global'' density-density interaction. It stems from the back-action of the quantized cavity field on the atoms. That is, as the atoms are strongly coupled to the dynamic cavity field, a local change in the atomic distribution affects significantly the cavity field. Since the cavity electromagnetic field is global (only restricted by the cavity mirrors), the modification of the cavity field is sensed by all the atoms in the cavity, regardless of their distance from the original atomic perturbation. This induces a density-density interaction among the atoms with the non-decaying, periodic strength $\mathcal{D}(\mathbf{r},\mathbf{r'})$, Equation~\eqref{eq:cavity-interaction-strength}. 

This global interaction can also be understood from the microscopic photon scattering processes: A pump-laser photon is scattered by an atom into the cavity mode, and then later scattered back into the pump mode by a second atom. These two atoms are, therefore, correlated and interact with each other via the exchange of a cavity photon. Since photons are delocalized over the entire cavity mode, the cavity-mediated interaction is of global range. The range of the cavity-mediated interaction can be tuned by exploiting multiple modes of a cavity, as will be discussed in Section~\ref{sec:multimode}. Furthermore, in addition to the density-density interaction, other forms of global interactions such as a spin-spin interaction can be mediated via cavity photons using more atomic internal states as described in Section~\ref{sec:spinor-selfordering}. 

The effective atom-only Hamiltonian can also be obtained by directly substituting the steady-state photonic field operator $\hat{a}_{\rm ss}$ in the effective Hamiltonian, Equation~\eqref{eq:H_eff_1comp}, as well as the Liouville superoperator, Equation~\eqref{eq:Liouville-op}. The resultant effective atom-only Hamiltonian will  have the same general form as $\hat{H}_{\text{eff-at}}$ \eqref{eq:H-eff-at-only}, but with a somewhat different cavity-mediated interaction strength. Nevertheless, within the realm of the validity of these effective atom-only Hamiltonians, namely, fast cavity-field dynamics, both effective atom-only Hamiltonians yield qualitatively the same results~\cite{Maschler2008Ultracold}.

In order to have a constructive photon scattering from the pump laser into the cavity mode, it is usually required to work on the red relative cavity detuning side, $\Delta_c<0$, as mentioned in the previous section. Therefore, for instance for an atom at the origin (i.e., $\mathbf{r}'=0$) the cavity-mediated interaction $\mathcal{D}(\mathbf{r},0)$ is maximally attractive (i.e., negative) for other atoms located at integer multiples of $\lambda_c$ along the $x$ and $y$ directions [$\mathbf{r}=(\ell,0)\lambda_c$ and $\mathbf{r}=(0,m)\lambda_c$, respectively, with $\ell$ and $m$ being integers] as well as at integer multiples of $\lambda_c/\sqrt{2}$ along the diagonal and off-diagonal directions [i.e., $\mathbf{r}=(\ell,\pm\ell)\lambda_c/2$]; see also Figure~\ref{fig:potentials}.
In general, the sign and the strength of the cavity-mediated interaction $\mathcal{D}(\mathbf{r},\mathbf{r}')$ can similarly be obtained for any two atoms located at arbitrary positions $\mathbf{r}$ and $\mathbf{r}'$.

The cavity-mediated interaction~\eqref{eq:cavity-interaction-strength} favors a $\lambda_c$-periodic checkerboard density pattern. When the energy gain due to the cavity-mediated global density-density interaction in the checkerboard density pattern dominates over other energy penalties arising from the local terms in the effective atom-only Hamiltonian $\hat{H}_{\text{eff-at}}$, Equation~\eqref{eq:H-eff-at-only}, the checkerboard density pattern is then stabilized in the system (in Section~\ref{subsec:thermal_nonthermal} we discuss the validity and the limitation of energy consideration in quantum-gas--cavity setups which are generically driven-dissipative systems). This so-called ``self-ordering'' of the atoms into the checkerboard density pattern in the effective atom-only description of the system coincides with the onset of the ``superradiance'' in the cavity---i.e., the appearance of a finite coherent component for the cavity field---as we will discuss in detail in Section~\ref{sec:SR_discrete}.
That said, in certain regimes the cavity-mediated interaction can induce phase transitions which are not accompanied by superradiance. Examples include cavity-fluctuation-induced superconductivity discussed in Section~\ref{subsec:cavity-induced pairing}, lattice magnetic orders in Section~\ref{sec:lattice-magnetic-orders-fermi}, and cavity-induced spin glasses with a single spin-1/2 atom per site~\cite{Gopalakrishnan2011Frustration}.

As discussed intuitively above, the cavity-mediated global atom-atom interactions favor an ordered, crystalline density structure. This can also be explained in a mathematically elegant way through the atom-only Heisenberg equation of motion for the atomic field operator [i.e., after substituting the steady-state photonic field operator $\hat{a}_{\rm ss}$~\eqref{eq:ss-a} in the Heisenberg equation of motion for the atomic field operator~\eqref{eq:Heisberg-eq-g}, which yields the effective atom-only Hamiltonian~\eqref{eq:H-eff-at-only} as discussed above],
\begin{align}
\label{eq:Heisberg-eq-g-at-only} 
i\hbar\frac{\partial \hat\psi(\mathbf{r},t)}{\partial t}&=
\left[-\frac{\hbar^2\boldsymbol{\nabla}^2}{2M}+V_{\rm ext}(\mathbf{r})
+\hbar V(\mathbf{r})
+\frac{2\hbar\Delta_c\eta_0}{|\tilde\Delta_c|^2} \eta(\mathbf{r}) \hat{\Theta}
+g_0\hat{n}(\mathbf{r})\right]\hat\psi(\mathbf{r},t).
\end{align}
Here we have kept the terms solely up to $1/\Delta_a^{2}$ (similar to the effective atom-only Hamiltonian $\hat{H}_{\text{eff-at}}$) and have introduced the operator
\begin{align}
\label{eq:order-paramter-op}
\hat{\Theta}=\frac{1}{\eta_0}\int \eta(\mathbf{r}')\hat{n}(\mathbf{r}')d\mathbf{r}'.
\end{align}
The operator $\hat{\Theta}$ can be envisaged as an order parameter for the system, characterizing the degree of the density crystallization in the checkerboard potential $\eta(\mathbf{r})$. The ground state of the system in the limit of fast cavity dynamics is obtained from Equation~\eqref{eq:Heisberg-eq-g-at-only} along with the self-consistent solution for the order parameter~\eqref{eq:order-paramter-op}, as will be discussed in Section~\ref{subsec:MF} for a mean-field approach. Note that $\hat{\Theta}$ is directly related to the steady-state cavity field operator~\eqref{eq:ss-a} as  
\begin{align}
\label{eq:ss-a-Theta}
  \hat a_\mathrm{ss} = \frac{\eta_0\hat \Theta}
  {\Delta_c -  \int U(\mathbf{r})\hat{n}(\mathbf{r})\mathrm{d}\mathbf{r}+ i\kappa}.
\end{align}
Through this review paper, we will see how one can engineer various novel self-consistent, emergent orders in ultracold atoms coupled to optical cavities.

We note that in the limit of fast cavity-field dynamics considered in this section, a great deal of interesting physics related to the retardation effects of the electromagnetic field has been already discarded. In Chapter~\ref{sec:dynamics}, we will encounter examples of collective phenomena which can only be described by taking into account the finiteness of the cavity timescale. One also has to note that such an adiabatically eliminated atom-only Hamiltonian does not always accurately capture all atomic dynamics~\cite{Damanet2019Atom}.

\subsection{Applicability of the Dicke model}
\label{subsubsec:DickeModel}

$N$ far-positioned two-level atoms prepared in their excited state decay independently from one another, incoherently emitting an electromagnetic field with an energy-density maximum proportional to $N$. However, if the atoms are trapped within a fraction of a wavelength, the emitted photons interfere constructively resulting in a transient superradiant pulse with an energy-density maximum proportional to $N^2$ as predicted by Dicke in 1954~\cite{Dicke1954Coherence}. Hepp and Lieb in 1973 predicted that a steady-state form of superradiance can occur in an ensemble of two-level atoms coupled identically to a quantized cavity mode~\cite{Hepp1973On, Hepp1973Equilibrium}. However, for atoms with two electronic states the superradiance is supposed to take place for atom-photon coupling in optical-frequency order, a situation beyond state-of-the-art experimental capabilities. Furthermore, the existence of the superradiant phase transition for such strong atom-photon couplings has been under an intense debate due to the ignored $\mathbf{A}^2$ (i.e., vector potential squared) term. 

Let us now examine the low-energy limit of the effective Hamiltonian $\hat{H}_{\rm eff}$, Equation~\eqref{eq:H_eff_1comp}, at zero temperature for \emph{bosonic} atoms in the absence of the external potential and the two-body interaction. The quantum potential $\eta(\mathbf{r})\propto\cos(k_cx)\cos(k_cy)$ defines the periodicity of the effective Hamiltonian: The system is $\lambda_c$ periodic along both $x$ and $y$ directions. That is, starting from the zero momentum state, only momentum states with integer multiples of $k_c$, $\mathbf{k}=(k_x,k_y)=(l,m)k_c$ with $l,m\in\mathbb{Z}$, can be occupied via the photon scattering processes. Therefore, the atomic field operator $\hat{\psi}(\mathbf{r})$ can be expanded in the momentum basis in the unit of $k_c$ using plane waves as,
\begin{align}
\hat{\psi}(\mathbf{r})=\frac{1}{\sqrt{A}}\sum_{l,m\in\mathbb{Z}}e^{ik_c(lx+my)}\hat{b}_{l,m},
\end{align} 
where $\hat{b}_{l,m}$ is the bosonic annihilation operator for momentum state $\mathbf{k}=(l,m)k_c$ and $A$ is the surface area of the Wigner-Seitz cell spanned by the elementary lattice vectors. Therefore, up to an immaterial constant term, the effective Hamiltonian~\eqref{eq:H_eff_1comp} in momentum space takes the form
\begin{align} 
\label{eq:H_eff_1comp-p}
\hat{H}_{\rm eff}=\sum_{l,m\in\mathbb{Z}}&\bigg[
\hbar\omega_r\left(l^2+m^2\right)\hat{b}_{l,m}^\dag\hat{b}_{l,m}
+\frac{\hbar\eta_0}{4}(\hat{a}^\dag+\hat{a})
\left(\hat{b}_{l+1,m+1}^\dag\hat{b}_{l,m}+\hat{b}_{l+1,m-1}^\dag\hat{b}_{l,m}+\text{H.c.}\right)
\nonumber\\
&+\frac{\hbar V_0}{4}\left(\hat{b}_{l,m+2}^\dag\hat{b}_{l,m}+\text{H.c.}\right)
+\frac{\hbar U_0}{4}\hat{a}^\dag\hat{a}\left(\hat{b}_{l+2,m}^\dag\hat{b}_{l,m}+\text{H.c.}\right)\bigg]
-\hbar\delta_c\hat{a}^\dag\hat{a},
\end{align}
where $\omega_r=\hbar k_c^2/2M$ is the recoil frequency and $\delta_c=\Delta_c-NU_0/2$ with $N$ being the total atom number\footnote{Note that $\delta_c=\Delta_c-NU_0/2$ as defined here is intimately related to the dispersively shifted cavity detuning $\delta_c\equiv\Delta_c- \int U(\mathbf{r})\hat{n}(\mathbf{r})d\mathbf{r}$ defined earlier in Section~\ref{sec:basic-model}. Throughout this review paper, we will use $\delta_c$ in a rather liberal way, but all ultimately related to the dispersively shifted cavity detuning.}. 

In the low-energy limit, one can restrict the momentum modes solely to $l,m=\{0,\pm1\}$. In particular, the momentum states $\mathbf{k}=(0,0)$ and $(\pm1,\pm1)$ form a closed coupled set with $\hat{b}_{0,0}^\dag\hat{b}_{0,0}+\sum_{l,m=\pm1}\hat{b}_{l,m}^\dag\hat{b}_{l,m}=N$ being a constant of motion. By defining the collective atomic raising and lowering $\hat{J}_+=\hat{J}_-^\dag=(\frac{1}{2}\sum_{l,m=\pm1}\hat{b}_{l,m}^\dag)\hat{b}_{0,0}$ and population-imbalance $\hat{J}_z=(\sum_{l,m=\pm1}\hat{b}_{l,m}^\dag\hat{b}_{l,m}-\hat{b}_{0,0}^\dag\hat{b}_{0,0})/2$ operators (see also Figure~\ref{fig:energy_scales}), the low-energy limit of the effective Hamiltonian, Equation~\eqref{eq:H_eff_1comp-p}, yields
\begin{align} 
\label{eq:H_eff_1comp-p-LE}
\hat{H}_{\rm LE}\simeq
\hbar\omega_0\hat{J}_z
-\hbar\delta_c\hat{a}^\dag\hat{a}
+\frac{1}{2}\hbar\eta_0(\hat{a}^\dag+\hat{a})(\hat{J}_++\hat{J}_-).
\end{align}
This low-energy Hamiltonian has the exact form of the Dicke Hamiltonian \cite{Garraway2011The, Kirton2018Introduction}, describing the dynamics of pseudospins with the transition frequency $\omega_0=2\omega_r$ coupled identically to a quantized bosonic field with the frequency $-\delta_c$.
The Dicke Hamiltonian is fully connected and shows a mean-field-type superradiant phase transition at the critical coupling $\sqrt{N}\eta_{0c}/2=\sqrt{-\omega_0\delta_c}/2$  breaking the $\mathbf{Z}_2$ parity symmetry of the system, $\hat{a}\to-\hat{a}$ and $\hat{J}_{\pm}\to-\hat{J}_{\pm}$, both at zero and finite temperature~\cite{Emary2003Chaos}. However, due to photon losses out of the cavity, the transition takes place instead at the critical coupling~\cite{Kirton2018Introduction}, 
\begin{align} \label{eq:eta-critical-Dicke-model}
\sqrt{N}\eta_{0c}=\sqrt{\frac{\omega_0(\delta_c^2+\kappa^2)}{-\delta_c}}. 
\end{align}
Since here the involved atomic states coupled via the radiation field are momentum states with an energy separation in a much lower recoil-energy regime $\hbar\omega_r$, the required atom-photon coupling for the superradiant phase transition reduces substantially and falls into the recoil-energy regime. As an important consequence, the no-go theorems concerning the superradiant phase transition do not apply in this case as well. Such a superradiant phase is generically present for \emph{both} driven bosonic and fermionic atoms inside optical cavities as we will see throughout this review article.

\subsection{Mean-field description}
\label{subsec:MF}

In this section we introduce the commonly-used self-consistent mean-field treatment of the system. This approach is justified since the cavity electromagnetic field is macroscopically occupied in the self-ordered phase and the cavity-mediated interactions in single-mode cavities are global, as discussed in the preceding section; mean-field approaches generically become accurate in globally interacting systems and fully connected models in the thermodynamic limit~\cite{Muehlschlegel1962Asymptotic, Stanley1971Introduction,Hepp1973On,Carollo2020Proof}. For the photonic field, this mean-field approach amounts to the assumption that the electromagnetic field is in a coherent state, and therefore $\alpha=\langle\hat{a}\rangle$ is the coherent field amplitude. By taking the quantum average from the Heisenberg equation of motion of the cavity operator, Equation~\eqref{eq:Heisberg-eq-a-loss}, one obtains a mean-field equation for the field amplitude,
\begin{align}
\label{eq:MF-eq-a-loss}
i\hbar\frac{\partial \alpha}{\partial t}&=
-\hbar\left(\Delta_c-\int U(\mathbf{r}) 
\langle\hat{n}(\mathbf{r}) \rangle d\mathbf{r}+i\kappa \right)\alpha
+\hbar \int \eta(\mathbf{r}) \langle\hat{n}(\mathbf{r})\rangle d\mathbf{r}.
\end{align}
The crucial approximation involved here is the factorization of the average $\langle\hat{n}(\mathbf{r})\hat{a}\rangle\simeq \langle\hat{n}(\mathbf{r})\rangle \alpha$. Combined with an equation for the quantum-averaged density $\langle\hat{n}(\mathbf{r})\rangle=n(\mathbf{r})=n(\mathbf{r},\alpha)$ with $\alpha=\alpha(n(\mathbf{r}))$, the set of self-consistent, coupled mean-field equations is closed. Below, we will consider the application of this mean-field approach to both bosonic and fermionic atoms.

We stress that this approximation only regards the atom-light interaction, ignoring any other interaction that might be present among the atoms. This includes intrinsic atom-atom interactions which affect the average density entering the above equations. In particular, within this mean-field approximation, the two-mode squeezing of the light-matter state is neglected~\cite{Emary2003Chaos}. However, due to the global cavity-mediated interactions, corrections to the above mean-field approximation are suppressed by a factor $1/V$ where $V$ is the volume of the atomic cloud~\cite{Piazza2013Bose}. Therefore, the mean-field description becomes exact in the thermodynamic limit $N,V\to\infty$, with $N/V=\text{const}$. 

Before proceeding, let us stress one important consequence of this approximation which limits its validity. The mean-field equation for the field amplitude, Equation~\eqref{eq:MF-eq-a-loss}, does not contain the loss-induced noise $\xi$, which in general leads to violation of fluctuation-dissipation relations characterizing the correct balance between damping and noise in a steady state. While this does not affect the coherent component of the cavity field $\alpha$, it does affect the atomic density and in general the atomic momentum distribution. This makes this mean-field approach inadequate for the description of cavity-induced redistribution of the atoms (see Section~\ref{subsec:thermal_nonthermal} below).

%..................................................................................
\subsubsection{Weakly interacting ultracold Bose gases}
\label{subsubsec:MF-bose}

Considering zero temperature, weakly-interacting bosons, and neglecting the quantum depletion of the Bose-Einstein condensate, one can approximate in Equation~\eqref{eq:MF-eq-a-loss} the quantum-averaged density as $n(\mathbf{r})=|\psi(\mathbf{r})|^2$, where $\psi(\mathbf{r},t)$ is the condensate wavefunction satisfying the time-dependent Gross-Pitaevskii equation~\cite{Dalfovo1999Theory, Nagy2008Self},
\begin{align}
\label{eq:MF-eq-g} 
i\hbar\frac{\partial \psi(\mathbf{r})}{\partial t}=\left[-\frac{\hbar^2}{2M}\boldsymbol{\nabla}^2+V_{\rm ext}(\mathbf{r})
+\hbar V(\mathbf{r})
+\hbar|\alpha|^2U(\mathbf{r})
+2\hbar\text{Re}[\alpha]\eta(\mathbf{r})
+g_0n(\mathbf{r})\right]\psi(\mathbf{r}).
\end{align}
This equation is derived from the Heisenberg equation of motion of the atomic field operator, Equation~\eqref{eq:Heisberg-eq-g}, assuming that the many-body atomic wavefunction is a coherent state such that the field operator $\hat{\psi}(\mathbf{r})$ can be substituted with its mean-field amplitude $\psi(\mathbf{r})=\langle\hat{\psi}(\mathbf{r})\rangle$~\cite{Horak2001Dissipative, Zhang2009Nonlinear}. The above Gross-Pitaevskii equation contains a dynamical (superradiant) potential 
$\hbar V_{\rm SR}(\mathbf{r})\equiv\hbar|\alpha|^2U(\mathbf{r})+2\hbar\text{Re}[\alpha]\eta(\mathbf{r})$ 
whose strength depends on the cavity field amplitude $\alpha$, which must be determined self-consistently from Equation~(\ref{eq:MF-eq-a-loss}).

The stationary state of the system can be found using the standard \textit{ansatz} for the condensate wavefunction $\psi(\mathbf{r},t)=\psi(\mathbf{r})e^{-i\mu t/\hbar}$, with $\mu$ being the atomic chemical potential, and correspondingly via the condition $\partial \alpha/\partial t=0$ for the cavity-field amplitude. The latter yields from Equation~(\ref{eq:MF-eq-a-loss}) the steady-state cavity-field amplitude,
\begin{align}
\label{eq:MF-eq-a-ss} 
\alpha_{\rm ss}=\frac{\int \eta(\mathbf{r}) n(\mathbf{r})d\mathbf{r}}
{\Delta_c-\int U(\mathbf{r})n(\mathbf{r})d\mathbf{r}+i\kappa}
=
\frac{\eta_0\Theta}
{\Delta_c-\int U(\mathbf{r})n(\mathbf{r})d\mathbf{r}+i\kappa},
\end{align}
where $\Theta=\langle \hat{\Theta}\rangle=(1/\eta_0)\int \eta(\mathbf{r}) n(\mathbf{r})d\mathbf{r}$ [see Equations~\eqref{eq:order-paramter-op} and~\eqref{eq:ss-a-Theta}] is the mean-field checkerboard density order parameter, which is to be determined self-consistently along with the dispersive cavity shift $-\int U(\mathbf{r})n(\mathbf{r})d\mathbf{r}$.

%..................................................................................
\subsubsection{Non-interacting ultracold Fermi gases}
\label{subsubsec:MF-fermi}

In the absence of an external (harmonic) trap $V_{\rm ext}(\mathbf{r})=0$, as the system is $\lambda_c$ periodic along both $x$ and $y$ directions, one can make a Bloch \emph{ansatz} for the single-particle states, $\psi_{m,\mathbf{q}}(\mathbf{r})=e^{i\mathbf{q}\cdot\mathbf{r}}u_{m,\mathbf{q}}(\mathbf{r})$, where $m$ is the band index, $\mathbf{q}=(q_x,q_y)$ is the quasimomentum in the first Brillouin zone $[-\pi/\lambda_c,\pi/\lambda_c]\times[-\pi/\lambda_c,\pi/\lambda_c]$, and $u_{m,\mathbf{q}}(\mathbf{r})$ is a periodic function with the same periodicity as the Hamiltonian, $u_{m,\mathbf{q}}(\mathbf{r}+\lambda_c\hat{e}_x+\lambda_c\hat{e}_y)=u_{m,\mathbf{q}}(\mathbf{r})$. Therefore, the fermionic field operator can be expanded in the basis of the Bloch functions, $\hat{\psi}(\mathbf{r})=\sum_{m,\mathbf{q}}\psi_{m,\mathbf{q}}(\mathbf{r})\hat{f}_{m,\mathbf{q}}$, where $\hat{f}_{m,\mathbf{q}}$ is the fermionic annihilation operator for the $m$th band at quasimomentum $\mathbf{q}$. The periodic Bloch functions $u_{m,\mathbf{q}}(\mathbf{r})$ then satisfy the eigenvalue equation,
\begin{align}
\label{eq:MF-eq-g-f} 
\left[\frac{\hbar^2}{2M}(-i\boldsymbol{\nabla}+\mathbf{q})^2+\hbar V(\mathbf{r})
+\hbar|\alpha|^2U(\mathbf{r})
+2\hbar\text{Re}[\alpha]\eta(\mathbf{r})\right]
u_{m,\mathbf{q}}(\mathbf{r})=\epsilon_{m,\mathbf{q}} u_{m,\mathbf{q}}(\mathbf{r}),
\end{align}
with Bloch-band energies $\epsilon_{m,\mathbf{q}}$. Recall that the local contact interaction is absent due to the Pauli exclusion principle as here we only consider spin-polarized fermions.  

Regarding the stationary state of the system, the cavity field amplitude is determined self-consistently via the mean-field equation (\ref{eq:MF-eq-a-ss}), with the atomic density now given by 
\begin{align}
\label{eq:density-f}
n(\mathbf{r})
=\sum_{m,\mathbf{q}}|u_{m,\mathbf{q}}(\mathbf{r})|^2n_{\rm FD}(\epsilon_{m,\mathbf{q}}),
\end{align}
where $n_{\rm FD}(\epsilon)=1/[1+e^{(\epsilon-\mu)/k_BT}]$ is the Fermi-Dirac distribution as a function of the energy $\epsilon$ at temperature $T$ (with $k_B$ being the Boltzmann constant). The chemical potential $\mu$ is determined self-consistently by fixing the total number of the atoms in the system,
\begin{align}
\label{eq:N-mu}
N=\int n(\mathbf{r}) d\mathbf{r}=\sum_{m,\mathbf{q}}\int 
\frac{|u_{m,\mathbf{q}}(\mathbf{r})|^2}{1+e^{(\epsilon_{m,\mathbf{q}}-\mu)/k_BT}}d\mathbf{r}.
\end{align}
Note that we have assume here that the steady-state distribution of the atoms is in a thermal equilibrium, with $T$ being fixed by the initial energy of the system. This is not strictly correct since energy is not conserved in this driven-dissipative system and the steady state of the system can be even a non-thermal type~\cite{Piazza2014Quantum} (see also Section~\ref{subsec:thermal_nonthermal}).

%----------------------------------------------------------------------------------------------------------------------------
\subsection{Analytical and numerical methods beyond the mean-field approach}
\label{subsec:analytical_methods}

A number of techniques have been employed for the description of the systems under consideration, which go beyond the mean-field approximation just introduced. A subgroup of them aims at a better description of correlations induced by the intrinsic atom-atom interactions, but still treating the atom-cavity correlations at a mean-field level and including only the coherent part $\alpha$ of the cavity field operator $\hat{a}$. These methods include dynamical mean-field approaches~\cite{Li2013Lattice,Bakhtiari2015Nonequilibrium}, multiconfigurational time-dependent Hartree~\cite{Alon2018Many}, exact diagonalization~\cite{Colella2018Quantum}, and quantum Monte Carlo~\cite{Habibian2013Bose}. On the other hand, other beyond-mean-field methods have been developed that also take into account the atom-cavity quantum correlations and therefore the role of cavity-induced noise in the steady state of the system. These include numerically exact methods applicable to small systems~\cite{Vukics2007C++QED} and more recently matrix-product-state-based methods allowing to describe larger systems~\cite{Halati2020Numerical}, as well as field-theoretical methods for real-time Green's functions~\cite{Gopalakrishnan2010Atom,Piazza2014Quantum} or perturbative expansions around the mean field~\cite{Bezvershenko2020Dicke}. 

%----------------------------------------------------------------------------------------------------------------------------
\subsection{Thermal versus non-thermal stationary states}
\label{subsec:thermal_nonthermal}

Coupling atoms to a lossy cavity field is known to lead to the redistribution of the atoms, referred to as cavity cooling or heating in quantum optics~\cite{Wolke2012Cavity}. The cavity-induced redistribution has recently been studied also for quantum gases~\cite{Piazza2014Quantum,Chiacchio2018Tuning,Nagy2018Quantum,Bezvershenko2020Dicke}.

Since atom-cavity setups are generically driven-dissipative systems, where a steady state results from the balance between energy inflow into the system from pump lasers and energy dissipation through the cavity mirrors, the steady state is in general not expected to be in thermal equilibrium.
Indeed, in the context of ultracold atoms, it has been shown that cavity-induced redistribution can lead to non-thermal steady states featuring broad power-law tails~\cite{Piazza2014Quantum}. In absence of short-range repulsion between the atoms, which generically favors a thermal steady state~\cite{Bezvershenko2020Dicke}, there is a crossover between a thermal and a non-thermal regime of the cavity-induced steady state of the atoms~\cite{Piazza2014Quantum}: When the cavity detuning $|\Delta_c|$ is much larger than the recoil frequency $\omega_r$, the atoms are cavity heated or cooled to a thermal state with temperature of order $\kappa$. 

One can estimate the impact of cavity-induced redistribution of atoms already by examining the mean-field approach of Section~\ref{subsec:MF}. There, the corrections to the noiseless mean-field result are suppressed by a factor of $1/N$. Therefore, the characteristic time for cavity-induced redistribution to set in scales with the atom number $N$ (or the volume $V$ in the thermodynamic limit). This is a common feature of long-range interacting systems, where the phase space for collisions leading to redistribution of particles is reduced with respect to a short-range interacting system~\cite{Campa2009Statistical}. 
A more precise estimation of the cavity-induced relaxation rate of the atoms gives~\cite{Piazza2014Quantum},
  \begin{align}
    \Gamma_{\rm rel}\simeq \frac{1}{N}\frac{\eta_0^2}{\kappa}\frac{\delta_c \omega_r}{\kappa^2}.
  \end{align}
  The relaxation rate is thus in general suppressed by the number of atoms $N$, as anticipated from above, which is due to the global nature of the light-matter coupling in this single-mode case. In cases like the ETH experiments, where the ratio $\omega_r/\kappa$ is in the order of $10^{-3}$ (see Table~\ref{tab:ExperimentalParameters}), the relaxation timescale is estimated to be of several seconds, which means that the atoms approximately remain in the initial thermal state for the whole duration of the experiment. This justifies in many occasions in this review article the use of thermal-equilibrium methods, i.e., the minimization of energy and the maximization of entropy, for studying many-body phases in quantum-gas--cavity setups.

The situation can be, however, quite different in experiments where $\kappa\sim\omega_r$  (see Hamb and T\"{u}b2 in Table~\ref{tab:ExperimentalParameters}). Important qualitative differences have indeed already been experimentally observed due to this fact (see Figure~\ref{fig:QuenchDynamics}). Though cavity cooling has been demonstrated~\cite{Wolke2012Cavity}, the cavity-induced relaxation and the emergence of non-thermal steady states have not yet been experimentally investigated, though the timescales should be within reach.

Let us conclude by noting that for other types of driving protocols, different than the continuous-wave, monochromatic driving pumps considered in this chapter, further non-thermal steady states and even non-steady states are expected to exist, as discussed in Section~\ref{sec:dynamics}.

%----------------------------------------------------------------------------------------------------------------------------
\subsection{Collective excitations}
\label{subsec:excitations_probing}

The excitations of a coupled system of matter and light are collective modes---so-called polaritons---that mix atomic and photonic degrees of freedom owing to the strong coupling between the two constituents. Polaritons govern the linear response of the atom-cavity system to perturbations and can be observed either via the fluctuation spectrum of the cavity field or via cavity probing with a coherent field. In the following we present two complementary approaches to theoretically describe these excitations. Their actual measurement is discussed in Section~\ref{subsec:excitation_measurement}.

%..................................................................................
\subsubsection{Polarizable atomic gases}
\label{sec:polarizable-atomic-gas}

The polariton spectrum can be computed by starting from the equilibrium mean-field equation~\eqref{eq:MF-eq-a-loss} for the cavity field amplitude $\alpha$ and expanding it around a self-consistent equilibrium solution. The mean-field equation for the cavity field amplitude, Equation~\eqref{eq:MF-eq-a-loss}, can be rewritten in a generic form as
\begin{equation}
  \label{eq:MF-eq-a-loss-eigen}
  i\partial_t\alpha=-(\Delta_c+i\kappa)\alpha+\sum_\ell\langle u_\ell(\alpha)|\alpha U+\eta| u_\ell(\alpha)\rangle n(\epsilon_\ell),
\end{equation}
using eigenstates $| u_\ell(\alpha)\rangle$ [and corresponding energies $\epsilon_\ell(\alpha)$] of the mean-field single-particle Hamiltonian $\mathcal{H}_{1,\rm eff}(\alpha)$, obtained from Equation~\eqref{eq:H_eff_density_1comp} by substituting $\hat{a}$ with its average $\alpha$.
Note that we have defined here the basis-independent first-quantized operators $U$ and $\eta$ such that  $U(\mathbf{r})=\langle\mathbf{r}|U|\mathbf{r}\rangle$ and $\eta(\mathbf{r})=\langle\mathbf{r}|\eta|\mathbf{r}\rangle$, with the position eigenstates $\ket{\mathbf{r}}$ forming a complete basis $1=\int d\mathbf{r}|\mathbf{r}\rangle\langle\mathbf{r}|$.
Here we do not yet specify $n(\epsilon_\ell)$ to be a fermionic or bosonic distribution, nor a choice of the external potential $V_{\rm ext}$ for the atoms.
Defining the field-amplitude fluctuation as $\delta\hat{a}(t)=\hat{a}(t)-\alpha$ and keeping terms in Equation~\eqref{eq:MF-eq-a-loss-eigen} up to linear order in $\delta\hat{a}$ and $\delta\hat{a}^\dag$, which requires computing perturbative corrections to the eigenvectors $| u_\ell(\alpha)\rangle$ and the eigenvalues $\epsilon_\ell(\alpha)$, we obtain the equation for the fluctuation:
\begin{equation}
  \label{eq:MF-eq-a-loss-linear}
 i\partial_t\delta\hat{a}=\left[-\delta_c(\alpha)-i\kappa+\chi_{\rm stat}(\alpha)\right](\delta\hat{a}+\delta\hat{a}^\dag).
\end{equation}
Here we have defined the dispersively shifted cavity detuning 
\begin{equation}
  \label{eq:disp_shift}
 \delta_c(\alpha)=\Delta_c-\sum_\ell n(\epsilon_\ell)\langle u_\ell(\alpha)|U| u_\ell(\alpha)\rangle,
\end{equation}
and the static polarization function of the atomic medium
\begin{equation}
  \label{eq:stat_polarization}
  \chi_{\rm stat}(\alpha)=\sum_{\ell,\ell'} \frac{n(\epsilon_\ell)-n(\epsilon_{\ell'})}{\epsilon_{\ell}-\epsilon_{\ell'}}\bigg|\langle u_{\ell'}(\alpha)|\alpha^*U+\eta| u_\ell(\alpha)\rangle\bigg|^2.
\end{equation}
The static polarization function characterizes the response of the medium to the cavity probe and can be immediately upgraded to include the energy $\hbar\omega$ of the probe photon, which gives the dynamical polarization function
\begin{equation}
  \label{eq:dyn_polarization}
  \chi_{\rm dyn}(\omega;\alpha)=\sum_{\ell,\ell'} \frac{n(\epsilon_\ell)-n(\epsilon_{\ell'})}{\hbar\omega+\epsilon_{\ell}-\epsilon_{\ell'}+i0^+}\bigg|\langle u_{\ell'}(\alpha)|\alpha^*U+\eta| u_\ell(\alpha)\rangle\bigg|^2,
\end{equation}
where the infinitesimally small imaginary part ensures the retarded character of this response function.
In the context of condensed-matter physics, this is also known as the Lindhard function~\cite{Fetter2012Quantum}, characterizing the density response of a material. Note that in the dispersive regime considered here, the cavity photons indeed couple to the atomic density.

To compute the frequency-resolved response of the system to a cavity probe we Fourier-transform Equation~\eqref{eq:MF-eq-a-loss-linear} and combine it with its complex conjugate to obtain the following system of equations:
\begin{align}
  \label{eq:MF-eq-aadag-loss-linear-fourier}
D^{-1}(\omega)  
\begin{bmatrix} 
\delta\hat{a} (\omega) \\ 
\delta\hat{a}^\dag (-\omega) 
\end{bmatrix}
=0,
\end{align}
with
\begin{align}
  \label{eq:ret-green-func}
D^{-1}(\omega)=    
\begin{bmatrix} 
\omega+\delta_c+i\kappa-\chi_{\rm dyn}(\omega) & -\chi_{\rm dyn}(\omega) \\ 
-\chi_{\rm dyn}^*(-\omega) &  -\omega+\delta_c-i\kappa-\chi_{\rm dyn}^*(-\omega)   
\end{bmatrix}.
\end{align}
The matrix $D(\omega)$ is nothing else than the retarded Green's function of the cavity electromagnetic field, including medium-polarization corrections, expressed through its positive- and negative-frequency components. The complex frequencies $\omega_{\rm pol}$ of the collective polariton modes correspond to the poles of the retarded Green's function, given by the condition $\mathrm{Det}[D^{-1}(\omega_{\rm pol})]=0$. This condition yields the following implicit equation for the polariton frequencies:
\begin{equation}
  \label{eq:poles-retarded-cavity}
 (\omega_{\rm pol}+i\kappa)^2- \delta_c^2+2\delta_c\chi_{\rm dyn}(\omega_{\rm pol})=0,
\end{equation}
where we have exploited the symmetry of the retarded polarization function $\chi_{\rm dyn}(\omega)=\chi_{\rm dyn}^*(-\omega)$.

The polariton frequencies $\omega_{\rm pol}$ obtained from Equation~\eqref{eq:poles-retarded-cavity} possess an imaginary part, since they always contain a photonic component. From the point of view of the atoms, this imaginary part arises from the non-adiabaticity of the cavity field. That is, the cavity field follows the changes in the atomic state with a delay on the order of $1/\kappa$. Therefore, the atomic excitations can be damped out through the cavity loss channel in appropriate parameter regimes, leading to cavity cooling of the atoms. 

\begin{figure}[t!]
  \begin{center}
    \includegraphics[width=\columnwidth]{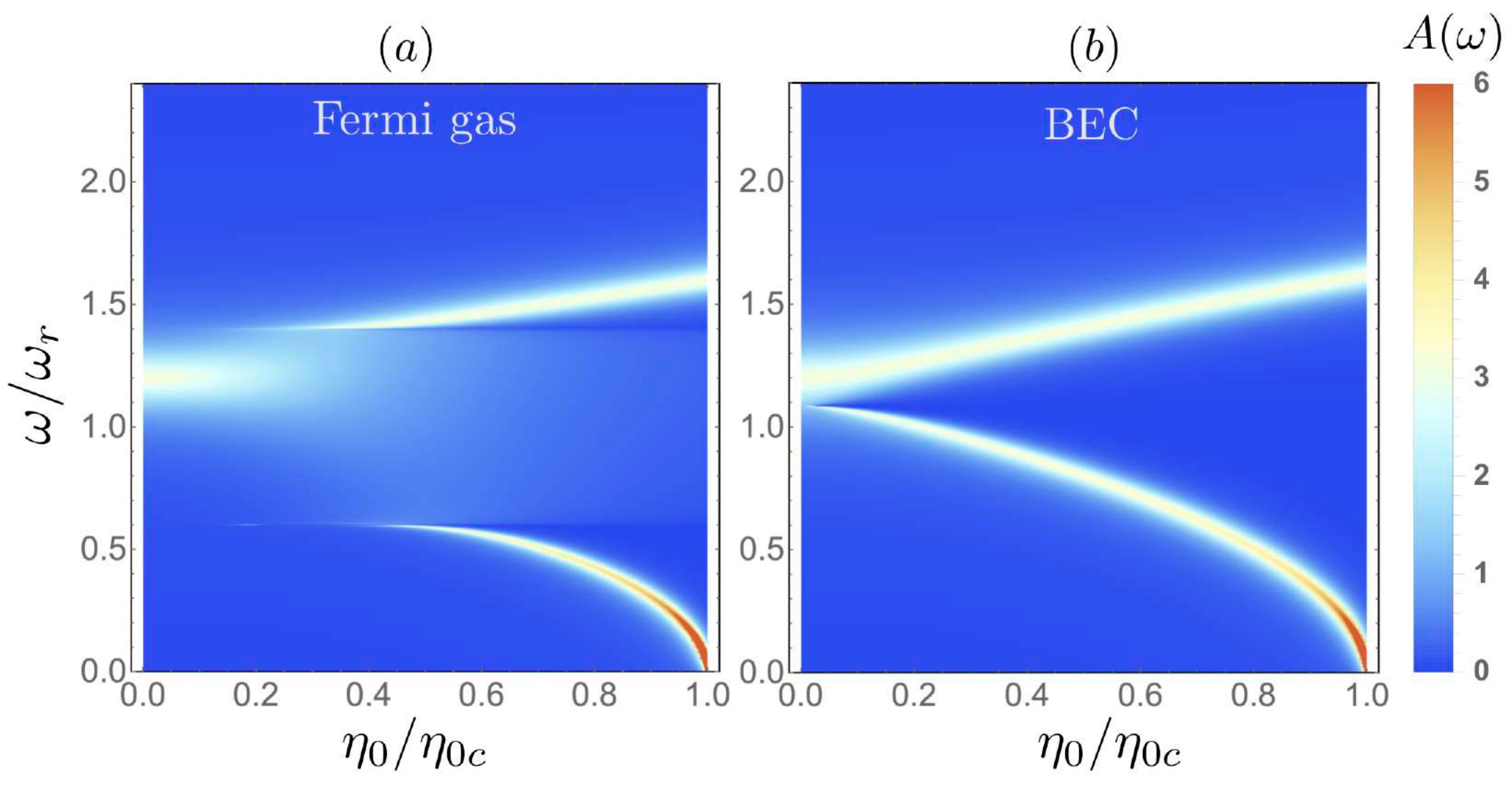}
    \caption{Spectrally resolved response of the system to a cavity probe. A fermionic (a) or a bosonic (b) cloud is coupled to a single standing-wave cavity mode of wavevector $k_c$. The cloud is taken to be homogeneous, one-dimensional and oriented along the cavity axis. The cloud average density is $n=0.127 k_c$ ($k_F=0.2k_c$) and $T=0$. Fermions are non interacting while the bosons have a contact interaction energy $g_0n=0.09 \hbar\omega_r$. The dispersively-shifted cavity detuning is $\delta_{\rm c}(0)=-1.2\omega_r$ and the loss rate is $\kappa=0.1\omega_r$. Two polariton collective modes are visible as isolated peaks. In the fermionic case, the particle-hole continuum of excitations is visible between the polaritons. We note that the parameters used here do not correspond to any of the current experiments and are chosen solely for illustrative purposes.
    }
    \label{fig:spectral_fermiVSbose}
  \end{center}
\end{figure}

The polariton resonances at $\omega_{\rm pol}$ appear as peaks in the cavity optical response function $A(\omega)=-2\pi\mathrm{Im}[D(\omega)]$. The resonant peaks have a finite width due to the decay of cavity photons. A typical cavity optical response $A(\omega)$ is shown in Figure~\ref{fig:spectral_fermiVSbose}, both for a fermionic and a bosonic atomic cloud in the absence of an external trap potential $V_{\rm ext}(\mathbf{r})=0$. Generically, two polariton modes appear, corresponding to mixing of the atomic density fluctuation and the single-mode photonic degree of freedom. By increasing the atom-cavity coupling $\eta_0$ the polariton peaks further separate.

In the absence of a coherent cavity field $\alpha=0$ [and in the absence of an external trap $V_{\rm ext}(\mathbf{r})=0$ as considered in Figure~\ref{fig:spectral_fermiVSbose}], the single-particle eigenstates in Equation~\eqref{eq:dyn_polarization} are plane waves $|u_{\mathbf{k}}\rangle=e^{i\mathbf{k}\cdot\mathbf{r}}/\sqrt{V}$ with energies $\epsilon_{\mathbf{k}}^0=\hbar^2k^2/2M$ and the dynamical polarization function reads

\begin{equation}
  \label{eq:dyn_polarization_homogeneous}
  \chi_{\rm dyn, hom}(\omega)=\frac{\eta_0^2}{2}\sum_{\mathbf{k}} \frac{n(\epsilon^0_{\mathbf{k+k_c}})-n(\epsilon^0_{\mathbf{k}})}{\hbar\omega+\epsilon^0_{\mathbf{k+k_c}}-\epsilon^0_{\mathbf{k}}+i0^+}.
\end{equation}
For (non-interacting) fermionic atoms at thermal equilibrium, $n(\epsilon)=n_{\rm FD}(\epsilon)$ and the dynamical polarization function has an imaginary part, i.e., finite absorption in the medium, corresponding to on-shell particle-hole excitations of the Fermi surface. This continuum of states is indeed visible in the optical response as a broad feature which shows sharp edges for low-temperature gas [see Figure~\ref{fig:spectral_fermiVSbose}(a)]. If the polariton energy lies within this particle-hole continuum, it does not correspond anymore to a resonant peak, rather to a locally increased spectral weight~\cite{Piazza2014Umklapp,Piazza2014Quantum,Sandner2015Self,Feng2017Single}. 

On the other hand, in a BEC with a finite two-body contact repulsion, the cavity photon excites Bogoliubov modes with dispersion $\epsilon_{\mathbf{k}}=\sqrt{\epsilon_{\mathbf{k}}^0(\epsilon_{\mathbf{k}}^0+2g_0n)}$ out of the condensate $n(\epsilon_{\mathbf{k}})=N\delta_{\mathbf{k},0}$. At zero temperature $T=0$ and neglecting condensate depletion as well as Beliaev damping of phonons~\cite{Piazza2013Bose,Konya2014Photonic,Konya2014Damping}, the dynamical polarization function of the BEC has no imaginary part, so that two resonant peaks are always present in the optical response as shown in Figure~\ref{fig:spectral_fermiVSbose}(b).

%..................................................................................
\subsubsection{Linear stability analysis and Bogoliubov spectrum}
\label{subsubsec:linear-stability-analysis}

Here we present an alternative, complementary approach for obtaining low-energy collective excitations of the system with bosonic atoms at zero temperature. It is based on a linear stability analysis around the mean-field steady-state solutions~\cite{Horak2001Dissipative, Nagy2008Self,Oztop2013Collective}. To this end, the Heisenberg equations of motion of the atomic and photonic field operators, Eqs.~\eqref{eq:Heisberg-eq-g} and \eqref{eq:Heisberg-eq-a-loss}, are linearized around the mean-field steady state $\psi(\mathbf{r})$ and $\alpha$ by keeping the quantum fluctuations $\delta\hat\psi(\mathbf{r},t)=e^{-i\mu t/\hbar}[\delta\hat\psi_+(\mathbf{r})e^{-i\omega_{\rm pol} t}+\delta\hat\psi_-^\dag(\mathbf{r})e^{i\omega_{\rm pol}^* t}]$ and $\delta\hat{a}(t)=\delta\hat{a}_+e^{-i\omega_{\rm pol} t}+\delta\hat{a}_-^\dag e^{i\omega_{\rm pol}^* t}$ up to first order,
\begin{subequations} 
\label{eq:linearized-eq-ga}
\begin{align}
\label{eq:linearized-eq-g} 
i\hbar\frac{\partial \delta\hat{\psi}}{\partial t}&=
\left[-\frac{\hbar^2}{2M}\boldsymbol{\nabla}^2+V_{\rm ext}(\mathbf{r})
+\hbar V(\mathbf{r})
+\hbar|\alpha|^2U(\mathbf{r})
+2\hbar\text{Re}[\alpha]\eta(\mathbf{r})
+2g_0n(\mathbf{r})-\mu\right]\delta\hat{\psi}\nonumber\\
&+g_0\psi^2(\mathbf{r})\delta\hat{\psi}^\dag
+\hbar\psi(\mathbf{r})
\left\{\left[U(\mathbf{r})\alpha^*+\eta(\mathbf{r})\right]\delta\hat{a}
+\left[U(\mathbf{r})\alpha+\eta(\mathbf{r})\right]\delta\hat{a}^\dag\right\},
\end{align}
\begin{align}
\label{eq:linearized-eq-a} 
i\frac{\partial \delta\hat{a}}{\partial t}&=
-\left(\Delta_c-\int U(\mathbf{r}) n(\mathbf{r})d\mathbf{r}+i\kappa \right)\delta\hat{a}
+\int \left[U(\mathbf{r})\alpha+\eta(\mathbf{r})\right]  
\left[\psi^*(\mathbf{r})\delta\hat{\psi}+\psi(\mathbf{r})\delta\hat{\psi}^\dag\right]
d\mathbf{r}.
\end{align}
\end{subequations}
The linearized equations~\eqref{eq:linearized-eq-ga} constitute Bogoliubov-type coupled equations for atomic and photonic quantum fluctuations. The Bogoliubov equations, Equation~\eqref{eq:linearized-eq-ga}, can be recast in a matrix form, 
\begin{align}
\label{eq:Bogoliubov-eq}
\mathbf{M}_B\mathbf{F}=\omega_{\rm pol}\mathbf{F},
\end{align}
where $\mathbf{F}=(\delta\hat\psi_+,\delta\hat\psi_-,\delta\hat{a}_+,\delta\hat{a}_-)^\top$ and $\mathbf{M}_B$ is a non-Hermitian matrix obtained from Equation~\eqref{eq:linearized-eq-ga}. 
The eigenvalues $\omega_{\rm pol}$ of the Bogoliubov equation~\eqref{eq:Bogoliubov-eq} yield the collective elementary excitation spectrum of the system above the mean-field steady state.

%----------------------------------------------------------------------------------------------------------------------------
\subsection{Basic experimental scheme and relevant parameters}
\label{subsec:basic-experimental-scheme}

The basic experimental scheme used to study self-organization of quantum gases in optical cavities is depicted in Figure~\ref{fig:coupling_scheme}. Ultracold atoms are trapped in an external potential $V_\mathrm{ext}(\mathbf{r})$ whose minimum is centered with respect to the mode waist of an optical cavity. Typically a near harmonic trapping potential is formed from two orthogonally intersecting, non-interfering laser beams (not shown) of sufficiently large waist, which are far red-detuned from atomic resonances~\cite{Grimm1999Optical}. The trapping laser frequencies  are chosen to be also far detuned from cavity resonances such that the effect of scattering trapping-light photons into the cavity can be neglected.

All bosonic experiments to date have used Bose-Einstein condensates of Rubidium-87 atoms~\cite{Steck2009Rubidium}, with typical atom numbers between a few thousands and a few hundred thousands. Besides the atom number $N$ and the temperature $T$ of the atomic cloud, the most relevant atomic parameters for these experiments are the atomic mass $M$, the atomic transition frequency $\omega_a$, the electric dipole matrix element $d_{ge}$ of the considered atomic transition and its decay rate $\gamma$, as well as the atomic s-wave scattering length $a^s_{gg}$ determining atomic two-body interactions. Feshbach resonances of Rubidium-87 are mostly very narrow, which makes it practically impossible to modify the scattering properties and the interaction strength, and limits the possibilities to realize phenomena requiring tunable atomic interactions. However, there is no fundamental difficulty in performing such experiments with atomic species having different properties.

\subsubsection{Optical resonators}\label{subsubsec:optical_resonators} 
\begin{figure}[t!]
  \begin{center}
    \includegraphics[width=1\columnwidth]{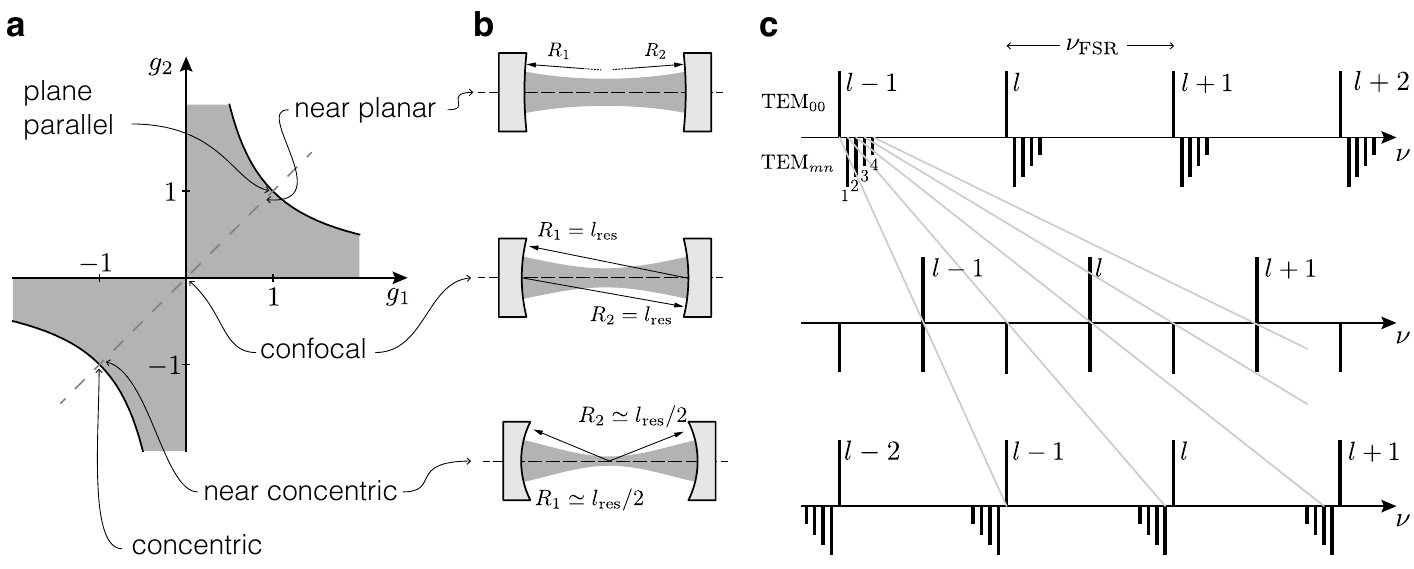}
    \caption{Stability diagram, typical resonator configurations, and mode spectra. 
    (a) The geometric cavity parameters $g_i$ can be used to identify the condition for the resonator stability, $0 \leq g_1 g_2 \leq 1$, shown as grey area. Special configurations for a symmetric resonator ($g_1=g_2$, the dashed line) are indicated in the stability diagram and schematically shown in (b). The mode spectra for the near-planar (top), confocal (middle), and near-concentric (bottom) configurations are shown in (c). In these spectra, the fundamental TEM$_{00}$ modes are depicted above the frequency axes, while the higher transverse modes indicated with the sum $(m+n)$ of the transverse mode indices are shown below. The longitudinal mode index $l$ refers to the TEM$_{00}$ modes which are spaced by $\nu_\mathrm{FSR}$. Moving from the planar to the concentric configuration, all modes move in frequency space as indicated by the grey lines. In the confocal configuration, degeneracy of modes with either even or odd sums of transverse mode indices is reached. Note that modes with different longitudinal indices are becoming degenerate in the confocal and concentric cases, while in the near planar case a (quasi-)degenerate mode family shares the same longitudinal mode index $l$.}
    \label{fig:CavityParameter}
  \end{center}
\end{figure}

Experiments have been carried out using linear cavities (groups at ETH Zurich, University of Hamburg, and Stanford University) as well as ring cavities (the group at University of T\"ubingen). Here we focus on the linear cavities and review the role of the most relevant parameters for their description~\cite{Kogelnik1966Laser,Siegman1986Lasers,Renk2012Basics,Haroche2006Exploring}. The geometry of a linear cavity is characterized by the radii of curvature $R^c_{1,2}$ of the two concave mirrors and by their separation, the length $l_\mathrm{res}$ of the resonator. The $g$-parameters $g_i = 1-l_\mathrm{res}/R^c_i$  are used to describe a resonator and its stability condition, $0\leq g_1 g_2 \leq 1$; see also Figure~\ref{fig:CavityParameter}. The two limiting and only marginally stable cases are the plane parallel Fabry-Pérot resonator ($R^c_1=R^c_2=\infty$) and the concentric configuration ($R^c_1+R^c_2=l_\mathrm{res}$). A third special, stable case is the confocal configuration ($R^c_1+R^c_2=2l_\mathrm{res}$).

In the paraxial approximation, the so-called transverse electromagnetic modes TEM$_{m n}$ are described by the Hermite-Gaussian wavefunctions, 
\begin{align}
  \label{eq:higher_order_gaussian_beam}
  \mathcal{E}_{mn}(\mathbf{r})&=\mathcal{E}_0\frac{w_0}{w(x)}H_m\left(\frac{\sqrt{2}y}{w(x)}\right) H_n\left(\frac{\sqrt{2}z}{w(x)}\right)e^{-(y^2+z^2)/w^2(x)}\nonumber\\
  &\times\cos\left[k_c\left(x+\frac{y^2+z^2}{2R(x)}\right)-\varphi^{\rm Gouy}_{mn}(x)+\varphi^{\rm offset}_{mn}\right].
\end{align}
Here, $m$ and $n$ are the integer transverse mode indices, the wave number $k_c$ is fixed by the mode frequency $\nu$ as $k_c=2\pi\nu/c$, $w(x)=w_0\sqrt{1+(\lambda_c x/\pi w_0^2)^2}$ is the mode waist with minimum  waist $w_0$ at longitudinal position $x=0$, $R(x)=x+\pi^2w_0^4/\lambda_c^2x$ is the local radius of curvature, and $H_m$ is a Hermite polynomial. The wave length $\lambda_c$ is set by $k_c=2\pi/\lambda_c$. Since the radius of curvature of the wavefront of the Gaussian beam is constantly evolving with the propagation distance, the beam acquires an additional phase shift with respect to the propagation of a plane wave. Indeed, the effective wave length depends on the distance $x$ from the beam waist. This spatially dependent Gouy phase shift $\varphi^{\rm Gouy}_{mn}(x)=(1+m+n) \arctan(\lambda_c x / \pi w_0^2)$ depends on the transverse mode indices. Finally, in order to satisfy the boundary conditions at the mirrors, one has to allow for a phase offset $\varphi^{\rm offset}_{mn}= (m+n) \arctan(\lambda_c l_{\rm res} / 2\pi w_0^2)$~\cite{Guo2019Emergent}.

In a cavity, the radii of curvature of the wavefronts at the position of the mirrors have to match the radii of curvature of the mirror surfaces. This condition determines the minimum mode-waist radius squared, $w_0^2=(\lambda_c l_{\rm res}/\pi)\sqrt{g_1g_2(1-g_1g_2)/(g_1+g_2- 2g_1g_2)^2}$, and therefore the mode volume $\mathcal{V}=\pi w_0^2 l_\mathrm{res}/4$ of the fundamental mode TEM$_{0 0}$ with a purely Gaussian transverse profile. 

In order to meet a resonance condition, the phase of the field must be an integer multiple $l$ of $2\pi$ after one resonator round trip, with $l$ thus counting the number of nodes of a mode in the longitudinal direction. 
This condition determines the resonance frequencies of the cavity modes
\begin{align}
  \label{eq:transverse_mode_spacing}
  \nu_{lmn} = \nu_\mathrm{FSR} \left[ l + (m+n+1)\frac{\arccos(\pm \sqrt{g_1 g_2})}{\pi} \right],
\end{align}
where we have used the fact that the Gouy phase shift across the resonator is given by $\varphi^{\rm Gouy}_{mn}(l_{\rm res}/2)-\varphi^{\rm Gouy}_{mn} (-l_{\rm res}/2)=\arccos(\pm\sqrt{g_1g_2})$, with the + (-) sign applying to the upper right $g_1,g_2>0$ (lower left $g_1,g_2<0$) quadrant. From Equation~\eqref{eq:transverse_mode_spacing}, one can directly derive the spacing $\Delta \nu_\mathrm{TEM}$ between adjacent transverse modes. The frequency separation between two adjacent longitudinal modes, i.e., $(lmn)$ and $((l+1)mn)$, is given by the free spectral range $\nu_\mathrm{FSR} = c/2 l_\mathrm{res}$.

Equation~(\ref{eq:transverse_mode_spacing}) indicates how the mode frequency separations between the longitudinal and transverse modes depend on the $g$-parameters of the cavity; see Figure~\ref{fig:CavityParameter}. For the two extreme cases of a plane parallel ($g_1=g_2=+1$) and a concentric ($g_1=g_2=-1$) resonator, all transverse modes TEM$_{mn}$ are degenerate and form mode families. In the plane parallel configuration, these mode families share the same longitudinal mode number $l$, and correspondingly a common longitudinal standing-wave profile $\cos(k_c x)$ with wavevector $k_c=\omega_c/c=2 \pi \nu_{lmn}/c$ (in a ring cavity the standing wave is substituted by two degenerate, counterpropagating plane waves). Moving away from the plane parallel situation, the transverse modes belonging to one such mode family separate increasingly. Reaching the confocal configuration ($g_1=g_2=0$), the initial mode families mix. At this point, all modes fall into either of two degenerate new mode families that alternate every half free spectral range, and where the sum of mode indices ($m+n$) is either an even or an odd number. Further decreasing the $g$-parameters, these sub-families of transverse modes split again until the concentric configuration is reached. In this situation, mode families of equal sum of mode indices ($l+m+n$) form.

The two experimentally most relevant cavity configurations are the near-planar ($R^c_i \gg l_\mathrm{res}, g_i\rightarrow+1$) and the confocal ($R^c_i \simeq l_\mathrm{res}, g_i\rightarrow 0$) geometries, since they are stable resonator configurations. Coupling to a single TEM$_{00}$ mode of a near-planar cavity has been considered in the preceding sections (as well as in the large parts of this review). Coupling of atoms to confocal (and to some extent also near-concentric) cavities will be discussed in Section~\ref{sec:multimode}.

Ideal mirror surfaces are assumed to be parabolic and rotationally symmetric. In practice, however, spherical mirrors are used and they might have different radii of curvature $\{R^c_x, R^c_y\}$ along different axes, resulting in an elliptical  cavity waist. In addition, the axes of the mirrors forming the cavity might also be misaligned with respect to each other. The solution of the Helmholtz equation beyond the paraxial approximation shows that the Gouy phase shift---ultimately determining resonance frequencies of the cavity modes---depends on the orientation of the field polarization with respect to the elliptical focus. To lowest order, this gives  rise to a birefringent splitting of the cavity modes along the ordinary and the extraordinary axes by~\cite{Uphoff2015Frequency},
\begin{align}
  \Delta\nu = -\frac{c \lambda_c}{8\pi^2 l_\mathrm{res}} \frac{R^c_x-R^c_y}{R^c_x R^c_y}\,.
\end{align}
This geometrically caused birefringence is, however, for macroscopic cavities usually very small and only becomes significant for microscopic cavities with very small radii of curvature.

More importantly for macroscopic cavities is a possible birefringence of dielectric coatings on the mirror surfaces. This birefringence can be a material property or due to mechanical stress caused during mounting or machining of the mirrors. The electric field penetrates the mirror surfaces typically by several wave lengths and is thus subject to a possible birefringence in the coatings. This leads to a splitting of the cavity resonance frequencies for the orthogonal polarization modes. All these effects lift the theoretically ideal mode degeneracies of multimode resonators to a certain degree in a realistic experiment.

The optical properties of the mirrors are captured by their reflectivity $\mathcal{R}_i$, transmissivity $\mathcal{T}_i$, and losses $\mathcal{L}_i$, where $\mathcal{R}_i+\mathcal{T}_i+\mathcal{L}_i=1$. The losses can be caused by scattering, diffraction, or absorption. The finesse $\mathcal{F}$ of a Fabry-Pérot resonator,
\begin{align}
\label{eq:finesse}
 \mathcal{F} = \frac{\pi \sqrt[4]{\mathcal{R}_1 \mathcal{R}_2}}{1-\sqrt{\mathcal{R}_1\mathcal{R}_2}},
\end{align}
characterizes the field enhancement inside the cavity. A photon traverses the resonator on average $\mathcal{F}/\pi$ times before it is lost, resulting in a power enhancement of  $\mathcal{F}/\pi$ with respect to the input coupled power. Finesse and free spectral range also determine the decay rate $\kappa$ of the electromagnetic field inside the cavity
\begin{align}\label{eq:kappa}
  \kappa = \pi \frac{\nu_\mathrm{FSR}}{\mathcal{F}},
\end{align}
and accordingly the full width at half maximum $\Delta \nu = \kappa/\pi$ of the transmission peak of a cavity.

\subsubsection{Cavity QED parameters}
The coupling between a single atom and a single quantized electromagnetic field inside an optical cavity is given by the vacuum-Rabi frequency
\begin{align}
  \label{eq:vacuum_Rabi_split}
  \mathcal{G}_0 = \frac{\mathbf{d}_{ge}\cdot \boldsymbol{\epsilon}_c \mathcal{E}_0}{\hbar} 
  = \frac{d_{ge}}{\hbar}\sqrt{\frac{\hbar \omega_c}{2 \epsilon_0 \mathcal{V}}} \cos(\angle(\mathbf{d}_{ge},\boldsymbol{\epsilon}_c)),
\end{align} 
where $\epsilon_0$ is the electric permittivity. We used the vector notations $\mathbf{d}_{ge}$ for the induced atomic electric-dipole moment for the transition $\ket{g}\leftrightarrow\ket{e}$ and $\boldsymbol{\epsilon}_c$  for the unit polarization vector of the intracavity field. Accordingly, $\angle(\mathbf{d}_{ge},\boldsymbol{\epsilon}_c)$ is the angle between the dipole moment and the polarization direction of the electric field. Here, $\mathcal{E}_0$ is the maximum electric field strength of a single photon in the peak intracavity field of the TEM$_{00}$ mode volume $\mathcal{V}$. The spatial-dependent Rabi frequency used in the previous sections is then defined as 
\begin{align}
  \label{eq:HermiteToRabi}
\mathcal{G}(\mathbf{r})=\mathcal{G}_0\frac{\mathcal{E(\mathbf{r})}}{\mathcal{E}_0},
\end{align}
where $\mathcal{E(\mathbf{r})}$ is the local electric field strength of a single intra-cavity photon.

Cavity-QED systems are often characterized by the cooperativity parameter $C=\mathcal{G}_0^2/\kappa \gamma$ (note that $\kappa$ and $\gamma$ are the cavity-field and atomic-dipole decay rates, i.e. $\kappa$ is half the cavity linewidth, see Equation (\ref{eq:kappa}), and $\gamma$ is half the atomic linewidth $\Gamma$), or by the Purcell factor $P_F = 24 \mathcal{F}/\pi k_c^2 w_0^2$~\cite{Tanji-Suzuki2011Interaction}. However, since self-organization experiments typically operate in the dispersive regime ($\Delta_a \gg \gamma$), where the occupation of the atomic excited state is negligible, the more relevant parameter is the collective coupling rate $\sqrt{N} \mathcal{G}_0$. The coupling strength $\mathcal{G}_0$ also determines the relevant parameters of the effective atom-cavity Hamiltonian,  Equation~(\ref{eq:H_eff_1comp}), i.e., the optical lattice depth $\hbar U_0= \hbar\mathcal{G}_0^2/\Delta_a$ created by a single intra-cavity photon and the two-photon Rabi frequency $\eta_0 = \mathcal{G}_0\Omega_0/\Delta_a$.

\subsubsection{Relevant energy scales} 
\begin{figure}[t!]
  \begin{center}
    \includegraphics[width=1\columnwidth]{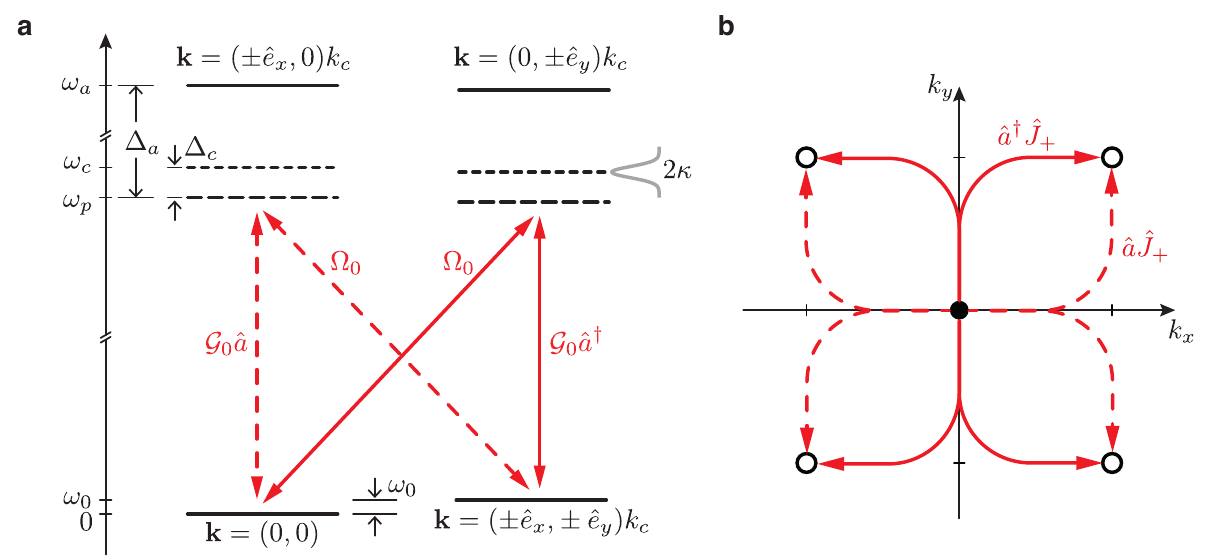}
    \caption{Basic atom-photon coupling scheme with relevant frequencies and coupling rates in self-ordering phenomena. (a) The relevant energy/frequency scales are given by the atomic kinetic energy $\hbar \omega_0=2\hbar\omega_r$, and the relative atomic $\Delta_a=\omega_p-\omega_a$ and cavity $\Delta_c=\omega_p-\omega_c$ detunings with respect to the pump frequency $\omega_p$ (frequency axis not to scale). The two pairs of red arrows (solid and dashed lines) show the two possible excitation paths that couple the zero momentum state $\mathbf{k}=(k_x,k_y)=(0,0)$ to the atomic excited momentum states $\mathbf{k}=(k_x,k_y)=(\pm k_c, \pm k_c)$. (b)  The two excitation paths (dashed and solid) corresponding to the two Raman channels are illustrated schematically in a momentum-space diagram. The operators $\hat a^\dag \hat{J}_+$ and $\hat a \hat{J}_+$ correspond to the creation and annihilation of a cavity photon while creating atoms in the excited momentum state, respectively, in the framework of the Dicke model; see Section~\ref{subsubsec:DickeModel}.}
    \label{fig:energy_scales}
  \end{center}
\end{figure}

The energy scales relevant for most of the self-organization physics described in this review article are shown in Figure~\ref{fig:energy_scales} together with the basic coupling scheme. Bosonic atoms are initially in the zero momentum state $\mathbf{k}=(k_x,k_y)=(0,0)$ and are coupled via cavity-assisted two-photon Raman transitions to the symmetric superposition of states  $\mathbf{k}=(k_x,k_y)=(\pm k_c, \pm k_c)$; see the discussion about the Dicke model in Section~\ref{subsubsec:DickeModel}. The energy difference $\hbar\omega_0 = 2 \hbar\omega_r$ between these levels is twice the recoil energy $\hbar\omega_r = \hbar^2 k_c^2/2M$. There are two possible excitation paths connecting these states. Either first a photon from the pump beam is absorbed by the atoms and then reemitted into the cavity mode, or these processes take place in the inverse order. The intermediate states are far detuned by $\Delta_a = \omega_p - \omega_a$ from the atomic excited state and thus only virtually populated.  The cavity resonance is detuned by $\Delta_c=\omega_p-\omega_c$ from the pump frequency $\omega_p$. The experiment operates in the sideband resolved regime if the cavity-field decay rate $\kappa$ is on the order of or smaller than the motional excitation frequency $\omega_0$.

Table~\ref{tab:ExperimentalParameters} summarizes the experimental parameters of the setups which have thus far performed experiments and published works on self-organization of quantum gases in optical cavities. 

\begin{sidewaystable}\sffamily
\begin{center}
\begin{tabular}{c | c c c c c c c} \hline
 Parameter & ETH1 \cite{Baumann2010Dicke} & ETH2 \cite{Leonard2017Supersolid} & Hamb \cite{Kessler2014Optomechanical} & Stan \cite{Kollar2015An} & EPFL \cite{Roux2020Strongly} & T\"ub1 \cite{Schmidt2014Dynamical}& T\"ub2 \cite{Wolf2018Observation}\\ \hline \hline
  & \multicolumn{5}{c|}{linear standing-wave cavity} & \multicolumn{2}{c}{ring cavity} \\ \hline
$R_{1,2}^c$ & \SI{75}{\milli \m} & \SI{75}{\milli \m}& \SI{25.06}{\milli \m} & \SI{9.97}{\milli \m} & \SI{25}{\milli \m} & -- & -- \\ \hline
$l_\mathrm{res}$ & \SI{176}{\micro \m} & \begin{tabular}{c} \SI{2.45}{\milli \m} \\\SI{2.80}{\milli \m} \end{tabular} & \SI{48.93}{\milli \m} & \begin{tabular}{c} adjustable \\ $\sim R_{1,2}^c \pm \epsilon$  \end{tabular}  & \SI{4.13}{\centi \m} & \SI{87}{\milli \m} & \SI{39}{\centi \m} \\ \hline
$w_0$ & \SI{25.3}{\micro \m} & \begin{tabular}{c} \SI{48.7}{\micro \m} \\ \SI{50.4}{\micro \m} \end{tabular} & \SI{31.2}{\micro \m} & \SI{35}{\micro \m} &  \SI{45.0}{\micro \m} & $(117 \times 88)$\si{\micro \m}  & \SI{170}{\micro \m}\\ \hline
$\nu_\mathrm{FSR}$ & \SI{852}{\giga \hertz} & \begin{tabular}{c} \SI{61.18}{\giga \hertz} \\ \SI{53.53}{\giga \hertz} \end{tabular} & \SI{3.063}{\giga \hertz} & \SI{14.95}{\giga \hertz} &  \SI{3.630}{\giga \hertz}  & \SI{3.45}{\giga \hertz} & \SI{0.77}{\giga \hertz}  \\ \hline
$\Delta \nu_\mathrm{TEM}$ & \SI{18.5}{\giga \hertz} & \begin{tabular}{c} \SI{4.99}{\giga \hertz} \\ \SI{4.67}{\giga \hertz} \end{tabular} & \SI{301.5}{\mega \hertz} &   \begin{tabular}{c} adjustable \\ $0 -100$\SI{}{\mega \hertz} \end{tabular}&  \SI{2.635}{\giga \hertz} & -- & --\\ \hline
$\mathcal{F}$ & \num{3.42e5} & \begin{tabular}{c} \num{2.08e5} \\ \num{3.34e4} \end{tabular} & \num{3.44e5} & \num{5.46e4} & \num{4.7e4} & \begin{tabular}{c} s: \num{1.3e5} \\ p: \num{2.8e3} \end{tabular} & \num{9.4e4}\\ \hline
$\kappa / (2\pi)$ & \SI{1.25}{\mega \hertz} & \begin{tabular}{c}  \SI{147}{\kilo \hertz} \\  \SI{800}{\kilo \hertz} \end{tabular} &  \SI{4.5}{\kilo \hertz} &  \SI{137}{\kilo \hertz} & \SI{77}{\kilo \hertz} & \begin{tabular}{c}  s: \SI{14}{\kilo \hertz} \\  p: \SI{650}{\kilo \hertz} \end{tabular} & \SI{5}{\kilo \hertz} \\ \hline
$\mathcal{G}_0 / (2\pi)$ &  \SI{15.5}{\mega \hertz} & \begin{tabular}{c}  \SI{2.14}{\mega \hertz}\\ \SI{1.94}{\mega \hertz} \end{tabular} & \SI{0.76}{\mega \hertz} &  \SI{1.47}{\mega \hertz}  & \SI{0.479}{\mega \hertz} & \SI{87.5}{\kilo \hertz} & \SI{24.3}{\kilo \hertz}  \\ \hline
$C$ & \num{63.1} & \begin{tabular}{c} \num{10.3} \\ \num{1.57} \end{tabular} & \num{42.3} & \num{5.2} & \num{2.1} & \num{0.2} & \num{0.04}  \\ \hline
$\lambda$ & \SI{780}{\nano \m} & \SI{785}{\nano \m} & \SI{803}{\nano \m} & \SI{780}{\nano \m} & \SI{670}{\nano \m} & \SI{780}{\nano \m} & \SI{780}{\nano \m}  \\ \hline
\end{tabular}
\end{center}
\caption{Parameters of the experimental setups discussed in this review (ETH1 and ETH2 are experiments operated at ETH Zurich, Hamb is operated at the University of Hamburg, Stan at the Stanford University, EPFL is operated at EPFL Lausanne, and T\"ub1 and T\"ub2 are operated at the University of T\"ubingen). $\lambda$ specifies the wave length at which the parameters were characterized and might be different from the wave length used in a specific experiment. The values for coupling strength and cooperativity for the linear cavities have been calculated based on the geometry of the resonator using the maximum dipole-allowed transition for the D2 line and assuming coupling to the TEM$_{00}$ mode. The setup ETH2 operates with two crossed cavities for which the parameters are given in consecutive lines. The setup at Stanford operates a cavity with adjustable length close to the confocal configuration.}
\label{tab:ExperimentalParameters}
\end{sidewaystable}

\section{Superradiant crystallization breaking a discrete symmetry}
\label{sec:SR_discrete}

In this chapter we explore the superradiant phase transition in laser-driven quantum gases coupled dispersively to single standing-wave modes of linear cavities and some other closely related variants. Fundamental aspects of these coupled systems were discussed in Section~\ref{sec:theory}.  Across such a superradiant phase transition, the atomic density self-orders into a crystalline pattern---spontaneously breaking the discrete $\mathbf{Z}_2$ symmetry of the systems---which constructively scatters the laser light into the cavity such that a coherent cavity field is built-up. Assuming that the system can be considered to be approximately in thermal equilibrium (see Section~\ref{subsec:thermal_nonthermal}), the self-organization transition occurs when the potential-energy gain due to the coupling of the atoms into the cavity field overcomes the kinetic-energy penalty of the density crystallization. Within an equivalent description based on cavity-mediated atom-atom interactions, presented in Section~\ref{subsubsec:cavity-induced-int}, it is these global interactions which favor the self-organization into the crystalline order. This chapter is devoted to the most fundamental self-ordering scenarios. In the following chapters, we will encounter several other instances of superradiant phase transition associated with different types of ordering.

\subsection{Self-organization phase transition}
\label{subsec:self-ordering_PT}

\subsubsection{Superradiant instability}
\label{subsec:SR_instability}

Consider a laser-driven quantum gas coupled dispersively to a single standing-wave mode of a lossy linear cavity, as discussed in detail in Section~\ref{sec:theory}. Such a system generically exhibits a Dicke-type superradiant phase transition where one of the collective modes becomes unstable at the onset of the superradiant threshold. In the present driven-dissipative system, this corresponds to a low-lying polariton mode switching from being damped to growing in time. That is, the imaginary part of the complex polariton frequency vanishes at the superradiant threshold; see Section~\ref{subsec:excitations_probing}.

\begin{figure}[t!]
  \begin{center}
    \includegraphics[width=0.7\columnwidth]{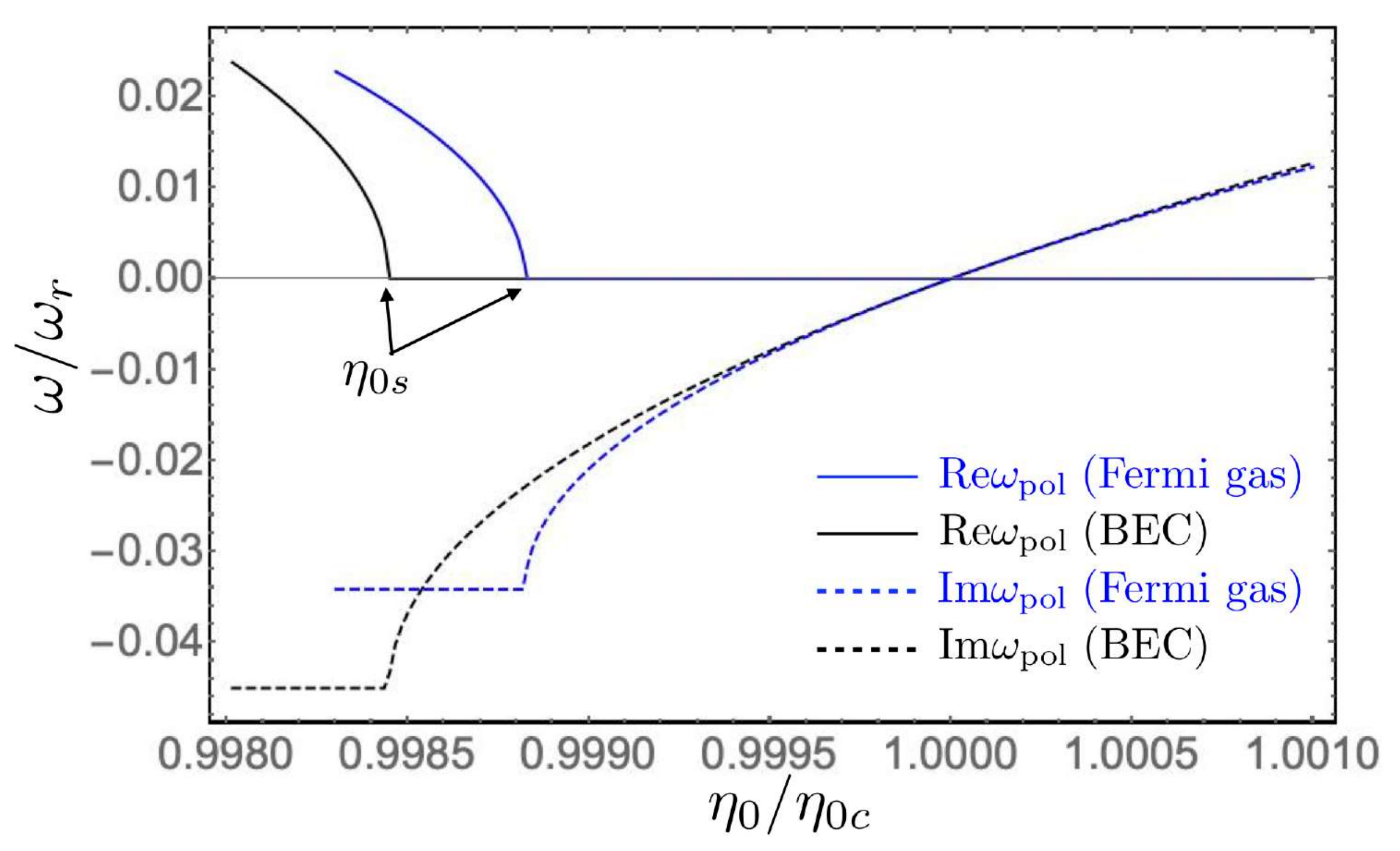}
    \caption{The low-lying (i.e., softening) complex polariton frequencies as a function of the pump strength $\eta_0$ for both bosons (black curves) and fermions (blue curves), similar to Figure~\ref{fig:spectral_fermiVSbose}, around the superradiant threshold $\eta_{0c}$.  After the real part of the frequency vanishes at $\eta_0=\eta_{0s}$, the imaginary branch (associated with the damping of the polariton) splits into two parts, where one (shown here) linearly vanishes approaching the pump-strength threshold $\eta_{0c}$. Beyond the threshold, this part of the imaginary branch becomes positive, signaling a dynamical instability toward the self-ordered, superradiant phase. }
    \label{fig:poles_fermiVSbose}
  \end{center}
 \end{figure}

At very small atom-photon coupling strengths $\eta_0$ (compared to the energy separation between polaritons), the polaritons are almost decoupled from one another and are still mainly either atom-like or photon-like. By increasing the coupling strength $\eta_0$, the polaritons are mixed more strongly. In the case that the photon-like polariton initially lies at a higher frequency than the atomic counterpart, the atom-like polariton starts then to soften---i.e., its real part approaches zero. As can be seen from Figure~\ref{fig:spectral_fermiVSbose}, this happens essentially in the same way for both bosonic and fermionic atoms, apart from the fact that in the latter case the polariton resonant peak becomes visible only when it exits the particle-hole continuum.
When the real part of the low-lying polariton frequency reaches zero at $\eta_0=\eta_{0s}$, the polariton splits into two branches differing only by their imaginary part. The branch with the smallest imaginary part is shown in Figure~\ref{fig:poles_fermiVSbose}, for both bosonic and fermionic atoms. At the critical pump strength $\eta_{0c}>\eta_{0s}$, one of the imaginary branches crosses zero and becomes positive for $\eta_0>\eta_{0c}$, signaling that the mean-field solution with $\alpha=0$ has become unstable towards a superradiant state where $\alpha\neq 0$. The threshold $\eta_{0c}$ is obtained from the characteristic equation \eqref{eq:poles-retarded-cavity} by setting the polariton frequency to zero, $\omega_{\rm pol}=0$. In the simple case of a spatially homogeneous 1D, zero-temperature BEC as considered in Figure~\ref{fig:spectral_fermiVSbose}(b), an explicit analytical expression for the threshold can be obtained,
\begin{align}
\label{eq:eta-critical_bose_hom}
\sqrt{N}\eta_{0c}=\sqrt{\frac{\delta_c^2+\kappa^2}{-2\delta_c}}
\sqrt{\omega_0+\frac{2g_0n}{\hbar}},
\end{align}
where $\delta_c=\Delta_c-NU_0/2$. This critical pump strength with $g_0=0$ (i.e., a non-interacting BEC) is consistent with the threshold for the Dicke-superradiance phase transition, Equation~\eqref{eq:eta-critical-Dicke-model}, except that in the 1D case discussed here there is an extra factor of $1/\sqrt{2}$ and $\omega_0 = \omega_r$.

Equivalently, the superradiant threshold $\eta_{0c}$, Equation~\eqref{eq:eta-critical_bose_hom}, can be obtained from the linear stability analysis (see Section~\ref{subsubsec:linear-stability-analysis}) of the trivial mean-field solution, i.e., a zero-temperature, homogeneous 1D BEC along the cavity axis and a vacuum cavity field $\alpha=0$. In the low-energy limit in the vicinity of the Dicke superradiance phase transition only the atomic condensate fluctuations $\delta\hat\psi_\pm(x)\propto\cos(k_cx)$ couples to the field fluctuations $\delta\hat{a}_\pm$ and vice versa~\cite{Nagy2008Self}. By restricting the atomic condensate fluctuations to $\propto\cos(k_cx)$ excitations, relevant to the photon scattering from the pump laser into the cavity on the onset of the self-ordeing, the excitations of the system in the restricted subspace is obtained via the fourth-order characteristic equation $\text{Det}(\mathbf{M}_B-\omega_{\rm pol}I_{4\times4})=0$,
\begin{align}
\label{eq:characteristic-eq}
\left(\omega_{\rm pol}^2- \frac{1}{\hbar^2}\epsilon_{k_c}^2\right)
\left[(\omega_{\rm pol}+i\kappa)^2-\delta_c^2\right]
+2N\eta_0^2\omega_r\delta_c=0,
\end{align}
where $\epsilon_{k_c}=\sqrt{\hbar\omega_r(\hbar\omega_r+2g_0n)}$ is the energy of the condensate's phonon mode with momentum corresponding to the cavity wave number $k_c$. The threshold $\eta_{0c}$ is the solution of the characteristic equation~\eqref{eq:characteristic-eq} with zero polarition frequency, $\omega_{\rm pol}=0$.

The solution of the characteristic equation~\eqref{eq:characteristic-eq} yields the low-energy polariton spectrum $ \omega_{\rm pol}(\eta_0)$ below the threshold $\eta_{0c}$.
For the case of $\epsilon_{k_c}^2\ll \hbar^2(\kappa^2+\delta_c^2)$, we can obtain simple analytical expressions. In particular, in the region $\eta_0<\eta_{0s}$, the real part of the lowest-lying polariton frequency can be approximated as
    \begin{align}
    \label{eq:softening_analytical_real}
    \text{Re}(\omega_{\rm pol})\simeq  \frac{\epsilon_{k_c}}{\hbar}\sqrt{1-\frac{\eta_0^2}{\eta_{0s}^2}},
    \end{align}
and the imaginary part can be approximated as
\begin{align}
      \label{eq:imaginary_analytical_noncritical}
    \text{Im}(\omega_{\rm pol})\simeq 
    -\frac{\kappa\epsilon_{k_c}^2}{\hbar^2(\delta_c^2+\kappa^2)}\frac{\eta_0^2}{\eta_{0c}^2}.
\end{align}
On the other hand, in the vicinity of the critical point where $\eta_0/\eta_{0c}\simeq 1$, the real part is zero and the imaginary part can be approximately written as
\begin{align}
      \label{eq:softening_analytical}
    \text{Im}(\omega_{\rm pol})\simeq 
    -\frac{\delta_c^2+\kappa^2}{2\kappa}\left(1-\frac{\eta_0^2}{\eta_{0c}^2}\right).
\end{align}
The range of pump strength $\eta_{0s}<\eta_0<\eta_{0c}$ where the critical polariton is purely damped, i.e., $\text{Re}(\omega_{\rm pol})=0$ and $\text{Im}(\omega_{\rm pol})<0$, is given by,
  \begin{align}
    \label{eq:overdamped_range}
    \eta_{0c}^2-\eta_{0s}^2=\eta_{0c}^2\left[\frac{\kappa\epsilon_{k_c}}{\hbar(\delta_c^2+\kappa^2)}\right]^2,
  \end{align}
  which is much smaller than $\eta_{0c}$ given the assumption $\hbar^2(\kappa^2+\delta_c^2)\gg \epsilon_{k_c}^2$.
If instead $\kappa,\delta_c\sim \epsilon_{k_c}/\hbar$, the overdamped range extends and occupies an appreciable fraction of $\eta_{0c}$ [note that the analytical expression~\eqref{eq:overdamped_range} is no longer valid in this regime].

By entering the superradiant phase with $\alpha\neq 0$, the system spontaneously breaks the $\mathbf{Z}_2$ parity symmetry of the Hamiltonian, Equation~\eqref{eq:H_eff_1comp}. This implies fixing the phase of the cavity field relative to the pump laser and correspondingly $\lambda_c$-periodic density ordering in either even or odd sites of the emergent optical lattice [see also experimental Figure~\ref{fig:DickePT}(f)]. 
The resultant $\lambda_c$-periodic density pattern optimizes indeed the constructive light scattering from the transverse pump laser into the cavity mode.

%----------------------------------------------------------------------------------------------------------------------------
\subsubsection{Criticality of the self-ordering phase transition}
\label{subsubsec:criticality-self-ordering}

From Figure~\ref{fig:poles_fermiVSbose} one can see that the critical behavior across the superradiant phase transition is the same for both bosons and fermions, and is characterized by the linear vanishing of the polariton damping as a function of the effective coupling strength $\eta_0$, as seen by linearizing Equation~(\ref{eq:softening_analytical}). In this simplest scenario for superradiance featuring a single cavity mode, the critical behaviour coincides with the one of a dissipative version of the Dicke model~\cite{Nagy2010Dicke,Garraway2011The,Kirton2018Introduction}, discussed in Section~\ref{subsubsec:DickeModel}. The static universality class is of the mean-field, classical Ising type. It is always at finite temperature due to cavity losses (the same holds in the absence of photon losses but at finite temperature of the atomic cloud). The dynamical universality class is a mean-field version of Model A in the Hohenberg-Halperin classification, since there is no conserved quantity~\cite{DallaTorre2013Keldysh}. The mean-field nature of the transition is generic to cases where the atoms are coupled to a finite (in the thermodynamic limit) number of cavity modes, resulting in a zero-dimensional effective theory for the order parameter~\cite{Piazza2013Bose}. In equivalent terms, the cavity-mediated atom-atom interactions in these cases are of infinite-range nature, so that the interacting models are fully connected.

Cavity losses render the superradiant phase transition effectively thermal. As discussed in Section~\ref{subsec:thermal_nonthermal}, our considered driven-dissipative systems are in general not in a global thermal-equilibrium state. However, the degree of freedom corresponding to the Dicke-type order parameter follows a critical dynamics in presence of an effective classical, thermal bath~\cite{DallaTorre2013Keldysh}. The corresponding temperature is thus always finite in the open system and depends on the character of the soft polaritonic mode which becomes undamped at the transition. In the regime where the cavity dynamics is much faster than the atomic one, $\kappa\gg \omega_r$ (experiments ETH1, ETH2, Stan, and T\"{u}b1 in Table~\ref{tab:ExperimentalParameters}), the soft polariton corresponds to an almost atomic excitation. This picture is supported by the measurement of the structure factor in the vicinity of the transition (see Figure~\ref{fig:DynamicStructureFactor}), where the occupation of the soft polariton is consistent with the temperature of the atomic gas ($\sim$kHz) before coupling to the cavity, rather than with the loss rate ($\sim$MHz). 
It is important to note, however, that this scenario neglects the effect of cavity-induced relaxation (the so-called cavity heating or cooling) of the atomic cloud, whereby the on-shell scattering processes involving cavity photons redistribute the atoms. This issue has been discussed in Section~\ref{subsec:thermal_nonthermal}.

We described here the superradiant instability in terms of a polariton mode becoming unstable. The atomic component of this polariton involves a single momentum corresponding to the cavity wave number $k_c$. This finite-momentum instability is analogous to a roton mode softening. In the standard situation of a closed equilibrium system, the softening is associated with an excitation energy approaching zero (the so-called energetic instability). In the present driven-dissipative case, however, the order-parameter degree of freedom is in contact with a classical thermal bath, as discussed above. What matters is thus not the excitation energy but its damping, which approaches zero and crosses it at the unstable point,  becoming positive. As a consequence, across the transition the mode grows exponentially in time, leading to a so-called dynamical instability. The critical regime of the self-organization phase transition has been experimentally explored; see Sections~\ref{subsubsec:BraggSpectroscopy} and \ref{subsubsec:fluctuations}. 

The non-equilibrium---or more precisely the driven---nature of this open atom-cavity system plays an essential role in the possibility of the experimental realization of the Dicke superradiant phase transition. The dispersively driven two-photon transitions between two momentum states lead to the ``paramagnetic-like'' term  proportional to $\hat{a}^\dag+\hat{a}$ in Equation~(\ref{eq:H_eff_1comp}) [and correspondingly in Equation~(\ref{eq:H_eff_1comp-p-LE})], thereby moving the superradiant transition from the ultrastrong coupling regime (as in the original Dicke model) to the strong coupling regime. 
That is, in the original Dicke model the strength of the atom-field coupling (i.e., paramagnetic term) has to be comparable to the bare cavity and atomic-transition frequencies, which lie in the optical domain of several hundred THz. In contrast, in the dispersively driven atom-cavity systems the strength of the atom-field coupling needs to be comparable to the cavity detuning/loss rate and the recoil frequency, i.e., to be in the several kHz to MHz regime. The validity of Dicke-type models and the fundamental possibility of the superradiant transition in the ultrastrong coupling regime is still a matter of active debate both in atomic and solid-state physics~\cite{Frisk2019Ultrastrong,Andolina2019Cavity}.

\subsubsection{Observation of the self-ordering phase transition} \label{subsubsec:SuperradiantPhaseTransition}
The first observation of the superradiant self-organization phase transition of a quantum gas made use of an almost pure BEC, such that $k_B T \ll \hbar \omega_r$~\cite{Baumann2010Dicke}. The relevant momentum states can thus be clearly resolved such that the low-energy description and the corresponding mapping to the Dicke model is applicable, as discussed in Section~\ref{subsubsec:DickeModel}. Therefore, the threshold for the superradiant self-ordering phase transition is set by the competition between the atomic quantum-kinetic energy, i.e., the energetic cost of spatially modulating the wave function, and potential energies. That is, the self-ordering is favored when the (negative) potential energy gain outweighs the kinetic energy penalty due to the density crystallization. This is different from the self-organization of thermal classical atoms~\cite{Black2003Observation}, where the phase transition from the normal to the self-organized phase is determined by a competition of thermal fluctuations and potential energy.

In typical experiments considered in this section, an atomic BEC is confined by an external trap to the  center of a single cavity mode, and illuminated transversally by a far red-detuned standing-wave pump field~\cite{Baumann2010Dicke}; see also Figure~\ref{fig:coupling_scheme}. The atomic detuning $\Delta_a$ between the pump field and the atomic resonance frequency is on the order of THz, such that the atoms are not electronically excited but only act as a dispersive medium scattering photons from the pump field into the cavity mode and vice versa. At the same time, the cavity detuning $\Delta_c$ between the pump field and the cavity resonance is comparatively small, on the order of $\sim$10 MHz (where the cavity decay rate $\kappa$ is $\sim$1MHz), such that photons are preferentially scattered via the atoms from the pump field into the cavity mode (and vice versa). Such two-photon scattering processes impart each a recoil kick onto the atoms along the pump and the cavity direction. The bare kinetic energy of this microscopic scattering process is thus $2\hbar \omega_r$. The transverse pump field induces a coupling between the atomic external degree of freedom and the cavity field, which leads to a softening of one of the polariton branches. This softening reduces the energetic cost of the photon scattering processes, and eventually leads to the superradiant instability described in Section~\ref{subsec:SR_instability}.

\begin{figure}[t!]
  \begin{center}
    \includegraphics[width=\columnwidth]{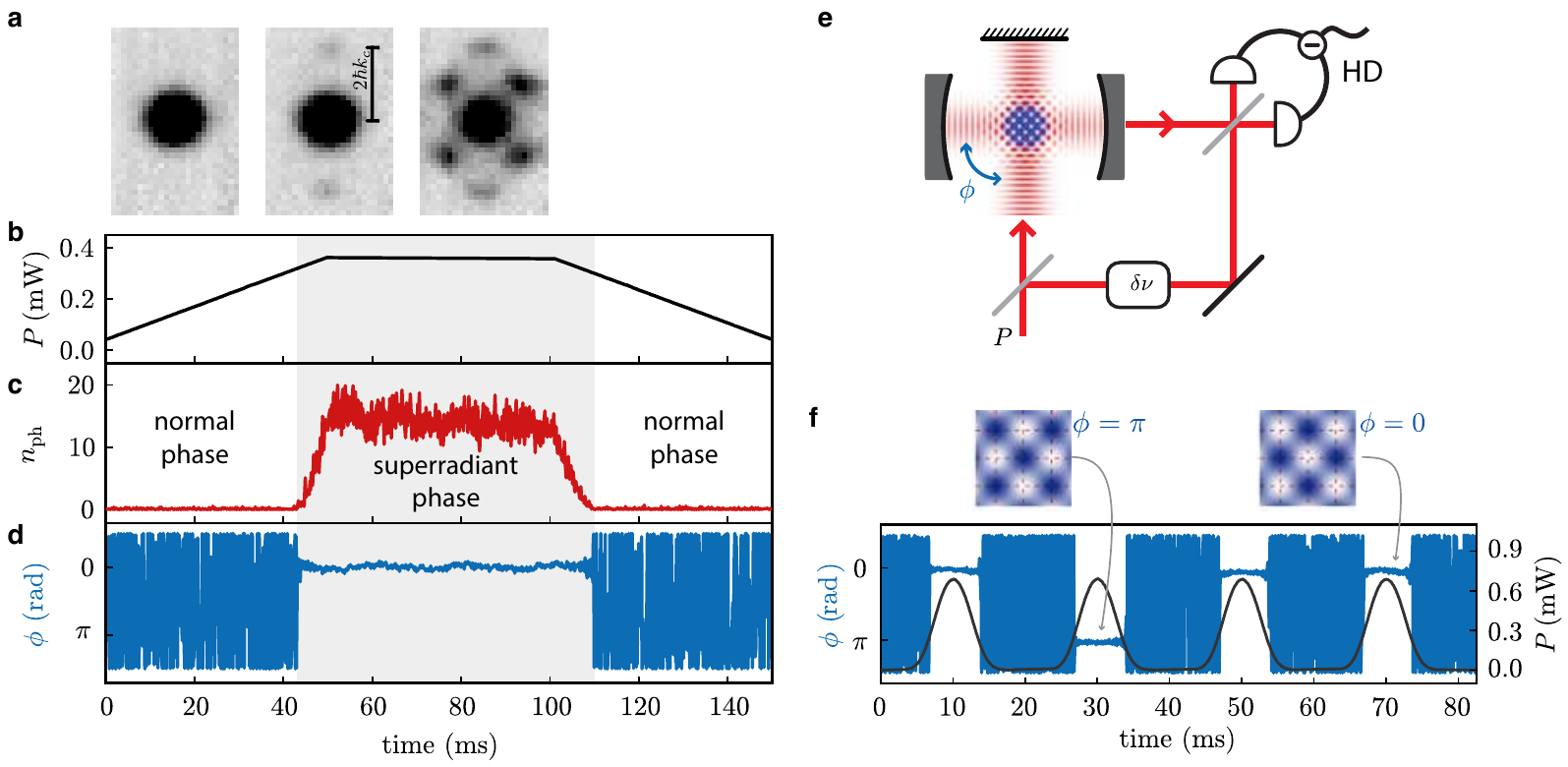}
    \caption{Self-organization of a driven BEC inside a single-mode standing-wave cavity, corresponding to the superradiant phase transition. (b) The transverse pump power $P$ is ramped up linearly. At the same time, the intensity (c) and the phase (d) of the intra-cavity field are measured. Initially, in the normal phase, the cavity field is close to the vacuum state with no well-defined phase. At the critical pump power, the system enters the superradiant phase with a nonzero cavity field possessing a well-defined phase over time. The process is reversible, i.e., by ramping down the pump field the system enters again the normal phase.  (a) Absorption images of the BEC after ballistic expansion reveal the occupation of various momentum states at different stages of the experiment. For increasing pump lattice depth, the peaks at $\pm2\hbar k_c\hat{e}_y$ due to the pump lattice become visible. In the self-organized phase, the four momenta at $\mathbf{k}=(k_x,k_y)=(\pm k_c, \pm k_c)$ become macroscopically populated. (e) The heterodyne detection (HD) scheme used to measure amplitude and phase of the cavity field: a fraction of the pump field is split, frequency shifted, and used as local oscillator. The output of the cavity is interfered with this field and sent to a balanced photo-detector.  (f) Symmetry breaking: If the pump power $P$ is repeatedly ramped across the critical point, the phase of the intra cavity light field takes either 0 or $\pi$, demonstrating the discrete $\mathbf{Z}_2$ parity symmetry breaking at the phase transition. The blue density plots symbolize the two possible atomic density configurations where either the even or the odd sites of the emergent checkerboard lattice are occupied. Figure adapted and reprinted with permission from Ref.~\cite{Baumann2010Dicke} published in 2010 by the Nature Publishing Group and Ref.~\cite{Baumann2011Exploring}~\textcopyright~2011 by the American Physical Society. 
    }
    \label{fig:DickePT}
  \end{center}
\end{figure}

The basic experimental sequence and main observations of this prototypical experiment are summarized in Figure~\ref{fig:DickePT}. The power $P$ of the transverse pump field inducing the coupling ($P\propto \eta_0^2$) is gradually increased. Above a critical pump power, the intra cavity light intensity raises sharply, indicating the superradiant phase transition point. At the same time, a superposition of four relevant momentum states $\mathbf{k}=(\pm k_c, \pm k_c)$ becomes macroscopically populated, as can be observed in absorption images after time of flight expansion of the BEC. In the normal phase, the atomic density is modulated by the transverse pump lattice but homogeneous along the cavity direction, leading to destructive interference of the light scattered off the atoms into the cavity mode. In contrast, in the superradiant phase, the spatial arrangement of the atomic system in the emergent optical lattice corresponds to the formation of a Bragg grating at which the pump field is efficiently scattered into the cavity mode and back. 

A good choice as an order parameter operator for the superradiant phase transition is the overlap integral $\hat{\Theta}$ between the atomic density operator and the checkerboard lattice, introduced in Equation~\eqref{eq:order-paramter-op}. It measures the localization of the atomic density on either the even ($\langle \hat \Theta \rangle > 0$) or odd ($\langle \hat \Theta \rangle < 0$) sublattice; see also Figure~\ref{fig:DickePT}. If the system is mapped to the Dicke model as in Section~\ref{subsubsec:DickeModel}, the expectation value of the order parameter is given by the expectation value of the atomic polarization, $\langle\hat\Theta \rangle = \langle \hat J_+ + \hat J_-\rangle/2$.

In the steady state, $\hat\Theta$ is directly linked to the cavity field operator $\hat a_\mathrm{ss}$ [see Equation~\eqref{eq:ss-a-Theta}], which can equivalently be used as the order parameter. Since the intra-cavity light field can be continuously detected in real time via the field leaking through the cavity mirrors, this is a particularly useful and versatile relation. In addition to monitoring the steady state of the system, this also enables access to fluctuations of the system as we discuss in Section~\ref{subsubsec:fluctuations}. Furthermore, this provides the fundamental tool for many cavity-based non-destructive measurement schemes; see Section~\ref{sec:quantmeasure}.

\subsubsection{Symmetry breaking}
Whether the even or odd sites of the underlying checkerboard lattice are predominantly occupied in the superradiant phase is determined by a spontaneous symmetry breaking process. After the phase transition, the phase of the intra-cavity field is locked with respect to the pump field's phase. Using a heterodyne detection scheme, this relative phase can be measured. It was experimentally confirmed that the phase locks either to 0 or $\pi$ relative to the pump phase; see Figure~\ref{fig:DickePT}(f)~\cite{Baumann2011Exploring}. This observation of two discrete values for the relative cavity-pump phase demonstrates the $\mathbf{Z}_2$ parity symmetry breaking expected from the basic Hamiltonian (\ref{eq:H_eff_1comp}), and corresponds to the self-organization of the atomic density with maxima at either the even or the odd sites of the emerging checkerboard lattice; see also Section~\ref{subsec:SR_instability}. An \emph{in-situ} observation of the atomic density distribution has so far not been possible. 

The symmetry at the phase transition is broken by a symmetry-breaking field rather than by a spontaneous process---as probably in all realistic experiments. The breaking is most likely caused by the finite size of the initially homogeneous atomic cloud having a slightly different overlap with either the even or the odd sites of the emergent optical lattice. This overlap integral depends $\lambda_c$-periodically on the relative position between the center of the atomic trap and the checkerboard lattice structure. The nonzero overlap difference then inhibits the complete destructive interference in the normal phase and a small coherent field scattered into the cavity acts as a symmetry breaking light field, such that the order parameter is renormalized by an additive constant. Measuring the distribution of the phase of the intra-cavity light field across the phase transition repeatedly allows to analyze the symmetry-breaking field. It is important to note that an equal distribution of the measurement outcomes does not prove the absence of symmetry-breaking fields, since it can as well be caused by technical fluctuations of a present symmetry-breaking field. In Ref.~\cite{Baumann2011Exploring}, a maximal symmetry-breaking field corresponding to $\sim$40 atoms distributed in excess on either the even or odd sites was estimated. When the phase-transition boundary was slowly crossed repeatedly, the symmetry was found to be broken and indeed one of the two possible measurement results was strongly favored. The symmetry breaking was also studied as a function of the quench time during which the phase-transition boundary was passed. For rapid quenches, the phase-transition boundary was crossed non-adiabatically and (thermal or quantum) fluctuations of the order parameter were frozen out analogous to the Kibble-Zurek mechanism~\cite{Zurek2005Dynamics}, such that a more equal distribution of the measurement results was found.

\begin{figure}[t!]
  \begin{center}
    \includegraphics[width=0.7\columnwidth]{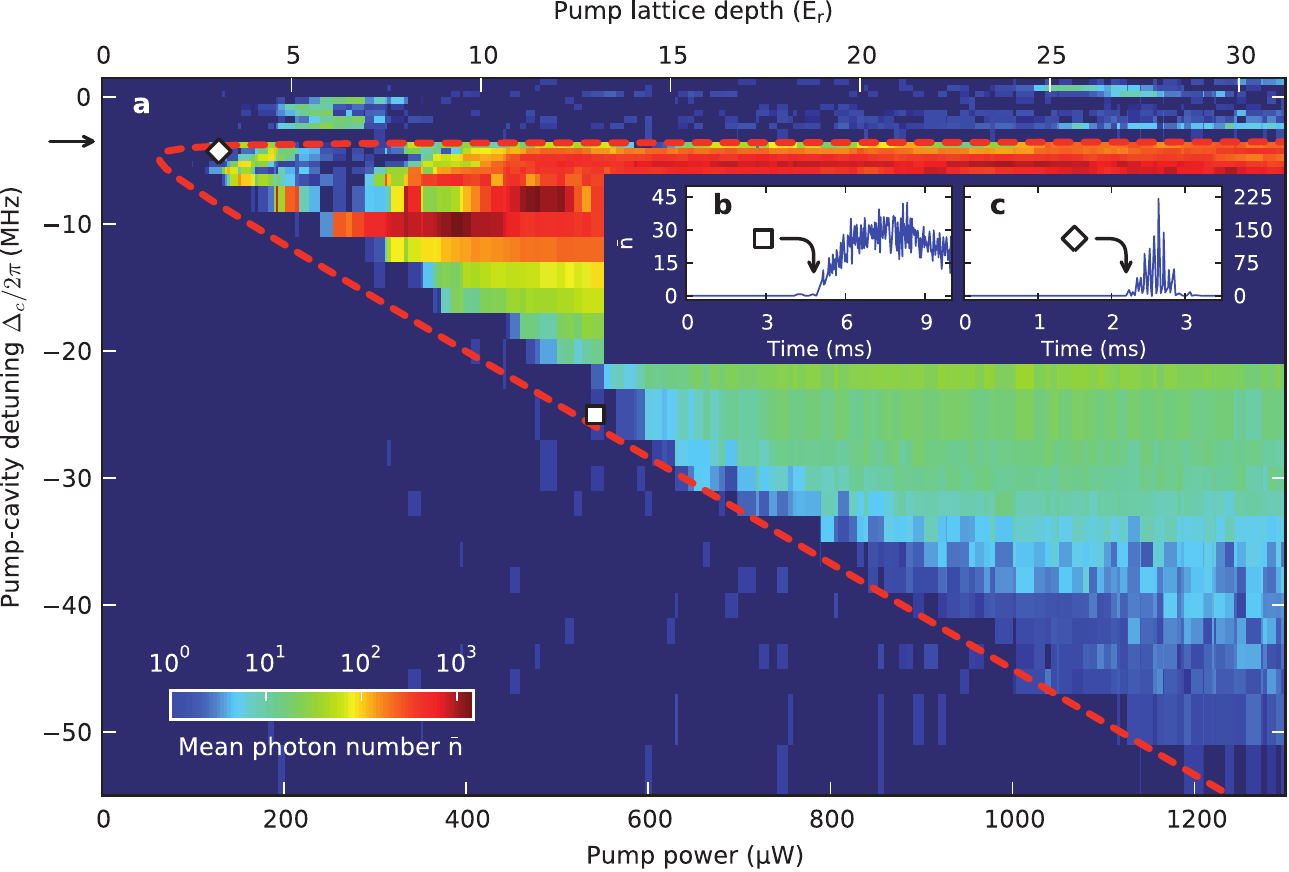}
    \caption{The non-equilibrium phase diagram of a transversely driven BEC dispersively coupled to a single standing-wave mode of a linear cavity. (a) The intra-cavity photon number is shown as a function of the pump-cavity detuning $\Delta_c$ and the pump power. A sharp phase boundary separating the normal and the Dicke-superradiant states is visible, which shifts for increasing absolute values of the detuning to increasing pump powers. The dashed curve is the mean-field prediction for the pump threshold, Equation~\eqref{eq:eta-critical_bose_hom}, reproducing the experimental phase boundary. The insets (b) and (c) show time traces of the intra-cavity photon number when entering the self-organized phase. For detunings close to the cavity resonance, an oscillatory behavior is observed since the dynamic dispersive shift of the cavity pushes the detuning $\delta_c$ to zero which interrupts the positive feedback driving the self-organization. Figure reprinted with permission from Ref.~\cite{Baumann2010Dicke}, published in 2010 by the Nature Publishing Group.}
    \label{fig:DickePhaseDiagram}
  \end{center}
\end{figure}

\subsubsection{Phase diagram} 
 The superradiant phase in this driven system is a stable steady state with lifetimes on the order of several 100~ms, limited by either atom loss or heating \cite{Baumann2010Dicke, Baumann2011Exploring}. This is different from observations of superradiance in free space \cite{Inouye1999Superradiant} or in longitudinally pumped ring cavities \cite{Slama2007Superradiant}, where the superradiant Rayleigh scattering is a transient phenomenon leading to a pulse of light rather than entering a stable quantum phase of matter. By scanning both the pump-cavity detuning $\Delta_c$ and the pump power $P$ ($\propto\eta_0^2$), the steady-state phase diagram of the system can be recorded as shown in Figure~\ref{fig:DickePhaseDiagram}. With increasing absolute values of the pump-cavity detuning $\Delta_c$, the critical point to enter the superradiant phase shifts to higher values of the pump power. The mean-field prediction for the phase boundary, Equation~\eqref{eq:eta-critical_bose_hom}, is in good agreement with the experimental data. The horizontal phase boundary at small detunings is dispersively shifted by $-\int U(\mathbf{r})\hat{n}(\mathbf{r})d\mathbf{r}$ with respect to the bare cavity resonance. This shift is a dynamic quantity, depending on the overlap between the cavity mode and the atomic density distribution, giving rise in addition to nonlinear effects in the system. For detunings close to the cavity resonance, the system can enter into an unstable regime, where the optomechanical interaction pushes the dispersively shifted cavity resonance $\delta_c$ to zero and the positive feedback that drives self-organization is interrupted. As a consequence, the intra-cavity field shows oscillations; see Figure~\ref{fig:DickePhaseDiagram}(c).

\subsubsection{Quench dynamics across the non-equilibrium phase transition}
The experiments considered so far operated in the regime where the time scale $1/\kappa$ for the cavity-field dissipation is much faster than the atomic motion, $\kappa \gg \omega_r$. In this regime, the intra-cavity field instantaneously follows the atomic dynamics and can thus be adiabatically eliminated. In contrast, if the cavity has a very narrow bandwidth, quench experiments allow access to the non-adiabatic regime, where both photonic and atomic fields are out of equilibrium and cannot be slaved to each other. Therefore, some of the features predicted for the non-equilibrium phase transition in Dicke-type models~\cite{Keeling2010Collective, Bhaseen2012Dynamics} can be observed. 

\begin{figure}[t!]
  \begin{center}
    \includegraphics[width=0.8\columnwidth]{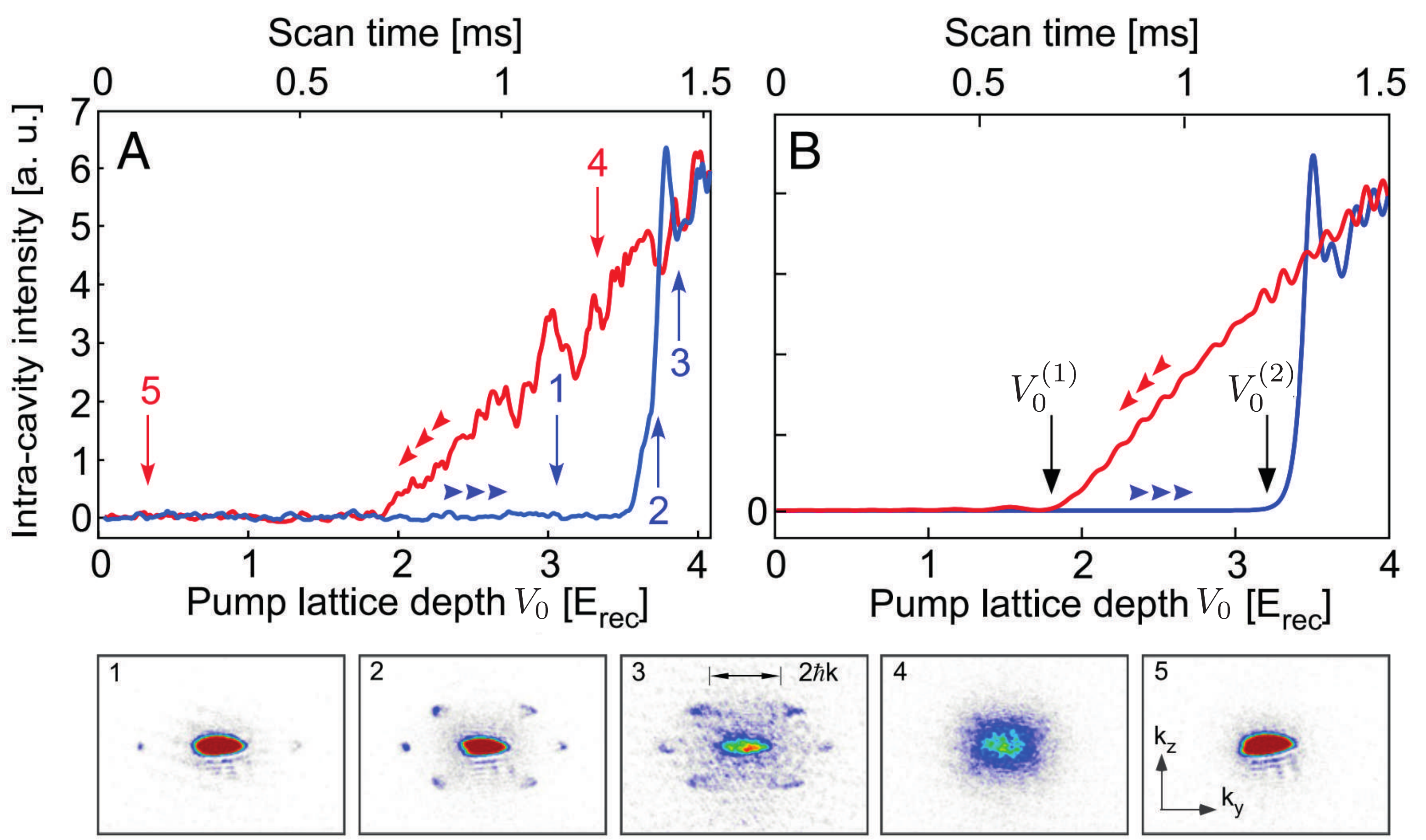}
    \caption{Quench dynamics accross the non-equilibrium Dicke-superradiant phase transition.
    (A) The intra-cavity intensity is shown for a quench of the pump lattice from 0 to $4E_\mathrm{rec}$ during $\SI{1.5}{\milli \second}$ (blue) and back again during $\SI{1.5}{\milli \second}$ (red). A clear hystersis loop is visible. The absorption images after time-of-flight expansion of the BEC shown below (A) were recorded at increasing times, indicated by the labels (1-5). During the jump of the intra-cavity field, higher atomic momentum states are occupied, then coherence is lost and finally re-gained within the experimental protocol. (B) A mean-field calculation of the quench dynamics for an infinite system without collisional interactions reproduces the observed dynamics. Figure adapted and reprinted with permission from Ref.~\cite{Klinder2015Dynamical}, published in 2015 by the United States National Academy of Science.}
    \label{fig:QuenchDynamics}
  \end{center}
\end{figure}

This sideband resolved regime has been experimentally realized \cite{Klinder2015Dynamical} with $\kappa=2\pi\times \SI{4.45}{\kilo \hertz} < 2 \omega_r = 2\pi \times \SI{7.1}{\kilo \hertz}$ (for detailed experimental parameters, see Table~\ref{tab:ExperimentalParameters}, Hamb. Also the experimental setup T\"{u}b2 operates in this regime). For $\Delta_c<0$, the resulting phase diagram is very similar to the phase diagram (Figure~\ref{fig:DickePhaseDiagram}) of the previously discussed system \cite{Baumann2010Dicke}. However, new features appear in the phase diagram for positive pump-cavity detunings, $\Delta_c>0$, as predicted by Refs.~\cite{Keeling2010Collective, Bhaseen2012Dynamics}. Namely, in this region of the phase diagram short superradiant pulses were emitted into the cavity, during which the atoms were irreversibly scattered into higher momentum states.

Features of non-adiabaticity are clearly visible in quench experiments shown in Figure~\ref{fig:QuenchDynamics}. Ramping from the normal phase into the superradiant phase and back each on a time scale of $\SI{1.5}{\milli \second}$, a sudden jump in the intra-cavity light field was observed when entering the superradiant phase, followed by a smooth decrease of the light field when exiting the superradiant phase. This hysteresis loop encloses an area that was found to scale with the quench time according to power laws, suggesting an interpretation in terms of the model discussed by Kibble and Zurek~\cite{Kibble1976Topology, Zurek2005Dynamics, Zurek1996Cosmological}. For larger cavity bandwidths, the hysteresis area is expected to vanish~\cite{Klinder2015Dynamical}.

\subsubsection{Phase diagram of bosons versus fermions}

\begin{figure}[t!]
  \begin{center}
    \includegraphics[width=0.65\columnwidth]{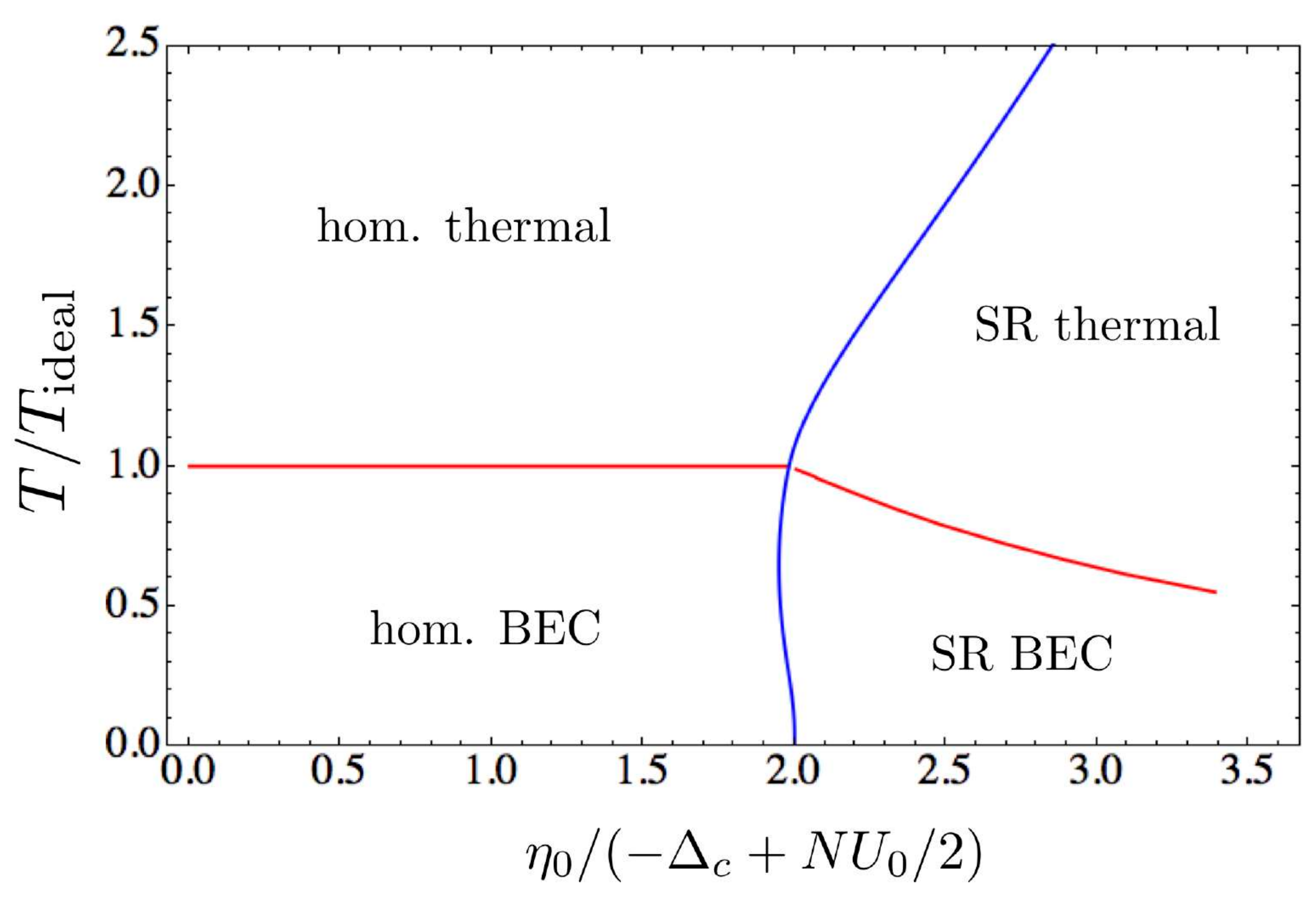}
    \caption{Phase diagram as a function of temperature vs.\ the driving pump strength for a three-dimensional, non-interacting cloud of driven bosonic atoms coupled dispersively to a single standing-wave mode of an optical cavity. A superradiant (SR) phase can exist both for thermal and Bose-condensed  atoms. There is an optimal temperature for which the pump-strength threshold for the superradiant phase transition (the blue curve) is the lowest. Adapted and reprinted with permission from Ref.~\cite{Piazza2013Bose}, published in 2013 by the Elsevier.}
    \label{fig:phasediagram_singlecav_BECtoThermal}
  \end{center}
\end{figure}

While the criticality of the superradiant phase transition in bosonic atoms is captured by the effective low-energy Dicke-type model of Equation~(\ref{eq:H_eff_1comp-p-LE}), some important qualitative effects, even in the simplest geometry considered so far where the driven atoms couple dispersively to a single cavity mode, cannot be reproduced by the low-energy Dicke Hamiltonian~\eqref{eq:H_eff_1comp-p-LE}. For instance, the superradiant threshold as a function of temperature, obtained from the full model of Equation~\eqref{eq:H_eff_1comp} for non-interacting bosons and shown in Figure~\ref{fig:phasediagram_singlecav_BECtoThermal}, behaves differently from the one of the Dicke model. The smallest pump-strength threshold appears at an optimal, finite temperature of the atomic cloud, while the Dicke Hamiltonian~\eqref{eq:H_eff_1comp-p-LE} has the lowest pump-strength threshold at zero temperature. This behavior results from the fact that the polarization function $\chi_{\rm dyn}$, Equation~(\ref{eq:dyn_polarization_homogeneous}), of the atomic medium at finite photon momentum initially increases if atomic momentum states in the direction perpendicular to the photon momentum become thermally occupied, simply because more states then participate in the scattering. This, however, holds only until the temperature broadening of the atomic distribution becomes an appreciable fraction of the cavity-photon momentum, which makes the numerator of the polarization function $\chi_{\rm dyn}$ smaller as longitudinal momenta of the order of the photon momentum become occupied.

The self-ordering of bosonic atoms can exhibit other interesting features, which are not shown in Figure~\ref{fig:phasediagram_singlecav_BECtoThermal}. For instance, it has been predicted that deep enough in the superradiant phase the BEC can enter a fragmented phase~\cite{Lode2017Fragmented}, where multiple atomic modes are macroscopically occupied but with no coherence among one another. The effect of such beyond-mean-field correlations is even stronger if an external (i.e., not cavity-induced), strong optical lattice is imposed in the system; see Section~\ref{sec:extended-BH}.

\begin{figure}[t!]
  \begin{center}
    \includegraphics[width=0.75\columnwidth]{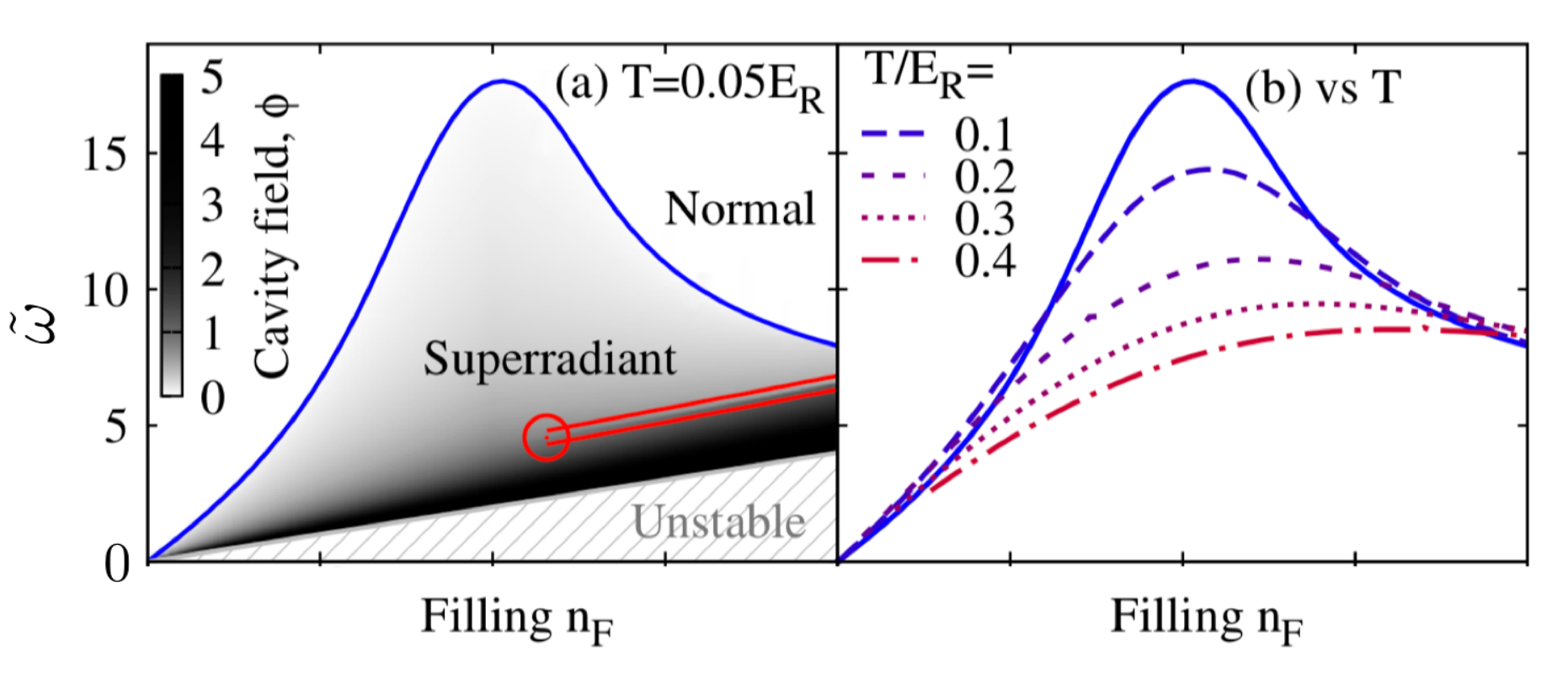}
    \caption{The superradiant phase transition in a two-dimensional, non-interacting cloud of driven fermionic atoms coupled dispersively to a single standing-wave mode of an optical cavity. (a) Phase diagram for a fixed pump intensity as a function of the dimensionless cavity-pump detuning  $\tilde{\omega}=-4\Delta_c/U_0N_l$ vs.\ the lattice filling factor $N/N_l$. The superradiant phase is characterized by a nonzero cavity field. The nesting  condition is fulfilled around the unit filling, visible as a peak around this filling. The empty circle in (a) is a critical point where the first order liquid-gas-like transition terminates. Panel (b) shows the effect of temperature, where the nesting effect is washed out by increasing temperature. Adapted and reprinted with permission from Ref.~\cite{Keeling2014Fermionic} \textcopyright~2014 by the American Physical Society.
    }
    \label{fig:phasediagram_singlecav_Fermi_Keeling}
  \end{center}
\end{figure}

Even more qualitative differences between the prediction of the full Hamiltonian~\eqref{eq:H_eff_1comp} and the Dicke model~\eqref{eq:H_eff_1comp-p-LE} appear in the self-ordering of fermionic atoms. Besides studies concerning optomechanical effects in Fermi gases inside cavities~\cite{Kanamoto2010Optomechanics,Padhi2013Cavity}, the self-organization of fermionic atoms has been first studied theoretically in Refs.~\cite{Keeling2014Fermionic, Piazza2014Umklapp, Chen2014Superradiance}, while the first experimental setup has recently become available~\cite{Roux2020Strongly, Roux2021Cavity}. The phase diagram of a driven 2D Fermi-gas coupled dispersively to a single standing-wave mode of a linear optical cavity as a function of the cavity-pump detuning versus the filling is shown in Figure~\ref{fig:phasediagram_singlecav_Fermi_Keeling}. 

One important additional feature for the fermionic atoms during the self-ordering process is the possibility of nesting between the photon momentum and the Fermi surface of the atoms, allowing for photon-induced Umklapp scattering processes transferring a fermion from one side of the Fermi surface to the other side~\cite{Chen2014Superradiance}. Nesting enhances the tendency towards the formation of a density modulation and thus superradiance. This is visible as a peak around unit filling (or equivalently $k_c=2k_F$) in the phase diagram shown in Figure~\ref{fig:phasediagram_singlecav_Fermi_Keeling}(a). The effect of nesting is more dramatic in one dimension~\cite{Piazza2014Umklapp}, where the pump-strength threshold for the self-ordering phase transition vanishes when the photon momentum is twice the Fermi momentum. 
This phenomenon is known in the condensed-matter context as the Peierls transition in electron-phonon models \cite{PeierlsQuantum1996}. By forming a density wave, a gap is opened in the Fermi surface so that the atoms in the superradiant cavity lattice form an insulator. Sufficiently away from the unit filling, the atoms are in a metallic state \cite{Piazza2014Umklapp, Fraser2019Topological,Rylands2020Photon}. Nesting effects require a sharp Fermi surface and are thus washed out with increasing temperature of the atoms, as illustrated in Figure~\ref{fig:phasediagram_singlecav_Fermi_Keeling}(b). This is especially true in one dimension, where the Peierls instability immediately disappears and the superradiant threshold is moved to a finite atom-photon coupling strength. The role of disorder in the self-organization of fermions has been also investigated in Ref.~\cite{Muller2012Quantum}.

 \subsection{Excitation spectra and fluctuations}\label{subsec:excitation_measurement}

In Section~\ref{subsec:excitations_probing} we developed the theory for obtaining collective excitations of atom-cavity systems in a general setting. Using this formalism, in Section~\ref{subsec:SR_instability} we obtained the low-energy polaritons of driven bosonic atoms coupled dispersively to a single standing wave of a linear cavity. In this section, we provide a summary of the experimental works that measured polaritons and related fluctuation spectra across the self-ordering phase transition in bosnic atoms. 

In the experimental situation of a finite-size atomic BEC with short-range collisional interactions and a transverse pump-lattice potential, the energy of the excitations have to be rescaled~\cite{Mottl2012Roton}. Contact interactions and the optical potentials affect the bare energy of the excitation mode already without cavity-mediated global interactions. The excited single-particle mode is replaced by a Bogoliubov mode, and the coupling between the ground-state wave function and that mode is rescaled by its spatial overlap integral weighted with the mode structure of the long-range interaction potential. In the normal phase, the excited Bogoliubov mode with momentum (respectively quasi-momentum along the transverse pump lattice) $(\pm k_c, \pm k_c)$ lies in the lowest-energy band of the transverse pump lattice potential, and its energy is shifted with respect to the bare kinetic energy by the mean-field shift. In the superradiant phase, the emergent checkerboard lattice halves the Brillouin zone (which is defined initially only along the pump direction, but now along both the pump and cavity directions), and the most strongly coupled excited Bogoliubov mode at $(\pm k_c, \pm k_c)$ lies in a higher-energy band at zero quasi-momentum of the optical checkerboard potential~\cite{Mottl2012Roton}. A nonzero energy gap between the ground state and the first excited state is expected to remain even at the critical point due to the finite size of the system.

\subsubsection{Cavity-enhanced Bragg spectroscopy}\label{subsubsec:BraggSpectroscopy}

\begin{figure}[t!]
  \begin{center}
    \includegraphics[width=0.65\columnwidth]{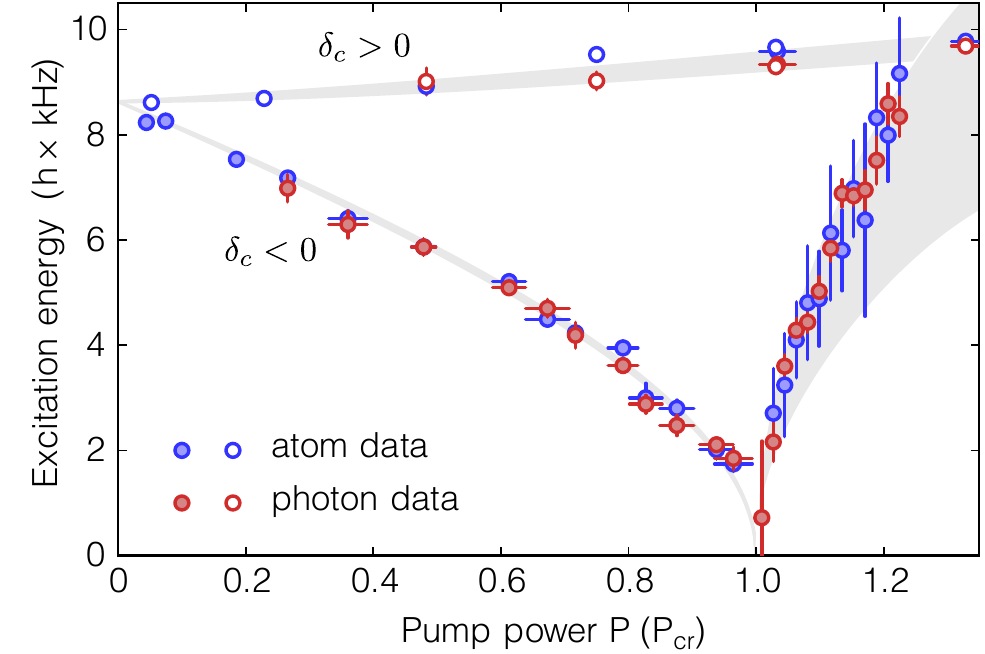}
    \caption{Measurement of the lowest excitation energy as a function of the transverse pump power across the self-ordering phase transition of a driven BEC inside a single-mode linear cavity. The excitation spectrum was measured via either the atomic population of the excited momentum state (blue), or via the light field leaking from the cavity (red). The observed mode softening (filled circles) for effective attractive cavity-mediated atom-atom interactions (corresponding to $\delta_c<0$) culminates around the critical pump power $P_{\rm cr}$. For the opposite blue detuning between pump and cavity ($\delta_c>0$), the excitation mode hardens instead since fluctuations are suppressed (open circles). Figure adapted and reprinted with permission from Ref.~\cite{Mottl2012Roton} published in 2012 by the American Association for the Advancement of Science.}
    \label{fig:RotonSoftening}
  \end{center}
\end{figure}

Experimental access to the polaritonic excitation spectrum can be gained by performing a variant of Bragg spectroscopy~\cite{Stenger1999Bragg, Steinhauer2002Excitation}. The system is prepared at a transverse pump power, corresponding to a certain coupling strength, and probed by a weak light pulse of amplitude $\sqrt{n_{\rm pr}}$ injected directly into the cavity mode~\cite{Mottl2012Roton}. The frequency of the probe light pulse has an adjustable difference $\delta_\mathrm{pr}$ relative to the transverse pump frequency, such that the interference between the two gives rise to a time-dependent amplitude-modulated checkerboard lattice potential $\hbar \sqrt{n_\mathrm{pr}} \eta(\mathbf{r}) \cos(\delta_\mathrm{pr} t +\varphi)$ acting on the atoms, where $\varphi$ is the relative phase between the two fields. If this perturbation is resonant with a collective excitation of the system, stimulated scattering of probe photons into the pump and vice versa will take place, leading to the occupation of the excited polariton. Subsequent to the probe pulse, all potentials are switched off, projecting the created atomic excitations onto the free-space momentum states $(\pm k_c, \pm k_c)$, which can be detected in absorption imaging after time-of-flight expansion of the BEC. Instead of measuring the atomic population,  the light field leaking from the cavity during the probe pulse can also be analyzed. If the atomic excited momentum state is populated by the perturbing potential, pump photons will be scattered into the cavity mode, leading to an enhanced cavity output field that oscillates at $\delta_\mathrm{pr}$. The amplitude of these oscillations is maximal on resonance with a collective excitation of the system. The resulting excitation spectrum is displayed in Figure~\ref{fig:RotonSoftening}.

A clear softening of the lowest-energy polariton mode is observed, consistent with the square-root vanishing of the real part of the polariton frequency [see Equation~(\ref{eq:softening_analytical_real})]. Note that the experimental parameters here do not allow one to resolve the overdamped critical regime $\eta_{0s}<\eta_0<\eta_{0c}$ [see Equation~(\ref{eq:overdamped_range})], where the polariton oscillation frequency (i.e., the real part of the polariton frequency) is zero but the damping (i.e., the imaginary part of the polariton frequency) is finite. We will return to the polariton damping in the next section. We note that if such a measurement would be performed in the Hamburg setup where $\kappa\sim\omega_r$ \cite{Klinder2015Dynamical}, the overdamped critical regime should become accessible experimentally.

\subsubsection{Measurement of fluctuation spectra and dynamic structure factor}\label{subsubsec:fluctuations}

The atom-photon coupling $\eta(\mathbf{r})$ induces checkerboard density correlations in the atomic gas with a $\lambda_c$ spatial periodicity in the $x$-$y$ plane. These correlated fluctuations, which correspond to the atomic component of the lowest polariton mode, are already present in the normal phase below the self-ordering critical point and are the precursors of the mean density modulation emerging in the superradiant phase. In a microscopic picture, these fluctuations correspond to the creation and annihilation of correlated atoms in an excited Bogoliubov mode. They are induced by photons being scattered by the atoms from the pump into the cavity mode and vice versa. These photon scatterings impart momentum kicks onto the atoms, and accordingly lead to a frequency shift of the involved photons, observable in the power spectral density $\mathcal{S}_{aa}(\omega)$ of the light field leaking from the cavity. When an atom is scattered from the BEC into the excited momentum state via a photon scattering process, the involved photon will be red shifted as it deposits energy in the system. For the opposite process of annihilating a mechanical excitation of the atomic gas, the photon is accordingly blue shifted. The energy shift depends on the distance to the critical point and is given by the real part of the lowest polariton energy $\hbar \omega_{\rm pol}$ [see Equation~(\ref{eq:softening_analytical_real})].

The density correlations are directly connected to the dynamic structure factor $S(\mathbf{k},\omega)$ of the atomic system, which is the spatial and temporal Fourier transform of the density-density correlations. The experimental setting allows direct access to this observable. While the cavity-enhanced Bragg spectroscopy described above is based on the stimulated scattering of photons from one coherent field to another, the measurement of fluctuations requires the analysis of spontaneously and inelastically scattered photons. With respect to free space, the probability to detect spontaneously scattered photons leaking from the cavity mode is increased by several orders of magnitude due to the enhanced vacuum field in the resonator. The density of states for scattering photons has to be high but flat in the energy range of interest, which requires the cavity to operate in the bad cavity limit with respect to the energy of the mechanical excitation, $\kappa \gg \omega_r$.

\begin{figure}[t!]
\centering
\includegraphics[width=0.6\columnwidth]{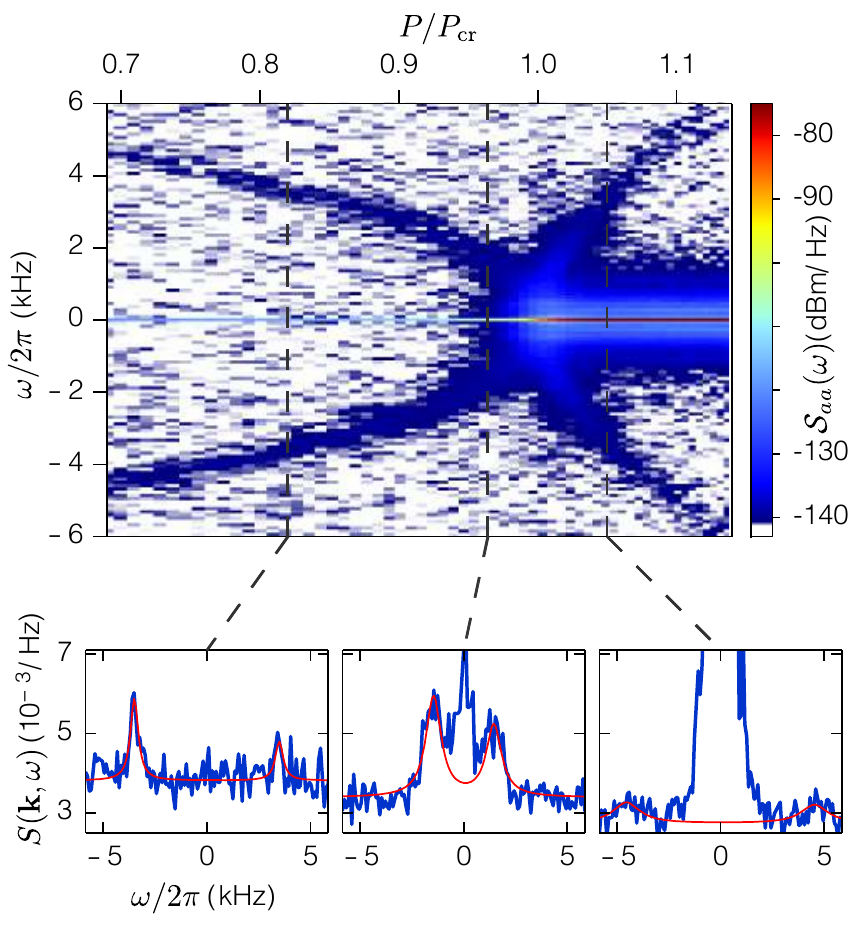}
\caption{Measured power spectral density of the cavity field across the self-ordering phase transition of a driven BEC inside a single-mode linear cavity. The power spectral density $\mathcal{S}_{aa}(\omega)$ of the light leaking from the cavity is shown as function of transverse pump power $P/P_{\rm cr}$ and frequency difference $\omega$ with respect to the transverse pump frequency. The two sidebands originate from the density fluctuations of the atomic gas, while the signal at $\omega=0$ stems from elastic scattering of pump photons at a density modulation. The cuts below the main figure display the extracted dynamic structure factor $S(\mathbf{k},\omega)$ at the wave vector given by $\mathbf{k}=(\pm k_c, \pm k_c)$. Figure adapted and reprinted with permission from Ref.~\cite{Landig2015Measuring} published in 2015 by the Nature Publishing Group.}
\label{fig:DynamicStructureFactor}
\end{figure}

A heterodyne detection scheme can be used to obtain a frequency-resolved measurement of the light field leaking from the cavity~\cite{Landig2015Measuring}. Figure~\ref{fig:DynamicStructureFactor} displays a measurement of the power spectral density $\mathcal{S}_{aa}(\omega)$ of the cavity field together with the extracted dynamic structure factor $S(\mathbf{k},\omega)$ as a function of the relative transverse pump power $P/P_{\rm cr}$ and the frequency $\omega$ relative to the frequency $\omega_p$ of the transverse pump. The geometric setup of pump and cavity fields determines the momentum $\mathbf{k}=(\pm k_c, \pm k_c)$  at which the dynamic structure factor is measured. Power in the frequency bin at $\omega=0$ corresponds to the elastic (Rayleigh) scattering of pump photons at an existing density modulation, while power at nonzero frequencies corresponds to inelastic (Raman) scattering of pump photons creating or annihilating density fluctuations. The signal at $\omega=0$ rises by 5 orders of magnitude, which is a measure of the density modulation  increasing dramatically, above the critical point at $P/P_{\rm cr}=1$. The red and blue detuned sidebands are a measure of the density fluctuations. Their frequency is decreasing towards the critical point in accordance with the expected mode-softening measurement of Figure~\ref{fig:RotonSoftening}. 

The width of the sidebands is a measure of the damping of the polaritonic mode. Since this mode has a dominant atomic character, the damping is mainly attributed to the decay of atomic momentum excitations, and the contribution of cavity decay to the damping is negligible (see the discussion at the beginning of Section~\ref{subsubsec:SuperradiantPhaseTransition}).  The behavior of the damping as a function of the atom-photon coupling strength has been investigated theoretically in detail and originates from Beliaev damping of the checkerboard density modulation and from finite temperature effects~\cite{Kulkarni2013Cavity, Konya2014Photonic, Konya2014Damping}. However, in the overdamped critical regime, where the polariton mode frequency is zero, the contribution to damping from Beliaev decay (or other collision-induced decays) also vanishes. This leaves the cavity loss as the only source of damping which cannot anymore be neglected in principle. As noted above, this regime was not accessible in the ETH Zurich experiments, but could be within reach in the Hamburg setup.

The lower panels in Figure~\ref{fig:DynamicStructureFactor} clearly show a sideband asymmetry, where the lower energetic red sideband has a larger amplitude than the blue sideband. Similar to thermometry in trapped ion experiments, the occupation of the relevant quasi-particle mode can be extracted from this data and was found to be on the order of a few quanta. Analyzing the system in terms of thermodynamic quantities further allowed to extract the irreversible entropy production rate of the system across the phase transition~\cite{Brunelli2016Experimental}.

The integrated power in the sidebands is a measure of the strength of the density fluctuations. In accordance with an earlier measurement~\cite{Brennecke2013Real}, the density fluctuations, which are the fluctuations of the order parameter of this phase transition, diverge when approaching the critical point. An analysis of the scaling of the order parameter fluctuations as a function of the distance from the critical point revealed critical exponents of $0.7(1)$ and $1.1(1)$ on the normal and the superradiant side, respectively. As discussed in Section~\ref{subsec:SR_instability}, the universality class of this mean-field phase transition predicts---due to the cavity losses---a critical exponent of 1 for the order parameter fluctuations~\cite{Nagy2011Critical, DallaTorre2013Keldysh, Piazza2013Bose, Kulkarni2013Cavity}. Howewver, this can be tuned by implementing non-Markovian baths~\cite{Nagy2015Nonequilibrium} or controlling the phase transition by feedback~\cite{Ivanov2020Feedback}.

\subsection{Variants and extensions}
The experimental schemes introduced in the previous sections have been extended to realize a number of  variants of the self-ordering phase transition breaking discrete symmetries, which we summarize in this section.
 
\subsubsection{Ring cavities}\label{subsubsec:RingCavities}
\begin{figure}[t!]
\centering
\includegraphics[width=1\columnwidth]{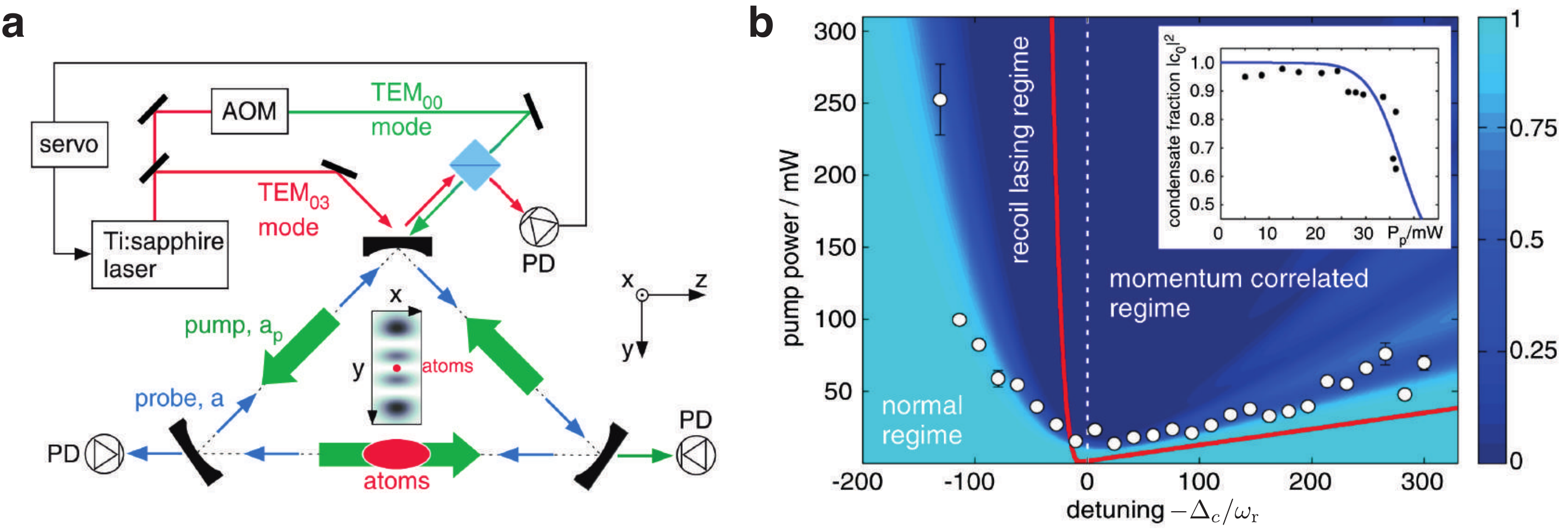}
\caption{Self-organization of a longitudinally driven BEC in a ring cavity with two degenerate, counterpropagating running-wave modes. (a) Atoms are placed within a ring cavity that is pumped on the TEM$_{00}$ mode (green) through one of the cavity mirrors. Photons scattered off the atoms populate the counterpropagating, degenerate mode (blue), leading to the emergence of a dipole potential from the interference of the two counterpropagating fields. (b) Stability diagram of the system. The white circles indicate the critical pump power above which the condensate zero-momentum state is significantly depopulated due to photon scattering. The background color shows the population of the condensate mode from a numerical simulation, and the red line indicates the numerical phase boundary obtained using a low-energy model without cavity damping. The inset shows an example from which a threshold can be deduced. Depending on the detuning between the cavity and pump field $\Delta_c$, the system either enters the recoil lasing regime or the momentum correlated regime separated by the white dashed line.  Figure adapted and reprinted with permission from Ref.~\cite{Schmidt2014Dynamical} \textcopyright\ 2014 by the American Physical Society.}
\label{fig:RingCavity}
\end{figure}

Atomic self-organization can also take place in ring cavities, which support a pair of degenerate, counterpropagating running-wave modes. First experiments investigated the phenomenon of collective atomic recoil lasing (CARL), which is an instability analogous to a free electron laser and has a similar origin as the Dicke phase transition. The atomic gas is trapped at the position of the cavity modes where one of them is pumped through one of the cavity mirrors; see Figure~\ref{fig:RingCavity}(a). Photons injected into the forward-propagating cavity mode can be scattered back via the atoms into the unpumped, counterpropagating cavity mode, while the atoms recoil. Exponential gain of this unpumped, counterpropagating field then triggers the so-called CARL instability. Specifically, the backscattering process leads to the buildup of an interference pattern of the two counterpropagating light fields which acts as a dynamic potential on the atoms. Correspondingly, the atoms form a matter-wave pattern further enhancing the photon scattering in a runaway process. This density wave is accelerated in space, synchronized with the co-moving optical potential generated from the interference of the two cavity fields. As a consequence, the frequency of the light scattered at the atoms is increasingly red-shifted with respect to the probe light due to the Doppler effect. In the bad cavity limit, this leads to the occupation of increasingly higher momentum states. For narrow band resonators in the good cavity limit, processes occupying increasingly higher momentum states are, however, suppressed,  effectively stabilizing the system. The CARL instability, originally predicted by Bonifacio et al.~\cite{Bonifacio1994Exponential}, has been demonstrated in both the thermal~\cite{Kruse2003Observation} and the ultracold regime~\cite{Slama2007Superradiant}.

The CARL instability sets in for positive, dispersively-shifted pump-cavity detuning, $\delta_c>0$, and is a runaway process. In an intuitive classical picture~\cite{Tomczyk2015Stability}, the atoms are located close to the maxima of the periodic dipole potential building up from the pumped and back-scattered fields, leading to the acceleration of the atomic system. A pulse of light is emitted while the atoms are transferred from the zero momentum state to higher momentum states in the direction of the pump mode.  In contrast, for negative pump-cavity detuning $\delta_c<0$, the phase of the field inside the resonator acquires a phase shift of $\pi$ with respect to the situation for $\delta_c>0$. The atoms thus are located close to the minima of the emerging dipole potential and can enhance scattering via constructive interference by building up a stationary density grating with a $\lambda_c/2$ periodicity for narrow-band resonators. Momentum states in the direction of the pump and in the opposite direction are thus symmetrically occupied, hence the name ``momentum correlated regime'' for  $\delta_c<0$~\cite{Schmidt2014Dynamical}. A stability diagram of this system is shown in Figure~\ref{fig:RingCavity}(b).

The position of the nodes of the emergent optical lattice and correspondingly the atomic density wave in the momentum correlated regime would be arbitrary and set by initial (quantum) fluctuations in an idealized ring cavity, breaking spontaneously the continuous symmetry of both the light-field phase and the origin of the spatial pattern. In a realistic ring cavity, some light is, however, always scattered into the unpumped counterpropagating mode at surface inhomogeneities of the mirrors. Depending on the chosen longitudinal mode of the resonator, this back scattering can be maximized or minimized~\cite{Krenz2007Controlling}. Exploiting the situation with maximized back scattering, this backscattered field was used to pin the atomic-density pattern in space and thus to act as a symmetry breaking field of the transition to the momentum correlated regime~\cite{Schuster2018Pinning}. Breaking of the continuous symmetry spontaneously can, nevertheless, be approximately realized by pumping the cavity with two non-interfering counterpropagating light fields as discussed in Section~\ref{sec:supersolid-ring-cavity-longitudinal-pump}, or by transversely pumping the atomic cloud inside the ring cavity as described in Section~\ref{sec:supersolid-ring-cavity-transverse-pump}.

Also the situation of a transversally pumped atomic cloud coupled to an initially empty ring resonator has been explored experimentally. Scattering from the transverse  pump beam under an angle into the ring cavity leads to population of higher momentum modes. The sideband resolved regime in which this resonator operates allows to control the population of  specific momentum states by varying the detuning between the cavity resonance and the pump frequency~\cite{Bux2011Cavity,Bux2013Control}.

\subsubsection{Spin-dependent self-organization}
\label{subsubsec:SpinDependentSO}

So far, we did not consider the spin degree of freedom of the atoms in the self-organization process. Tilting the quantization axis of the atoms with respect to the polarization of the driving electric fields, and considering scattering of light into both polarization modes of the cavity allows to extend the single-mode self-organization to schemes involving two photonic modes, as discussed in Ref.~\cite{Landini2018Formation, Morales2019Two}.

To this end, we need to rephrase the atom-field interaction Hamiltonian in terms of the atomic polarizability, taking into account that the atom has a more complex internal structure than the two-level scheme assumed in Section~\ref{sec:theory}.  The Hamiltonian describing the interaction between an atomic dipole operator $\hat{\mathbf{d}}$ and an electric field $\hat{\mathbf{E}}$ in the dipole approximation $\hat{H}_\mathrm{A-L}=-\hat{\mathbf{d}} \cdot \hat{\mathbf{E}}$ can be rewritten in the dispersive limit as~\cite{Deutsch2010Quantum, Kien2013Dynamical,Landini2018Formation},
\begin{align}
  \hat{H}_\mathrm{A-L} = -\alpha_s (\hat{\mathbf{E}}^\dag \cdot \hat{\mathbf{E}}) \hat{I}_{\hat{\mathbf F}}
  + i \alpha_v (\hat{\mathbf{E}}^\dag \times \hat{\mathbf{E}}) \cdot \frac{\hat{\mathbf{F}}}{2F} 
  - \alpha_t \sum_{i,j=1}^3 \frac{3(\hat{F}_i \hat{F}_j +\hat{F}_j \hat{F}_i) - 2 F^2\delta_{ij}}{2F(2F-1)}\hat{\mathbf{E}}_i \cdot \hat{\mathbf{E}}_j, 
  \label{eq:H_AL}
\end{align}
where $\hat{\mathbf{F}}$ is the total angular momentum operator with maximum eigenvalue $F$, $\hat{I}_{\hat{\mathbf F}}$ is a $(2F+1)\times(2F+1)$ identity operator in the internal (angular-momentum) space, and $\alpha_s$, $\alpha_v$, $\alpha_t$ are the scalar, vector, and tensor polarizabilities of the atom, respectively~\cite{Kien2013Dynamical}. They depend on the electronic structure of the atom and the frequency of the driving electric field. The scalar part captures the induced atomic dipole oscillating in phase and in the direction of the polarization of the driving electric field, which we assume to be linear for the moment. Light scattered by the scalar part thus maintains the polarization of the driving light field. In contrast, the vector part is imaginary and can change the polarization of the scattered light field through the Faraday effect. Light emitted due to the vector polarizability has a polarization direction orthogonal to both the atomic spin and the pump polarization, and oscillates $\pi/2$ out of phase with respect to the driving electric field. The tensor part finally describes non-linear spin processes, and will be neglected here since it is irrelevant at the detunings considered in the following. 

For driven atoms inside a cavity, the total electric field operator $\hat{\mathbf{E}}$ at the position of the atoms contains contributions from both the (classical) pump fields $\mathbf{E}_\mathrm{p}$ and the (quantum) cavity fields $\hat{\mathbf{E}}_\mathrm{c}$,
\begin{align}
  \hat{\mathbf{E}} = \sum\limits_{\substack{\mathrm{pump}\\\mathrm{fields}}}\mathbf{E}_\mathrm{p} + \sum\limits_{\substack{\mathrm{cavity}\\\mathrm{modes}}}\hat{\mathbf{E}}_\mathrm{c},
\end{align}
and each electric field is described by its unit polarization vector, amplitude, spatial mode profile, and time dependence. Inserting the pump and cavity modes assumed in Section~\ref{sec:basic-model} into Equation~\eqref{eq:H_AL}, one recovers the potentials derived in Equation~\eqref{eq:Heisberg-eq-ga}, now in term of the scalar polarizability. Specifically, one finds the pump lattice depth $V_0=\Omega_0^2/\Delta_a = -\alpha_s E_\mathrm{0p}^2$, the cavity lattice depth per photon $U_0 = \mathcal{G}_0^2/\Delta_a = -\alpha_s \mathcal{E}_0^2$, and the two-photon scattering rate $\eta_0=\Omega_0\mathcal{G}_0/\Delta_a=-\alpha_s E_\mathrm{0p} \mathcal{E}_0$. Here, $E_\mathrm{0p}$ and $\mathcal{E}_0$ denote the electric field amplitudes of the pump and the cavity fields, respectively.

\begin{figure}[t!]
\centering
\includegraphics[width=1\columnwidth]{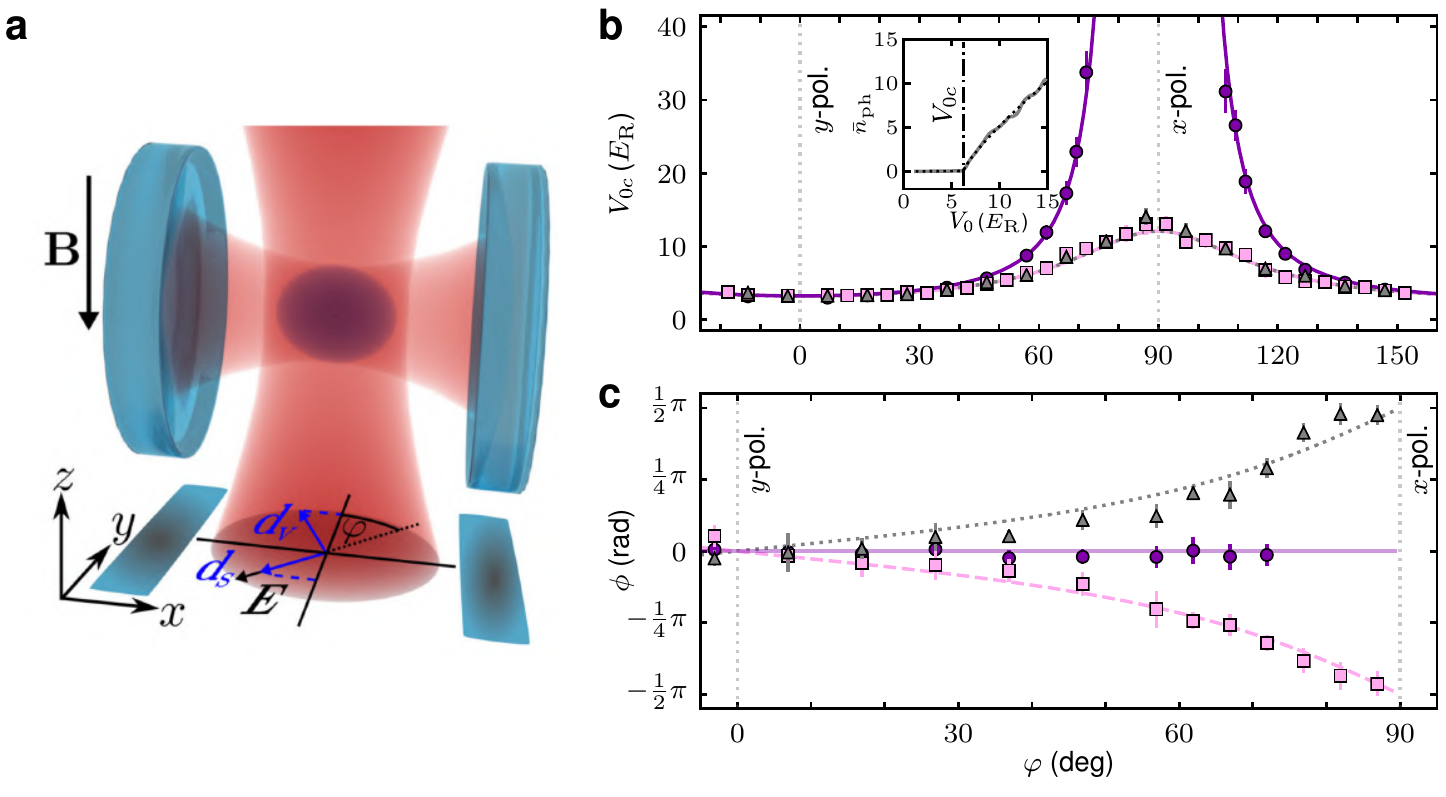}
\caption{Spin-dependent self-organization of a transversely driven BEC in a single-mode optical cavity. (a)~The BEC is illuminated by a pump field along the $z$ direction whose linear polarization can be rotated in the $x$-$y$ plane by an angle $\varphi$. The induced atomic dipole has a scalar $\hat{\mathbf d}_s$ and a vector $\hat{\mathbf d}_v$ component which oscillate $\pi/2$ out of phase. This leads to a phase shift of the light scattered into the cavity which depends on $\varphi$. (b) Critical value of the pump-lattice depth $V_{0c}\propto\eta_{0c}^2$ for self-organization as a function of the polarization angle $\varphi$ for the different Zeeman states (circles: $m_F=0$, triangles and squares: $m_F=\pm1$). The inset shows a typical measurement of the critical pump-lattice depth. (c) Phase of the light field scattered into the cavity with respect to the phase of the pump field for self-organization of the different Zeeman states. 
Figure adapted and reprinted with permission from Ref.~\cite{Landini2018Formation} \textcopyright~2018 by the American Physical Society.}
\label{fig:FormationSpinTexture}
\end{figure}

\paragraph{Spin texture formation} \label{par:SpinTextureFormation}The term proportional to the spin-dependent vector polarizability $\alpha_v$ becomes important if the cross product $(\hat{\mathbf{E}}^\dagger \times \hat{\mathbf{E}})$ in Equation~(\ref{eq:H_AL}) becomes non-zero. This can be achieved for example by rotating the polarization  of the pump field (illuminating the atoms in the $z$ direction) with respect to the polarization of the cavity field by an angle $\varphi$, as shown in Figure~\ref{fig:FormationSpinTexture}(a)~\cite{Landini2018Formation}. That is, the polarization of the pump field is rotated to make an angle $\pi/2-\varphi$ with respect to the cavity axis, since the polarization of the cavity mode is parallel to cavity mirrors. The induced atomic dipole can then be decomposed into a scalar and a vector component $\hat{\mathbf d}=\hat{\mathbf d}_s+i\hat{\mathbf d}_v=-\alpha_s\hat{\mathbf E}/2+i\alpha_v\hat{\mathbf F}\times\hat{\mathbf E}/4F$ [where the orientation of $\hat{\mathbf F}$ is fixed by the bias magnetic field $\mathbf{B}$; see Figure~\ref{fig:FormationSpinTexture}(a) and also Equation~(\ref{eq:H_AL})], such that light scattered into the cavity mode will be phase shifted with respect to the phase of the pump field. This can also be seen from the many-body Hamiltonian for $N$ atoms---mapped into a Dicke model with macroscopic spin operators as introduced in Section~\ref{subsubsec:DickeModel}---in the Zeeman state $m_F$~\cite{Landini2018Formation},
\begin{align}\label{eq:DickeHamiltonianZeeman}
  \hat{H}_{m_F} = -\hbar \delta_c \hat{a}^\dagger \hat{a}+ 2\hbar \omega_r \hat{J}_{z, m_F} + \frac{\hbar}{2} \left[ \eta_{s0} (\hat{a}^\dagger+\hat{a})\cos\varphi + i \eta_{v0} (\hat{a}^\dagger - \hat{a}) \sin\varphi \right] \hat{J}_{x, m_F}.
\end{align}
Here, $\cos\varphi$ ($\sin\varphi$) appears due to the projection of the scalar $\hat{\mathbf d}_s$ (vector $\hat{\mathbf d}_v$) dipole component along the polarization of the cavity field in the $y$ direction; see Figure~\ref{fig:FormationSpinTexture}(a). The polarization angle $\varphi$ of the pump electric field controls the ratio between the scalar coupling $\eta_{s0}\propto\alpha_s$ and the vectorial coupling $\eta_{v0}\propto\alpha_v$. Since the scalar and vector polarizabilities have different strengths, the critical point for the self-organization phase transition depends on the angle $\varphi$. Figures~\ref{fig:FormationSpinTexture}(b) and (c) display the threshold for the self-organization and the phase shift of the light scattered into the cavity as a function of the polarization angle of the pump field, respectively. The behavior depends on the Zeeman state the atoms are prepared in. Note that we assume here the Zeeman splitting of the atomic levels to be much larger than both the coupling strength and the detuning $\delta_c$, such that spin changing processes are suppressed; spin changing processes are considered and discussed in Section~\ref{sec:spinor-selfordering}. For $m_F=0$, only the scalar polarizability is present. Accordingly, the time phase of the cavity field is independent of $\varphi$, and the threshold for self-organization diverges if the polarization of the pump field becomes parallel to the cavity axis (i.e., $\varphi=\pi/2$). In contrast, for the Zeeman states $m_F=\pm1$ the vector polarizability can become nonzero allowing self-organization even for $\varphi=\pi/2$;  the field scattered into the cavity experiences either a positive or a negative phase shift.

In a more general case of a balanced spin mixture of atoms prepared in the Zeeman states $m_F=+1$ and $m_F=-1$, the light fields scattered by the individual spin components interfere, leading to competition between the different self-organized states. Depending on the ratio of real and imaginary components of the field scattered into the cavity, it can become energetically more favorable for the system to arrange in a spin modulated texture than in a checkerboard density modulation. Such a phase transition was observed in Ref.~\cite{Landini2018Formation}, where the phase of the light field scattered into the cavity with respect to the phase of the pump field jumped from 0 to $\pi/2$ at a critical angle of the pump field polarization. Accordingly, a spin texture formed with a $\lambda_c$-periodic checkerboard modulation of the atomic spins. The phase transition between the normal and either of the self-organized states (spin modulated or density modulated) has the same behavior as the Dicke model. However, for sufficiently small detuning with respect to the cavity resonance, an intriguing dynamical instability can be induced driving the system to oscillate between the different atomic patterns (see Section~\ref{sec:dissipation-induced-instability-spinor-BEC}).

\paragraph{Polarization-dependent self-organization} 
So far, we have considered superradiance of atoms in only one of the two (theoretically) degenerate modes of a linear cavity with orthogonal polarizations. Coupling to the other orthogonally polarized mode of the cavity is, however, also possible if the atom has a non-zero vector polarizability at the frequency of the transverse pump. Self-organization can thus take place involving both polarization modes, as has been experimentally demonstrated~\cite{Morales2019Two}.

\begin{figure}[t!]
\centering
\includegraphics[width=1\columnwidth]{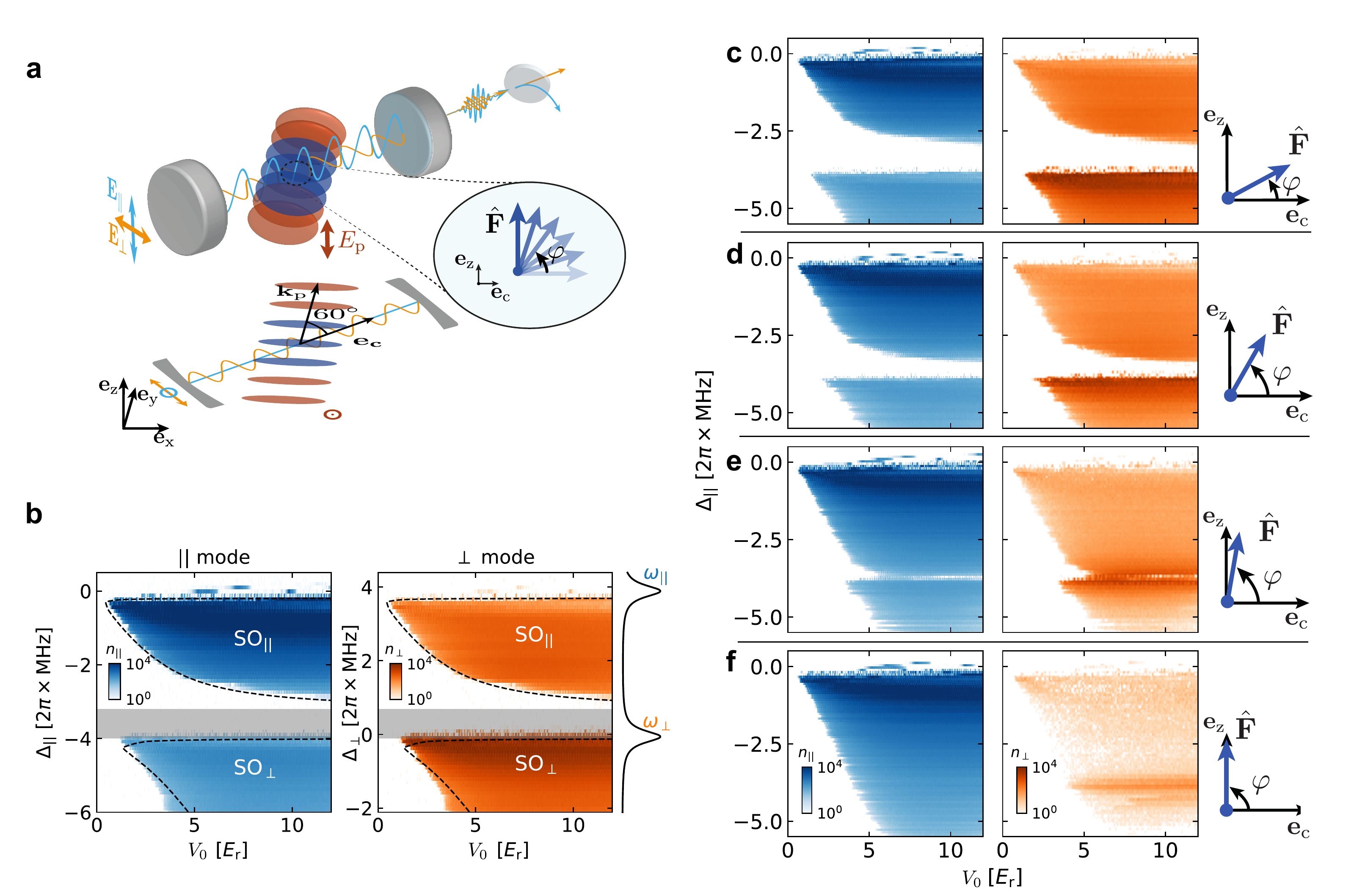}
\caption{Two-mode self-organization of a driven BEC inside a linear cavity via competing of two mutually orthogonal, polarized cavity modes. (a) The electric field $\mathbf{E}_\mathrm{p}$ of the pump laser is polarized along $z$ while the atomic spin $\hat{\mathbf F}$ can be rotated by an angle $\varphi$ in a plane spanned by the cavity axis and the $z$ axis. The vector polarizability can lead to scattering of photons from the pump into both polarization modes of the cavity. (b) The population of both polarization modes is simultaneously measured and shown in blue and orange, respectively, for an angle of $\varphi=0(4)^\circ$. Two regions of self-organization are visible in the resulting phase diagram. The self-organized phase SO$_\parallel$ is completely suppressed for detunings close to the cavity resonance of the orthogonally polarized mode, as indicated by the grey region. For detunings effectively red with respect to the orthogonal polarized pump, the self-organized phase SO$_\perp$ emerges. The dashed lines show the theoretically predicted phase boundaries. Panels (c)-(f) show neasurements of this phase diagram for different angles $\varphi = 30(4)^\circ, 60(4)^\circ, 80(4)^\circ, 90(2)^\circ$. Figure adapted and reprinted with permission from Ref.~\cite{Morales2019Two} \textcopyright~2019 by the American Physical Society.}
\label{fig:TwoModeDickeModel}
\end{figure}

Figure~\ref{fig:TwoModeDickeModel}(a) shows the basic experimental scheme. The polarization of the pump electric field $\mathbf{E}_p$ is oriented along the $z$ axis, and an externally applied bias magnetic field defining the atomic quantization axis allows to rotate the orientation of the atomic pseudospin $\hat{\mathbf{F}}$ by an angle $\varphi$ with respect to the cavity axis. Therefore, the induced dipole moment $\hat{\mathbf d}=\hat{\mathbf d}_s+i\hat{\mathbf d}_v=-\alpha_s\hat{\mathbf E}/2+i\alpha_v\hat{\mathbf F}\times\hat{\mathbf E}/4F$ can have components along two orthogonally polarized cavity modes designated by the annihilation operators $\hat{a}_\parallel$ 
(for the mode with a parallel polarization with respect to the pump polarization) and $\hat{a}_\perp$ (for the mode with an orthogonal polarization with respect to the pump polarization). In particular, $\hat{\mathbf d}_s$ is aligned along the polarization of the $\hat{a}_\parallel$ mode, while $\hat{\mathbf d}_v$ is parallel to the polarization of the $\hat{a}_\perp$ mode. The degeneracy of the two mutually orthogonal cavity modes is lifted in the experiment due to birefringence (see Section~\ref{subsubsec:optical_resonators}), and hence they have different resonance frequencies. Depending on the detunings $\Delta_\parallel$ and $\Delta_\perp$ between the pump frequency and the respective cavity-mode frequency, the pump field might thus be red-detuned to one cavity mode while it is blue-detuned with respect to the other cavity mode. 

Using the general atom-field interaction formalism described above [see Equation~\eqref{eq:H_AL}] and the mapping of Section~\ref{subsubsec:DickeModel}, the low-energy Hamiltonian of the system for atoms in the Zeeman substate $m_F$ is~\cite{Morales2019Two}, 
\begin{align}
\hat{H}&=-\hbar\Delta_\parallel\hat{a}_\parallel^\dagger\hat{a}_\parallel-\hbar\Delta_\perp\hat{a}_\perp^\dagger\hat{a}_\perp+\hbar\omega_0\hat{b}^\dagger\hat{b}
+\alpha_s\frac{E_\textrm{0p}\mathcal{E}_0}{2\sqrt{2}}\left(\hat{a}_\parallel^\dagger+\hat{a}_\parallel\right)
\left(\hat{b}^\dagger\hat{b}_0+\hat{b}_0^\dagger\hat{b}\right)\notag\\
&+i\alpha_v\left[\frac{E_\textrm{0p}\mathcal{E}_0}{2\sqrt{2}}
\left(\hat{a}_\perp^\dagger-\hat{a}_\perp\right)\left(\hat{b}^\dagger\hat{b}_0+\hat{b}_0^\dagger\hat{b}\right)
+\frac{\mathcal{E}_0^2}{2}\left(\hat{a}_\parallel^\dagger\hat{a}_\perp-\hat{a}_\perp^\dagger\hat{a}_\parallel\right)\hat{b}^\dagger_0\hat{b}_0\right]
\frac{m_F}{2F}\cos\varphi,
\label{spinHamiltonian}
\end{align}
where $\hat{b}$ ($\hat{b}^\dagger$) is the operator annihilating (creating) an atom in the momentum superposition state with energy $\omega_0$ resulting from the scattering of photons between the pump and the cavity, and $\hat{b}_0$ ($\hat{b}_0^\dagger$) is the operator annihilating (creating) an atom in the zero-momentum BEC.  The term proportional to the scalar polarizability $\alpha_s$ describes coupling of the BEC to the real quadrature ($\hat{a}_\parallel+\hat{a}_\parallel^\dagger$) of the parallel polarized cavity mode. The first term proportional to the vector polarizability $\alpha_v$ captures the coupling of the BEC to the imaginary quadrature $i(\hat{a}_\perp^\dagger-\hat{a}_\perp)$ of the orthogonal polarized cavity mode, while the second term describes a direct scattering of photons between the two mutually orthogonal cavity modes via the atoms in the BEC. 

Since for certain pump frequencies the signs of the detunings $\Delta_\parallel$ and $\Delta_\perp$ can be opposite, the coupling to the two polarization modes competes, and self-organization can be fully suppressed in a certain parameter range, as can be seen in Figure~\ref{fig:TwoModeDickeModel}(b). Specifically, if the pump frequency is blue detuned with respect to the resonance frequency $\omega_\perp$ of the orthogonal polarized mode but red detuned with respect to the resonance frequency $\omega_\parallel$ of the parallel polarized mode, the phase diagram shows a sliver without self-organization also for large pump fields. In this region, a mode hardening due to the effective blue detuning competes with a mode softening due to the effective red detuning. Accordingly, the critical point in the grey shaded area is pushed to infinity. 

\subsubsection{Self-organization with repulsive optical potentials}
\label{subsubsec:blue_detuned_SR}

Self-organization can also take place for low-field seeking atoms with $\Delta_a>0$, i.e., in the repulsive regime of blue-detuned pump and cavity lattices. This was predicted theoretically early on in Ref.~\cite{Keeling2010Collective}. In particular, the possible existence of dynamical instabilities in this parameter regime has attracted more interest recently~\cite{Bhaseen2012Dynamics, Piazza2015Self, Kessler2019Emergent, Lin2020Pathway}; see Section~\ref{sec:dynamics} for discussions regarding dynamical aspects.

At first sight self-organization in repulsive fields might seem counterintuitive, since the atoms will localize in the intensity minima of the pump field, suppressing the scattering of photons into the cavity and thus inhibiting self-organization. However, this purely classical picture of point-like particles does not take into account the finite extent of the BEC wavefunction around the potential minima and that the localization of particles comes at the cost of kinetic energy. As a result, the atomic gas still has a finite overlap with the blue detuned pump field. Since the induced atomic dipoles are driven above their resonance, the light field scattered by the atoms into the cavity oscillates out of phase with respect to the pump field. At the position of the atoms, these two fields thus interfere destructively, again creating potential minima for the atoms; see also Figure~\ref{fig:potentials}. This process allows the atoms to lower their potential energy via increased scattering of photons into the resonator, and hence self-organization in a repulsive potential takes place.

\paragraph{Coupling to the $P$ band of the pump lattice}

\begin{figure}[t!]
\centering
\includegraphics[width=1\columnwidth]{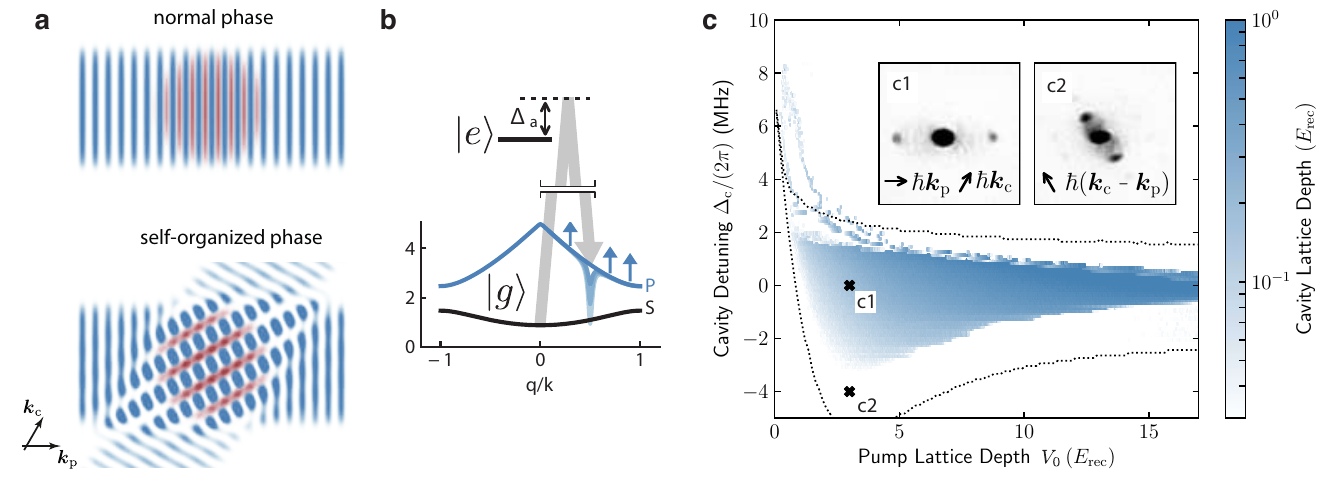}
\caption{Self-organization of a driven BEC in a single-mode linear cavity in the repulsive lattice regime. (a)~The BEC (red) is exposed to a repulsive, blue-detuned optical pump lattice. In the self-organized phase, the field scattered by the atoms into the cavity interferes at the position of the atoms destructively with the pump field. This way the atoms can lower their potential energy by self-organization. (b) The two-photon process for the blue atomic detuning (grey arrows) couples the BEC in the ground S band to a higher momentum state in the $P$ band of the pump lattice. With increasing pump-lattice depth the coupling term leads to a mode softening at the according wave vector (pale blue lines) and finally to self-organization when this mode touches zero relative to the ground state energy. At the same time, the energy of the $P$ band increases with the pump-lattice depth (blue arrows). For deeper lattices, this effect becomes dominant and suppresses self-organization. (c) Phase diagram showing the intra-cavity lattice depth as a function of the cavity detuning and the pump-lattice depth. The self-orderd phase has only limited extent for increasing pump lattice depth. Dotted lines are results from a numerical mean-field calculation showing the phase boundaries. The insets show absorption images of the atomic cloud after ballistic expansion for parameters indicated by the crosses in the main plot.  Figure adapted and reprinted with permission from Ref.~\cite{Zupancic2019P} \textcopyright~2019 by the American Physical Society.}
\label{fig:PBandOrganization}
\end{figure}

Although the basic process for self-ordering of low-field seeking atoms is very similar to self-organization of atoms in red-detuned, attractive light fields, significant differences exist. In a blue-detuned pump lattice, the atoms in the ground state localize at the nodes of the pump lattice potential. The maxima of the atomic density are thus shifted by $\lambda_c/4$ with respect to the maxima of the pump field. Taking the atomic density maxima as origin of the coordinate system, the term $\eta(\mathbf{r})$ in Equation~(\ref{eq:H_eff_density_1comp}) describing the coupling between the pump and the cavity thus changes parity and becomes $\propto \sin(k_c y)$ instead of $\propto \cos(k_c y)$. Accordingly, while for atomic red detuning the BEC is coupled from the zero-momentum state of the S band to finite quasi-momenta within the S band, for blue detuned lattices the BEC is coupled to quasi-momenta in the P band, localized at the maxima of the pump potential. The coupling leads to a mode softening at the wave vector of the scattering process [see Figure~\ref{fig:PBandOrganization}(b)], eventually enabling self-organization for a sufficiently strong pump field. However, for a further increased pump-lattice depth, the energy gap between the $S$ and $P$ bands increases, and the self-organization is finally terminated at some point. As a result, the extent of the self-organized phase as a function of the pump-lattice depth is limited, as can be seen from the experimentally recorded phase diagram in Figure~\ref{fig:PBandOrganization}(c)~\cite{Zupancic2019P}. This is in stark contrast to the self-organization in the red-detuned regime, cf.\ Figure~\ref{fig:DickePhaseDiagram}.

The self-organized atomic density favours the emergence of stripes [see Figure~\ref{fig:PBandOrganization}(a)] rather than a checkerboard pattern for two reasons. First, in the experiment the pump field is tilted with respect to the cavity mode by 60$^\circ$, such that two atomic momentum modes with different energies are addressed by the two-photon scattering. The lower energetic one is predominantly occupied. Second, the structure of the coupling disfavors for $\Delta_a>0$ a simultaneous population of both modes, different from the situation for $\Delta_a<0$.

The phase diagram in Figure~\ref{fig:PBandOrganization}(c) extends also to positive detunings $\Delta_c$ between the pump and the cavity since also the dispersive shift, $-\int U(\mathbf{r}) n(\mathbf{r})d\mathbf{r}$, has a flipped sign with respect to the case for $\Delta_a<0$. For increasing pump-lattice depth at positive cavity detunings, cyclic spiking of the cavity field and population of very high atomic momenta was observed. Uniquely for $\Delta_a, \Delta_c>0$, self-organization can become a metastable state, and the system eventually lowers its energy by letting the photon number in the cavity diverge.

\paragraph{Structural phase transition} Applying a pump field with a standing-wave and a running-wave component allows to realize for low-field seeking atoms, $\Delta_a>0$, a situation where a simultaneous coupling to the $P$ band and to the $S$ band of the pump lattice can be brought to competition~\cite{Li2020Measuring}. Since the self-organized density patterns of these two possibilities are not compatible with each other, a first order structural phase transition between them can be induced.

\begin{figure}[t!]
\centering
\includegraphics[width=1\columnwidth]{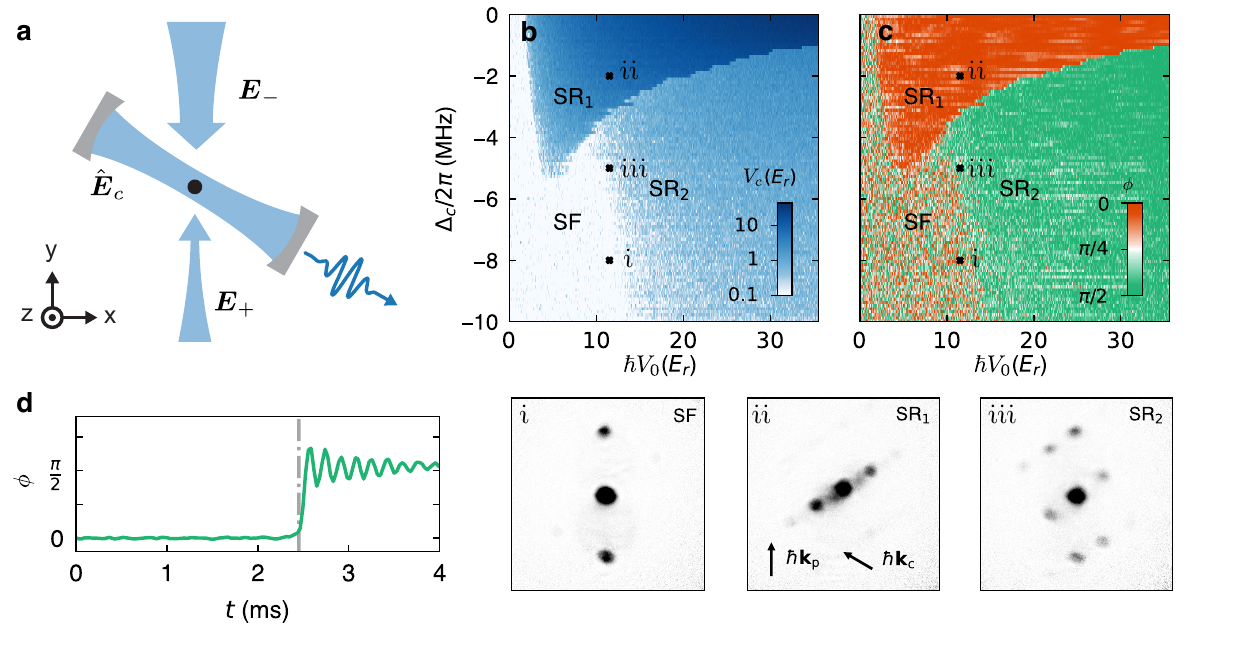}
\caption{Self-organization of a driven BEC inside a single-mode cavity with imbalanced pump fields. (a)~A BEC inside a linear cavity is illuminated by imbalanced pump fields $\mathbf{E}_+$, $\mathbf{E}_-$, effectively leading to the coupling of the quantum gas to a standing-wave and a running-wave pump field. The interference of these two components with the cavity field gives rise to two incompatible patterns the atoms might self-organize to. Phase diagram showing the intra-cavity lattice depth $V_c=\hbar U_0 \langle\hat{a}^\dag \hat{a}\rangle$ (b) and the phase mapped to the quadrant $\phi \in [0,\pi/2]$ of the cavity field (c) as a function of the pump-lattice depth $\hbar V_0$  and the pump-cavity detuning $\Delta_c$. Besides the normal superfluid phase (SF), two self-organized superradiant phases SR$_1$ and SR$_2$ can be distinguished. The cavity field has different amplitude as well as a $\pi/2$ phase shift in the two phases. In addition, the two phases can be distinguished by the momentum distribution revealed from absorption images after ballistic expansion of the quantum gas ($i$-$iii$). (d) A first order structural phase transition between the SR$_1$ and SR$_2$ phases can be induced by ramping up the pump-lattice depth to $\hbar V_0 = 25\,E_r$ at a detuning of $\Delta_c/2\pi =-1.75$~MHz. The phase of the cavity field shows a sudden jump of $\pi/2$, followed by a damped oscillation.
Figure adapted and reprinted with permission from Ref.~\cite{Li2020Measuring}. }
\label{fig:StruturalPhaseTransition}
\end{figure}

Experimentally this was implemented by applying imbalanced pump fields with wave vectors $\pm\mathbf{k}_p$ and electric field amplitudes $E_\pm$ to a BEC inside a cavity with the mode wave vector $\mathbf{k}_c$; see Figure~\ref{fig:StruturalPhaseTransition}(a). The according pump electric field is then given by $\mathbf{E}_p=(\mathbf{E}_+ + \mathbf{E}_-)/2$ with $\mathbf{E}_\pm=E_\pm e^{\pm i \mathbf{k}_p \mathbf{r}} \mathbf{e}_z$, which results in a pump-lattice depth $V_p=-\alpha_s E_+ E_-$.  The interference between the pump electric field and the cavity electric field generates two potential energy terms with different symmetry. This can be seen from inserting the total electric field $\hat{\mathbf{E}}=\mathbf{E}_p+\hat{\mathbf{E}}_c$ into the atom-field interaction Hamiltonian~(\ref{eq:H_AL}), where $\hat{\mathbf{E}}_c=\mathcal{E}_0 \boldsymbol{\epsilon}_c \hat{a} \cos(\mathbf{k}_c \cdot \mathbf{r})$ is the positive frequency component of the cavity field. Ignoring the term proportional to the vectorial polarization in Equation~\eqref{eq:H_AL}, this results in 
\begin{align}
\hat{H}_{A-L} (\mathbf{r})&= \hbar V_0 \cos^2(\mathbf{k}_p\cdot \mathbf{r}) +\hbar U_0 \hat{a}^\dagger \hat{a} \cos^2(\mathbf{k}_c\cdot \mathbf{r})
+ \hbar \eta_{01} (\hat{a}+\hat{a}^\dagger ) \cos(\mathbf{k}_p\cdot \mathbf{r})\cos(\mathbf{k}_c\cdot \mathbf{r})\nonumber\\
&+i \hbar \eta_{02} (\hat{a}-\hat{a}^\dagger ) \sin(\mathbf{k}_p\cdot \mathbf{r})\cos(\mathbf{k}_c\cdot \mathbf{r}),
 \label{eq:HamiltonianAR}
\end{align}
where the potential depths $\hbar \eta_{01}, \hbar \eta_{02}$ depend on the tunable imbalance between the pump fields $E_\pm$.

From the resultant atom-field interaction Hamiltonian, Equation~(\ref{eq:HamiltonianAR}), it becomes obvious that the two different spatial structures involving $\cos(\mathbf{k}_p \cdot \mathbf{r})$ and $\sin(\mathbf{k}_p\cdot \mathbf{r})$ couple to the real and the imaginary quadratures of the cavity field, respectively. This allows to distinguish them experimentally by measuring the phase $\phi$ of the cavity field, as shown in Figures~\ref{fig:StruturalPhaseTransition}(b) and (c). Two superradiant phases, termed SR$_1$ and SR$_2$, are observed which differ in amplitude and phase of the cavity field. In addition, absorption images of the atomic cloud after ballistic expansion reveal that also the spatial structure of the two self-ordered states is different. SR$_1$ shows momentum components indicating a dominantly striped density modulation, while SR$_2$ shows momentum components corresponding to a checkerboard density pattern.

When the system is driven from SR$_1$ to SR$_2$ by increasing the pump-lattice depth $V_0$, a structural phase transition takes place. This transition is of first order as was shown theoretically by investigating the energy landscape. In addition, the real-time observation of the phase of the cavity field shows an abrupt jump by $\pi/2$, followed by a damped oscillation. The oscillation frequency is in agreement with the curvature of the free energy landscape minimum, as calculated numerically from a mean-field model~\cite{Li2020Measuring}.

\section{Superradiant crystallization breaking a continuous symmetry: supersolids}
\label{sec:supersolid}

In the preceding section, the self-ordering of ultracold atoms inside single-mode, standing-wave linear cavities was discussed. We saw that the crystallization in these setups breaks spontaneously the \emph{discrete} $\mathbf{Z}_2$ parity symmetry of the Hamiltonian, leading to a two-fold degenerate ground state with gapped collective excitations. In this section, we consider scenarios where the superradiant crystallization of a superfluid BEC breaks spontaneously an external \emph{continuous} $U(1)$ symmetry [in addition to the spontaneously broken $U(1)$ superfluid gauge symmetry of the BEC], giving rise to an infinitely degenerate ground state with a gapless Goldstone mode (in addition to the gapless condensate phonon mode associated with the broken superfluid gauge symmetry). This state fulfills the criteria for the minimal version of a ``supersolid'', an enigmatic state of matter characterized by the coexistence of crystalline and superfluid orders which thus supports frictionless flow of particles in a crystal. 
It is important to note that the existence of a BEC does not directly imply superfluidity and by this dissipationfree flow. However, ultracold gases with finite collisional interactions have been proved to support superfluid flow~\cite{Onofrio2000Observation}, and the observation of phase coherence is considered as sufficient indication for superfluidity~\cite{Greiner2002Quantum}.

Already the previously discussed self-organized phases of superfluid BECs breaking the discrete $\mathbf{Z}_2$ symmetry~\cite{Baumann2010Dicke, Mottl2012Roton} can be regarded as a variant of supersolids, often termed ``lattice supersolids''~\cite{Matsuda1970Off}. However, the underlying discrete $\mathbf{Z}_2$ symmetry does not allow for the continuous ground-state degeneracy envisioned in the original discussions of supersolidity.  As a consequence, lattice supersolids do not sustain dissipationless particle currents.

More precisely, during the formation of a supersolid two continuous symmetries must be spontaneously broken, namely, a continuous spatial translational symmetry and an  internal $U(1)$ superfluid gauge invariance. This paradoxical state of matter was predicted almost 50 years ago to exist in solid helium-4~\cite{Andreev1969Quantum, Thouless1969The, Leggett1970Can}. Despite intensive experimental efforts, there has been so far no conclusive evidence for the observation of supersolidity in helium. Dilute bosonic quantum gases had been proposed as possible candidates for the realization of a supersolid state of matter, if appropriate long-range interactions which favor the superfluid gas to form a crystal could be engineered~\cite{ODell2003Rotons, Santos2003Roton}. These interactions might stem from induced or permanent electric dipole moments of the atoms, or can exist for atoms with strong magnetic dipole moments. Cavity-mediated global interactions turn out to be a suitable, alternative route to realize a supersolid, if the discrete $\mathbf{Z}_2$ symmetry of composite atom-cavity systems can be replaced by a continuous $U(1)$ symmetry~\cite{Gopalakrishnan2009Emergent}. This was achieved via symmetry enhancement using two crossed standing-wave linear cavities~\cite{Leonard2017Supersolid}, as we detail below in Section~\ref{sec:supersolidity-two-cavities}. Supersolids have since then also been experimentally demonstrated in quantum gases making use of spin-orbit coupling~\cite{Li2017A}, of magnetic dipolar interactions~\cite{Tanzi2019Observation, Bottcher2019Transient, Chomaz2019Long}, and of ring cavities~\cite{Schuster2020Supersolid}. Supersolids in systems with finite-range interactions exhibit a continuous dispersion relation and thus have finite rigidity. In contrast, systems with exclusively global-range interactions show (in the thermodynamic limit) a mode softening at isolated points in momentum space and thus form defect-free and perfectly rigid crystals. The resulting supersolid in the latter systems thus can be seen as the minimal possible version of a supersolid.

\subsection{Supersolidity in two-crossed linear cavities}
\label{sec:supersolidity-two-cavities}

\subsubsection{Symmetry enhancement using two linear cavities}\label{subsubsec:TwoCavities}
A continuous symmetry can be engineered through symmetry enhancement with competing order parameters, as has been discussed in the context of high-temperature superconductors, high-energy physics, and cosmology~\cite{Eichhorn2013Multicritical}. The concept is based on fine-tuning Hamiltonian parameters in order to combine the symmetry groups of multiple order parameters into a single group with a higher symmetry. Several schemes enhancing symmetries based on generalized Dicke models exploiting the atomic  internal degrees of freedom have been discussed~\cite{Fan2014Hidden, Baksic2014Controlling, Moodie2018Generalized}.  The idea can also be applied to the external degree of freedom of a driven quantum gas coupled dispersively to two standing-wave linear cavities. This allows to realize a supersolid in the self-organized superradiant phase, if the two systems each breaking the discrete $\mathbf{Z}_2$ parity symmetry individually are combined to break a continuous $U(1)$ symmetry in real space and cavity field space.

\begin{figure}[t!]
\centering
\includegraphics[width=1\columnwidth]{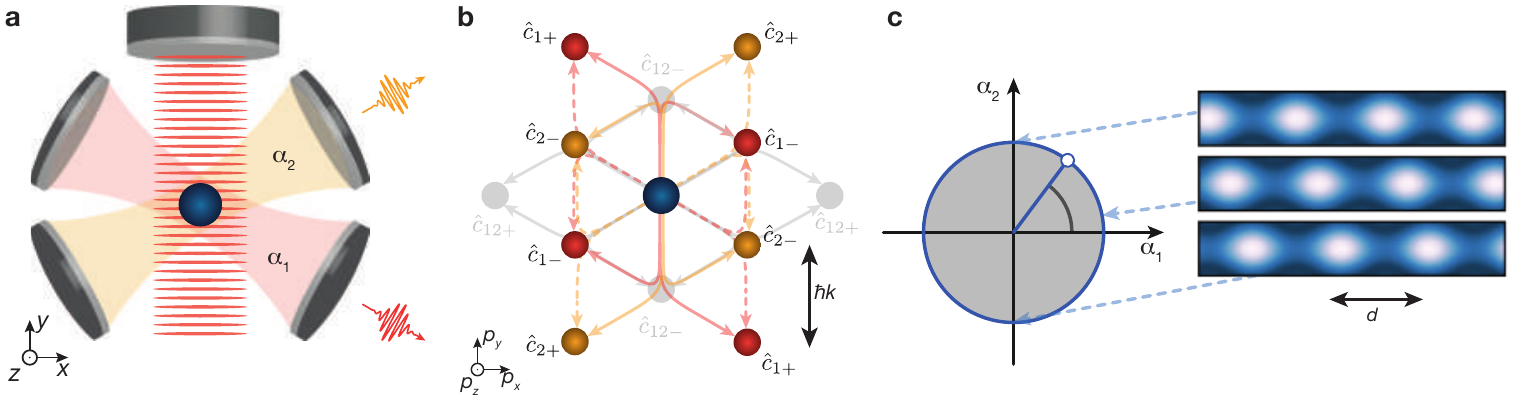}
\caption{Self-organization of a driven BEC in two crossed standing-wave cavities, giving rise to a supersolid state. (a) A BEC (blue) is symmetrically coupled to two linear cavities that cross at a 60\textdegree\ angle. The atomic gas is dispersively illuminated by a standing-wave transverse pump in the plane of the two cavities. (b) Momentum space representation of the involved two-photon scattering processes. Processes scattering photons from the pump field into either cavity are colored in red and yellow, respectively. Since pump and cavity fields are not orthogonal, a higher and a lower energetic momentum state ($\hat{c}_{j\pm}$) can be occupied. Scattering processes between the two cavities shown in grey  ($\hat{c}_{12\pm}$) are negligible for large atomic detunings. (c) A vector order parameter $\alpha = \alpha_1+i\alpha_2 = |\alpha| \exp(i\phi)$ can be constructed from the two scalar field-amplitude order parameters $\alpha_1$ and $\alpha_2$. Different ratios $\alpha_1/\alpha_2$ are energetically degenerate but lead to a spatial translation of the crystallized atomic density along the $x$ axis.
 Figure adapted and reprinted with permission from Ref.~\cite{Leonard2017Supersolid} published at 2017 by the Nature Publishing Group.}
\label{fig:CrossedCavities}
\end{figure}

In the experimental realization of a supersolid at the ETH Zurich group, a single BEC was coupled dispersively to two linear optical cavities crossed at a 60\textdegree\ angle, and subject to a standing-wave transverse pump beam~\cite{Leonard2017Supersolid}; see Figure~\ref{fig:CrossedCavities}(a). Self-organization of the atoms can take place in either of the cavities, each breaking a discrete parity symmetry. However, if the two cavity modes are tuned to be degenerate, photons from the pump field can be scattered into either of the two cavity modes without an energetic bias (neglecting scattering between the cavities for the moment), and simultaneous self-organization populating both modes in the two cavities can take place. The optical potential acting on the atoms is formed from the interference of all three involved fields~\cite{Safaei2015Triangular}. As a consequence, the $x$ position of the emergent lattice the atoms arrange into depends on the ratio of the field amplitudes $\alpha_1$ and $\alpha_2$ in the two cavities.  Conversely, the position of the atomic ordering along the $x$ axis determines the cavity-field amplitudes $\alpha_1$ and $\alpha_2$. Since the cavities are degenerate, the energy of the system is independent of the ratio $\alpha_1/\alpha_2$, which can take any value. Thus the self-ordered atomic crystal can be continuously translated along $x$ while the mode occupations are accordingly redistributed without energetic cost. The light-field amplitudes $\alpha_{1,2}$ are the individual scalar order parameters characterizing the BEC crystallization  according to either cavity-mode field; see Section~\ref{subsubsec:SuperradiantPhaseTransition}. They can be combined into a vector order parameter $\alpha = \alpha_1+i\alpha_2 = |\alpha| \exp(i\phi)$ that exhibits a continuous rotational $U(1)$ symmetry; see Figure~\ref{fig:CrossedCavities}(c). This idea is in close analogy to constructing the XY model from two Ising models with equal couplings.

The basic model introduced in Section~\ref{sec:basic-model} can be extended to two cavities ($j=1,2$) with photon annihilation operators $\hat{a}_j$, wave vectors $\mathbf{k}_{cj}$, detunings $\Delta_{cj}$ between pump frequency and cavity resonances, and maximum two-photon coupling rates $\eta_{0j}$. The resultant many-body Hamiltonian of the system reads
\begin{align}\label{eq:2cavitiesMB}
  \hat{H}_\mathrm{eff} &= -\hbar\sum_{j=1,2} \Delta_{cj} \hat{a}_j^\dagger \hat{a}_j
  + \int \hat{\psi}^\dagger (\mathbf{r}) \biggl\{ -\frac{\hbar^2 \nabla^2}{2M} 
  + \hbar V_0 \cos^2(\mathbf{k}_p \cdot \mathbf{r} + \varphi) 
  \nonumber\\
  & + \sum_{j=1,2} \left[ \hbar U_{0j} \cos^2(\mathbf{k}_{cj} \cdot \mathbf{r}) \hat{a}_j^\dagger \hat{a}_j
  + \hbar \eta_{0j}  \cos(\mathbf{k}_p \cdot \mathbf{r} + \varphi)\cos(\mathbf{k}_{cj} \cdot \mathbf{r}) ( \hat{a}_j^\dagger + \hat{a}_j)  \right] 
  \nonumber\\
   & + \hbar \sqrt{U_{01} U_{02}}  \cos(\mathbf{k}_{c1} \cdot \mathbf{r}) \cos(\mathbf{k}_{c2} \cdot \mathbf{r}) 
   ( \hat{a}_1^\dagger\hat{a}_2 + \hat{a}_2^\dagger\hat{a}_1) 
   \biggr\} \hat{\psi}(\mathbf{r}) d\mathbf{r}.
  \end{align}
Here, the wave vector of the pump is $\mathbf{k}_p$, and $\varphi$ is the experimentally tunable spatial phase of the pump lattice with respect to the crossing point of the two cavity modes. The last line in Equation~(\ref{eq:2cavitiesMB}) describes the scattering of photons between the two cavities.  

Following the procedure of Section~\ref{subsubsec:DickeModel}, we expand the atomic field operator $\hat{\psi}(\mathbf{r})$ in the momentum basis. The angle of 60\textdegree ~between the different modes implies that two-photon scattering processes with a bare energy of $\hbar \omega_- = 1\hbar \omega_r$ and of $\hbar  \omega_+ = 3\hbar \omega_r$ are possible as depicted in Figure~\ref{fig:CrossedCavities}(b). The resulting annihilation operators are labeled $\hat{c}_{j-}, \hat{c}_{j+}$ for the momenta involved in photon scattering processes between the pump field and cavity $j$, and $\hat{c}_{12-}, \hat{c}_{12+}$ for the momenta involved in photon scattering process between the two cavities. The annihilation operator for the zero-momentum state of the BEC is designated by $\hat{c}_0$. Neglecting terms that do not involve the macroscopic occupation of the BEC zero-momentum state ($\langle \hat{c}_0 \rangle \simeq \sqrt{N}$), the low-energy limit of the Hamiltonian~\eqref{eq:2cavitiesMB} takes the form,
\begin{align}\label{eq:SSHamiltonian}
\hat{H}_{\rm LE} &= -\hbar\sum_{j=1,2} \delta_{cj} \hat{a}_j^\dagger \hat{a}_j 
+\sum_{j=1,2}\sum_{s=-,+}
\left[\hbar \omega_s \hat{c}_{js}^\dagger\hat{c}_{js}
 +\frac{\hbar\eta_{0j}}{2\sqrt{2}} 
  \left(\hat{c}_{js}^\dagger \hat{c}_0 + \mathrm{H.c.}\right)\left( \hat{a}_j^\dagger + \hat{a}_j\right)\right]
\nonumber\\
  &+  \hbar \frac{\sqrt{U_{01} U_{02}}}{2\sqrt{2}} \left(\hat{c}_{12-}^\dagger \hat{c}_0  + \hat{c}_{12+}^\dagger \hat{c}_0+ \mathrm{H.c.}\right)  
  \left( \hat{a}_1^\dagger\hat{a}_2 + \hat{a}_2^\dagger\hat{a}_1\right),
\end{align}
where the dispersive shifts are incorporated in the detunings $\delta_{cj}$. The two-photon scattering rate between the pump and either cavity is proportional to $\eta_{0j}=\Omega_0 \mathcal{G}_{0,j}/\Delta_a$, while the two-photon scattering between the two cavities is proportional to $\sqrt{U_{01}U_{02}}=\mathcal{G}_{0,1} \mathcal{G}_{0,2}/\Delta_a$. The pump Rabi frequency $\Omega_0$ and the atomic detuning $\Delta_a$ are freely tunable experimental parameters. This allows to go to a regime of large detunings, where photon-scattering processes from one cavity to the another cavity become negligible. In this case, the Hamiltonian of the system reduces to the first line of Equation~(\ref{eq:SSHamiltonian}).

Besides the $U(1)$ superfluid gauge symmetry of the BEC, the system in the regime of negligible inter-cavity photon scattering exhibits a continuous $U(1)$ symmetry for the fine-tuned situation of  equal couplings ($\eta_{01} = \eta_{02}$) and equal detunings ($\delta_{c1} = \delta_{c2}$): The Hamiltonian [i.e., the first line of Equation~(\ref{eq:SSHamiltonian})] remains unchanged under a simultaneous rotation by an arbitrary angle $\theta$,
\begin{align}
\mathcal{R}=
\begin{pmatrix}
\cos\theta & -\sin\theta \\
\sin\theta & \cos\theta
\end{pmatrix},
\end{align}
in the space of photonic and atomic-momentum field operators,
\begin{align} \label{U1-transformation-2cavities}
\begin{pmatrix}
 \hat{a}_1\\
\hat{a}_2
\end{pmatrix}
\to
\mathcal{R}
\begin{pmatrix}
 \hat{a}_1\\
\hat{a}_2
\end{pmatrix},\quad
\begin{pmatrix}
\hat{c}_{1\pm}\\
\hat{c}_{2\pm}
\end{pmatrix}
\to
\mathcal{R}
\begin{pmatrix}
\hat{c}_{1\pm}\\
\hat{c}_{2\pm}
\end{pmatrix}.
\end{align}
This transformation shifts photon occupations between the two cavities and simultaneously redistributes the momentum excitations. The according continuous $U(1)$ symmetry $\hat{\mathcal U}(\theta)=e^{i\theta \hat{C}}$ has the generator
\begin{align}
  \hat{C} = -i\left[ \hat{a}_1^\dagger \hat{a}_2 -\hat{a}_2^\dagger \hat{a}_1
  +\sum_{s=\pm} \left(\hat{c}_{1s}^\dagger \hat{c}_{2s} - \hat{c}_{2s}^\dagger \hat{c}_{1s} \right) \right].
\end{align}

%----------------------------------------------------------------------------------------------------------------------------
\subsubsection{Observation of supersolidity}\label{subsubsec:TwoCavitieySuperSolid}
In the experiment~\cite{Leonard2017Supersolid}, a BEC was symmetrically coupled to the two crossed cavity modes, and the spatial phase $\varphi$ of the pump field was set to $\pi/2$. An atom-pump detuning $\Delta_a=-2\pi \times 2.4$~THz was chosen to minimize the cavity-cavity photon scattering processes. At a certain transverse pump lattice depth, photons leaking from both cavities were simultaneously recorded as a function of the detunings $\Delta_{cj}$ of the two cavities to construct a phase diagram of the system. Three extended phases were observed: a normal phase without self-organization and with no coherent field in either cavity for large enough detunings $\Delta_{cj}$, and two self-organized phases where either the cavity 1 or the cavity 2 was populated. In addition, self-organization where both cavities were simultaneous populated was only observed for fine-tuned parameters where the effective cavity detunings were identical, $\delta_{c1} = \delta_{c2}$.

\begin{figure}[t!]
\centering
\includegraphics[width=\columnwidth]{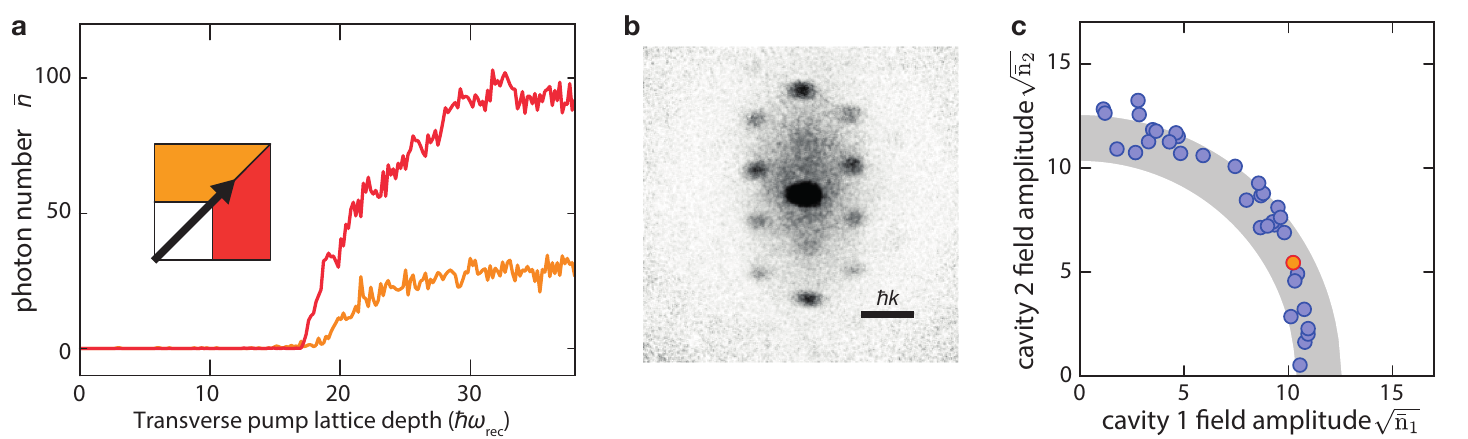}
\caption{Observation of supersolid properties across self-organization of a driven BEC in two crossed standing-wave cavities. (a) For degenerate cavity modes, the mean intra-cavity photon numbers were recorded while the transverse pump-lattice depth was increased. Both cavity modes are populated, however, with different amplitudes. (b) Absorption image after ballistic expansion of the atomic cloud. The observation of clear momentum peaks shows the phase coherence of the BEC and is a strong indication of the superfluid character. The observed momentum peaks also indicate a crystallized atomic density in accord with the pump and both cavity fields. (c) Field amplitudes in both cavities for repeated experimental runs under identical conditions signal the breaking of a continuous $U(1)$ symmetry. Along with the phase coherence of the BEC in (b), this is an indication of the supersolidity of the system. The data point highlighted in red corresponds to the trace shown in (a). Figure adapted and reprinted with permission from Ref.~\cite{Leonard2017Supersolid} published at 2017 by the Nature Publishing Group.}
\label{fig:Supersolid}
\end{figure}

An experimental trace of both mean intra-cavity photon numbers as a function of transverse pump power is shown in Figure~\ref{fig:Supersolid}(a) for the case of degenerate cavities (i.e., $\delta_{c1} = \delta_{c2}$). Light was detected in both cavities, however, with different amplitudes. The absorption image of the atomic cloud taken after ballistic expansion indeed showed momentum peaks corresponding to simultaneous self-organization in both cavities; see Figure~\ref{fig:Supersolid}(b). The clear observation of this matter-wave interference is a strong indication of the superfluid nature of the BEC.

In order to prove that the system across the double-superradiance crystallization breaks spontaneously the fine-tuned continuous $U(1)$ symmetry [see Equation~\eqref{U1-transformation-2cavities}] of the system, the experiment was repeated multiple times. Each repetition showed a different ratio of the field amplitudes in the two cavities, while all parameters were kept constant; see Figure~\ref{fig:Supersolid}(c). From this observation one can conclude that the superfluid crystallizes at a different location in each realization, thereby breaking a continuous spatial symmetry. The observed combination of a broken continuous spatial symmetry with the superfluid phase coherence of the atomic gas is a sufficient requirement for realizing a spontaneously formed crystal with dissipation-free flow, i.e. the minimal version of a supersolid.

\subsubsection{Excitation spectrum of the supersolid}
\label{sec:excitation-spectrum-supersolid-two-cavities}

The presence of an infinitely degenerate ground state has a direct consequence on the excitation spectrum of a system. In the Landau picture, the free energy of a supersolid resembles a sombrero hat. Accordingly, the fluctuations of the order parameter $\alpha=\alpha_1+i\alpha_2=|\alpha|\exp(i \phi)$ reveal two different excitations in the supersolid phase: An amplitude mode (or Higgs or massive Goldstone mode) originating from amplitude fluctuations $\delta |\alpha|$ at constant phase $\phi$ with a finite excitation energy due to the radial curvature of the Landau potential, and a phase mode (or Goldstone mode) with vanishing excitation energy which stems from phase fluctuations $\delta \phi$ at constant amplitude $|\alpha|$.

\begin{figure}[t!]
\centering
\includegraphics[width=\columnwidth]{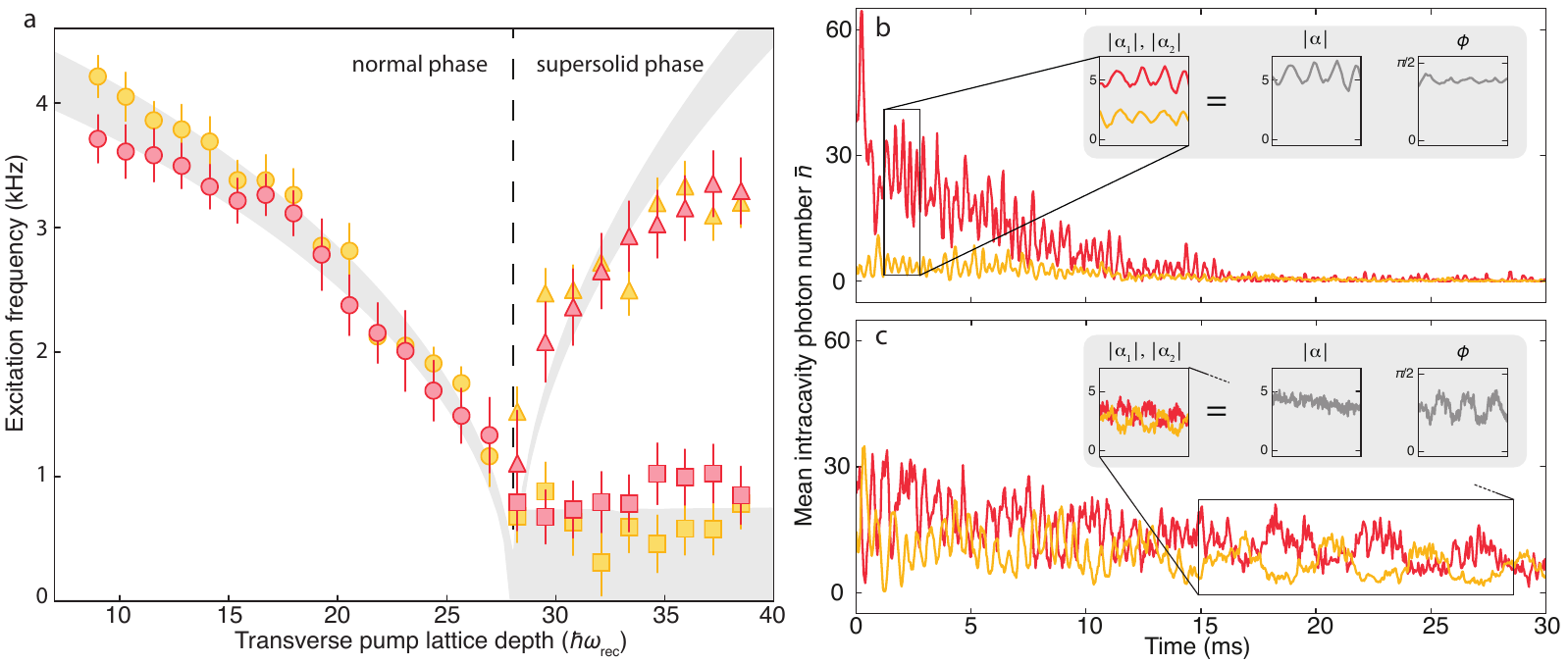}
\caption{Excitations of the supersolid state realized in the two crossed cavity setup. (a) Excitation spectrum recorded using cavity-enhanced Bragg spectroscopy shows a mode softening twoards the critical point at the normal-to-supersolid phase transition. Above the critical point, two excitation branches appear: the lower polariton remains (almost) gapless while the higher polariton hardens. Investigating real-time dynamics after a short excitation pulse at higher (b) and lower (c) polaritons reveals that these excitation branches can be identified as a gapped amplitude (Higgs) and a gapless phase (Goldstone) mode, respectively. Figure adapted and reprinted with permission from Ref.~\cite{Leonard2017Monitoring} published in 2017 by the American Association for the Advancement of Science.}
\label{fig:SupersolidExcitations}
\end{figure}

The excitation spectrum was probed in the two-crossed cavity system described above using cavity-enhanced Bragg spectroscopy (see Section~\ref{subsubsec:BraggSpectroscopy}) measuring the response of the system to a weak probe field injected into either cavity. Figure~\ref{fig:SupersolidExcitations}(a) shows the resulting excitation spectrum as a function of transverse pump-lattice depth. The second order phase transition from the normal phase to the supersolid is accompanied by the softening of the lowest polariton that vanishes at the critical point. In the supersolid phase, two excitation branches emerge. One polariton mode hardens again, similar to the rising excitation branch of the lattice supersolid, cf.\ Figure~\ref{fig:RotonSoftening}. The second polariton, however, remains at low energies. The observed non-zero excitation gap of the lower excitation branch in the supersolid phase is likely caused by imperfect fine-tuning of the degeneracy of the two cavities, and/or by residual cavity-cavity scattering as discussed below. In principle, photon dissipation from the cavities might also have a contribution to this non-zero excitation gap, as it mixes the amplitude and phase modes.

This separation of excitation frequencies inside the ordered phase suggests an interpretation as Goldstone and Higgs modes. The nature of these modes can, however, also be directly tested by observing real-time dynamics of the order parameter after exciting the system with a strong pulse. Since the supersolid order parameter is composed of the two cavity field amplitudes, $\alpha=\alpha_1+i\alpha_2$, both amplitude and phase response of the system can be directly reconstructed. Figure~\ref{fig:SupersolidExcitations}(b) and~(c) show real-time dynamics after a 1-ms-long excitation pulse quasi-resonant with the amplitude mode and the phase mode, respectively. While the amplitude (or Higgs) mode captures changes of the strength of the atomic density modulation, the phase (or Goldstone) mode is associated to changes in the position of the otherwise perfectly rigid crystal. The reconstructed phase and amplitude in the insets clearly confirm the expected character of the two modes.

\subsubsection{Coupled order parameters}
\label{sec:coupled-order-parameters}

The system with two crossed cavities was theoretically analyzed in Ref.~\cite{Lang2017Collective}, showing that the $U(1)$ symmetry is an approximate symmetry that holds in the limit of vanishing cavity fields. The atom-mediated scattering of photons between the two cavities favors a state with equal cavity populations for degenerate cavities and reduces the $U(1)$ symmetry to a $\mathbf{Z}_2\otimes\mathbf{Z}_2$ symmetry. An analysis of the system in terms of a Ginzburg-Landau picture was carried out~\cite{Lang2017Collective}  including  finite temperature effects and finite cavity-decay rates. The estimated non-zero effective mass of the  Goldstone mode due to higher order scattering processes slightly breaks the $U(1)$ symmetry (which is exactly restored only at the critical point~\cite{Wu2018Emergent}), giving rise to shallow minima in the Landau free-energy landscape.  However, the high probability to escape these shallow minima due to the added noise from cavity losses restores the continuous symmetry.

The rate of photon scattering processes between the two cavities can be tuned relative to the rate of photon scattering processes from the pump to either cavity by changing the atom-pump detuning $\Delta_a$; see Section~\ref{subsubsec:TwoCavities}. The inter-cavity photon scattering processes involve coupling of the zero-momentum BEC to additional higher momentum states. If one cavity is populated with photons, this cavity field can then act as an additional pump field for the second cavity and influence the critical point for self-organization~\cite{Morales2018Coupling}. Such inter-cavity photon scattering processes break the degeneracy and either favor a superposition of both cavity modes or a symmetry breaking between them.

\begin{figure}[t!]
\centering
\includegraphics[width=\columnwidth]{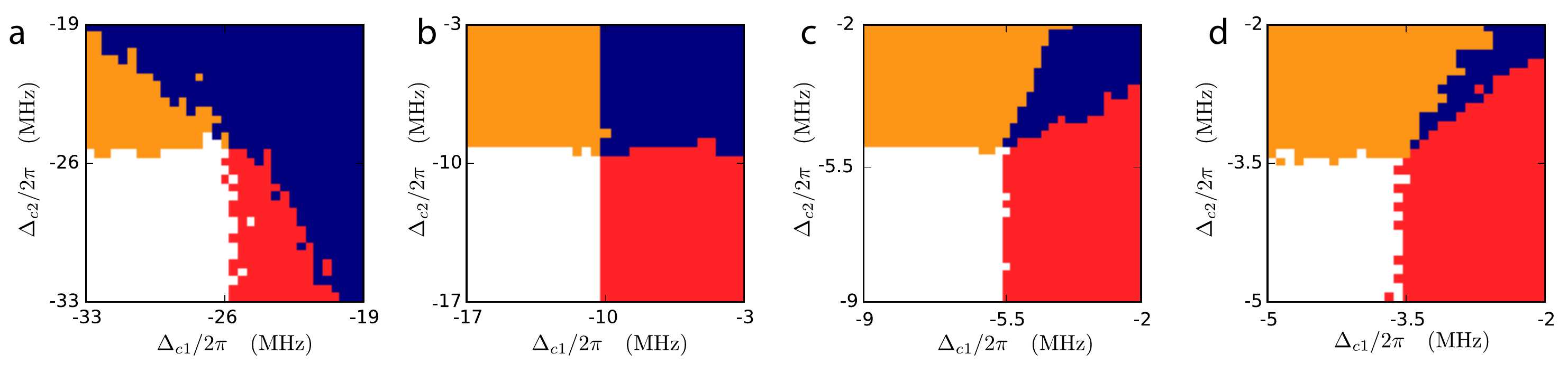}
\caption{Mutual enhancement and competition of self-organization in two crossed cavity modes. Experimental phase diagrams as a function of the two cavity detunigns $\Delta_{cj}$ for $\Delta_a/(2\pi)=$\SI{-73}{\giga \hertz}~(a), \SI{-400}{\giga \hertz}~(b), \SI{-1}{\tera \hertz}~(c), and \SI{-2.4}{\tera \hertz}(d). Four phases are visible, characterized by no occupation of either cavity (white region),  occupation of cavity 1 (red) , or of cavity 2 (yellow), and simultaneous occupation of both cavities (blue). By setting the atomic detuning closer to the atomic resonance, photon scattering processes between the two cavities are enhanced and occupation of one cavity lowers the critical point for self-organization in the other cavity. Figure adapted and reprinted with permission from Ref.~\cite{Morales2018Coupling} published at 2018 by the Nature Publishing Group.}
\label{fig:CouplingOPs}
\end{figure}

Figure~\ref{fig:CouplingOPs} shows experimental phase diagrams as a function of the two cavity detunings $\Delta_{cj}$ for different atomic detunings $\Delta_a$. For small absolute atomic detunings, inter-cavity photon scattering processes are strongly enhanced, and self-organization in one cavity favors self-organization in the other cavity. Thus, for a wide range of parameters, the system organizes simultaneously in both cavity modes, giving rise to a broken  $\mathbf{Z}_2\otimes\mathbf{Z}_2$ symmetry; see Figure~\ref{fig:CouplingOPs}(a). The opposite limit of large absolute atomic detunings has already been discussed in detail in Section~\ref{subsubsec:TwoCavitieySuperSolid}. In this case, the simultaneous superradiance in both cavities is suppressed due to the absence of photon scattering between the two cavities and only possible for degenerate cavity modes [i.e., along the diagonal in Figure~\ref{fig:CouplingOPs}(d)], resulting in a broken $U(1)$ symmetry.

In the limit of very strong inter-cavity photon scattering, it was theoretically discussed that an exotic vestigially ordered phase, located between the normal and the superradiant phases, can be entered~\cite{Gopalakrishnan2017Intertwined}. In this vestigially ordered phase, the atomic cloud acquires a density modulation, but neither cavity mode is macroscopically populated. In order to reach this phase, its kinetic energy cost has to be lower than the kinetic energy cost of the density wave from the standard self-ordered phase. This can be achieved for small angles between the two cavities (or in multi-mode cavities), since the atomic kinetic energy cost is given by the momentum difference between the two cavity-mode wave vectors. 

\begin{figure}[t!]
\centering
\includegraphics[width=0.9\columnwidth]{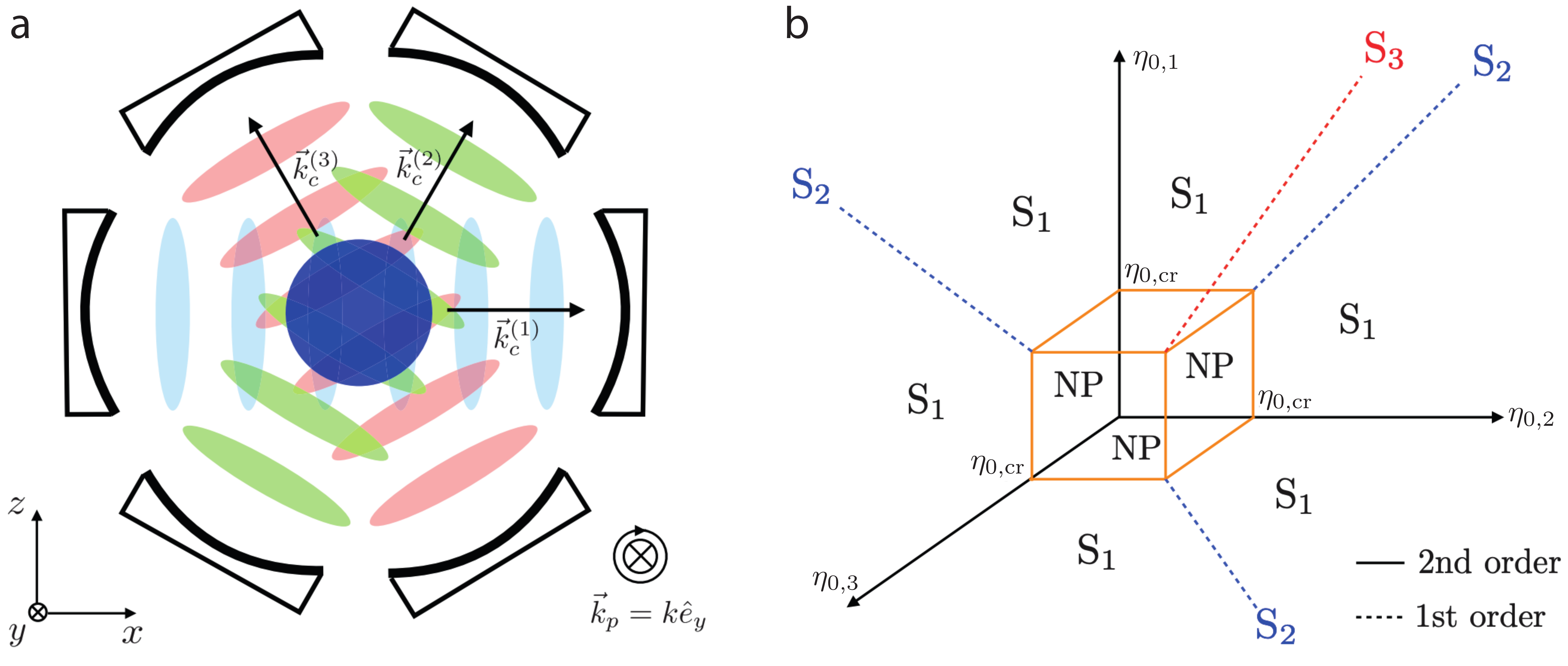}
\caption{Self-ordering of a transversely-driven BEC in three crossed cavities. (a) A BEC (blue circle) is placed at the intersection of three single-mode linear cavities, which are all located in the $x$-$z$ plane and aligned at the same angle of 60$^\circ$ from each other. An external, circularly polarized laser dispersively drives the atoms from the transverse ($y$) direction, resulting in two-photon scattering processes between the pump and the cavity modes. (b) Mean-field ground-state phase diagram of the system as a function of the effective couplings $\eta_{0,j}$ to the individual cavities. The orange cube indicates the normal phase with no photon in any cavity. Phases S$_1$ with a single superradiant cavity breaking a $\mathbf{Z}_2$ symmetry, phases S$_2$ with double superradiance breaking a $U(1)$ symmetry, and the phase S$_3$ with triple superradiance breaking a continuous $SO(3)$ symmetry are indicated. Figure adapted and reprinted with permission from Ref.~\cite{Chiacchio2018Emergence} \textcopyright~2018 by the American Physical Society.}
\label{fig:Chiacchio2018Emergence_Fig2}
\end{figure}

\subsubsection{Further enhancing symmetries}
The double crossed linear-cavity setup can be extended to more cavities with higher geometrical complexity, in order to realize higher continuous symmetries with more intriguing phase diagrams. By arranging three cavities in a plane symmetrically such that they make a 60\textdegree\ angle with one another and placing a transversely-driven BEC in their common intersection [see Figure~\ref{fig:Chiacchio2018Emergence_Fig2}(a)], one can obtain a much richer phase diagram as shown in Figure~\ref{fig:Chiacchio2018Emergence_Fig2}(b)~\cite{Chiacchio2018Emergence}. In particular, a continuous $SO(3)$ rotational symmetry is realized in the low-energy physics for symmetric coupling of the BEC to all cavities [the red dashed line denoted by S$_3$ in Figure~\ref{fig:Chiacchio2018Emergence_Fig2}(b)]. Here the continuous $SO(3)$ symmetry corresponds to the redistribution of photons among the three cavities at a given fixed total intensity. The continuous $SO(3)$ rotational symmetry is spontaneously broken at the onset of the triple superradiance, where all the cavity modes are populated. Correspondingly, two gapless Goldstone modes appear in the collective excitations of the system. This can be understood based on the fact that $SO(3)$ is a higher dimensional symmetry, related to three different continuous translations of the atomic positions along the directions perpendicular to each one of the cavity wave vectors. Therefore, the emergent superradiant potential can be located anywhere in the plane of the cavities.

\subsection{Supersolidity in a transversely pumped ring cavity}
\label{sec:supersolid-ring-cavity-transverse-pump}

Another scheme to engineer a supersolid has been proposed based on a BEC within a ring cavity with a pair of degenerate counterpropagating running electromagnetic modes $e^{\pm i k_cx}$ with the corresponding annihilation operators $\hat{a}_\pm$~\cite{Gopalakrishnan2009Emergent, Gopalakrishnan2010Atom, Mivehvar2018Driven}. This hinges quite intuitively on the ring geometry of the cavity~\cite{Nagy2006Self}, which does not impose any hard-wall boundary on the electromagnetic fields, thereby respecting the continuous translational symmetry of the space. 

\subsubsection{Intrinsic continuous $U(1)$ symmetry in ring-cavity geometries}
\label{sec:model-MF-supersolid-ring-cavity-transverse}

\begin{figure}[t!]
\centering
\includegraphics [width=\textwidth]{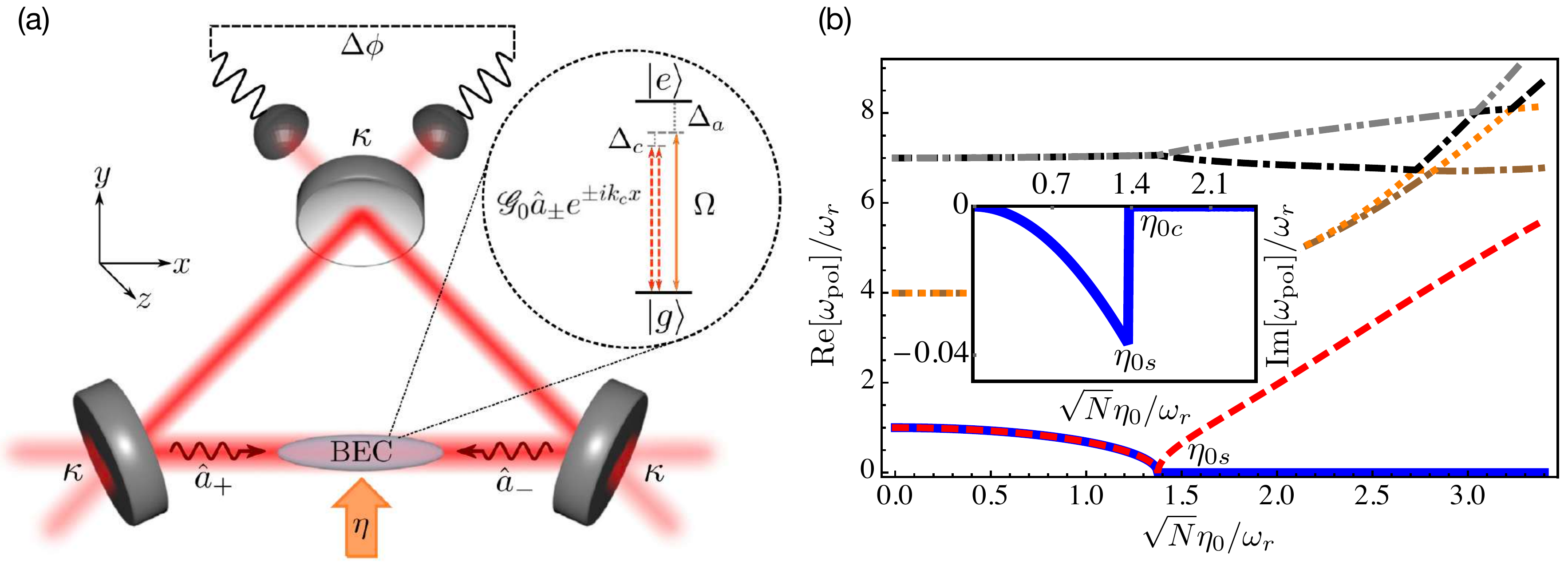}
\caption{Self-organization of a transversely-pumped BEC in a ring cavity, resulting in a supersolid state.
(a)~Schematic sketch of the system. A transversely-driven 1D BEC is dispersively coupled to a
pair of degenerate, counterpropagating modes of a ring cavity. 
Thanks to the ring-geometry of the cavity the system possesses a continuous $U(1)$ (simultaneous translation and phase rotation) symmetry,
in addition to the continuous $U(1)$ superfluid gauge symmetry.  (b)~The low-lying excitations of the system show the appearance of 
a gapless Goldstone mode (the blue solid curve) at $\eta_{0s}$ due to the spontaneously broken $U(1)$ symmetry, 
signaling the formation of a supersolid state. 
This Goldstone mode corresponds to the center-of-mass motion of the entire modulated BEC along the cavity axis 
dragging the superradiant optical lattice $V_{\rm SR}(x)$ with itself. Interestingly, as illustrated in the inset 
the damping (i.e., the imaginary part) of the Goldstone mode vanishes at the critical pumping 
strength $\eta_{0c}$ (slightly higher than $\eta_{0s}$),
resulting in a dissipationless particle current.
Figure adapted and reprinted with permission from Ref.~\cite{Mivehvar2018Driven} \textcopyright~2018 by the American Physical Society.} 
\label{Fig:Mivehvar2018Driven_Fig1andFig3}
\end{figure}

A one-dimensional BEC is dispersively coupled to two cavity modes $\hat{a}_\pm$ with strengths $\mathcal{G}_\pm(x)=\mathcal{G}_0 e^{\pm i k_cx}$ and illuminated coherently by a standing-wave pump laser with the Rabi frequency $\Omega_0$ in the transverse direction as depicted in Figure~\ref{Fig:Mivehvar2018Driven_Fig1andFig3}(a)~\cite{Mivehvar2018Driven}. Following a procedure similar to Section~\ref{sec:basic-model}~\cite{Moore1999Quantum}, in the rotating frame of the pump laser the system is described by the effective Hamiltonian 
$\hat{H}_{\rm eff}=
\int \hat\psi^\dag(x)\hat{\mathcal{H}}_{1,\rm eff}\hat\psi(x)dx
-\hbar\Delta_c(\hat{a}_+^\dagger\hat{a}_++\hat{a}_-^\dagger\hat{a}_-),$ 
with the effective single-particle atomic Hamiltonian density:
\begin{align} \label{eq:1-eff-H-ring}
\hat{\mathcal{H}}_{1,\rm eff}=&
-\frac{\hbar^2}{2m}\frac{\partial^2}{\partial x^2}
+\hbar U_0 \left(\hat{a}_+^\dag\hat{a}_++\hat{a}_-^\dag\hat{a}_-
+\hat{a}_+^\dag\hat{a}_- e^{-2ik_cx}
+\hat{a}_-^\dag\hat{a}_+ e^{2ik_cx}\right)\nonumber\\
&+\hbar\eta_0 \Big(\hat{a}_+ e^{ik_cx}+\hat{a}_- e^{-ik_cx}
+\text{H.c.}
\Big).
\end{align} 
The photon losses are accounted for by the master equation \eqref{eq:master-eq}, where the Liouvillean now includes both cavity modes,
$\mathcal{L} \hat\rho= \kappa\sum_{\ell=+,-}
(2 \hat{a}_\ell\hat\rho \hat{a}_\ell^\dagger - \{\hat{a}_\ell^\dagger \hat{a}_\ell, \hat\rho \}).$

In addition to the continuous $U(1)$ gauge symmetry, the system possesses another $U(1)$ symmetry, as the effective Hamiltonian $\hat{H}_{\rm eff}$ and the Liouvillean are \emph{both} invariant under a simultaneous spatial translation $x\to\mathcal{T}_{X}x=x+X$ and cavity-phase rotations $\hat{a}_\pm\to\mathcal{U}_X\hat{a}_\pm=\hat{a}_\pm e^{\mp i k_c X}$. This continuous $U(1)$ symmetry is ultimately related to the ring geometry of the cavity which does not impose a hard wall boundary on the electromagnetic fields and respects the continuous spatial translational symmetry. In contrast to the two crossed linear cavities where the external  $U(1)$ symmetry---i.e., the invariance under simultaneous continuous spatial translation and field rotation---is merely an approximate and fine-tuned symmetry as discussed in Section~\ref{sec:supersolidity-two-cavities}, the external $U(1)$ symmetry in the ring-cavity setup is an exact symmetry, independent of parameter regimes. 

The external $U(1)$ symmetry is spontaneously broken at the onset of the superradiant phase transition, where the field amplitudes $\alpha_\pm=\langle\hat{a}_\pm\rangle=|\alpha_\pm|e^{i\phi_\pm}$ acquire non-zero values with equal absolute values $|\alpha|=|\alpha_+|=|\alpha_-|$ and arbitrary phases $\phi_\pm$. That is, in the superradiant state the relative phase $\Delta\phi\equiv(\phi_+-\phi_-)/2$ of the two cavity modes is fixed to an arbitrary value between $0$ and $2\pi$, spontaneously breaking the the continuous external $U(1)$ symmetry. A spontaneously chosen value of $\Delta\phi$ fixes the position of the minima of the emergent superradiant potential
$V_{\rm SR}(x)=2U_0|\alpha|^2
\cos(2k_cx+2\Delta\phi)+4\eta_0|\alpha|\cos(k_cx+\Delta\phi)\cos(\Phi)$, 
with $\Phi\equiv(\phi_++\phi_-)/2$ being the total phase. In accordance with the emergent superradiant lattice, the BEC crystallizes (i.e., the BEC density is modulated), spontaneously breaking the continuous translational invariance. The crystalline state, nevertheless, inherits the superfluidity of the BEC with a long-range phase coherence. Therefore, the resultant steady state in the superradiant phase comprises a supersolid~\cite{Mivehvar2018Driven}.

In fact, full quantum considerations show that at the onset of the superradiance a superposition of field amplitudes with different phases $e^{\pm i k_cx}$ correlated with the corresponding density fluctuations peaked at $x$ emerges, forming a highly entangled atom-field state: $\ket{\Psi}= \int_0^{\lambda_c} dx \ket{\psi_x}\otimes\ket{\alpha e^{ik_cx}} \otimes\ket{\alpha e^{-ik_cx}}$~\cite{Gietka2019Supersolid}. This highly entangled state is very fragile, susceptible to quantum fluctuations and noises. This state subsequently collapses to a state with a certain random relative phase via quantum jumps induced by cavity photon losses, forming the supersolid state: $\ket{\Psi_0}=\ket{\psi_{x_0}}\otimes\ket{\alpha e^{i k_cx_0}} \otimes\ket{\alpha e^{-ik_cx_0}}$.

\subsubsection{Collective excitations and supersolidity}
\label{sec:collective-excitations-ring-cavity}

The collective excitations of the system shown in Figure~\ref{Fig:Mivehvar2018Driven_Fig1andFig3}(b) confirm the supersolidity of the system~\cite{Mivehvar2018Driven}. In particular, a gapless Goldstone mode (the solid blue curve) appears beyond the pump strength $\eta_{0s}$ (cf.\ Figure~\ref{fig:poles_fermiVSbose}), corresponding to the spontaneously broken continuous $U(1)$ symmetry. The Goldstone mode corresponds to the center-of-mass motion of the entire modulated BEC along the cavity axis dragging the superradiant optical lattice $V_{\rm SR}(x)$ with itself. The lowest gapped polariton branch (the dashed red curve) instead corresponds to a Higgs amplitude mode. These are reminiscent of the observed Goldstone and Higgs modes in the crossed-cavity experiment, cf.\ Figure~\ref{fig:SupersolidExcitations}.  Note that the self-ordering in the ring cavity (and the two crossed linear cavities) is in sharp contrast to the self-organization in a single linear cavity, where only a discrete $\mathbf{Z}_2$ symmetry is spontaneously broken and the first excitation gap closes at the critical pump strength but then re-opens again; cf.\ Figure~\ref{fig:RotonSoftening}.

The imaginary parts of the polaritons exhibit a peculiar behavior. Above the critical pump strength $\eta_{0c}$, all the collective excitations except the gapless Goldstone mode acquire imaginary parts. The imaginary part of the lowest polariton mode as a function of $\sqrt{N}\eta_0/\omega_r$ is shown in the inset of Figure~\ref{Fig:Mivehvar2018Driven_Fig1andFig3}(b). We note that although this mode is damped for small pump strengths, the imaginary part vanishes at the critical pump strength $\eta_{0c}$ (note that the corresponding real part vanishes at the slightly lower pump strength $\eta_{0s}$, where the damping reaches its maximum value; see also Figure~\ref{fig:poles_fermiVSbose}). This means that the center of mass of the entire modulated BEC can move freely along the cavity axis dragging the superradiant optical lattice with itself without experiencing any friction. The fact that supersolidity survives even in the presence of
dissipation is due to the fact that the corresponding Lindblad operators also respect the $U(1)$ symmetry of the system. Such a dissipationless supersolid particle flow in a ring cavity---an open quantum system---can be used as a very sensitive tool for a precise, non-destructive measurement of various forces including the gravitational force~\cite{Gietka2019Supersolid}; see Section~\ref{sec:quantum-sensing} for more details.

\subsection{Supersolidity in a longitudinally pumped ring cavity}
\label{sec:supersolid-ring-cavity-longitudinal-pump}
\begin{figure}[t!]
\centering
\includegraphics[width=\columnwidth]{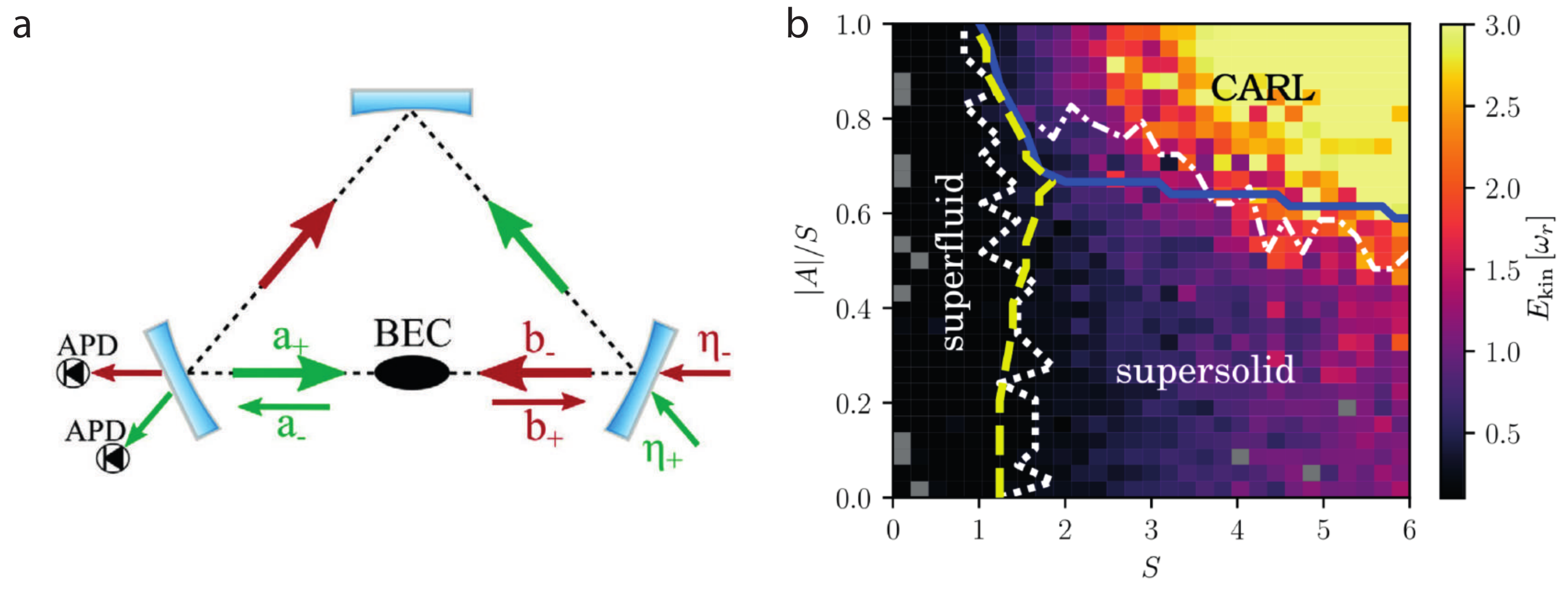}
\caption{Supersolid state in a longitudinally pumped ring-cavity--BEC geometry. (a)~A BEC placed inside a ring cavity dispersively couples to two pairs of transverse running-wave modes $\{\hat{a}_\pm,\hat{b}_\pm\}$, from which the counterpropagating modes $\hat{a}_+$ and $\hat{b}_-$ are pumped at rates $\eta_+$ and $\eta_-$. (b) Phase diagram of the system as a function of the sum $S$ of the two pump-field strengths and their normalized asymmetry $|A|/S$. Shown as color scale is the kinetic energy transferred to the atomic system within an interaction time of \SI{1}{\milli \second}. A superfluid, a supersolid, and an instable phase of collective atomic recoil lasing (CARL) are visible. 
Figure adapted and reprinted with permission from Ref.~\cite{Schuster2020Supersolid} \textcopyright~2020 by the American Physical Society.}
\label{fig:RingSuperSolid}
\end{figure}

In a related ring-cavity--BEC geometry, signatures of supersolidity have been experimentally observed  in a BEC of $^{87}$Rb atoms~\cite{Schuster2020Supersolid}. This scheme exploited \emph{two} pairs of running-wave transverse modes (TEM$_{00}$ and TEM$_{20}$), where the modes in each pair are degenerate (i.e., propagating in opposite directions) and do not interfere with the modes in the other pair because of a large frequency offset between the pairs. Two counterpropagating modes, one mode from each pair, were longitudinally pumped by coherent laser fields through a cavity mirror; see Figure~\ref{fig:RingSuperSolid}(a). For symmetric and strong enough pumping, the uniform BEC was self-ordered into a stationary crystalline state, breaking the continuous spatial symmetry of the system. The crystalline state maintains the long-range phase coherence of the BEC; hence, it is a supersolid state. Superfluidity of the atomic crystal was inferred from the observation of distinct momentum peaks after ballistic expansion. The breaking of the continuous spatial symmetry was, however, not proved in the experiment.

The system also remains in the stable self-ordered phase when the pump fields are not fully symmetric. The atomic system adapts to a stable momentum distribution such that the net force on its center of mass vanishes. This stabilizing mechanism compensating asymmetric pumping is related to cavity cooling, and makes this system more robust. For strongly asymmetric pumping, however, there is no stationary state. In particular, above a certain pump asymmetry the system enters the collective atomic recoil laser (CARL) regime; see Section~\ref{subsubsec:RingCavities}. Figure~\ref{fig:RingSuperSolid}(b) shows the phase diagram of the system as a function of the sum of the two pump strengths, $S$, and the relative pump asymmetry, $|A|/S$. Three phases are visible: the normal superfluid phase, the stable supersolid phase, and a runaway CARL-like state.

\subsection{Supersolidity in other systems}
The supersolids induced by the interaction with dynamical light fields as described in this section crystallize with a lattice constant that is determined by the wave length of the light. As discussed in Section~\ref{subsubsec:cavity-induced-int}, the light fields induce an effective interatomic interaction of global  range with a particular spatial structure stemming from the interfering light fields. The crystallization is thus also expected to be defect-free and homogeneous, because all atoms couple equally to the cavity modes. This results in a perfectly rigid crystalline structure, which inhibits the presence of phonons at non-zero wave numbers. Similar arguments also apply for supersolid states induced by spin-orbit coupling~\cite{Li2017A}.

More recently, supersolidity has also been engineered using quantum gases with strong magnetic dipole moments~\cite{Tanzi2019Observation, Bottcher2019Transient, Chomaz2019Long}, exploiting the interplay between trap shapes and collisional and dipolar interactions. In contrast to the light-induced supersolids in cavities, dipolar supersolids can have phononic excitations. System sizes have so far, however, been limited to a handful of lattice sites. Light-induced supersolids can, in principle, also show phononic excitations, provided a continuum of electromagnetic modes is available. This is for instance possible in confocal cavities, as discussed in Section~\ref{sec:multimode}, and even in free space~\cite{Ostermann2016Spontaneous}.

\section{Multimode cavities: supermode polaritons, quantum crystalline phases, and droplets}
\label{sec:multimode}

In this section, we shall consider the case where atoms are coupled to a large number of modes of an optical cavity. We will discuss novel phenomena emerging in this case with respect to the few-mode cases discussed in the preceding sections, and illustrate consequences for the superradiant self-organization and corresponding crystalline phases. In the absence of a multifrequency or a Floquet drive \cite{Kraemer2014Self,Clark2019Interacting,Johansen2020Multimode}, coupling to a large number of cavity modes is achieved by tuning the cavity to a degenerate point, where several tens or hundreds of modes are almost resonant with the driving laser (see Figure~\ref{fig:CavityParameter}). In this situation, the collective polaritonic excitation, whose coherent occupation corresponds to superradiance, is a superposition of a large number of cavity modes. The superposition of many cavity modes generically leads to destructive interference which makes the scattered light spot more local. As an important consequence, cavity-mediated atom-atom interactions become finite range, offering a promising route to observe interaction-induced fluctuations and beyond-mean-field physics. As the largest constructive interference is obtained by a localized atomic cloud, the cavity-mediated interaction is attractive and strongest at short distances. Unless counteracted by sufficiently strong Van-der-Waals interatomic repulsion, this leads to the formation of superradiant, self-bound crystalline droplets instead of extended crystals.

\subsection{Cavity-mode structure and stable degenerate points}
\label{subsec:degenerate_cavity_modes}

The generalization of the theoretical model of Equation~\eqref{eq:H_eff_1comp} to the multimode case is straightforward and reads
\begin{align} 
\label{eq:H_eff_1comp_multimode}
\hat{H}_{\rm eff}=\int \hat\psi^\dag(\mathbf{r}) 
\bigg\{&-\frac{\hbar^2}{2M}\boldsymbol{\nabla}^2+V_{\rm ext}(\mathbf{r}) 
+\frac{\hbar}{\Delta_a} \Big[\left|\Omega(\mathbf{r})\right|^2 
+\sum_{\nu,\nu'}\mathcal{G}_{\nu'}^*(\mathbf{r}) \mathcal{G}_\nu(\mathbf{r})\hat{a}_{\nu'}^\dag\hat{a}_\nu
\nonumber\\
& +\sum_\nu\big(\Omega^*(\mathbf{r}) \mathcal{G}_\nu(\mathbf{r})\hat{a}_\nu+\text{H.c.}\big) \Big]
+\frac{g_0}{2}\hat{n}(\mathbf{r}) \bigg\}\hat{\psi}(\mathbf{r})d\mathbf{r}
-\hbar\sum_\nu\Delta_\mu\hat{a}_\nu^\dag\hat{a}_\nu,
\end{align}
where we have allowed for complex electromagnetic mode functions. The functions $\mathcal{G}_\nu(\mathbf{r})$ are defined as in Equation~\eqref{eq:HermiteToRabi}. Note that the contact interaction term proportional to $g_0$ is identically zero for fermions due to the Pauli exclusion principle. We assume a single frequency $\omega_p$ for the pump laser, defining the relative detuning of cavity modes $\Delta_\nu=\omega_p-\omega_\nu$. Here $\nu=(l,m,n)$ is a multi-index composed of the longitudinal mode index $l$ and the transverse mode indices $m,n$ of a given TEM$_{mn}$ cavity mode (see Figure~\ref{fig:CavityParameter}).
In addition, we define a loss rate $\kappa_\nu$ for each mode inducing the Lindblad dynamics [cf.\ Equation~\eqref{eq:Liouville-op}]
\begin{align}
\label{eq:Liouville-op-multimode}
\mathcal{L}\hat{\rho}=\sum_\nu \kappa_\nu
\left(2\hat{a}_\nu\hat{\rho}\hat{a}_\nu^\dag
-\hat{a}_\nu^\dag\hat{a}_\nu\hat{\rho}
-\hat{\rho}\hat{a}_\nu^\dag\hat{a}_\nu\right).
\end{align}

As discussed in Section~\ref{subsubsec:optical_resonators}, in the near-planar case all the modes of a given family [see Figure~\ref{fig:CavityParameter}(c)] share the same longitudinal index $l$ as well as the same longitudinal standing-wave modulation $\cos(k_c x)$. That is, the whole family possesses the same number of nodes in the longitudinal direction. The higher-frequency modes within the family are then characterized by an increasing transverse index $m+n$.

On the other side of the stability diagram in the concentric case (see Figure~\ref{fig:CavityParameter}), modes belonging to the same family instead share the same total number of nodes $l+m+n$. These degenerate modes thus can have a different number of longitudinal nodes, captured by the index $l$. Sufficiently away from the cavity center $x\gg \lambda_c$, all the modes in a given family still show a longitudinal periodic modulation with wave number $k_c$. Indeed, due to the small beam waist $w_0\sim\lambda_c$, the Gouy phase $\varphi^{\rm Gouy}_{mn}(x)$ [see Equation~\eqref{eq:higher_order_gaussian_beam}] quickly saturates to a spatially independent constant away from the center. Differently from the near-planar case, however, in the near-concentric case the Gouy phase is crucial in determining the mode frequency as well as in inducing a phase offset $\varphi^{\rm offset}_{mn}$ which differs for modes with different total transverse index $m+n$.

Requiring that the electric field satisfies the boundary conditions at the mirrors $x=\pm\ell_{\rm res}/2$ fixes the phase offset $\varphi^{\rm offset}_{mn}$. Equation~\eqref{eq:higher_order_gaussian_beam} then implies that modes with different $m+n$ are offset with respect to each other by a phase $\varphi^{\rm offset}_{mn}=(m+n)\arctan(\lambda_c\ell_{\rm res}/2\pi w_0^2)$. Let us consider the oscillating part (i.e., without the Gouy phase shift) of the longitudinal modulation in combination with the offset just discussed. In the concentric limit, $\arctan(\lambda_c\ell_{\rm res}/2\pi w_0^2)=\pi/2$, so that different transverse modes come with a periodic longitudinal modulation $\cos(k_c x)$ for $m+n=0$, $\sin(k_c x)$ for $m+n=1$, $-\cos(k_c x)$ for $m+n=2$, $\dots$. In the confocal case instead, $\arctan(\lambda_c\ell_{\rm res}/2\pi w_0^2)=\pi/4$, so that different transverse modes of a family come with a modulation $\cos(k_c x)$ for $m+n=0$, $\sin(k_c x)$ for $m+n=2$, $-\cos(k_c x)$ for $m+n=4$, $\dots$, and similarly for an odd family $m+n=1,3,5,\dots$. 

Approaching both the plane-parallel and the concentric limit in principle allows to reach full mode degeneracy. The concentric degeneracy is preferable over the plane-parallel one since the latter is characterized by a diverging mode volume, hence a vanishing light-matter coupling [see Equation~\eqref{eq:vacuum_Rabi_split}]. We will return to the concentric case in Section~\ref{subsec:multimode_droplets}.
However, the extent to which degeneracy can be achieved in both limits is affected by resonator stability. Experiments realizing multimode cavity QED with ultracold atoms have for this reason so far utilized the confocal degeneracy \cite{Kollar2015An,Kollar2017Supermode,Vaidya2018Tunable,Guo2019Sign}. As shown in Figure~\ref{fig:CavityParameter}, the latter lies indeed in the middle of the stability diagram and can successfully be reached, as we will discuss in the next two sections.

Before proceeding, however, let us stress one difference between confocal and concentric degeneracies which is especially relevant for the purpose of studying many-body physics. In the concentric case, a degenerate family includes modes with all possible numbers of nodes, so that the full destructive interference resulting from their superposition allows for light fields which can be local over the characteristic distance between atoms. On the other hand, in the confocal case, a degenerate family excludes every second mode, so that the destructive interference is only partial and revivals of cavity fields at large distances appear (see Section~\ref{subsec:tuning_interaction_range}).

\subsection{Multimode superradiance}
\label{subsec:multimode_superradiance}

A few years after the first observation of steady-state superradiance with a single-mode cavity discussed in Section~\ref{sec:SR_discrete}, the Stanford group observed the superradiant transition with Bose-condensed atoms also in the multimode regime, by tuning a cavity around the confocal point~\cite{Kollar2017Supermode}. The main novelty in this respect is the emergence of cavity supermode polaritons. This mixes the bare cavity modes by the intra-cavity atomic medium. If this system becomes superradiant, it forms a supermode polariton condensate.

To understand this concept let us revisit the computation of collective polaritonic excitations of Section~\ref{subsec:excitations_probing} in the present multimode case described by Equations~\eqref{eq:H_eff_1comp_multimode} and~\eqref{eq:Liouville-op-multimode}. The linearized equation of motion for the field amplitude of mode $\nu$ reads [cf.\ Equation~\eqref{eq:MF-eq-a-loss-linear}],
\begin{equation}
  \label{eq:MF-eq-amu-loss-linear}
  i\partial_t\delta\hat{a}_\nu=\sum_{\nu'}\left[-(\Delta_\nu+i\kappa_\nu)\delta_{\nu,\nu'}
  +\chi_{{\rm stat},\nu\nu'}\right](\delta\hat{a}_{\nu'}+\delta\hat{a}_{\nu'}^\dag),
\end{equation}
with the static polarization function given by
\begin{equation}
  \label{eq:stat_polarization_multimode}
  \chi_{{\rm stat},\nu\nu'}=\sum_{\ell,\ell'} 
  \frac{n(\epsilon_\ell)-n(\epsilon_{\ell'})}{\epsilon_{\ell}-\epsilon_{\ell'}}
  \langle u_{\ell}|\eta_\nu| u_{\ell'}\rangle \langle u_{\ell'}|\eta_{\nu'}| u_{\ell}\rangle.
\end{equation}
Here $|u_{\ell}\rangle$ are the single-particle eigenstates of the mean-field atomic Hamiltonian with the corresponding eigenvalues $\epsilon_\ell$, and $\eta_\nu(\mathbf{r})=\Omega(\mathbf{r})\mathcal{G}_\nu(\mathbf{r})/\Delta_a=\langle\mathbf{r}|\eta_{\nu}|\mathbf{r}\rangle$. Furthermore, we have for simplicity assumed to be in the non-superradiant phase, $\alpha_{\nu,0}=0$, neglected the terms quadratic in the photon operators, and also assumed real laser and cavity-mode functions, $\{\Omega(\mathbf{r}),\mathcal{G}_\nu(\mathbf{r})\}\in\mathbb{R}$.

The main observation is that the polarization function is in general a non-diagonal matrix in the cavity-mode space. This means that the presence of the atoms leads to scattering of photons between the cavity modes, thus mixing them to form a supermode polariton. An additional source of coupling between different modes besides the atomic cloud is mirror aberrations~\cite{Kollar2017Supermode}.

As discussed in relation to Equations~\eqref{eq:MF-eq-aadag-loss-linear-fourier} and~\eqref{eq:poles-retarded-cavity}, the polariton excitations are obtained as poles of the Green's function of the electromagnetic field. Since the polarization function is not diagonal, the original two-by-two matrix structure resulting from the positive and negative frequency components of the Green's function in Equation~\eqref{eq:ret-green-func} acquires now a block structure, where each of the four entries is a matrix in mode space with the non-diagonal part coming from $\chi_{\rm{stat},\nu\nu'}$. It is clear that each polaritonic pole acquires in general contributions from all cavity-mode sectors, i.e., becomes a supermode polariton.

\begin{figure}[t!]
  \begin{center}
    \includegraphics[width=0.7\columnwidth]{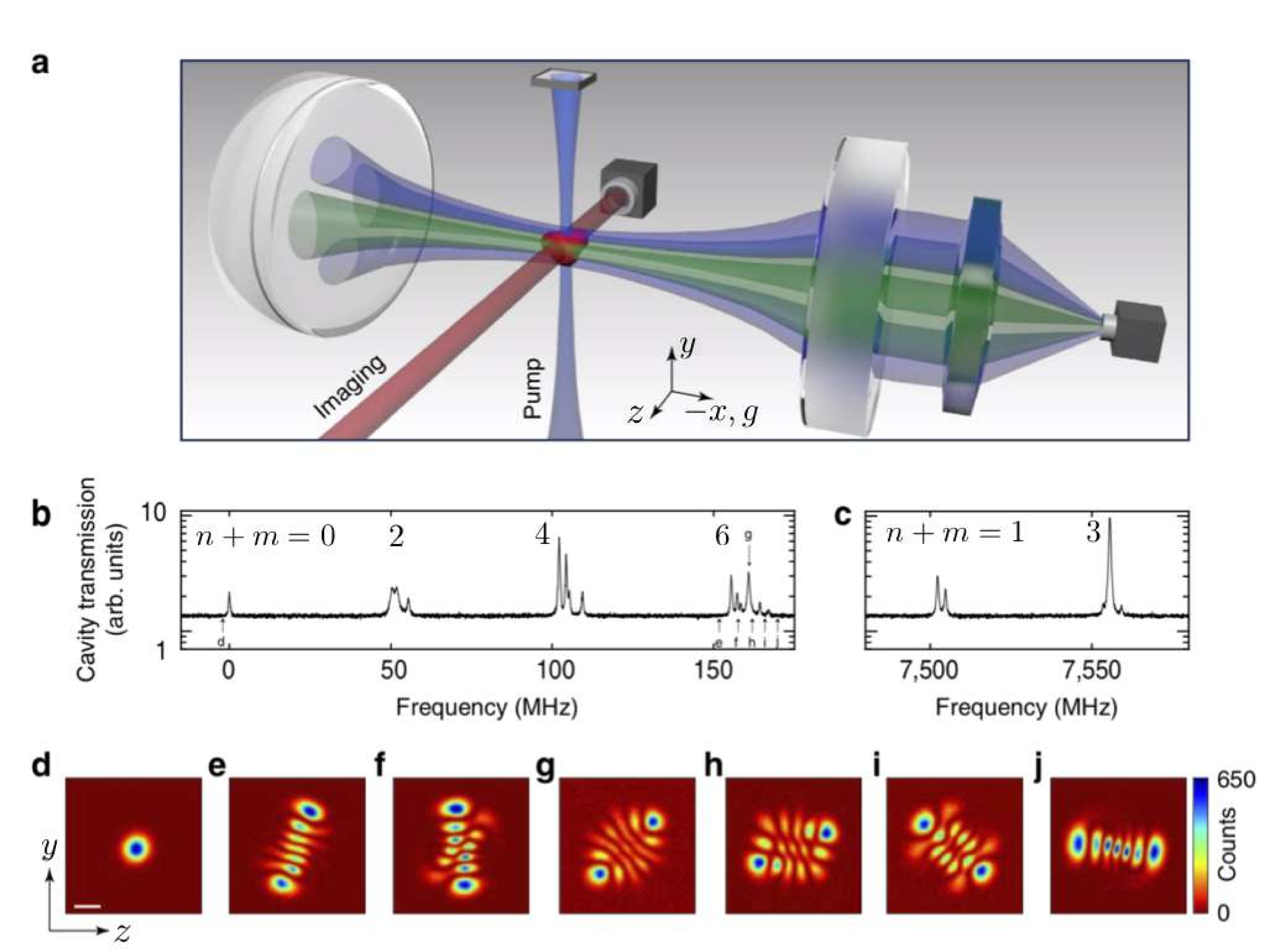}
    \caption{Observation of supermode-polariton superradiance with a BEC inside a near-confocal cavity. (a)~Experimental configuration with the BEC at the center of the cavity. (b,c) Cavity transmission spectrum showing the quasi-degenerate mode families, even in (b) and odd in (c), in a near-confocal cavity. (d-e) Superradiant emission above the superradiant-threshold coupling for a laser frequency indicated by the corresponding arrows in panel (b). A varying admixture from different transverse TEM$_{mn}$ modes is observed by changing the laser frequency. Figure adapted and reprinted with permission from Ref.~\cite{Kollar2017Supermode} published at 2017 by the Nature Publishing Group.}
    \label{fig:SupermodePolaritonCondensation}
  \end{center}
\end{figure}

At the superradiant instability point where one of the complex polariton frequencies vanishes, the admixture of each bare cavity mode to the unstable supermode polariton can be read out from the eigenvectors of the inverse electromagnetic Green's function at $\omega=0$.
This phenomenology has been experimentally analyzed in Ref.~\cite{Kollar2017Supermode} for a BEC trapped inside a near-confocal cavity such that the atomic-cloud's extension was significantly smaller than the mode waist, leading to an almost mode-independent polarization function as discussed above. The experimental results are summarized in Figure~\ref{fig:SupermodePolaritonCondensation}. The resonator length is tuned to a near-confocal point with the corresponding mode frequencies visible in the transmission spectrum of Figure~\ref{fig:SupermodePolaritonCondensation}(b), where the even transverse-mode families are still separated by few tens of MHz. Referring to Section~\ref{subsubsec:optical_resonators}, we recall that a given family can be labelled by just $m+n$, as the longitudinal index $l$ is fixed in the near-confocal condition by $l+(m+n)/2$ being equal to a constant. By tuning the laser frequency to different values within the $m+n=6$ sub-family, a superradiant emission containing a different admixture of cavity modes is observed.

When the atomic cloud is coupled to a single mode of the cavity, the superradiant phase transition breaks spontaneously the discrete $\mathbf{Z}_2$ symmetry of the system, by fixing the phase of the cavity field relative to the pump laser (see Section~\ref{sec:SR_discrete}). This is the case since the longitudinal modulation has a single possible phase offset, and so does the corresponding density modulation $\propto\cos(k_cx)$ of the atomic cloud in the superradiant phase. As we have seen in Section~\ref{subsec:degenerate_cavity_modes}, the situation changes at the confocal degeneracy, where an additional longitudinal modulation $\propto\sin(k_cx)$ is also possible. This can effectively lead to a continuous $U(1)$ symmetry \cite{Guo2019Emergent}, as we will discuss in the next section. For now, we just note that in the near-confocal case of Figure~\ref{fig:SupermodePolaritonCondensation} the distance from confocality is large enough to allow for only a single possible longitudinal modulation and thus realizes only the standard $\mathbf{Z}_2$ symmetry breaking.

\subsection{Cavity-tuning of the atom-atom interaction in the confocal regime}
\label{subsec:tuning_interaction_range}

\subsubsection{The interatomic potential at the confocal point}
\label{subsubsec:confocal_potential}

Following the procedure of Section~\ref{subsubsec:cavity-induced-int} for many cavity modes, the cavity-mediated interactions between two atoms in the multimode case can be obtained as,
\begin{align}
  \label{eq:multimode_interaction_potential} 
  \mathcal{D}(\mathbf{r},\mathbf{r'})=\frac{2\mathcal{G}_0^2\Omega(\mathbf{r})\Omega(\mathbf{r'})}{\Delta_a^2}
  \sum_{m,n}
  \frac{\Delta_{mn}\mathcal{E}_{mn}(\mathbf{r}) \mathcal{E}_{mn}(\mathbf{r'})}
  {\mathcal{E}_0^2\left(\Delta_{mn}^2+\kappa_{mn}^2\right)}
  \equiv 
  \frac{2\mathcal{G}_0^2\Omega(\mathbf{r})\Omega(\mathbf{r'})}
  {\Delta_a^2|\Delta_0|}\tilde{\mathcal{D}}(\mathbf{r},\mathbf{r'}),
\end{align}
where the last equality defines the dimensionless interaction potential $\tilde{\mathcal{D}}$ solely due to the cavity field. The Hermite-Gaussian functions $\mathcal{E}_{{mn}}(\mathbf{r})$ are given in Equation~\eqref{eq:higher_order_gaussian_beam}.
Having restricted to a single almost-degenerate family, the sum over modes in Equation~(\ref{eq:multimode_interaction_potential}) runs only over the transverse indices $\{m,n\}$, since the longitudinal index $l$ for given transverse indices $m$ and $n$ is determined by the resonance condition~\eqref{eq:transverse_mode_spacing}. As discussed already previously, having chosen a confocal degeneracy, only half of the transverse modes contribute, either with even or odd $m+n$.

\begin{figure}[t!]
  \begin{center}
    \includegraphics[width=\columnwidth]{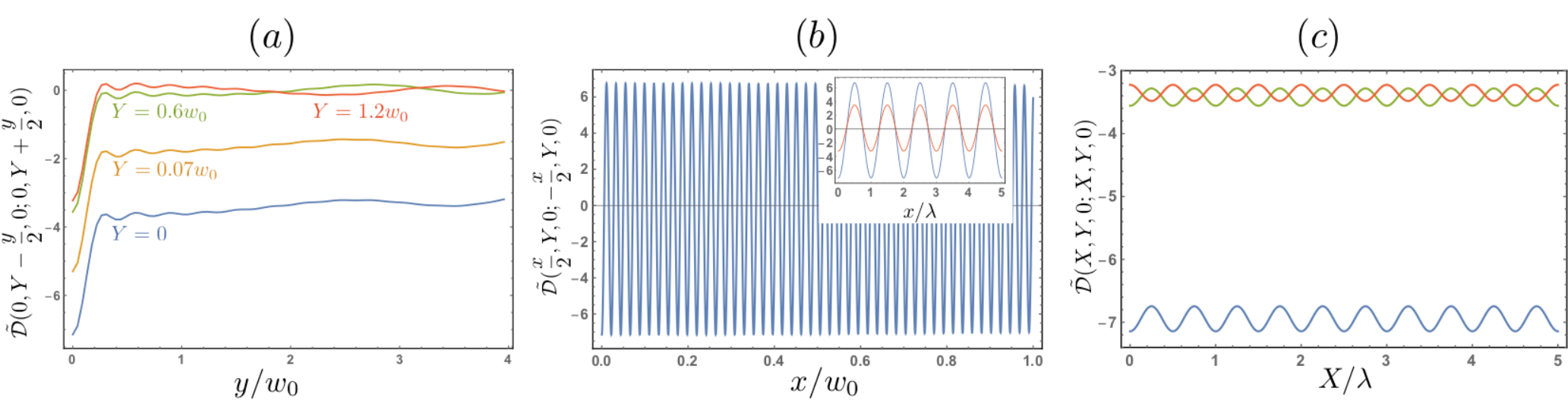}
    \caption{Dimensionless interaction potential  $\tilde{\mathcal{D}}(\mathbf{r}_1, \mathbf{r}_2)$ [defined in Equation~\eqref{eq:multimode_interaction_potential}] between two atoms at positions $\mathbf{r}_1$ and $\mathbf{r}_2$ mediated by a cavity close to the confocal degeneracy point. (a)~The dimensionless interaction potential as a function of the relative interatomic distance $y$ along the transverse direction on the mid-plane of the cavity, for different values of the center-of-mass position $Y$ along the transverse direction. Even for the atoms off the cavity center, $Y\neq0$, a residual long-range oscillating interaction remains as a function of $y$. (b)~$\tilde{\mathcal{D}}(\mathbf{r}_1, \mathbf{r}_2)$ as a function of the relative interatomic distance $x$ along the longitudinal direction. Inset: The same quantity shown over a smaller range and for two different values of the center-of-mass position $Y$ along the transverse direction. (c)~$\tilde{\mathcal{D}}(\mathbf{r}_1, \mathbf{r}_2)$ as a function of the center-of-mass position $X$ along the longitudinal direction, for different values of the center-of-mass position $Y$ along the transverse direction. The distance from the confocal point is $\delta l_{\rm res}=l_{\rm res}-R^c=0.4~\mu$m, i.e., $\epsilon=1.1$~MHz (see text). The parameters are taken from Ref.~\cite{Vaidya2018Tunable}. A laser is red-detuned by $102$~GHz from the atomic resonance and by $\Delta_{0}=-30$~MHz from the lowest mode of an even family. The curvature of both mirrors is $R^c=1$~cm, with the loss rate $\kappa=2\pi\times 167$~kHz. The mode waist is $w_0=35~\mu$m and the number of supported transverse modes is such that $n+m<100$.
    }
    \label{fig:ConfocalPotential}
  \end{center}
\end{figure}

The dimensionless interaction potential is shown in Figure~\ref{fig:ConfocalPotential} for a cavity tuned slightly away from the confocal point, so that a large number of modes contributes. This is measured by the mode-spacing parameter $\epsilon=c\delta l_{\rm res}/l_{\rm res}^2$ with $\delta l_{\rm res}=l_{\rm res}-R^c$, which enters the mode detuning: $\Delta_{mn}=\Delta_0-(m+n)\epsilon$. Following the notation of Ref.~\cite{Vaidya2018Tunable}, we quantify the amount of confocality by the parameter $M^*=|\Delta_0|/\epsilon$, whose square can be loosely associated with the number of modes participating. In the numerically computed potential $\tilde{\mathcal{D}}(\mathbf{r},\mathbf{r'})$ of Figure~\ref{fig:ConfocalPotential}, one can observe all the main qualitative features of cavity-mediated interactions around a confocal point. In the next section, an approximate analytical form will be derived~\cite{Vaidya2018Tunable}.

We first note that for a laser which is red-detuned from the lowest mode of the family, the interaction potential has an attractive minimum at zero interatomic distance. This is a generic feature of cavity-mediated interactions in the multimode case, resulting from the fact that the largest constructive interference in the scattering of laser photons into the cavity happens always when the atoms are at the same spot. This attractive minimum has important consequences for the many-body physics, which will be discussed in Section~\ref{subsec:multimode_droplets}.

The width of the attractive minimum as a function of the interatomic distance along the transverse direction is roughly given by $\xi=w_0/\sqrt{2M^*}$, i.e., it is set by the amount of confocality $M^*$. In the ideal confocal case $M^*\to\infty$, this width goes to zero. However, this does not happen in reality mainly due to the finite number of modes supported by the cavity, which limits the destructive interference between the modes.
In the following, we argue that the length $\xi$ plays the role of the interaction range. This is, however, not strictly true since in the confocal case only either even or odd transverse modes contribute to a given degenerate family.
One consequence of this is that the interaction potential in general does not fall to zero at large interatomic distances, as shown in Figure~\ref{fig:ConfocalPotential}(a),
which can also be explained as resulting from the presence of a mirror image for each atom. However, moving the center of mass of the atoms transversally away from the cavity axis, the image-induced background interaction decays to zero over a length set again by $\xi$, as shown by the green and red curves in Figure \ref{fig:ConfocalPotential}(a). 
A further consequence of having only half of the modes involved is the presence of residual oscillations around zero, also shown in Figure~\ref{fig:ConfocalPotential}(a). 
In the limit of infinitely many modes the ratio between the amplitude of the oscillations and the global minimum approaches zero. The oscillations have a period set by the beam waist $w_0$ and essentially no decay (the much faster wiggles visible in the figure are instead a spurious effect of using a hard cutoff in the cavity-mode sum instead of the more physical smooth cutoff). In the equivalent atom-image picture, these residual oscillations correspond to the interference fringes generated by the atom and its image. As discussed in Ref.~\cite{Guo2019Emergent}, this long-range oscillating part can, however, be eliminated by bringing the neighboring even family into play using a second driving laser tuned one free spectral range away.

Figure~\ref{fig:ConfocalPotential}(b) shows instead the behavior of the interaction potential in the longitudinal direction as a function of the interatomic distance. Here the potential oscillates with a period set by the wavelength $\lambda_c$. In the confocal case the dependence of the Gouy phase $\varphi_{mn}^{\rm Gouy}(x)$ on the longitudinal coordinate is slow since $w_0\gg\lambda_c$, so that every transverse mode has essentially the same longitudinal spatial dependence over several periodic oscillations. As discussed in Section~\ref{subsec:degenerate_cavity_modes}, different transverse modes can still have a different phase offset of the longitudinal oscillations. The consequence of this on the interaction potential is observable in its dependence on the center-of-mass position along the longitudinal direction, shown in Figure~\ref{fig:ConfocalPotential}(c). The longitudinal oscillations quickly change from a cosinusoidal to a sinusoidal modulation by moving the center of mass transversally beyond a distance $w_0$ from the cavity center. Along the longitudinal direction the interaction potential does not decay as a function of interatomic distance on experimentally relevant length scales [see Figure~\ref{fig:ConfocalPotential}(b)]. 

\subsubsection{Experimental characterization}

The features of the cavity-mediated interaction potential around the confocal degeneracy point have been explored in a series of recent experiments performed by the Stanford group~\cite{Vaidya2018Tunable,Guo2019Sign}. The experimental approach consisted in using movable BECs of a small size compared to the beam waist, and measuring the interaction strength as a function of position. Based on a theoretical model, which we will discuss in the following, the interaction strength as well as the length scales characterizing the spatial form of the interaction have been inferred from the experimental value of the superradiant threshold as a function of the BEC position.

The theoretical model describing the experiments is discussed in Refs.~\cite{Vaidya2018Tunable,Guo2019Emergent}. It is based on a multimode version of the mean-field approach introduced in Section~\ref{subsec:MF}. In the particular case of a BEC, the mean-field approximation corresponds to the self-consistent solution of the two coupled equations~\eqref{eq:MF-eq-a-loss} and~\eqref{eq:MF-eq-g}, which can be directly generalized to the present multimode case starting from the Hamiltonian~\eqref{eq:H_eff_1comp_multimode}. We are interested in the regime close to the superradiant threshold. We thus consider the linearized version of the coupled equations about the homogeneous, non-superradiant phase with $\alpha_{mn}=0$. The equation of motion for the cavity fields is thus given by Equation~(\ref{eq:MF-eq-amu-loss-linear}), specified to the BEC case.

In the single-mode self-ordering discussed in Section~\ref{subsec:self-ordering_PT}, it was sufficient to consider the following \textit{ansatz} for the BEC wavefunction $\psi(\mathbf{r})=\psi_{\rm el}(x)\psi_{\rm etr}(y,z)[\psi_0+\sqrt{2}\Theta\cos(k_cx)\cos(k_cy)]$, where $\psi_{\rm el}, \psi_{\rm etr}$ are fixed envelope functions describing the density modulation along the longitudinal and transverse directions from an external trap, and $\Theta$ is the $\mathbf{Z}_2$ order parameter of the superradiant phase transition [see Equation~\eqref{eq:order-paramter-op}]. The modulation $\cos(k_cy)$ with period $\lambda_c=2\pi/k_c$ along the transverse direction $y$ accounts for the imposed transverse standing-wave pump laser $\Omega(\mathbf{r})=\Omega_0\cos(k_cy)$, while the similar modulation $\cos(k_cx)$ along the longitudinal direction $x$ accounts for the spatial dependence of the cavity mode function in the longitudinal direction for a near-planar cavity. The combination of the two modulations corresponds to the total chequerboard interaction potential of Equation~\eqref{eq:cavity-interaction-strength}.

The analysis of the confocal interaction potential of the previous section implies the following generalization of the superradiant \textit{ansatz} for the BEC wavefunction:
\begin{align}
  \psi(\mathbf{r})=\psi_{\rm el}(x)\psi_{\rm etr}(y,z)
  \left\{\psi_0+\sqrt{2}\left[\Theta_{\rm c}\cos(k_cx+\gamma)+\Theta_{\rm s}\sin(k_cx+\gamma)\right]\cos(k_cy)\right\}.
\label{eq:SR_BECansatz_confocal}  
\end{align}
Apart from the common phase shift $\gamma$ that has been introduced for later convenience~\cite{Guo2019Emergent}, the only difference with respect to the \textit{ansatz} for the single-mode case is the introduction of a second component $\Theta_{\rm s}$ of the order parameter. This accounts for the fact discussed in the previous section that the longitudinal modulation of modes belonging to the same confocal degenerate family can be either cosinusoidal or sinusoidal [see also Figure~\ref{fig:ConfocalPotential}(c)]. These two components of the order parameter are, however, only independent under fine-tuned conditions, which will be discussed later.

The experimental characterization focused on the behavior of the interaction potential along the transverse direction $\mathbf{r}_\perp=(y,z)$. The linearized equations for $\alpha_{mn}$ and $\Theta_{\rm c,s}$ must, therefore, be integrated along the longitudinal direction $x$. Since a tight tweezer-trap confines the BEC such that both envelopes $\psi_{\rm el,etr}$ are centered around the position $\mathbf{R}_0=(X_0,Y_0,Z_0)$ and have characteristic size $w_{\rm BEC}\sim10~\si{\mu m} \ll w_0^2/\lambda_c$, one can neglect the non-periodic longitudinal dependence of the cavity mode functions. At the same time, since $w_{\rm BEC}\gg \lambda_c$, one can set all oscillating terms to zero in the integrated equations.
The cavity fields $\alpha_{mn}$ can be adiabatically eliminated by substituting their steady-state solution [see Equation~\eqref{eq:MF-eq-a-ss} for the single-mode case] into the equations of motion for the atomic order parameters $\Theta_{\rm c,s}$. This then leads to effective coupled equations of motion for the atomic order parameters, 
\begin{align}
  i\partial_t\Theta_{\tau}&=(\mu+2\omega_r)\psi_{\rm etr}^{\rm avg} \Theta_\tau
  +\frac{g_0 }{2}\psi_{\rm etr}^{\rm avg}\psi_0^2\Theta_{\tau}^*
  +\frac{\mathcal{G}_0^2\Omega_0^2N}{\Delta_a^2|\Delta_0|}\nonumber\\
  &
  \times\int d\mathbf{r}_\perp d\mathbf{r}_\perp'\sum_{\tau' =\rm c,s}
  \tilde{\mathcal{D}}_{\tau\tau'}^{\rm eff}(\mathbf{r}_\perp, \mathbf{r}_\perp';X_0) 
  \psi_{\rm etr}(\mathbf{r}_\perp) \psi_{\rm etr}(\mathbf{r}_\perp')|\psi_0|^2\left(\Theta_{\tau'}+\Theta_{\tau'}^*\right),
  \label{eq:confocal_orderparams}
\end{align}
where $\tau,\tau'=\{{\rm c,s}\}$, $g_0$ is the short-range atom-atom interaction strength, $\psi_{\rm etr}^{\rm avg}=\int d\mathbf{r}_\perp \psi_{\rm etr}(\mathbf{r}_\perp)$, and 
\begin{align}
  \label{eq:multimode_interaction_potential_effective} 
  \tilde{\mathcal{D}}_{\tau\tau'}^{\rm eff}(\mathbf{r}_\perp, \mathbf{r}_\perp';X_0)
  =|\Delta_0|\mathrm{Re}\sum_{m,n}\frac{\Xi_{mn}(\mathbf{r}_\perp;X_0) \Xi_{mn}(\mathbf{r}_\perp';X_0)}
  {\Delta_{mn}+i\kappa}\mathcal{O}_{mn}^\tau(X_0) \mathcal{O}_{mn}^{\tau'}(X_0),
\end{align}
is the dimensionless, effective cavity-mediated interaction potential [cf.\ Equation~\eqref{eq:multimode_interaction_potential}]. Here the dimensionless transverse Hermite-Gaussian functions are defined as 
\begin{align}
\Xi_{mn}(\mathbf{r}_\perp;X_0)=\frac{w_0}{w(X_0)}
H_m\left(\frac{\sqrt{2}y}{w(X_0)}\right) H_n\left(\frac{\sqrt{2}z}{w(X_0)}\right)e^{-\mathbf{r}_\perp^2/w^2(X_0)},
\end{align}
and the overlap integrals (resulting from the integration along the longitudinal direction) as $\mathcal{O}_{mn}^{c}(X_0)=\cos[(m+n)\phi_0(X_0)]$ and $\mathcal{O}_{mn}^{s}(X_0)=\sin[(m+n)\phi_0(X_0)]$, with
\begin{align}
  \label{eq:effective_gouy}
  \phi_0(X_0)=\frac{\pi}{4}+\arctan{\frac{\lambda_c X_0}{\pi w_0^2}}.
\end{align}

An analytical expression for the effective cavity-mediated interaction potential can be written in the following matrix form in the order-parameter space, specified for simplicity to an even degenerate family at perfect confocality $\epsilon=0$ and at the cavity midplane $X_0=0$, neglecting losses $\kappa=0$~\cite{Guo2019Emergent}:
\begin{align}
  \label{eq:multimode_interaction_potential_effective_analytical} 
  \mathrm{sgn}(\Delta_0)\tilde{\mathcal{D}}^{\rm eff}(\mathbf{r}_\perp, \mathbf{r}_\perp';X_0)\bigg|_{\epsilon=\kappa=X_0=0}=\mathbb{I}\tilde{\mathcal{D}}_{\rm loc}^{\rm eff}(\mathbf{r}_\perp, \mathbf{r}_\perp')+\sigma^z \tilde{\mathcal{D}}_{\rm nonloc}^{\rm eff}(\mathbf{r}_\perp, \mathbf{r}_\perp'),
\end{align}
where $\mathbb{I}$ and $\sigma^i$ are the identity and Pauli matrices in the order-parameter space. The potential is split into a local part, 
\begin{align}
  \label{eq:multimode_interaction_potential_effective_analytical_local}
  \tilde{\mathcal{D}}_{\rm loc}^{\rm eff}(\mathbf{r}_\perp, \mathbf{r}_\perp')\bigg|_{\epsilon=0}
  =\frac{1}{4}\left[\delta\left(\frac{\mathbf{r}_\perp-\mathbf{r}_\perp'}{w_0}\right)
  +\delta\left(\frac{\mathbf{r}_\perp+\mathbf{r}_\perp'}{w_0}\right)\right],
\end{align}
and a nonlocal part, 
\begin{align}
  \label{eq:multimode_interaction_potential_effective_analytical_nonlocal}
  \tilde{\mathcal{D}}_{\rm nonloc}^{\rm eff}(\mathbf{r}_\perp, \mathbf{r}_\perp')\bigg|_{\epsilon=\kappa=X_0=0}
  =\frac{1}{\pi}\cos\left(\frac{2\mathbf{r}_\perp\cdot \mathbf{r}_\perp'}{w_0^2}\right).
\end{align}
The former originates from the peak of size $\xi$ visible in the actual interatomic potential in Figure~\ref{fig:ConfocalPotential}(a). The Dirac-delta form of $\tilde{\mathcal{D}}_{\rm loc}^{\rm eff}(\mathbf{r}_\perp, \mathbf{r}_\perp')$ is a peculiarity of the ideal confocal case where an infinite number of modes contributes to the cavity-induced interatomic potential. Slightly away from confocality, one has instead~\cite{Vaidya2018Tunable},
\begin{align}
  \label{eq:multimode_interaction_potential_effective_analytical_local_nearconf}
 \tilde{\mathcal{D}}_{\rm loc}^{\rm eff}(\mathbf{r}_\perp, \mathbf{r}_\perp') 
 =\frac{M^*}{4\pi}\sum_{s=\pm1}
 K_0\left(\frac{\mathbf{r}_\perp-s\mathbf{r}_\perp'}{\xi}
 \sqrt{1+\frac{(\mathbf{r}_\perp+s\mathbf{r}_\perp')^2}{8M^*w_0^2}}\right),
\end{align}
with $K_0$ being the modified Bessel function of the second kind. In both Equations~(\ref{eq:multimode_interaction_potential_effective_analytical_local}) and~(\ref{eq:multimode_interaction_potential_effective_analytical_local_nearconf}), there is a contribution to the local interaction potential from the mirror image located at $-\bm{r}_\perp'$. The nonlocal term~(\ref{eq:multimode_interaction_potential_effective_analytical_nonlocal}) instead oscillates without decay, with a period set by the beam waist $w_0$. This corresponds to the residual long-range oscillations in the interatomic interaction potential in Figure~\ref{fig:ConfocalPotential}(a).

Let us now turn to the actual experimental characterization of the interaction potential. As mentioned, this relies on inferring its strength from the value of the superradiant threshold. For a BEC of small transverse size $w_{\rm BEC}$ compared to the beam waist $w_0$, we see from Equation~(\ref{eq:confocal_orderparams}) that close to the confocal point the local part of the interaction potential is much more important than the nonlocal part in determining the threshold, which would depend on the self-interaction $\tilde{\mathcal{D}}^{\rm eff}(\mathbf{R}_{\perp 0}, \mathbf{R}_{\perp 0};X_0)$ around the center of the small atomic cloud $\mathbf{R}_0=(X_0, \mathbf{R}_{\perp 0})=(X_0,Y_0,Z_0)$. The contribution of the nonlocal part to the self-interaction is indeed of order one, while the contribution from the local part is of order $M^*\gg 1$ near confocality. For the determination of the superradiant threshold, the two coupled equations for the order-parameter components $\Theta_{\rm c,s}$ can be thus considered approximately decoupled. This allows to obtain a simple relation between the critical value $\Omega_{0\rm c}$ of the laser strength and the strength of the local part of the effective self-interaction potential~\cite{Vaidya2018Tunable}:
\begin{align}
  \label{eq:critical_relation_laser_interaction}
\tilde{\mathcal{D}}_{\rm loc}^{\rm eff}(\mathbf{R}_{\perp 0}, \mathbf{R}_{\perp 0})\simeq -\frac{\omega_r\Delta_a^2|\Delta_0|}{N\mathcal{G}_0^2\Omega_{0 \rm c}^2}.
\end{align}
It is evident that the self-interaction has to be negative (i.e., the short-range part of the interatomic interaction potential must be attractive) in order for the superradiant threshold to be reachable. We note at this point the different sign convention between the present review article and Ref.~\cite{Vaidya2018Tunable}. 

\begin{figure}[t!]
  \begin{center}
    \includegraphics[width=\columnwidth]{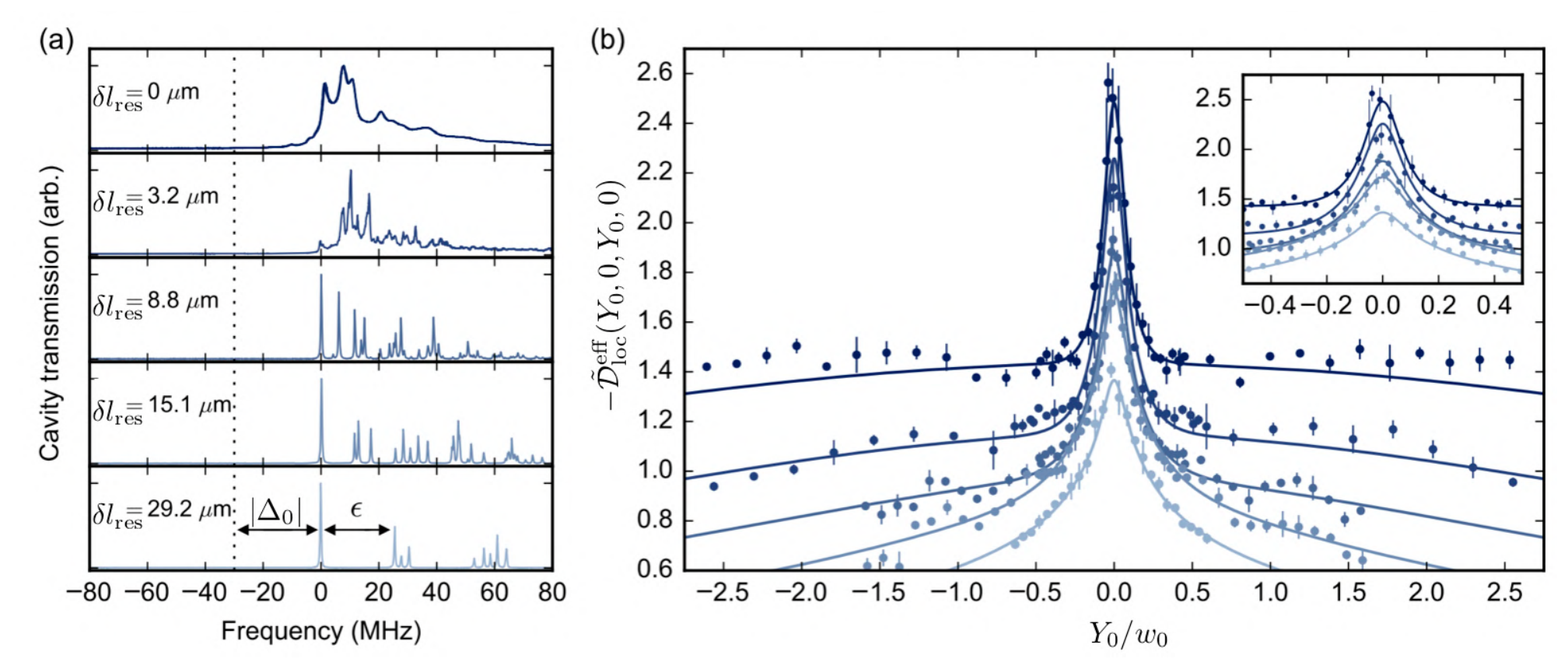}
    \caption{Experimental characterization of the cavity-mediated interatomic potential close to and away from confocality. (a) Cavity transmission spectrum for five different cavity lengths, from confocal $\delta l_{\rm res}=l_{\rm res}-R^c=0$ to a regime similar to the one presented in Figure~\ref{fig:SupermodePolaritonCondensation}, where the even mode families are well separated. For all the different cavity lengths in (a), panel (b) shows the absolute value of the strength of the effective self-interaction potential for a small-sized BEC located at different positions on the cavity midplane: $X_0=0$, measured using the theoretical model of Equation~(\ref{eq:critical_relation_laser_interaction}). The change in the width of the peak at the cavity center as a function of the degree of confocality is appreciable. The inset shows a zoom of the data close to the cavity center. Figure adapted and reprinted with permission from Ref.~\cite{Vaidya2018Tunable} \textcopyright~2018 by the American Physical Society.}
    \label{fig:ConfocalPotentialEffective_measurement}
  \end{center}
\end{figure}

The relation of Equation~(\ref{eq:critical_relation_laser_interaction}) has been used in the experiment~\cite{Vaidya2018Tunable} to reconstruct the spatial dependence of the local effective interaction~(\ref{eq:multimode_interaction_potential_effective_analytical_local_nearconf}). More precisely, due to the divergence of the Bessel function at zero argument, Equation~\eqref{eq:critical_relation_laser_interaction}, which assumes the transverse envelope of the BEC to be infinitely localized $\psi_{\rm etr}\propto\delta(\mathbf{r}_\perp-\mathbf{R}_{\perp 0})$, cannot be directly used. The transverse spatial average in Equation~(\ref{eq:confocal_orderparams}) needs instead to be performed over a finite BEC size. The experimental results are presented in Figure~\ref{fig:ConfocalPotentialEffective_measurement} for different cavity lengths. As clear from Equation~(\ref{eq:multimode_interaction_potential_effective_analytical_local_nearconf}), even though
the experimental data of Figure~\ref{fig:ConfocalPotentialEffective_measurement}(b) only measure the self-interaction as a function of the center-of-mass position $Y_0$, they already show the relevant scales characterizing the local interaction potential: the width $\xi$ of the (attractive) peak and a background value decaying on a scale much slower than $w_0$. Those features of the effective potential are also found in the microscopic interatomic interaction potential shown in Figure~\ref{fig:ConfocalPotential}. Measurements involving two separate BECs have also been performed, which allowed to access the finite-distance behavior of the local part of the effective interaction potential. This first Stanford experiment~\cite{Vaidya2018Tunable} did focus only on the local part of the interaction potential. The nonlocal part was the object of a second experiment~\cite{Guo2019Sign}, to which we turn next.

\begin{figure}[t!]
  \begin{center}
    \includegraphics[width=\columnwidth]{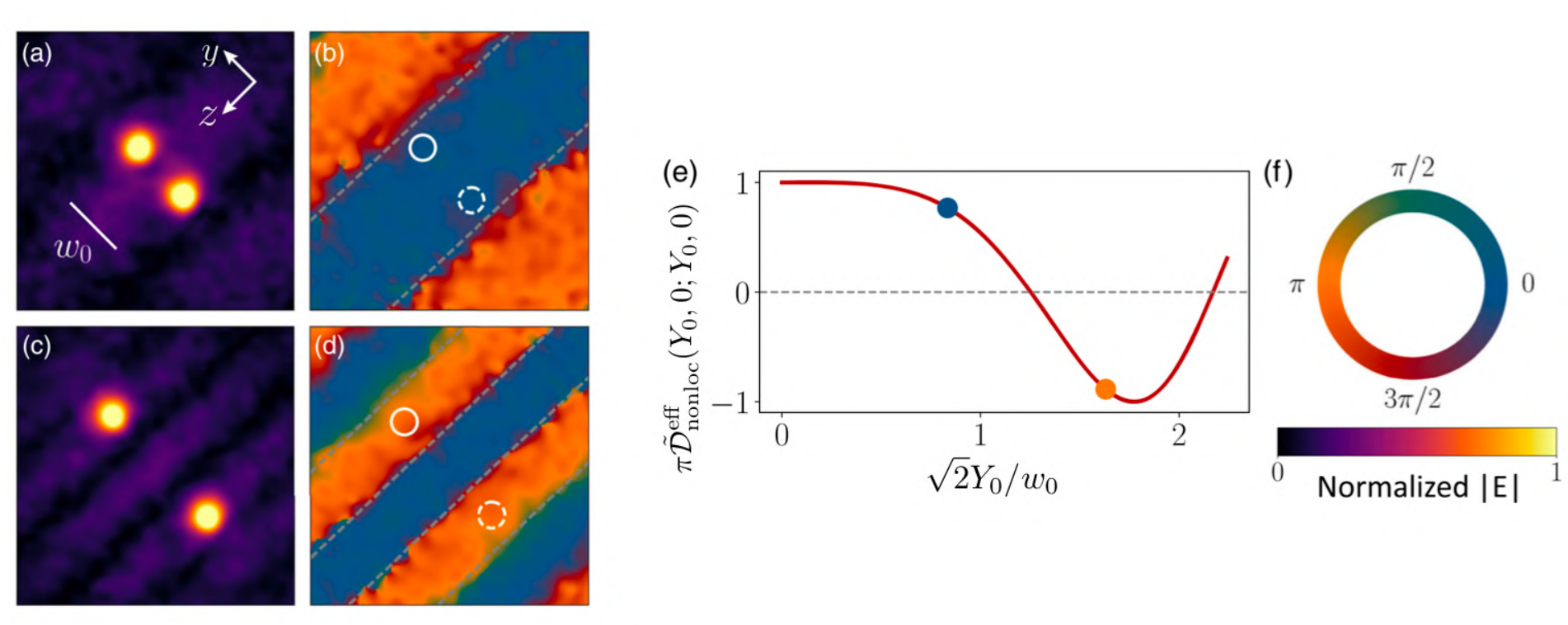}
    \caption{Experimental characterization of the superradiant density wave via the cavity output field at the confocal point. For two different positions of a single BEC in the transverse direction, the light intensity, (a) and (c), and phase, (b) and (d), are shown. Panel (e) shows the theoretical prediction (see text) for the relative phase between the fringes and the bright spot, with circles marking the values used in panels (b) and (d). Figure adapted and reprinted with permission from Ref.~\cite{Guo2019Sign} \textcopyright~2019 by the American Physical Society.}
    \label{fig:measurement_cavityoutput_confocal}
  \end{center}
\end{figure}

While the nonlocal part does not appreciably modify the superradiant threshold, it is crucial in determining the associated density modulation of the BEC. This is already evident from Equation~(\ref{eq:multimode_interaction_potential_effective_analytical}), where the nonlocal interaction favors one of the two components $\Theta_{\rm c,s}$ of the order parameter depending on the transverse position. This means that, depending on the BEC location the superradiant longitudinal density modulation would be either cosinusoidal or sinusoidal. In fact, as discussed in Ref.~\cite{Guo2019Emergent}, away from the cavity midplane the nonlocal part of the effective interaction is not a real quantity anymore, which results in a coupling between the two order parameters via an additional term in Equation~(\ref{eq:multimode_interaction_potential_effective_analytical}) proportional to $\sigma^x\mathrm{Im}\tilde{\mathcal{D}}_{\rm nonloc}^{\rm eff}$. This implies that in general the superradiant density modulation along the longitudinal direction is not either cosinusoidal or sinusoidal, but rather a superposition of the two with a phase dependent on the position of the BEC. The behavior of  this relative phase has also been characterized experimentally by means of the cavity output field~\cite{Guo2019Sign}. The principle of such measurement can be understood in the adiabatic cavity limit, which we used to derive Equation~(\ref{eq:confocal_orderparams}). In this limit, the cavity field components $\alpha_{mn}$ are locked to the order-parameter components $\Theta_{\rm c,s}$, i.e., the cavity fields are fully determined by the instantaneous value of the atomic order parameters. The total cavity field in real space is then given by a sum over the Hermite-Gaussian modes, weighted by the coefficients $\alpha_{mn}$. The measured cavity output corresponds to the forward-travelling component of the total cavity field passing through one of the mirrors. Choosing the parameters as in Equation~(\ref{eq:multimode_interaction_potential_effective_analytical}), it reads~\cite{Guo2019Emergent}:
\begin{align}
  \label{eq:cavityfield_forward_confocal}
  \alpha^{\rm F}(\mathbf{r}_\perp)\propto \int d\mathbf{r}_\perp'\psi_{\rm etr}(\mathbf{r}_\perp') \tilde{\mathcal{D}}_{\rm loc}^{\rm eff}(\mathbf{r}_\perp, \mathbf{r}_\perp')+e^{-2i\text{arg}\left(\Theta_{\rm c}+i \Theta_{\rm s}\right)}\int d\mathbf{r}_\perp'\psi_{\rm etr}(\mathbf{r}_\perp') \tilde{\mathcal{D}}_{\rm nonloc}^{\rm eff}(\mathbf{r}_\perp, \mathbf{r}_\perp'),
\end{align}
where we have assumed the BEC wavefunction to be real.
In the cavity output field, the two terms in Equation~(\ref{eq:cavityfield_forward_confocal}) can be clearly separated, as shown in Figure~\ref{fig:measurement_cavityoutput_confocal}. In panels (a) and (c), for two different positions of the BEC the light intensity is presented, showing two bright spots corresponding to the BEC and its mirror image, together with interference fringes. In accordance with the discussion above, the bright spots localized around the BEC and its image originate from the local part of the effective interaction potential, while the periodic fringes originate from the nonlocal part. Having separated the two contributions, a measurement of the local phase of the output field, as shown in panels (b) and (d), allows then to extract the relative phase between the local and nonlocal parts of the interaction potential, and thus the relative weight of the cosinusoidal and sinusoidal modulations of the superradiant BEC density. Panel (e) shows the theoretical prediction for the relative phase as a function of the BEC position, with the two circles marking the values used for the experimental output of panels (b) and (d). The measured sign flip is consistent with the theoretical prediction.

We conclude this section by discussing the symmetries of the effective model of Equation~(\ref{eq:confocal_orderparams}). In the generic case just discussed, where the nonlocal part of the interaction potential shifts and also couples the two components $\Theta_{\rm c,s}$ of the order parameter, one is left with the usual single-mode scenario where the superradiant transition corresponds to the spontaneous breaking of a discrete $\mathbf{Z}_2$ symmetry [see Equation~\eqref{eq:order-paramter-op}]. In the present notation, this symmetry corresponds to the choice of the global shift $\gamma=0,\pi$ in Equation~(\ref{eq:SR_BECansatz_confocal}), while $\Theta_{\rm s,c}$ are not independent so that a superposition of fixed relative phase is chosen, as shown in Figure~\ref{fig:measurement_cavityoutput_confocal}. However, under certain conditions the two components of the order parameter can become independent and equivalent, so that a continuous $U(1)$ symmetry emerges in the system~\cite{Guo2019Emergent}. Assuming perfect confocality and choosing the cavity midplane $X_0=0$, the effective interaction is given by Equation~(\ref{eq:multimode_interaction_potential_effective_analytical}). Taking the transverse BEC size $w_{\rm BEC}$ much smaller than $\lambda_c$, one can approximately consider only the self-interaction $\mathbf{r}_\perp=\mathbf{r}_\perp'$
in Equation~(\ref{eq:confocal_orderparams}) for the order-parameter components. By choosing the position of the BEC such that $\tilde{\mathcal{D}}_{\rm nonloc}^{\rm eff}(\mathbf{R}_{\perp 0}, \mathbf{R}_{\perp 0})=0$, the equations for $\Theta_{\rm c,s}$ are not only decoupled but also equivalent, so that their relative weight can be chosen arbitrarily, thus a $U(1)$ symmetry emerges. Such special positions exist also out of the cavity midplane, and the $U(1)$ symmetry can exist approximately also for small deviations from confocality or finite BEC size~\cite{Guo2019Emergent}.

\subsection{Superradiant quantum crystals versus self-bound crystalline droplets}
\label{subsec:multimode_droplets}

Having discussed the characteristics of the cavity-mediated interaction potential around the confocal degeneracy, let us now explore the consequences for the many-body physics. In particular, we focus on the types of crystalline phases and phase transitions that can appear in this case. As already observed in Section~\ref{subsubsec:confocal_potential}, the main feature of the cavity-mediated interaction in or around the confocal point is the presence of an attractive minimum at zero interatomic distance [see Figure~\ref{fig:ConfocalPotential}(a) and~(b)]. This feature is quite generic to laser-driven atoms coupled to multiple cavity modes, as the strongest constructive light scattering occurs typically when all the atoms are located at the same position. For a laser which is red detuned from a degenerate mode family, the resulting potential is attractive. As indicated by Equation~(\ref{eq:critical_relation_laser_interaction}), red detuning is necessary to observe superradiance and the accompanying crystallization of the atoms. Moreover, for a blue-detuned laser with respect to the cavity modes, heating of the atomic cloud might become an issue.

The consequence of an attractive interaction minimum at zero interatomic distance for an atomic bosonic cloud is the instability of the latter towards collapse. This collapse is counteracted by the finite temperature or the quantum kinetic energy, as well as by the short-range atomic repulsion, therefore leading to the formation of a stable, self-bound droplet state. Considering that the cavity-mediated interaction generically features also an oscillatory part, depending on the period of such oscillations compared to the droplet size, the latter may exhibit a density modulation associated with superradiance. The density crystallization breaks spontaneously a (typically discrete) translation symmetry of the system, accompanied with a phase-locking of the superradiant cavity field with respect to the pump laser. Noting Equation~(\ref{eq:multimode_interaction_potential}), we stress that the oscillatory part of the interaction might stem from the cavity modes, i.e., from $\tilde{\mathcal{D}}$, and/or from the spatial profile of the driving laser $\Omega(\mathbf{r})$.

\begin{figure}[t!]
  \begin{center}
    \includegraphics[width=0.7\columnwidth]{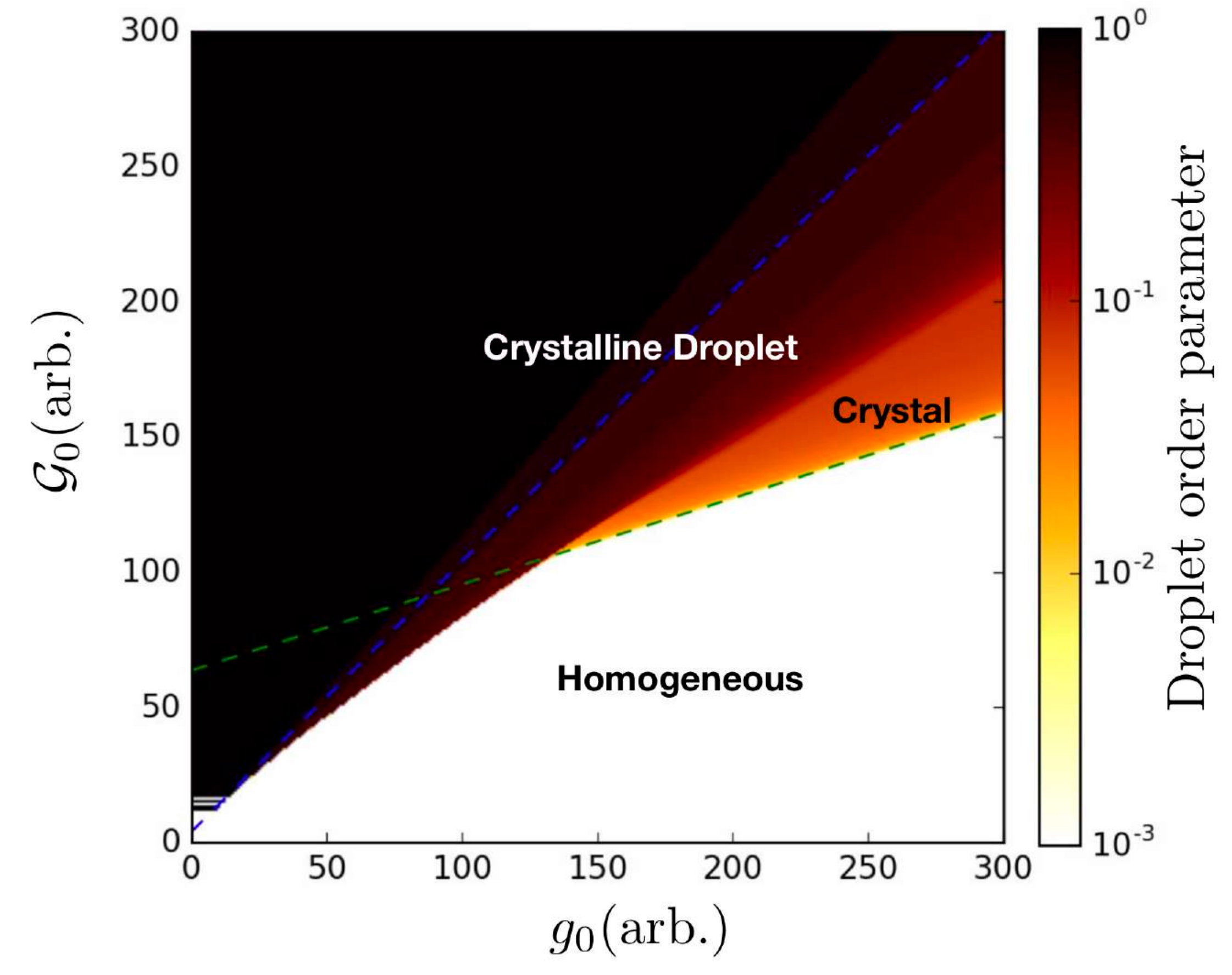}
    \caption{Qualitative phase diagram for a BEC of atoms interacting via a cavity-mediated interaction potential with an attractive minimum of finite width $\xi$ at zero distance, as is the case in confocal cavities (see Figure~\ref{fig:ConfocalPotential}). The axes correspond to the strength of the atom-cavity coupling $\mathcal{G}_0$ and the strength of the short-range interatomic repulsion $g_0$. The color scale indicates the value of the droplet order parameter or the inverse participation ratio $\int d\mathbf{r} \rho^2(\mathbf{r})/\left[\int d\mathbf{r} \rho(\mathbf{r})\right]^2$. Adapted from Ref.~\cite{Karpov2021Quantum}.}
    \label{fig:multimode_phase_diagram_karpov}
  \end{center}
\end{figure}

Such superradiant droplets have been predicted to appear for multimode cavities~\cite{Karpov2019Crystalline,Karpov2021Quantum} and also for a related geometry involving a retroreflected beam passing through the atomic cloud~\cite{Zhang2018Long}.
The generic zero-temperature phase diagram, as a function of the strength of the atom-cavity coupling $\mathcal{G}_0$ versus the interatomic-repulsion strength $g_0$, is shown in Figure~\ref{fig:multimode_phase_diagram_karpov}. As expected, a strong enough interatomic repulsion is necessary to stabilize an extended crystalline phase against the droplet. Otherwise, the superradiant instability of the homogeneous phase always leads to the formation of a self-bound, crystalline droplet state through a first-order phase transition~\cite{Karpov2019Crystalline,Karpov2021Quantum}. Taking further into account the static optical lattice created by the pump laser, a strong enough repulsion $g_0$ suppresses number fluctuations and brings in the Mott-insulator physics, as discussed in Section~\ref{sec:extended-BH} for extended superradiant states. As shown in Ref.~\cite{Karpov2021Quantum}, this can lead to different types of quantum liquid-like droplets. In Section~\ref{sec:qc}, we will see another instance of transition between a superradiant localized state and a superradiant extended quasicrystalline phase due to the competition between the atom-cavity coupling (i.e., the cavity-mediated attractive atomic interaction) and the interatomic repulsion in a different geometry involving few cavity modes.

The qualitative phase diagram of Figure~\ref{fig:multimode_phase_diagram_karpov} is generic to systems where the attractive minimum of the cavity-mediated interaction at zero distance has a finite width $\xi$, as is the case close to a confocal degeneracy (see Figure~\ref{fig:ConfocalPotential}). The attractive minimum of size $\xi$ has been indeed observed from the superradiant light emission using a BEC inside a confocal cavity as discussed in the previous section. Due to the tight tweezer trap, the BEC had a characteristic size $w_{\rm BEC}\sim 10~\mu\text{m}\sim \text{min}\xi\sim w_0/2$ (see Figure~\ref{fig:ConfocalPotentialEffective_measurement}). Therefore, even though the droplet size can be appreciably smaller than $\xi$ (depending on the interatomic repulsion), the cavity-output spatial resolution available in Refs.~\cite{Vaidya2018Tunable, Guo2019Sign} would not have allowed to identify the droplet. However, the use of a shallower trap creating a BEC whose transverse extension is above $\xi$ should allow to distinguish the superradiant droplet from the extended crystalline phase.

In the phase diagram of Figure~\ref{fig:multimode_phase_diagram_karpov}, the nature of the transition between the homogeneous and the crystalline phase has not been specified. The reason is that this depends on the actual range of the interaction and not strictly on the width $\xi$ of the attractive minimum at zero distance. A clear example is the confocal interaction potential of Figure~\ref{fig:ConfocalPotential}, where the actual range of the interaction is infinite (compared to available atom-cloud sizes) while $\xi$ can be small. As already discussed, this is a consequence of the imperfect destructive interference due to the missing odd or even modes in a confocal family. In this situation, the crystallization transition is of second-order and of the mean-field type, i.e., the same as in the single-mode case (see Section~\ref{subsec:SR_instability}). However, as already mentioned above, by keeping the cloud away from the cavity axis and eliminating the long-range oscillating background via an additional laser, a confocal cavity can be used to engineer a genuine short-range interaction with a scale given by $\xi$, in which case beyond-mean-field effects would become accessible.

A further option is to approach the concentric limit. As discussed in Sections~\ref{subsubsec:optical_resonators} and~\ref{subsec:degenerate_cavity_modes}, a degenerate family in the concentric case possesses both even and odd transverse modes, which allows for completely destructive interference at long distances and thus a finite cavity-mediated interaction range. As already mentioned  before, the perfect concentric limit can never be achieved as it corresponds to an unstable point, so that only a near-concentric regime is accessible. As shown in Figure~\ref{fig:CavityParameter}(c), this regime has the peculiarity that higher-order transverse modes lie lower in frequency, which means that for a red-detuned pump laser the highest-order transverse modes are the  closest ones to resonance and thus mostly coupled with the atoms. Therefore, the resultant cavity-mediated interaction potential strongly depends on the mode cutoff, i.e., on the details of the cavity mirrors beyond the simple curvature and reflectivity. A sharp mode cutoff is certainly not a good approximation. While proper mode-dependent loss rates $\kappa_\nu$ in Equation~(\ref{eq:Liouville-op-multimode}) can work well in concentric cavities, where among others diffraction losses due to finite mirror sizes need to be accurately modeled~\cite{Nguyen2018Operating}. Therefore, qualitative features of the photon-mediated interaction in near-concentric cavities will depend on the actual cavity employed. While a large portion of the phase diagram should in general feature a droplet phase due to the attractive minimum at zero distance expected also in near-concentric cavities, the properties of the transition between the homogeneous and the extended crystalline phase have to be discussed case by case.

\begin{figure}[t!]
  \begin{center}
    \includegraphics[width=\columnwidth]{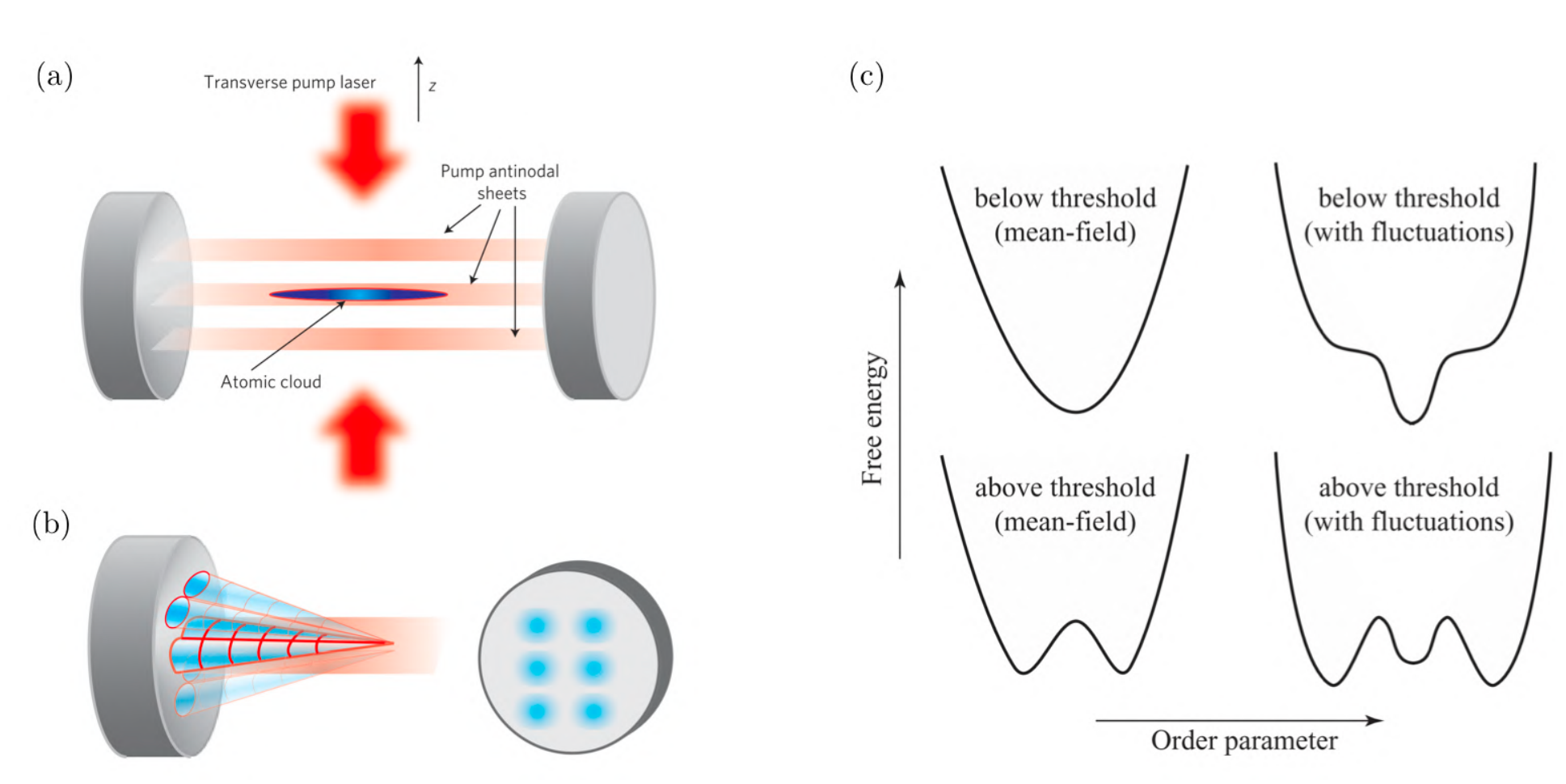}
    \caption{Brazovskii-type superradiant crystallization. (a)~Proposed setup involving a quasi-two-dimensional cloud of bosons trapped on a longitudinal plane around the cavity center. (b)~Nodal structure of one of the modes of the concentric cavity (TEM$_{12}$ with 5 longitudinal nodes in the picture). (c)~Sketch of the free energy landscape as a function of the superradiant order parameter. The right column corresponds to a fluctuation-induced first-order transition of the Brazovskii-type. Panels (a) and (b) have been adapted and reprinted with permission from Ref.~\cite{Gopalakrishnan2009Emergent} published at 2009 by the Nature Publishing Group, and panel (c) from Ref.~\cite{Gopalakrishnan2010Atom}~\textcopyright~2010 by the American Physical Society.}
    \label{fig:concentric_Brazovskii}
  \end{center}
\end{figure}

Here we limit ourselves to briefly discuss a particular example investigated theoretically in Ref.~\cite{Gopalakrishnan2009Emergent}. Though considering an idealized case, it well illustrates the promise of multimode cavities for realizing interesting many-body physics of crystallization. The proposed setup is illustrated in Figure~\ref{fig:concentric_Brazovskii}. The basic idea is that (thermal or quantum) fluctuations, which become important for a short-enough interaction range, can modify the order of the transition as illustrated by the free energy sketched in Figure~\ref{fig:concentric_Brazovskii}(c). In Refs.~\cite{Gopalakrishnan2009Emergent,Gopalakrishnan2010Atom}, starting from the microscopic description of the atom-cavity system, an effective free energy for the superradiant order parameter is derived. We will not present the calculation here and limit ourselves to describe the qualitative picture instead. Assuming one can approximately realize a radial symmetry on the atomic plane for a perfectly concentric cavity, the superradiant order parameter corresponds to a density modulation with a wave vector whose modulus is set by the laser wavelength but its direction is arbitrary chosen in the plane. This plane-wave-like picture relies on an approximate momentum conservation valid in the limit of very large clouds with respect to the wavelength, and also on excluding (via a sharp cutoff) modes which have a more radial than longitudinal nodes. Within this description, getting closer to the superradiant threshold the order parameter fluctuations grow. Since they can have an arbitrary direction in the plane, they form a quasi-continuum which impacts the critical behavior very strongly, so that the order of the transition is changed from the second-order mean-field type to a fluctuation-induced weak first order. This mechansim is known as the Brazovskii mechanism~\cite{Brazovski1996Phase,Chaikin1995Principles}. Besides relying on a quasi-continuum of order-parameter components with radial symmetry, it also requires the free energy not to contain cubic terms, so that at the mean-field level the crystallization transition can remain of second order (which is actually not the typical case in most materials~\cite{Chaikin1995Principles}). In the specific geometry of Figure~\ref{fig:concentric_Brazovskii}(a) and within the approximations employed in Ref.~\cite{Gopalakrishnan2009Emergent}, one can show that the cubic terms are indeed absent from the effective free energy for the superradiant order parameter~\cite{Gopalakrishnan2009Emergent,Gopalakrishnan2010Atom}. We note that the instability towards the self-bound droplets discussed above, which corresponds to a strongly first-order transition, is automatically excluded if the cubic terms are absent.

\section{Multi-component quantum gases inside cavities: density and spin self-ordering}
\label{sec:spinor-selfordering}

So far, we have mainly focused on many-body cavity QED systems where the internal dynamics of atoms are not relevant (i.e., they evolve much faster than the external degrees of freedom) and are omitted. That is, ultracold atomic gases are approximated as polarizable quantum media. Nonetheless, it is natural to wonder what will happen if the internal dynamics of the atoms also become important. This results in a new degree of freedom---``pseudospin''---which follows the same algebra as a genuine ``spin'', and allows in turn for studying combined density and spin self-ordering dynamics~\cite{Safaei2013Raman, Mivehvar2017Disorder, Kroeze2018Spinor, Colella2018Quantum, Mivehvar2019Cavity, Colella2019Antiferromagnetic, Ostermann2019Cavity, Masalaeva2020Spin}.\footnote{We will just use spin instead of pseudospin to refer to internal states and internal dynamics of the atom throughout this review paper.} In particular, the exchange of cavity photons by the atoms in this case can also result in long-range interactions between atomic spins independent of the temperature of the atomic cloud, similar to dipolar interactions between polar molecules. This is in contrast to spin exchange interactions  stemming from the Hubbard contact interactions in atomic gases trapped in free-space optical lattices, which only come into play at very low temperatures. The cavity-mediated spin interactions can be modified and tuned experimentally, allowing one to study different magnetic orders. Therefore, many-body cavity-QED systems provide a promising, versatile platform to implement various spin-model Hamiltonians and explore quantum magnetism in a controlled way. 

\subsection{Multi-component ultracold atoms coupled to cavities: Fundamentals}
\label{subsec:multi-com-atoms-cavity}

\begin{figure}[t!]
\centering
\includegraphics [width=0.95\textwidth]{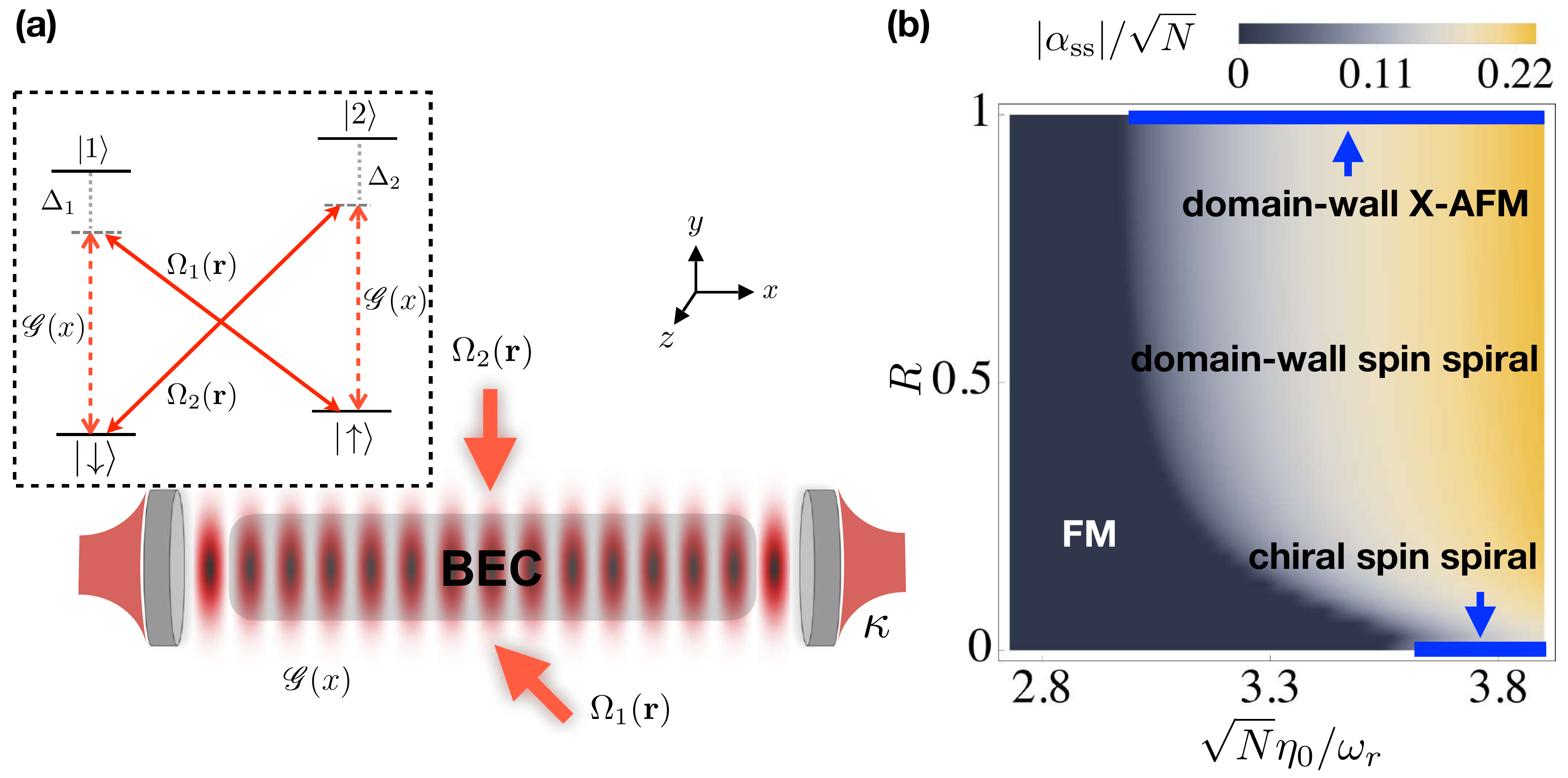}
\caption{Scheme and phase diagram for the cavity-induced density and spin self-ordering.
(a) Four-level bosonic atoms are coupled to a single standing-wave mode of a linear cavity and pumped in the transverse direction by two pump lasers. As can be seen from the inset, the atom-photon couplings 
comprise a double $\Lambda$ scheme, with far off-resonant, red-detuned transitions.
(b) The phase diagram of the system in the $R$-$\eta_0$ plane  for $\text{Re}(\tilde{\delta}_c)<0$ 
exhibits four fundamentally different cavity-induced spin orders. 
Below the pump-laser strength threshold $\eta_{0c}(R)$, the ground state of the system is a spin-polarized ferromagnetic (FM) normal state.
Increasing the pump-laser strength beyond the threshold, the system undergoes the Dicke superradiant phase transition 
and exhibits an emergent magnetic order: domain-wall $X$-antiferromagnetic ($X$-AFM), domain-wall spin-spiral, or
chiral spin-spiral order depending on the competition among different cavity-induced long-range spin-spin interactions. The structure of the density self-ordering is different in each phase. 
Figure adapted and reprinted with permission from Ref.~\cite{Mivehvar2019Cavity}~\textcopyright~2019 by the American Physical Society.} 
\label{Fig:Mivehvar2019Cavity_Fig1}
\end{figure}

Spin-dependent self-organization of a spinor BEC was already discussed in Section~\ref{subsubsec:SpinDependentSO}, where the population of each spin component was conserved. Here we instead focus on more general processes, where the cavity-assisted photon scattering drives spin-changing Raman processes and therefore the population of individual spin components is not conserved across the superradiant phase transition. This leads to the concept of combined, stationary density and spin self-ordering introduced in Ref.~\cite{Mivehvar2017Disorder}, where they studied the Dicke superradiance phase transition for a generalized atomic system with both internal and external quantized degrees of freedom. The nonequilibrium dynamics of a similar, related setup was also studied in Ref.~\cite{Safaei2013Raman}. The considered setup consists of four-level ultracold atoms with the ground-state manifold $\{\downarrow,\uparrow\}$ and excited states $\{1,2\}$ with energies $\{\hbar\omega_\downarrow=0, \hbar\omega_\uparrow,\hbar\omega_{1},\hbar\omega_{2}\}$, tightly confined into the $x$-$y$ plane containing the axis of a standing-wave linear cavity~\cite{Dimer2007Proposed, Kroeze2018Spinor, Mivehvar2019Cavity}. The atoms are pumped in the transverse direction using two sufficiently red-detuned, non-interfering pump lasers and also strongly coupled to an off-resonance empty mode of the cavity. Namely, the first (second) laser with the frequency $\omega_{p1}$ ($\omega_{p2}$) drives the transition $\uparrow\>\leftrightarrow1$ ($\downarrow\>\leftrightarrow2$) with the position-dependent Rabi rate $\Omega_{1}(\mathbf{r})$ [$\Omega_{2}(\mathbf{r})$], while the transitions  $\downarrow\>\leftrightarrow 1$ and $\uparrow\>\leftrightarrow 2$ are both induced by the cavity mode with the frequency $\omega_c$ and the same coupling strength $\mathcal{G}(x)=\mathcal{G}_0\cos(k_cx)$ as depicted in Figure~\ref{Fig:Mivehvar2019Cavity_Fig1}(a). For instance, for an atomic species with Zeeman sublevels with corresponding magnetic quantum numbers $m_\downarrow=m_1$ and $m_\uparrow=m_2=m_\downarrow+1$, the polarization of the cavity field is chosen to be linear along the quantization axis $z$, while the pump lasers have in-plane polarizations.\footnote{Alternatively, the pump lasers $\Omega_{1,2}(\mathbf{r})$ can be aligned in the $z$ direction with left/right circular polarizations, where now the physics in the $x$-$z$ plane is identical to the one presented in the text for given choices.} The scheme constitutes a double $\Lambda$ configuration with large atomic detunings $\Delta_{1}\equiv(\omega_{p1}+\omega_{p2})/2-\omega_1$ and
$\Delta_2\equiv\omega_{p2}-\omega_2$, where the ground states $\{\downarrow,\uparrow\}$ are the relevant spin states and form a spin-1/2.

Adiabatic elimination of the atomic excited states yields an effective Hamiltonian 
$\hat{H}_{\rm eff}=\int \hat\Psi^\dag(\mathbf{r})\hat{\mathcal H}_{1,\rm eff}\hat\Psi(\mathbf{r})d\mathbf{r}-\hbar\Delta_c\hat{a}^\dag\hat{a}$,
where $\Delta_c\equiv(\omega_{p1}+\omega_{p2})/2-\omega_c$, and 
$\hat{a}$ and $\hat\Psi=(\hat\psi_\uparrow,\hat\psi_\downarrow)^\top$ are
the photonic and two-component (spinor) atomic annihilation field operators, respectively.
The single-particle atomic Hamiltonian density reads~\cite{Mivehvar2019Cavity},
\begin{align} \label{eq:single-particle-H-den-spinor}
\hat{\mathcal H}_{1,\rm eff}=\hbar
\begin{pmatrix}
-\frac{\hbar}{2M}\boldsymbol{\nabla}^2+\delta+V_\uparrow(\mathbf{r})+U_\uparrow(\mathbf{r})\hat{a}^\dag\hat{a} & 
\eta_1^*(\mathbf{r})\hat{a}+\eta_2(\mathbf{r})\hat{a}^\dag
\\
\eta_1(\mathbf{r})\hat{a}^\dag+\eta_2^*(\mathbf{r})\hat{a} & 
-\frac{\hbar}{2M}\boldsymbol{\nabla}^2+V_\downarrow(\mathbf{r})+U_\downarrow(\mathbf{r})\hat{a}^\dag\hat{a}
\end{pmatrix},
\end{align}
with the classical pump potentials $V_{\uparrow(\downarrow)}(\mathbf{r})=|\Omega_{1(2)}(\mathbf{r})|^2/\Delta_{1(2)}$, the quantum cavity potentials 
$U_{\uparrow(\downarrow)}(\mathbf{r})=|\mathcal{G}(x)|^2/\Delta_{2(1)}$, and the two-photon Raman-Rabi coupling strength 
$\eta_j(\mathbf{r})=\mathcal{G}^*(x)\Omega_j(\mathbf{r})/\Delta_j$ with $j=1,2$. Here, $\delta\equiv\omega_\uparrow-(\omega_{p2}-\omega_{p1})/2+\delta_{\rm Zeeman}$ is the effective frequency difference between the two spin states, where an external magnetic field $B_{\rm ext}$ has been applied to fix the quantization axis (along $z$) and tune the effective two-photon detuning via the Zeeman shift $\delta_{\rm Zeeman}(B_{\rm ext})$. The two-body contact interactions are assumed to be small and are omitted. Note that for \emph{bosonic} atoms by restricting to an appropriate low-energy manifold including both internal and external states, one can map the effective Hamiltonian $\hat{H}_{\rm eff}$, corresponding to Equation~\eqref{eq:single-particle-H-den-spinor}, into the Dicke Hamiltonian~\cite{Li2010Extended, Kroeze2018Spinor}. 

From the steady-state photonic field operator, 
\begin{align} \label{eq:ss-a--spinor}
\hat{a}_{\rm ss}=\frac{\int \left[\eta_1(\mathbf{r})\hat{s}_-(\mathbf{r})
+\eta_2(\mathbf{r})\hat{s}_+(\mathbf{r})\right]d\mathbf{r}}
{\Delta_c+i\kappa-\int \left[U_\downarrow(\mathbf{r})\hat{n}_\downarrow(\mathbf{r})
+U_\uparrow(\mathbf{r})\hat{n}_\uparrow(\mathbf{r})\right]d\mathbf{r}},
\end{align}
where $\hat{n}_\tau(\mathbf{r})=\hat{\psi}^\dag_\tau(\mathbf{r})\hat{\psi}_\tau(\mathbf{r})$ and $\hat{\mathbf s}(\mathbf{r})=\hat\Psi^\dag(\mathbf{r})\pmb\sigma\hat\Psi(\mathbf{r})$ (with $\pmb\sigma$ being the vector of Pauli matrices) are the local density and spin operators, respectively, one can see that the cavity field is derived by pump-photon scattering processes involving the atomic spin operators $\hat{s}_+(\mathbf{r})=\hat{s}_-^\dag(\mathbf{r}) =\hat{\psi}^\dag_\uparrow(\mathbf{r})\hat{\psi}_\downarrow(\mathbf{r})$, rather than the atomic density as in the normal self-ordering of single component atoms; cf.\ for instance Equation~\eqref{eq:ss-a}. That is, the scattering of pump-laser photons into the cavity mode is accompanied by the atomic spin flips $\uparrow\leftrightarrow\downarrow$ and corresponding momentum kicks to the atom depending on the pump lasers' spatial profiles $\Omega_{1,2}(\mathbf{r})$.

\subsubsection{Cavity-induced long-range spin-spin interactions and effective spin Hamiltonians}
\label{subsubsec:cavity-induced-spin-int}

For the cavity-field dynamics evolving on a much faster time scale compared to the internal (i.e., spin) and external (i.e., center of mass) atomic dynamics, the former is locked to the latter ones in the sense of the Born-Oppenheimer approximation. To this end, the steady-state photonic field operator \eqref{eq:ss-a--spinor} can be adiabatically eliminated in close analogy to the single-component atoms described in Section~\ref{subsubsec:cavity-induced-int}. This leads to an effective atom-only Hamiltonian consisting of a (local) spin-independent non-interacting part for the center-of-mass motion, plus a long-range interaction part for the spin degree of freedom~\cite{Mivehvar2019Cavity}---cf.\ the penultimate term in the single-component effective atom-only Hamiltonian~\eqref{eq:H-eff-at-only},
\begin{align}\label{eq:eff-spin-H}
\hat{H}_{\rm spin}=\iint \Big\{
\sum_{\beta=x,y}&
J_{\rm Heis}^{\beta}(\mathbf{r}',\mathbf{r})\hat{s}_\beta(\mathbf{r}')\hat{s}_\beta(\mathbf{r})
+J_{\rm DM}^{z}(\mathbf{r}',\mathbf{r})
\left[\hat{s}_x(\mathbf{r}')\hat{s}_y(\mathbf{r})-\hat{s}_y(\mathbf{r}')\hat{s}_x(\mathbf{r})\right]
\nonumber\\
+&J_{\rm cc}^{xy}(\mathbf{r}',\mathbf{r})
\left[\hat{s}_x(\mathbf{r}')\hat{s}_y(\mathbf{r})+\hat{s}_y(\mathbf{r}')\hat{s}_x(\mathbf{r})\right]
\Big\}d\mathbf{r}d\mathbf{r}'
+\int B_z(\mathbf{r})\hat{s}_z(\mathbf{r}) d\mathbf{r}.
\end{align}
The first term in the effective spin Hamiltonian $\hat{H}_{\rm spin}$, Equation~\eqref{eq:eff-spin-H}, corresponds to the $x$ and $y$ components of a Heisenberg-type interaction $\hat{\mathbf s}(\mathbf{r}')\cdot\hat{\mathbf s}(\mathbf{r})$. The second term corresponds to the $z$ component of a Dzyaloshinskii-Moriya-type (DM) interaction $\hat{\mathbf s}(\mathbf{r}')\times\hat{\mathbf s}(\mathbf{r})$. The third term is a long-range coupling between the $x$ and $y$ components of the spins, hence referred to as the cross-component spin interaction. Finally, the last term serves as a local magnetic bias field, fixing the spin quantization axis. Depending on the sign of $J_{\rm Heis}^{\beta}$, the Heisenberg interaction favors ferromagnetic (FM) or antiferromagnetic (AFM) ordering. On the other hand, the DM and cross-component interactions favor chiral spin states such as spin spiral and skyrmion. It is the competition among these cavity-induced spin-spin interactions as well as the local magnetic bias field that determines the nature of the emergent magnetic order in the superradiant phase. The magnetic field and the spin-coupling coefficients are position dependent and are related to the cavity mode function and the pump-field spatial profiles as [cf.\ the density-density interaction strength $\mathcal{D}(\mathbf{r}',\mathbf{r})$ in Equation~\eqref{eq:cavity-interaction-strength}]: 
$B_z(\mathbf{r})=\hbar\delta/2+\hbar|\Omega_1(\mathbf{r})|^2/2\Delta_1-\hbar|\Omega_2(\mathbf{r})|^2/2\Delta_2$,
$J_{\rm Heis}^{x/y}=\text{Re}(c_1)\pm\text{Re}(c_2)$, $J_{\rm DM}^{z}=-\text{Im}(c_1)$, and $J_{\rm cc}^{xy}=-\text{Im}(c_2)$ with
\begin{align}
\label{eq:c1-c2}
c_1(\mathbf{r}',\mathbf{r})
&=2\hbar\mathcal{G}(x')\mathcal{G}(x)
\left[\frac{1}{\Delta_1^2\tilde\delta_c}\Omega_1(\mathbf{r}')\Omega_1^*(\mathbf{r})
      +\frac{1}{\Delta_2^2\tilde\delta_c^*}\Omega_2^*(\mathbf{r}')\Omega_2(\mathbf{r})\right],\nonumber\\
c_2(\mathbf{r}',\mathbf{r})
&=\frac{2\hbar\mathcal{G}(x')\mathcal{G}(x)}{\Delta_1\Delta_2}
\left[\frac{1}{\tilde\delta_c}\Omega_2(\mathbf{r}')\Omega_1^*(\mathbf{r})
      +\frac{1}{\tilde\delta_c^*}\Omega_1^*(\mathbf{r}')\Omega_2(\mathbf{r})\right].
\end{align}
The dispersively-shifted complex cavity detuning $\tilde\delta_c\equiv\delta_c+i\kappa=\Delta_c-\int [U_\downarrow(\mathbf{r})\hat{n}_\downarrow(\mathbf{r})+U_\uparrow(\mathbf{r})\hat{n}_\uparrow(\mathbf{r})]d\mathbf{r}+i\kappa$ has been introduced for the shorthand. We note that the coupling coefficients are periodic and sign changing, reminiscent of the Ruderman–Kittel–Kasuya–Yosida interaction in some heavy-fermion materials and metallic spin glasses~\cite{Gopalakrishnan2011Frustration}.

One can implement other spin models by utilizing different atom-photon coupling schemes. For instance, in a single $\Lambda$ scheme where one leg of the transition is driven by a pump laser and the second leg is coupled to another pump laser and a linear-cavity mode, in addition to the long-range $XXZ$ Heisenberg interactions and the cross coupling between the $x$ and $z$ components of the spins, cavity photons also induce long-range density-density and density-spin interactions~\cite{Masalaeva2020Spin}. Therefore, this cavity-QED system offers an alternative approach for simulating $t$-$J$-$V$-$W$-like models implemented via polar molecules in optical lattices~\cite{Gorshkov2011Tunable}. It has also been recently predicted that long-range, cavity-mediated interactions of the antiferromagnetic Heisenberg type can frustrate magnetic order enough to stabilize spin liquid phases, even in absence of geometric frustration~\cite{Chiocchetta2020Cavity}. We also note that the range of the cavity-mediated spin-spin (and density-spin) interactions can be tuned and become finite range by utilizing multimode cavities similar to the tunable density-density interactions in the single-component atoms coupled to multimode cavities as discussed in Section~\ref{subsec:tuning_interaction_range}~\cite{Gopalakrishnan2011Frustration}.

\subsection{Spinor Bose gases}
\label{subsec:spinor-BEC-self-ordering}

In this section, we focus on bosonic atoms with the effective Hamiltonian~\eqref{eq:single-particle-H-den-spinor}~\cite{Mivehvar2019Cavity}. The steady state of the system can be obtained for any given laser profiles $\Omega_{1,2}(\mathbf{r})$. An instructive and interesting scenario is to consider the spatial profiles $\Omega_{1,2}(y)=\Omega_{01,02}(e^{\pm ik_cy}+Re^{\mp ik_cy})/(1+R)$, where $0\leqslant R\leqslant1$ is the reflectivity of mirrors retroreflecting the pump lasers. $R=0$ and 1 correspond to pure running-wave and pure standing-wave pump lasers, respectively, while $0<R<1$ yields a mixture of running and standing waves. The amplitudes $\Omega_{01,02}$ and $\mathcal{G}_0$ are taken to be real, with the balanced Raman condition $\eta_0\equiv\mathcal{G}_0\Omega_{01}/\Delta_1=\mathcal{G}_0\Omega_{02}/\Delta_2$. 
Once again we note that the pump lasers do not interfere with each other, and the pump lattices $V_{\uparrow/\downarrow}(\mathbf{r})$ do not add up or cancel one another as they act on different internal atomic spin states. The two extreme limits of the above considered laser profiles corresponding to $R=1$ and $0$ have been experimentally realized and will be discussed in Sections~\ref{subsec:ExperimentSpinSO} and \ref{subsubsec:synthetic-SOC-experiment}, respectively. 

The mean-field approach for spinor bosonic atoms is identical to the single-component bosonic atoms discussed in Section~\ref{subsubsec:MF-bose}, saving that here $\Psi(\mathbf{r})=\langle\hat{\Psi}(\mathbf{r})\rangle=(\psi_\uparrow,\psi_\downarrow)^\top$ is a two-component condensate wave function. The mean-field phase diagram of the system in the $R$-$\eta_0$ plane is shown in Figure~\ref{Fig:Mivehvar2019Cavity_Fig1}(b) for $\text{Re}(\tilde{\delta}_c)<0$, to ensure the possibility of the superradiance phase transition. Below the pump-laser strength threshold $\eta_{0c}(R)$, the ground state of the system is a spin-polarized FM normal state with no photon in the cavity. At the threshold $\eta_{0c}(R)$, the system undergoes a Dicke superradiant phase transition into a magnetically ordered state. The nature of the emergent magnetic order depends on the spatial profiles of the lasers through $R$. 

\begin{figure}[t!]
\centering
\includegraphics [width=\textwidth]{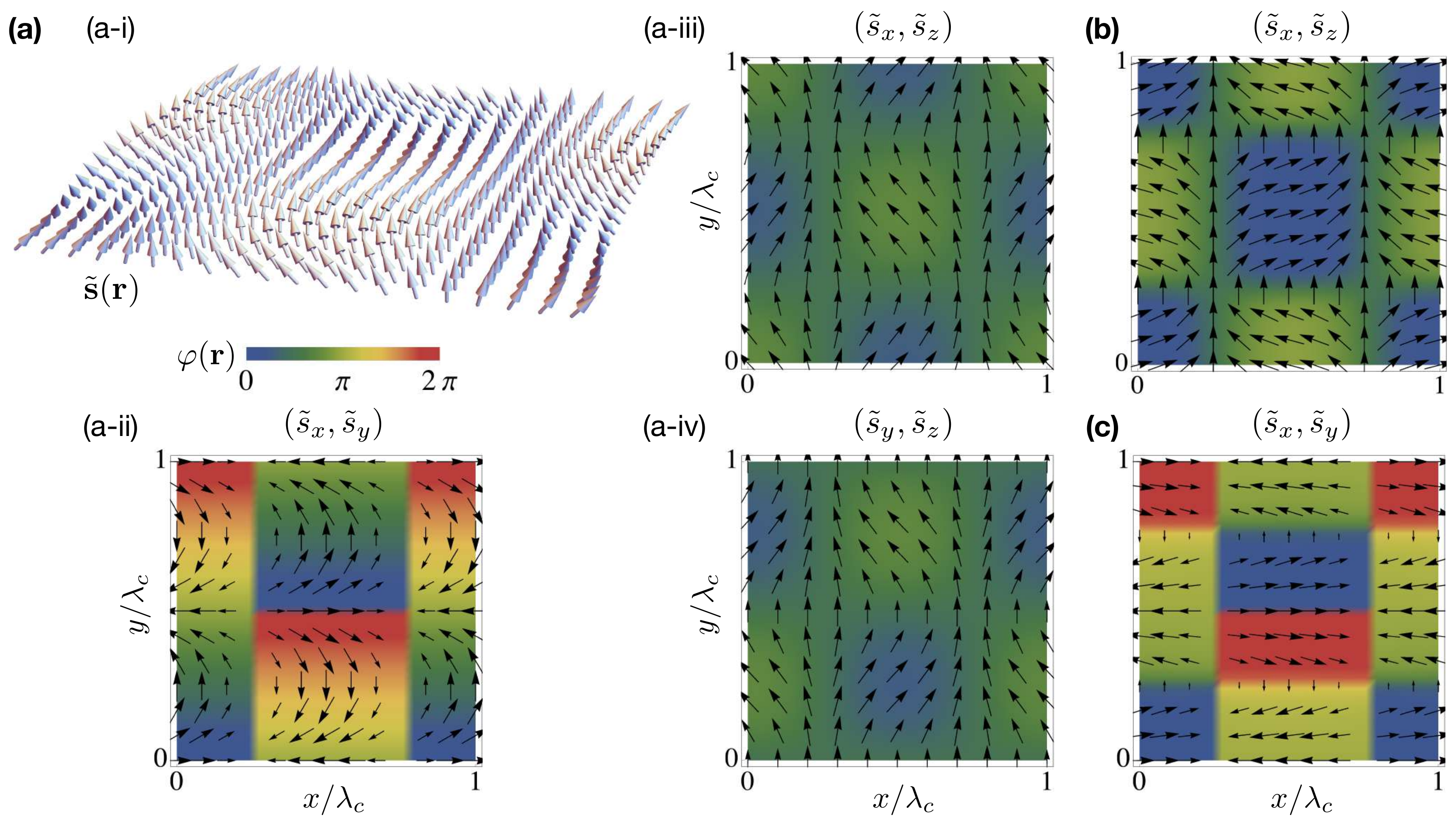}
\caption{
Typical cavity-induced chiral spin-spiral, domain-wall $X$-antiferromagnetic ($X$-AFM), and domain-wall spin-spiral textures.
(a) Shown is a chiral spin-spiral order (a-i) and its projections in different mutual spin planes (a-ii)-(a-iv).
The spin does a full $2\pi$ rotation in the $\tilde{s}_x$-$\tilde{s}_y$ plane along the $y$ direction over one $\lambda_c$, 
as can be seen clearly in the projection of $\tilde{\mathbf{s}}(\mathbf{r})$ into the $\tilde{s}_x$-$\tilde{s}_y$ spin plane (a-ii).
 The projection of the normalized spin texture into the 
(b) $\tilde{s}_x$-$\tilde{s}_z$ spin plane for the domain-wall $X$-AFM order and
(c) $\tilde{s}_x$-$\tilde{s}_y$ spin plane for the domain-wall spin-spiral texture.
 The color code in each figure indicates the respective spin angle, e.g., $\varphi=\tan^{-1}(\tilde{s}_y/\tilde{s}_x)$ in (a-ii).
Figure adapted and reprinted with permission from Ref.~\cite{Mivehvar2019Cavity}~\textcopyright~2019 by the American Physical Society.} 
\label{Fig:Mivehvar2019Cavity_Fig3}
\end{figure}

For pure standing-wave pumps $\Omega_{1,2}(y)=\Omega_{01,02}\cos(k_cy)$ corresponding to $R=1$, a spin wave is stabilized in the superradiant phase; see Figure~\ref{Fig:Mivehvar2019Cavity_Fig3}(b) which shows the projection of the normalized local spin $\tilde{\mathbf s}(\mathbf{r})\equiv\mathbf{s}(\mathbf{r})/s_n(\mathbf{r})$, with 
$\mathbf{s}(\mathbf{r})=\langle \hat{\mathbf s}(\mathbf{r})\rangle$ and 
$s_n(\mathbf{r})=\sqrt{s_x^2(\mathbf{r})+s_y^2(\mathbf{r})+s_z^2(\mathbf{r})}$,
in the $\tilde{s}_x$-$\tilde{s}_z$ plane as a function of $\mathbf{r}$. Note that different spin domains are separated by domain-wall lines. 
In strong pump strengths, the optical potentials $V_\tau(\mathbf{r})+U_\tau(\mathbf{r})|\alpha|^2$ (with $\tau=\{\downarrow,\uparrow\}$) and the Raman coupling $\eta_1^*(\mathbf{r})\alpha+\eta_2(\mathbf{r})\alpha^*=2\eta_0\text{Re}(\alpha)\cos(k_cx)\cos(k_cy)$  localize the atoms in a $\lambda_c/2$-periodic density pattern. Hence, the spin wave evolves into a $\lambda_c$-periodic checkerboard $X$-AFM spin lattice order in a more conventional sense.

In contrast, for pure running-wave lasers $\Omega_{1,2}(y)=\Omega_{01,02}e^{\pm ik_cy}$ corresponding to $R=0$, transverse conical, chiral spin-spiral order is favored. The spirals appear in the $\tilde{s}_x$-$\tilde{s}_y$ plane as the spin performs a full $2\pi$ rotation in this plane over one wave length $\lambda_c$ along the $y$ direction; see Figure~\ref{Fig:Mivehvar2019Cavity_Fig3}(a). For $0<R<1$, the system exhibits ``domain-wall spin-spiral'' texture; see Figure~\ref{Fig:Mivehvar2019Cavity_Fig3}(c). In this phase, although the projection of the spin in the $\tilde{s}_x$-$\tilde{s}_y$ plane still sweeps a full $2\pi$ angle over one wave length $\lambda_c$ along the $y$ direction, the $y$ component of the spin is considerably suppressed compared to the chiral spin-spiral order in the $R=0$ case. 

We note that in the case of two pure running-wave pumps $\Omega_{1,2}(y)=\Omega_{01,02}e^{\pm ik_cy}$, a dynamical spin-orbit coupling is induced for the BEC on the superradiant phase and the emergent spin-spiral magnetic order is indeed intimately connected to this dynamical spin-orbit coupling; this will be discussed in more detail in Section~\ref{subsec:synthetic-SOC}. By choosing more general pump configurations $\Omega_{1,2}(\mathbf{r})$, more exotic spin textures can be obtained.

\subsubsection{The effective spin Hamiltonian and the spin self-ordering}
\label{subsubsec:effective-spin-H_spin-slefordering}

The various emergent magnetic orders in the superradiant phase can be described and accounted for in an elegant way based on cavity-mediated spin-spin interactions and the spin Hamiltonian $\hat{H}_{\rm spin}$, Equation~\eqref{eq:eff-spin-H}. For the pure standing-wave pumps $\Omega_{1,2}(y)=\Omega_{01,02}\cos(k_cy)$ corresponding to $R = 1$, the coefficients $c_1$ and $c_2$ are identical (i.e., $c_1=c_2$) and real. Therefore, all the long-range spin-spin interactions vanish except the $x$ component of the Heisenberg Hamiltonian [see Equation~\eqref{eq:eff-spin-H} and the paragraph below it]. The spin Hamiltonian $\hat{H}_{\rm spin}$ then reduces to a $X$-type long-range Heisenberg model
$\iint J_{\rm Heis}^{x}(\mathbf{r}',\mathbf{r})\hat{s}_x(\mathbf{r}')\hat{s}_x(\mathbf{r}) d\mathbf{r}d\mathbf{r}'$ 
with the periodically modulated coupling strength 
$J_{\rm Heis}^{x}\propto\text{Re}(\tilde\delta_c)\cos(k_cx')\cos(k_cx)\cos(k_cy')\cos(k_cy)$. For the red-detuned dispersively shifted cavity frequency $\text{Re}(\tilde\delta_c)<0$, the cavity-induced Heisenberg interaction, therefore, leads to the checkerboard domain-wall $X$-AFM spin texture shown in Figure~\ref{Fig:Mivehvar2019Cavity_Fig3}(b). On the other hand for the pure running-wave pumps $\Omega_{1,2}(y)=\Omega_{01,02}e^{\pm ik_cy}$ corresponding to $R=0$, the coefficients $c_1$ and $c_2$ are both complex and different from each other, $c_1\neq c_2$. As a consequence, all the long-range spin-spin interactions in the effective spin Hamiltonian $\hat{H}_{\rm spin}$ are nonzero. They all have the same $\text{Re}(\tilde\delta_c)\cos(k_cx')\cos(k_cx)$ position dependence along the $x$ direction, but different modulations along the $y$ direction: $J_{\rm Heis}^x\propto\cos(k_cy')\cos(k_cy)$, $J_{\rm Heis}^y\propto\sin(k_cy')\sin(k_cy)$, $J_{\rm DM}^{z}\propto-\sin k_c(y'-y)$, and $J_{\rm cc}^{xy}\propto\sin k_c(y'+y)$. The DM and cross-component spin interactions stabilize transverse, conical spin-spiral order, while the magnetic domains are favored by the Heisenberg interactions as before; see Figure~\ref{Fig:Mivehvar2019Cavity_Fig3}(a). However, these cavity-mediated interactions are compatible with one another and do not lead to a frustration for this chosen geometry of the pump lasers and the cavity. The spirals appear solely in the $\tilde{s}_x$-$\tilde{s}_y$ plane as the DM interaction has only the $z$ component and the cross-component spin interaction only couples the $x$ and $y$ components of the spins. Note also that the $\lambda_c/4$ shift of the magnetic domains in the $\tilde{s}_x$-$\tilde{s}_z$ and $\tilde{s}_y$-$\tilde{s}_z$ planes along the $y$ axis is accurately accounted for by the $y$ dependence of $J_{\rm Heis}^{x/y}$ given above. For $0<R<1$, all the cavity-mediated spin-spin interactions are again nonzero. However, the coupling coefficients $J_{\rm Heis}^{y}$ and $\{J_{\rm DM}^{z}, J_{\rm cc}^{xy} \}$ approach zero, respectively, as $(1-R)^2$ and $1-R^2$ when $R\to1$. Whereas, the coupling coefficient $J_{\rm Heis}^{x}$ is independent of $R$ and remains the dominant coupling coefficient. This is indeed the reason for the suppression of the $y$ component of the spin in the domain-wall spin-spiral phase compared to the spin-spiral state where all the spin-spin interactions are in the same order of magnitude.

\subsubsection{The origin of the self-ordering in the generalized Dicke model}
\label{subsubsec:disorder-induced-slefordering}

For chosen pump lasers $\Omega_{1,2}(y)=\Omega_{01,02}(e^{\pm ik_cy}+Re^{\mp ik_cy})/(1+R)$, the system possesses a $\mathbf{Z}_2\otimes U(1)$ symmetry, where $\mathbf{Z}_2$ is the parity symmetry $\hat{a}\to\hat{a}'=-\hat{a}$ and $\mathbf{r}\to \mathbf{r}'=\mathbf{r}+(\lambda_c/2)\hat{e}_x$, and the continuous $U(1)$ symmetry represents the freedom of the total phase of the two bosonic atomic fields $\hat{\psi}_\tau(\mathbf{r})$. The $U(1)$ symmetry associated with the freedom of the relative phase of the two atomic fields is explicitly broken on the onset of the superradiant phase transition by the cavity-induced position-dependent Raman coupling term with the zero quantum average below the superradiant threshold $\langle \eta_1(\mathbf{r})\hat{a}^\dag+\eta_2^*(\mathbf{r})\hat{a} \rangle=0$ and with the always zero spatial average $\int[\eta_1(\mathbf{r})\hat{a}^\dag+\eta_2^*(\mathbf{r})\hat{a}]d\mathbf{r}=0$; see Equation~\eqref{eq:single-particle-H-den-spinor}. As a consequence, in the self-ordered phase the relative condensate phase varies in space to minimize the Raman coupling energy, leading to the spin orders. For instance, for the pure standing-wave pump lasers with $R=1$, the Raman energy is $2\hbar\eta_0(\alpha+\alpha^*)\int\sqrt{n_\uparrow n_\downarrow}\cos(k_cx)\cos(k_cy)\cos(\Delta\phi)d\mathbf{r}$, where $n_\tau(\mathbf{r})=\langle \hat{n}_\tau(\mathbf{r})\rangle$ are the condensate densities and $\Delta\phi(\mathbf{r})$ is the relative phase between the two condensate wave functions. As $\cos(k_cx)$ and $\cos(k_cy)$ change sign along the $x$ and $y$ directions, respectively, the relative phase $\Delta\phi(\mathbf{r})$ then jumps (discontinuously) along both $x$ and $y$ directions to minimize the energy, resulting in domain-wall AFM spin order. On the other hand, for the pure running-wave pump lasers with $R=0$, the Raman energy is $2\hbar\eta_0(\alpha+\alpha^*)\int\sqrt{n_\uparrow n_\downarrow}\cos(k_cx)\cos(k_cy-\Delta\phi)d\mathbf{r}$. As before the relative phase $\Delta\phi(\mathbf{r})$ jumps (discontinuously) along the $x$ direction to minimize the energy. However, to minimize the energy along the $y$ direction, it changes smoothly in a screw-like manner due to the term $\cos(k_cy-\Delta\phi)$, resulting in the chiral spin-spiral texture along the $y$ direction. Owing to the explicit breaking of the $U(1)$ relative-phase symmetry by the zero quantum- (below the superradiant threshold) and spatial-averaged Raman field, one may envisage the self-organization and the magnetic ordering in this generalized Dicke model (with both internal and external quantized degrees of freedom of the atoms included) as an order-by-disorder process~\cite{Mivehvar2017Disorder}, with disorder stemming from strong fluctuations of the cavity field around the superradiant threshold. This can be considered reminiscent of the spontaneous ordering in a two-dimensional classical ferromagnetic $XY$ spin model with a uniaxial random magnetic field~\cite{Minchau1985Two,Feldman1998Exact,Wehr2006Disorder,Niederberger2008Disorder}. The relative condensate phase is related to the spin angle, and the position-dependent Raman field with zero quantum (below the superradiant threshold) and spatial average plays the role of the random magnetic field. However, one has to note that quantum fluctuations here might not necessarily lower the free energy as in the original concept of order-by-disorder processes~\cite{Villain1980Order, Green2018Quantum}. 

\subsection{Spinful Fermi gases}

\subsubsection{The density and spin self-ordering}
\label{subsubsec:fermionic-density-spin-selfordering}

Closely related schemes to the model discussed in Section~\ref{subsec:multi-com-atoms-cavity} have been considered with ultracold fermionic atoms~\cite{Feng2017Quantum, Colella2018Quantum, Colella2019Antiferromagnetic}. Reference~\cite{Feng2017Quantum} has considered a similar double-$\Lambda$ atom-photon-coupling scheme as in the inset of Figure~\ref{Fig:Mivehvar2019Cavity_Fig1}(a) with a 2D Fermi gas perpendicular to the cavity axis (i.e., in the $y$-$z$ plane), where the pumps are also aligned along the cavity axis. In the mean-field limit, above a certain critical pump strength the Fermi gas undergoes a phase transition from a normal state into a superradiant state with partial polarization. Depending on the effective energy separation between the two spin states, the Dicke phase transition is either first order or second order. A variant of the scheme discussed in Section~\ref{subsec:multi-com-atoms-cavity} has been considered for a 1D Fermi gas within a ring cavity~\cite{Colella2019Antiferromagnetic}. In particular, the transitions  $\downarrow\>\leftrightarrow 1$ and $\uparrow\>\leftrightarrow 2$ are coupled to a pair of degenerate, counterpropagating modes of the ring cavity with the coupling strength $\mathcal{G}_\pm(x)=\mathcal{G}_0e^{\pm ik_cx}$. Below the Dicke superradiant phase transition, the system is in a normal mixed or polarized Fermi-gas state depending on the effective energy separation between the two spin states. A spin wave is stabilized in the superradiant phase, which evolves into an $X$-AFM lattice order in the strong pump limit. Due to the ring geometry of the cavity the system possesses a $U(1)\times\mathbf{Z}_2$ symmetry, where the former is spontaneously broken by the emergent density order and the latter by the $X$-AFM spin order at the onset of the superradiant phase transition. Furthermore, in moderate pump strengths the self-ordered states exhibit unexpected positive momentum pair correlations between fermions with opposite spin.

\subsubsection{Self-organization across the BEC-BCS crossover}

\begin{figure}[t!]
  \begin{center}
    \includegraphics[width=0.55\columnwidth]{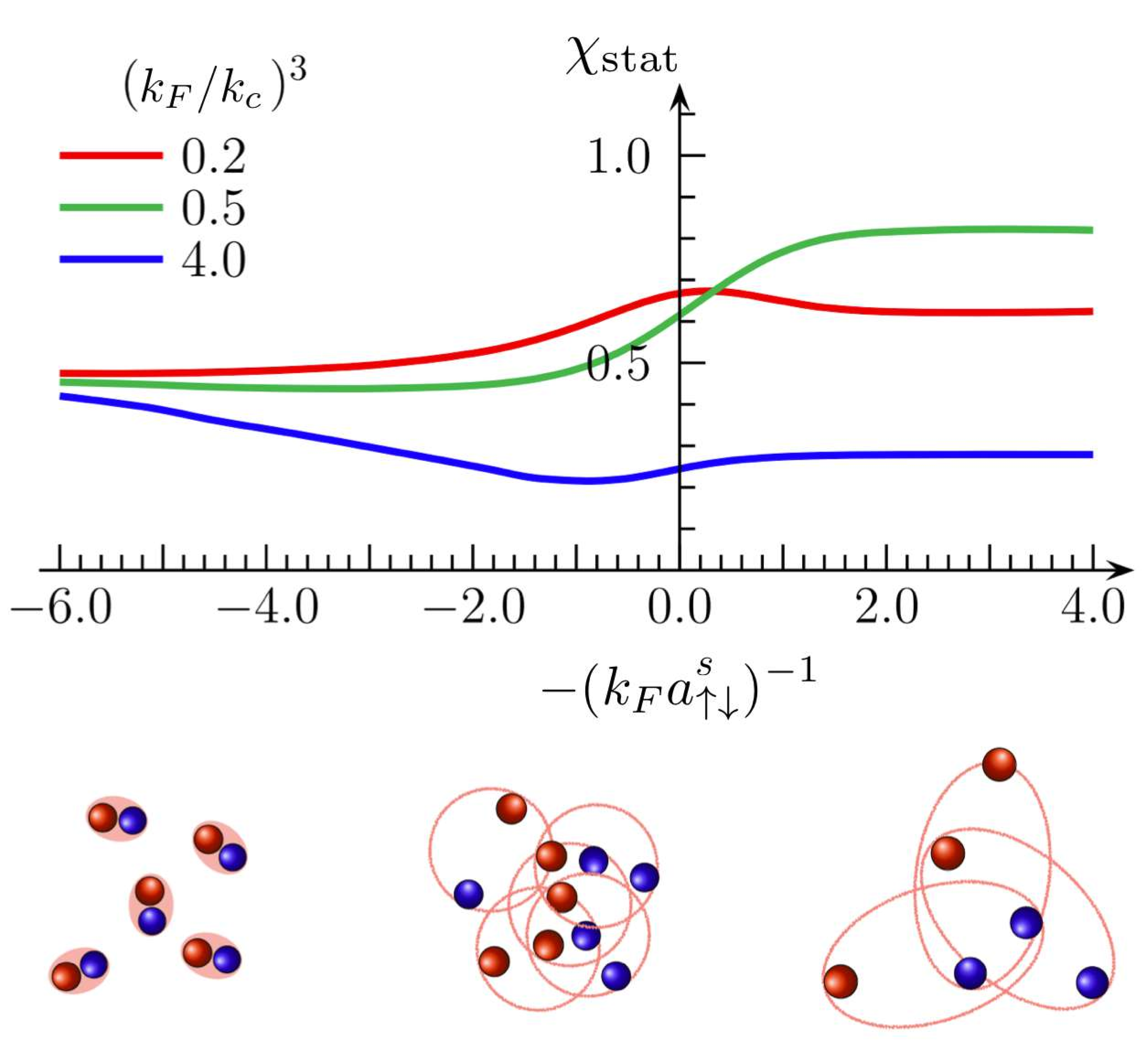}
    \caption{Behavior of the density response function $\chi_{\rm stat}$ at the cavity wave vector, 
    for interacting fermions in the BEC-BCS crossover. The latter is parametrized by the inter-species scattering length $a_{\uparrow\downarrow}^s$. The images below the plot illustrate pictorially how the size of the Cooper pairs changes in the BEC-BCS crossover. As discussed below Equation~\eqref{eq:stat_polarization}, $\chi_{\rm stat}$ quantifies the response of the atoms to a cavity photon and thus the tendency of the system to self-organize into the superradiant phase. Figure adapted and reprinted with permission from Ref.~\cite{Chen2015Superradiant}~\textcopyright~2015 by the American Physical Society.}
    \label{fig:superradiant_BECBCS}
  \end{center}
\end{figure}

The recent experimental realization of a tunable Fermi gas inside an optical cavity~\cite{Roux2020Strongly} has opened up interesting avenues of research with interacting spinful Fermi gases. In the experiments performed at the EPF Lausanne the scattering length governing the interactions between fermions in different spin states can be tuned using a standard Feshbach resonance. This allows to control the pairing between the atoms within the cavity, interpolating between the BEC limit (small positive scattering length), where tightly-bound pairs are well described as a condensate of localized bosons, and the BCS limit (small negative scattering length), where weakly-bound pairs are formed. Between these two limits the so-called unitary point is crossed, at which the scattering length diverges. The behavior of the cavity-induced self-organization transition across the BEC-BCS crossover has been studied theoretically in Ref.~\cite{Chen2015Superradiant}. The main finding, shown in Figure~\ref{fig:superradiant_BECBCS}, is that self-organization is maximally enhanced in the unitary regime for low densities, in the BCS regime for moderate densities close to Fermi surface nesting, or in the BEC regime for high densities.

\subsubsection{Cavity-induced superfluid pairing without superradiance}
\label{subsec:cavity-induced pairing}

In the scenario just discussed in the preceding section, the pairing is induced by the short-range atom-atom interactions, while the cavity-mediated interactions favor the self-organization of the atomic cloud into a density and spin pattern. Another interesting question is whether and how pairing can be instead induced by the cavity-mediated interactions. This question has been addressed in Ref.~\cite{Colella2018Quantum}, where $\Lambda$-type fermionic atoms are coupled to a far red-detuned transverse pump laser with the Rabi rate $\Omega(\mathbf{r})$ and a standing-wave cavity mode with strength $\mathcal{G}_0\cos(k_cx)$. In the bad cavity limit, cavity photons mediate global-range interaction between the atomic ground spin states,
\begin{align}
  \label{eq:colella_potential}
  \iint \mathcal{D}(\mathbf{r},\mathbf{r'})
  \hat{\psi}_\downarrow^\dag(\mathbf{r}) \hat{\psi}_\uparrow(\mathbf{r}) 
  \hat{\psi}_\uparrow^\dag(\mathbf{r'}) \hat{\psi}_\downarrow(\mathbf{r'})d\mathbf{r}d\mathbf{r'},
\end{align}
where $\mathcal{D}(\mathbf{r},\mathbf{r'})=(\mathcal{D}_0/\Omega_0^2)\Omega(\mathbf{r})\Omega(\mathbf{r}')\cos(k_cx)\cos(k_cx')$ with $\mathcal{D}_0=\hbar\Delta_c\Omega_0^2\mathcal{G}_0^2/\Delta_a^2|\tilde\Delta_c|^2$; cf.\ Equation~\eqref{eq:cavity-interaction-strength}. Restricting to 1D along the cavity axis, i.e., $\Omega(\mathbf{r})=\Omega_0$, for $\mathcal{D}_0<0$ one finds the equivalent of the $X$-AFM spin texture discussed above and shown in Figure~\ref{Fig:Mivehvar2019Cavity_Fig3}(b). One the other hand, for $\mathcal{D}_0>0$ a superfluid phase of paired fermions is favored. In Ref.~\cite{Zheng2020Cavity}, the possibility of generating the more exotic Fulde-Ferrel-Larkin-Ovchinnikov type of pairing---where the fermion pair has a net center-of-mass momentum---is discussed.
As we discussed previously, for positive $\mathcal{D}_0>0$ corresponding to $\Delta_c>0$, superradiance is strongly suppressed. The superfluid paired phase is indeed not of the superradiant type, i.e., coherent macroscopic occupation of the cavity mode does not occur in any pump strength. In this case,  superradiant and pairing processes, therefore, strongly exclude each other, and in particular for $\mathcal{D}_0<0$ only superradiance is possible. In Section~\ref{sec:cavity-induced-lattice-pairing} we discuss another scheme for cavity-induced pairing in an external optical lattice inside ring cavities in the red-detuned regime, $\Delta_c<0$.

\subsection{Experimental realizations of spin self-organization}
\label{subsec:ExperimentSpinSO}

Spin self-organization of atomic systems has been experimentally explored using schemes similar to the one discussed in Sections~\ref{subsec:multi-com-atoms-cavity} and~\ref{subsec:spinor-BEC-self-ordering}. While first experiments were based on thermal atoms~\cite{Zhiqiang2016Nonequilibrium, Zhang2018Dicke}, we here discuss a more recent realization using a BEC~\cite{Kroeze2018Spinor}. Operating in the quantum degenerate regime implies that experiments take place at effectively zero temperature, that all atoms couple identically to light fields as they share the same spatial wave function, and that the self-organization of the matter wave can be clearly observed in the appearance of sharp momentum peaks. In the spin-changing superradiant phases as discussed in Section~\ref{subsec:spinor-BEC-self-ordering}, the atomic spins are flipped due to the cavity-assisted Raman scattering processes [see Figure~\ref{Fig:Mivehvar2019Cavity_Fig1}(a)]. Correspondingly, recoil momenta associated with the Raman processes are imparted to the atoms. As a result, the spinor superradiant phases are spin-decorated patterns (see Figure~\ref{Fig:Mivehvar2019Cavity_Fig3}).

\begin{figure}[t!]
\centering
\includegraphics[width=0.85\textwidth]{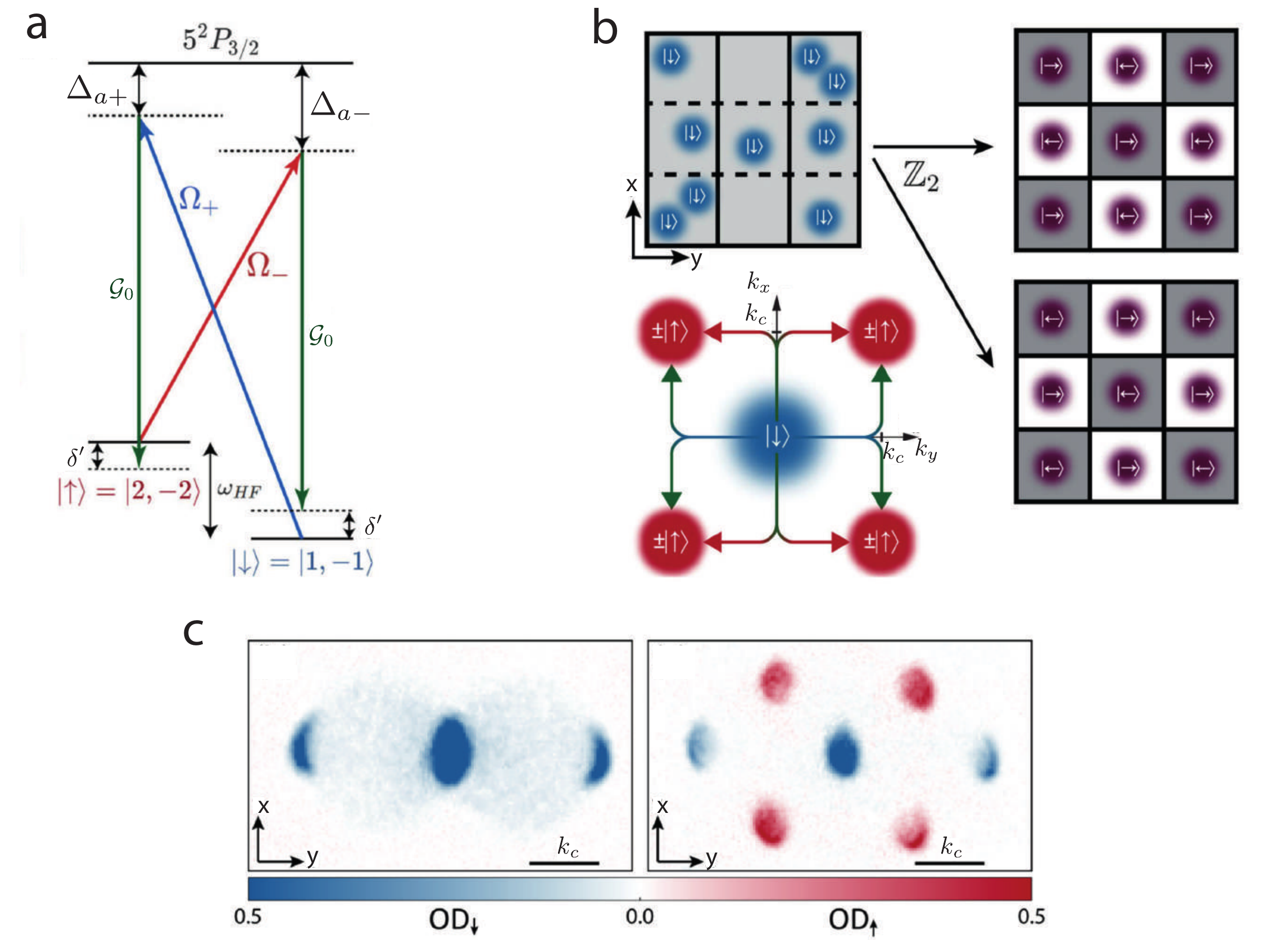}
\caption{Observation of the density and spin self-organization of a spinor BEC. (a)~Experimental scheme using two hyperfine states of $^{87}$Rb as the effective spin-1/2 system. These states are coupled via two separate cavity-assisted Raman transitions, where each transition involves the cavity mode and an additional coherent pump field. (b) The Raman scattering processes flip the spin of the atoms while transferring them to a superposition of momentum states (bottom). The system undergoes a transition from the normal phase with no spin ordering to a spin-decorated checkerboard superradiant state breaking a $\mathbf{Z}_2$ symmetry. (c) Spin-selective absorption images of the atomic cloud after ballistic expansion in the normal (left) and the superradiant phase (right). The images reveal that the excited momentum states in the superradiant phase are only populated by spin flipped atoms. Figure adapted and reprinted with permission from Ref.~\cite{Kroeze2018Spinor}~\textcopyright~2018 by the American Physical Society.}
\label{fig:SpinSO}
\end{figure}

The experimental scheme used in Ref.~\cite{Kroeze2018Spinor} is shown in Figure~\ref{fig:SpinSO}(a), which corresponds to the $R=1$ case discussed in Section~\ref{subsec:spinor-BEC-self-ordering}. Two different hyperfine ground states of $^{87}$Rb are chosen as the effective spin-1/2 system ($\ket{\uparrow}$, $ \ket{\downarrow}$) and coupled via two different Raman processes, each involving a coherent pump field and the TEM$_{00}$ cavity mode. The amplitudes of the Raman processes are balanced, such that an effective Dicke model is realized. The standing-wave pump beams illuminate the atoms from the side and are arranged such that the resulting pump lattices are in phase at the location of the atoms. The frequency of the light field scattered into the cavity by either Raman process is the mean frequency of the two coherent Raman pump fields illuminating the atoms. Since the frequencies of the pump fields and the cavity field are different, the potentials for the densities are square lattices while the spin potential has a checkerboard shape [see Equation~\eqref{eq:single-particle-H-den-spinor}]. During self-organization, an according spin-decorated checkerboard pattern forms. 

Figure~\ref{fig:SpinSO}(b) shows the underlying scattering processes together with illustrations of the normal and possible superradiant phases breaking a $\mathbf{Z}_2$ symmetry. In the organized phase, coherent Raman scattering creates a superposition of atoms in the initial zero-momentum state with the spin state $\ket{\downarrow}$ and of atoms in the recoiled momentum state with the spin state $\ket{\uparrow}$. Spin-selective absorption imaging of the atomic cloud after ballistic expansion allows to distinguish both the momentum states and the spin states. According images in the normal and the superradiant phases are shown in Figure~\ref{fig:SpinSO}(c).

The experiment was repeated by utilizing the TEM$_{10}$ mode instead of the TEM$_{00}$ mode. The node in the field of the TEM$_{10}$ mode is equivalent to a $\pi$ phase shift in the plane perpendicular to the cavity axis. This spatial phase shift of the cavity field leads to the creation of a spinor domain wall across the superradiant self-organization. In the experiment, the phase of the field leaking from the cavity was analyzed spatially by interfering it with a local oscillator in a camera, thus realizing a spatially resolved homodyne detection. Indeed, a $\pi$ phase shift across the nodal line was observed in the cavity field. This nodal line acts as a domain wall in the atomic systems. Accordingly, also the momentum distribution of the atomic cloud---derived from absorption images after ballistic expansion---shows a node in the Bragg peaks of the self-organized atomic cloud.

\section{Lattice superradiance: generalized extended Hubbard models}
\label{sec:extended-BH}

So far for single cavity setups, we have considered scenarios where the atoms are initially free along the cavity axis below the superradiance phase transition. Only beyond the superradiant phase transition, photons scattered from the pump laser into the cavity create an optical lattice along the cavity axis for the atoms. The atoms then self-order in this emergent potential into a Bragg density grating which maximizes the photon scattering from the pump into the cavity. We now consider lattice scenarios where the atoms are already initially trapped in a strong 2D ``external'', static optical lattice below the superradiant phase transition. As before, photons scattered by the atoms from a transverse pump field into the cavity result in cavity-mediated long-range interactions, competing directly with the kinetic energy and the local interactions of the strongly correlated atoms~\cite{Maschler2005Cold, Maschler2008Ultracold, Fernandez-Vidal2010Quantum, Li2013Lattice, Bakhtiari2015Nonequilibrium, Caballero-Benitez2015Quantum, Dogra2016Phase, Chen2016Quantum, Sundar2016Lattice, Flottat2017Phase, Panas2017Spectral, Liao2018Theoretical, Chiacchio2018Tuning, Nagy2018Quantum, Blass2018Quantum, Igloi2018Quantum, Himbert2019Mean, Lin2019Superfluid, Chen2020Extended, Chanda2020Self-organized, Kubala2020Ergodicity,Sicks2020Haldane}. Here, for instance, the cavity-mediated long-range interactions can be incommensurate with respect to the external static lattice spacing, leading to frustration.   

For ultracold atoms in a static optical lattice, the competition between the kinetic energy and the local contact interaction leads to the quantum phase transition between a superfluid (SF) state and a Mott-insulator (MI) phase \cite{Jaksch1998Cold, Greiner2002Quantum}. The Mott-insulator state is favored in the strongly interacting regime where the atoms are localized in potential minima and the coherence between the atoms is lost. Adding the cavity-mediated long-range interactions into this system enriches significantly the physics resulting, in addition to the superfluid and Mott-insulator phases, in new states in the superradiant phase including lattice-supersolid (LSS), charge-density-wave (CDW), and Bose-glass (BG) states. The lattice supersolid is the lattice variant of the supersolid discussed in Section~\ref{sec:supersolid}, where here the crystalline order is stabilized through a discrete spatial symmetry breaking (cf.\ Section~\ref{sec:supersolid}). The charge-density wave is an insulating state with a modulated periodic density. The Bose glass is a compressible state, containing disjoint superfluid islands in the Mott-insulator background. It is a characteristic state of disordered Bose-Hubbard models.

\subsection{One-component extended Bose-Hubbard model}
\label{sec:1comp-extended-BH}

\begin{figure}[t!]
\centering
\includegraphics [width=.5\textwidth]{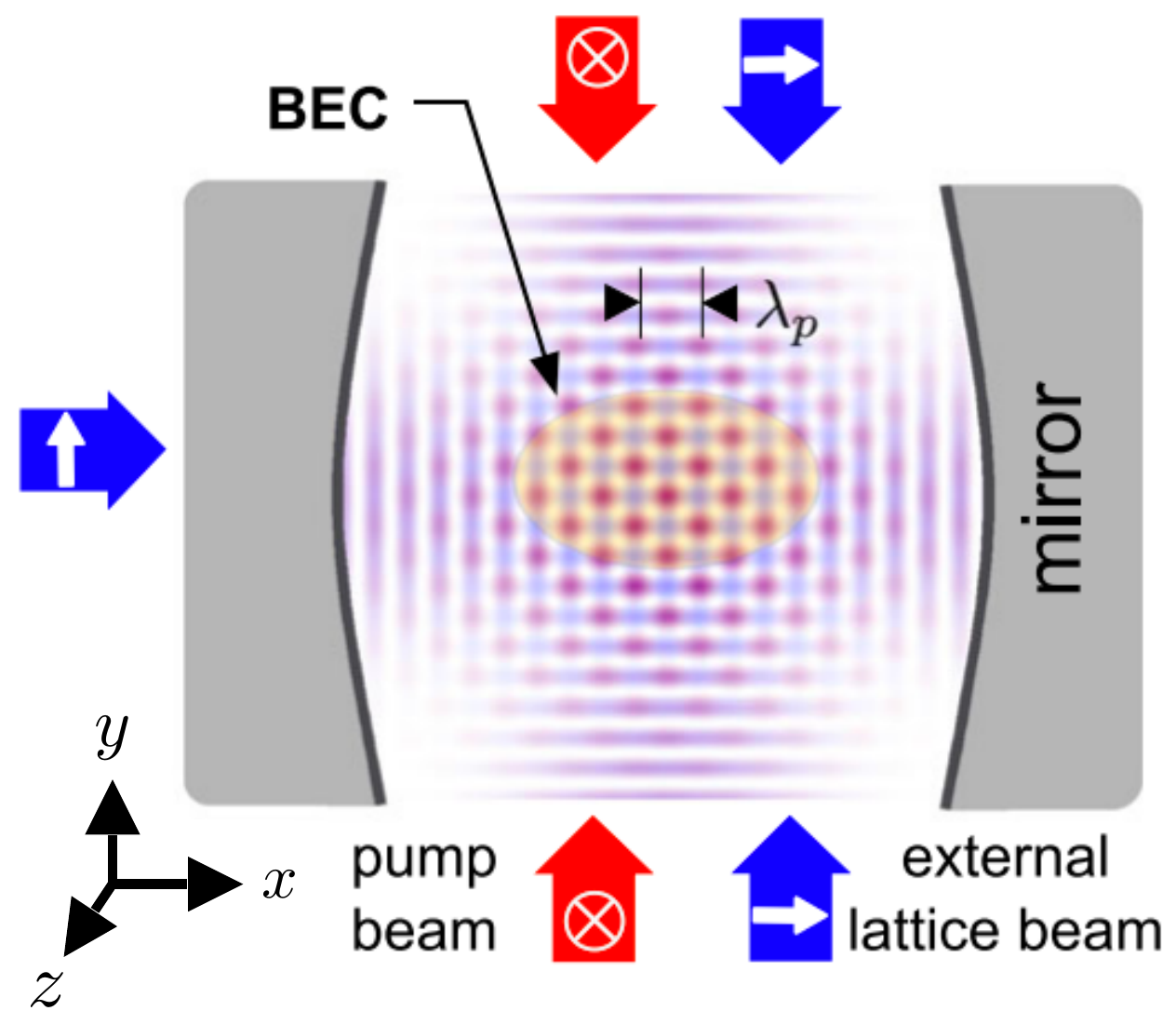}
\caption{Schematic sketch of a setup for realizing a generic lattice model with cavity-mediated global interactions.
Bosonic atoms in a 2D pancake in the $x$-$y$ plane are trapped by a strong external, statice square lattice generated by two standing-wave fields (indicated in blue) inside a cavity.
The cavity mode is populated via photon scattering of the pump laser (indicated in red) off the atoms. The cavity photons mediate global, all-to-all interactions among the atoms, described by an extended Bose-Hubbard-type Hamiltonian. Likewise, an extended Hubbard-type Hamiltonian can be realized for fermionic atoms.
Figure adapted and reprinted with permission from Ref.~\cite{Bakhtiari2015Nonequilibrium}~\textcopyright~2015 by the American Physical Society. 
} 
\label{Fig:Bakhtiari2015Nonequilibrium_Fig1}
\end{figure}

Let us now consider a generic lattice model inside a linear cavity~\cite{Maschler2008Ultracold, Li2013Lattice, Bakhtiari2015Nonequilibrium, Dogra2016Phase, Chen2016Quantum, Sundar2016Lattice, Flottat2017Phase, Panas2017Spectral, Liao2018Theoretical, Chiacchio2018Tuning, Nagy2018Quantum, Himbert2019Mean}. Ultracold bosonic atoms are trapped in a three dimensional external \emph{static} lattice, generated by three mutually orthogonal standing-wave laser fields not interfering with each other. The BEC is split into a stack of identical 2D pancakes along the $z$ direction by the lattice, and we focus our description into one of these pancakes. In the $x$-$y$ plane, the atoms experience the external, static square optical lattice 
\begin{align}
\label{eq:Vext-latt}
V_{\rm ext}(\mathbf{r})=V_{\rm latt}^{(0)}\left[\cos^2(k_{\rm latt}x)+\cos^2(k_{\rm latt}y)\right], 
\end{align}
with the lattice spacing $\lambda_{\rm latt}/2=\pi/k_{\rm latt}$ and lattice depth $V_{\rm latt}^{(0)}$. The polarizations and frequencies of the lasers generating this lattice potential are chosen such that the photon scattering from these lasers into the cavity by the atoms are either not possible or strongly suppressed. As depicted in Figure~\ref{Fig:Bakhtiari2015Nonequilibrium_Fig1}, the atoms are further driven in the $y$ direction by a standing-wave pump laser with wavenumber $k_c=2\pi/\lambda_c$ and strongly coupled to an empty mode of the cavity with the mode function $\cos(k_cx)$, similar to the basic model described in Section~\ref{sec:basic-model}. The initially empty cavity mode is populated by photons scattered off the atoms from the transverse pump field. This results in an additional, \emph{dynamical} potential on top of the classical external static lattice. In the 2D $x$-$y$ plane, the system is still described by the Hamiltonian of Equation~\eqref{eq:H_eff_1comp}, with $V_{\rm ext}(\mathbf{r})$ given by the lattice potential of Equation~\eqref{eq:Vext-latt}. The external static lattice and the dynamical potentials can be commensurate or incommensurate with each other, depending on the ratio $\lambda_c/\lambda_{\rm latt}$. These two cases will be considered separately in Sections~\ref{sec:ceBH} and~\ref{sec:ieBH}.

For a very strong external classical lattice $V_{\rm ext}(\mathbf{r})$, as a lowest-order approximation one can ignore the modification of $V_{\rm ext}(\mathbf{r})$ by the transverse pump and dynamical potentials and expand the atomic field operator in the basis of Wannier functions of the lowest Bloch band of the classical lattice as $\hat{\psi}(\mathbf{r})=\sum_{\mathbf{j}}W_\mathbf{j}(\mathbf{r})\hat{b}_\mathbf{j}$, where $W_\mathbf{j}(\mathbf{r})$ is the lowest-band Wannier function localized at  lattice site $\mathbf{r}_\mathbf{j}=\mathbf{j}\lambda_{\rm latt}/2=(j_x,j_y)\lambda_{\rm latt}/2$ with $\{j_x,j_y\}\in\mathbb{Z}$ and $\hat{b}_\mathbf{j}$ is the corresponding atomic annihilation operator satisfying the canonical bosonic commutation relation. By doing so, the transverse pump and the dynamical potentials are basically treated as perturbations; see Section~\ref{sec:corrections-cavity-eBHm} for a brief discussion regarding beyond perturbative treatments. Therefore, in the lowest-order approximation the system is described by the lattice Hamiltonian,
\begin{align}
\label{eq:H-latt}
\hat{H}_{\rm latt}&=
-J\sum_{\langle \mathbf{j},\mathbf{j}'\rangle}(\hat{b}_\mathbf{j}^\dag\hat{b}_{\mathbf{j}'}+\text{H.c.})
+\frac{1}{2}U_s \sum_{\mathbf{j}}\hat{n}_\mathbf{j}(\hat{n}_\mathbf{j}-1)
+\sum_{\mathbf{j}}\left(\hbar V_0B_\mathbf{j}^{(y)}-\mu\right)\hat{n}_\mathbf{j}\nonumber\\
&+\hbar\sum_{\mathbf{j}}\left[U_0\hat{a}^\dag\hat{a}B_\mathbf{j}^{(x)}
+\eta_0(\hat{a}^\dag+\hat{a})A_\mathbf{j}\right]\hat{n}_\mathbf{j}
-\hbar\Delta_c\hat{a}^\dag\hat{a},
\end{align} 
where $J$ is the hopping amplitude, $U_s$ the strength of the short-range contact interaction, $\hat{n}_\mathbf{j}=\hat{b}_\mathbf{j}^\dag\hat{b}_\mathbf{j}$ the local particle number operator at site $\mathbf{j}$, $\mu$ is the chemical potential included explicitly to fix the number of the particles, and $\{V_0,U_0,\eta_0\}$ are defined as before (see Section~\ref{sec:basic-model}). Like $\{J,U_s\}$, the coefficients $A_\mathbf{j}$ and $B_\mathbf{j}^{(x/y)}$ are related to overlap integrals of the Wannier functions (i.e., matrix elements of the spatial part of the pump and dynamic cavity potentials): 
\begin{align}
\label{eq:A-B-coeff-cBHM}
A_\mathbf{j}&\equiv\int |W_\mathbf{j}(\mathbf{r})|^2\cos(k_cx)\cos(k_cy)d\mathbf{r},\nonumber\\
B_\mathbf{j}^{(x)}&\equiv\int |W_\mathbf{j}(\mathbf{r})|^2\cos^2(k_cx)d\mathbf{r},\nonumber\\ 
B_\mathbf{j}^{(y)}&\equiv\int |W_\mathbf{j}(\mathbf{r})|^2\cos^2(k_cy)d\mathbf{r}. 
\end{align}
Note that off-site matrix elements of the potentials have been ignored here, as these contributions are normally much smaller than on-site matrix elements, Equation~\eqref{eq:A-B-coeff-cBHM}. The contribution of these higher-order corrections will be discussed briefly in Section~\ref{sec:corrections-cavity-eBHm}. We note that in the absence of the cavity field [the second line in Equation~\eqref{eq:H-latt}], the lattice Hamiltonian $\hat{H}_{\rm latt}$ reduces to the common Bose-Hubbard Hamiltonian describing the dynamics of locally interacting bosonic atoms in the static lattice $V_{\rm ext}(\mathbf{r})$ with a site-dependent chemical potential $\mu-\hbar V_0B_\mathbf{j}^{(y)}$ due to the transverse pump-field lattice potential $\hbar V(\mathbf{r})$ that can have a different wave vector than the original static lattice. 

 The distinction between phenomena arising due to the dynamical (semiclassical) nature of cavity-light potentials and effects due to their quantumness was underlined in Refs.~\cite{Caballero-Benitez2015Quantum, Caballero-Benitez2016Quantum}. Here we focus on the large cavity decay-rate limit, where the cavity field reaches very fast its steady state,
\begin{align}
\label{eq:ss-a-lattice}
\hat{a}_{\rm ss}=\frac{\eta_0\sum_\mathbf{j}A_\mathbf{j}\hat{n}_\mathbf{j}}
{\Delta_c+i\kappa-U_0\sum_\mathbf{j}B_\mathbf{j}^{(x)}\hat{n}_\mathbf{j}}.
\end{align}
The cavity field acquires a nonzero value only if the ``weighted'' average density operator [cf.\ Equation~\eqref{eq:order-paramter-op}],
\begin{align}
\label{eq:theta}
\hat\Theta\equiv\frac{1}{N_\Lambda}\sum_\mathbf{j}A_\mathbf{j}\hat{n}_\mathbf{j},
\end{align}
with $N_\Lambda$ being the total number of lattice sites, is nonzero. Adiabatically eliminating the steady-state photonic field operator $\hat{a}_{\rm ss}$~\eqref{eq:ss-a-lattice} under the assumption that $\Delta_c,\kappa\gg U_0\sum_\mathbf{j}B_\mathbf{j}^{(x)}\hat{n}_\mathbf{j}$ and retaining terms up to $1/\Delta_a^2$ yields a generalized, extended Bose-Hubbard model, 
\begin{align}
\label{eq:eBH}
\hat{H}_{\rm eBH}&=
-J\sum_{\langle \mathbf{j},\mathbf{j}'\rangle}\left(\hat{b}_\mathbf{j}^\dag\hat{b}_{\mathbf{j}'}+\text{H.c.}\right)
+\frac{1}{2}U_s \sum_{\mathbf{j}}\hat{n}_\mathbf{j}(\hat{n}_\mathbf{j}-1)
+\sum_{\mathbf{j}}\left(\hbar V_0B_\mathbf{j}^{(y)}-\mu\right)\hat{n}_\mathbf{j}
\nonumber\\
&+N_\Lambda U_l\Big(\frac{1}{N_\Lambda}\sum_\mathbf{j}A_\mathbf{j}\hat{n}_\mathbf{j}\Big)^2,
\end{align} 
where the last term represents the cavity-mediated global interaction with the strength $U_l=2\hbar N_\Lambda\Delta_c\eta_0^2/(\Delta_c^2+\kappa^2)$~\cite{Dogra2016Phase}; cf.\ Equations~\eqref{eq:H-eff-at-only} and~\eqref{eq:cavity-interaction-strength}. 
A negative cavity-mediated long-range interaction  $U_l<0$ favors a state which maximizes the absolute value of $\Theta=\langle\hat\Theta\rangle$. The parameter $\Theta$ is zero in the superfluid and Mott-insulator states with average uniform densities. The onset of nonzero $\Theta$ signals a quantum phase transition into other ordered states with modulated densities, driven by the cavity-mediated interaction $N_\Lambda U_l\hat\Theta^2$. The nonzero $\Theta$ also corresponds to the superradiant phase with nonzero cavity field amplitude $\alpha_{\rm ss}=\langle \hat{a}_{\rm ss}\rangle\neq0$; see Equation~\eqref{eq:ss-a-lattice}. Therefore, $\Theta=\langle\hat\Theta\rangle$ can be considered as an order parameter, the lattice analog of the continuum order parameter introduced in Equation~\eqref{eq:order-paramter-op}. The nature of the states with $\Theta\neq0$ depends crucially on the ratio $\lambda_c/\lambda_{\rm latt}$, as we will see shortly in Sections~\ref{sec:ceBH} and \ref{sec:ieBH}.    

The extended Bose-Hubbard Hamiltonian~\eqref{eq:eBH} can be diagonalized exactly for small system sizes. Other numerical techniques such as quantum Monte Carlo~\cite{Flottat2017Phase}, variational Monte Carlo~\cite{Bogner2019Variational}, multi-configurational time-dependent Hartree method~\cite{Lin2019Superfluid}, and even exact diagonalization of an infinite system in the hardcore limit (in 1D)~\cite{Blass2018Quantum,Igloi2018Quantum} have also been utilized to find the ground state of this system. Due to the fully-connected, long-range nature of cavity-mediated interactions, the mean-field approach provides, however, a faithful description of the system in a wide range of parameters in the thermodynamics limit which we focus on in the following section (see also Section~\ref{subsec:MF}). That said, one has to note that only the long-range interaction part of the Hamiltonian is treated appropriately by the mean-field approximation. When the short-range hopping terms and/or the on-site repulsions are also included, the dimensionality plays an important role---implying, e.g., substantial shifts of phase boundaries and/or even completely vanishing of some phases. Before discussing the mean-field approach, we finally also note that such extended Bose-Hubbard models can also be obtained in the deep superradiant phase without an external optical lattice~\cite{Fernandez-Vidal2010Quantum, Lin2019Superfluid} or even in longitudinally pumped cavities due to the optomechanical interaction~\cite{Maschler2005Cold, Larson2008Mott, Larson2008Quantum}.

\subsubsection{Mean-field approximation}

The extended Bose-Hubbard Hamiltonian, Equation~\eqref{eq:eBH}, can be studied most readily using a mean-field method. This mean-field approach decouples the off-site terms in the Hamiltonian $\hat{H}_{\rm eBH}$~\cite{Sheshadri1993Superfluid}:
\begin{align}
\label{eq:MF-decoupling}
\hat{b}_{\mathbf{j}}^\dag\hat{b}_{\mathbf{j}'}+\hat{b}_{\mathbf{j}'}^\dag\hat{b}_{\mathbf{j}}
&\simeq\beta_{\mathbf{j}'}\hat{b}_{\mathbf{j}}^\dag+\beta_{\mathbf{j}'}^*\hat{b}_{\mathbf{j}}
+\beta_{\mathbf{j}}\hat{b}_{\mathbf{j}'}^\dag+\beta_{\mathbf{j}}^*\hat{b}_{\mathbf{j}'}
-\left(\beta_{\mathbf{j}}^*\beta_{\mathbf{j}'}+\beta_{\mathbf{j}}\beta_{\mathbf{j}'}^*\right),\nonumber\\
\hat\Theta^2&\simeq2\Theta\hat\Theta-\Theta^2,
\end{align}
where $\beta_\mathbf{j}=\langle\hat{b}_{\mathbf{j}}\rangle$ is the superfluid order parameter and $\Theta=\langle\hat\Theta\rangle$ is the ``Bragg density'' order parameter, assumed to be independent of the superfluid order parameter. This assumption is used frequently in the literature, but its validity has not yet been checked thoroughly.

\paragraph{Commensurate: Superfluid, Mott-insulator, density-wave, and lattice-supersolid states}
\label{sec:ceBH}
We first consider the simplest case corresponding to $\lambda_c=\lambda_{\rm latt}$. One then has $B=B_\mathbf{j}^{(x)}=B_\mathbf{j}^{(y)}$ independent of site $\mathbf{j}$ and $A_\mathbf{j}=(-1)^{j_x+j_y}A$ (with $A\equiv |A_\mathbf{j}|$) alternating its sign between $\ell\equiv j_x+j_y\in$ even (e) and odd (o) sites. The Hamiltonian~\eqref{eq:eBH} can then be recast as,
\begin{align}
\label{eq:eBH-commensurate}
\hat{H}_{\rm ceBH}&=
-J\sum_{\langle e,o\rangle}\left(\hat{b}_e^\dag\hat{b}_o+\hat{b}_o^\dag\hat{b}_e\right)
+\frac{1}{2}U_s \sum_{\ell\in e,o}\hat{n}_\ell(\hat{n}_\ell-1)
-\left(\mu-\hbar V_0B\right)\sum_{\ell\in e,o}\hat{n}_\ell\nonumber\\
&+\frac{A^2}{N_\Lambda}U_l\Big(\sum_{e}\hat{n}_e-\sum_{o}\hat{n}_o\Big)^2,
\end{align} 
where the total number of atoms $N=\langle \hat{N}\rangle=\sum_j\langle \hat{n}_j\rangle$ has to be determined self-consistently. Let us first heuristically discuss the extended Bose-Hubbard Hamiltonian~\eqref{eq:eBH-commensurate} and consider in particular the effect of the cavity-mediated long-range interactions $N_\Lambda U_l\hat\Theta^2\propto U_l (\sum_{e}\hat{n}_e-\sum_{o}\hat{n}_o)^2$. In the absence of the cavity-mediated interaction, $U_l=0$, the Hamiltonian~\eqref{eq:eBH-commensurate} reduces to the common Bose-Hubbard model, where for $J/U_s>J_c(\mu/U_s)/U_s$ with $J_c(\mu/U_s)$ being the critical hopping amplitude the system is in the superfluid state.  For $J/U_s<J_c(\mu/U_s)/U_s$, the system enters into the Mott-insulator state, where the atoms are localized on lattice sites with an equal, integer number of atoms per site and the quantum coherence between the atoms is lost. Both states respect the symmetry between even and odd sites. However, a negative cavity-mediated long-range interaction  $U_l<0$ favors a population imbalance between even and odd sites, which breaks the $\mathbf{Z}_2$ symmetry of even and odd sites [originating from the independence of the long-range interaction energy $\propto(\sum_{e}\hat{n}_e-\sum_{o}\hat{n}_o)^2$ on the sign of the population imbalance]. For large enough cavity-mediated interaction strength, this results in a $\lambda_c$-periodic checkerboard density structure which acts as a Bragg grating and maximizes the photon scattering from the pump lattice to the cavity field, signaling the superradiant phase transition. The resultant state in the superradiant phase with a population imbalance is either a charge-density wave or a lattice supersolid, depending on if the quantum coherence between the atoms is lost or not.

The phase diagram of the system can be obtained using the decoupling mean-field approach, Equation~\eqref{eq:MF-decoupling}. Application of the mean-field decoupling~\eqref{eq:MF-decoupling} to the Hamiltonian $\hat{H}_{\rm ceBH}$~\eqref{eq:eBH-commensurate} leads to an effective two-site mean-field Hamiltonian
\begin{align}
\label{eq:MFeBH-commensurate}
\hat{H}_{\rm ceBH}^{\rm MF}&\simeq \sum_{\ell\in e,o}\left[-J\tilde\beta_\ell\left(\hat{b}_\ell^\dag+\hat{b}_\ell-\beta_\ell\right)
+\frac{U_s}{2} \hat{n}_\ell(\hat{n}_\ell-1)
-\left(\mu-\hbar V_0B\right)\hat{n}_\ell\right]\nonumber\\
&+2AU_l\Theta\Big(\sum_{e}\hat{n}_e-\sum_{o}\hat{n}_o\Big)
-N_\Lambda U_l\Theta^2,
\end{align}
where, without loss of generality, we have assumed that the superfluid order parameters are real $\beta_{e/o}\in\mathbb{R}$ (as there is no gauge potential), $\tilde\beta_e=\sum_{n.n.o}\beta_o=4\beta_o$ (assuming an isotropic situation in this 2D model each even site has four identical, nearest-neighbor odd sites and vice versa), $\tilde\beta_o=4\beta_e$, and $\Theta=\langle\hat\Theta\rangle=A(\sum_en_e-\sum_on_o)/N_\Lambda$ with $n_{e/o}=\langle \hat{n}_{e/o}\rangle$. The Bose-Hubbard model with the cavity-mediated infinite-range interaction in the mean-field level, represented by the Hamiltonian $\hat{H}_{\rm ceBH}^{\rm MF}$ in Equation~\eqref{eq:MFeBH-commensurate}, is equivalent to the extended Bose-Hubbard Hamiltonian with the nearest-neighbor density-density interaction $\propto\sum_{\langle \mathbf{j},\mathbf{j}'\rangle}\hat{n}_\mathbf{j}\hat{n}_{\mathbf{j}'}$~\cite{Dogra2016Phase}. The mean-field Hamiltonian $\hat{H}_{\rm ceBH}^{\rm MF}$ can be diagonalized in a unit cell consisting of an even and an odd lattice site, with the order parameters $\beta_{e/o}$ and $\Theta$ to be determined self-consistently.

\begin{figure}[t!]
\centering
\includegraphics[width=1\columnwidth]{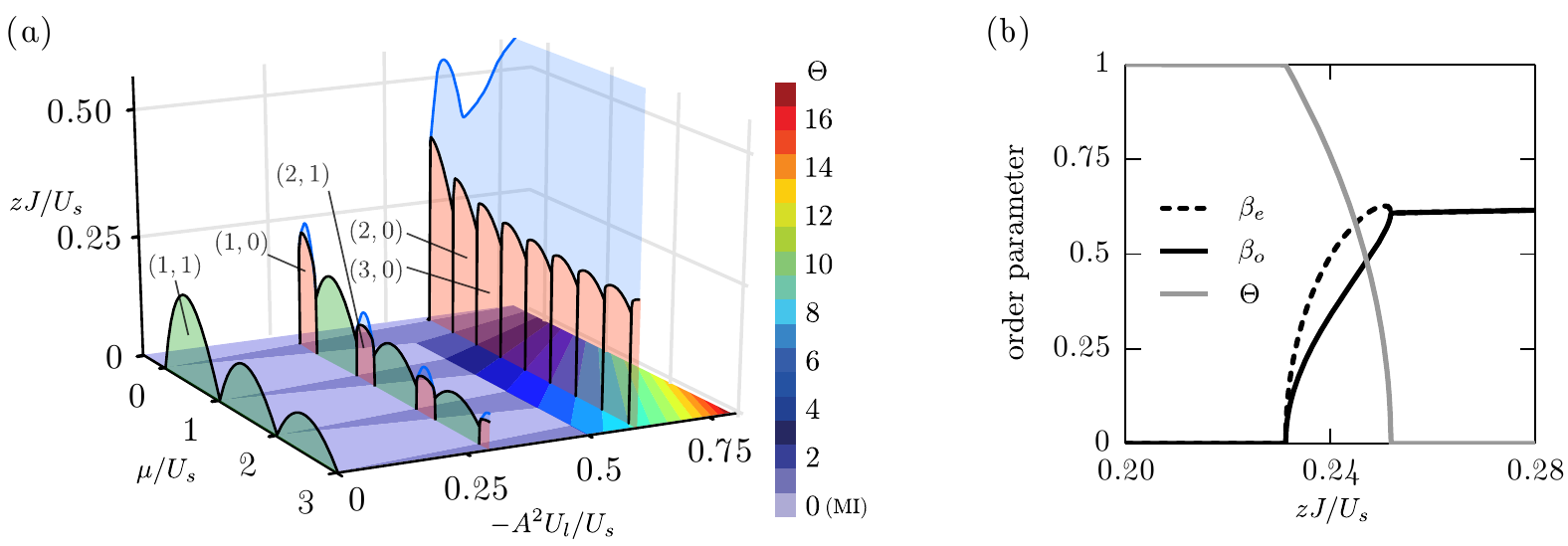}
\caption{Phase diagram of a cavity-induced commensurate extended Bose-Hubbard model with $\lambda_c=\lambda_{\rm latt}$. (a) Phase diagram as function of rescaled parameters $\mu/U_s$, $zJ/U_s$, and $-A^2 U_l/U_s$ with coordination number $z$.  The colorbar refers to $zJ/U_s=0$ and indicates the average imbalance $\Theta$ between the number of atoms on an even and an odd site. The transparent colors show the different phases for $zJ/U_s> 0$: MI (green), CDW (red), and LSS (blue). The SF phase is not indicated, but fills the remaining space. The labels correspond to the site populations ($n_e$,$n_o$). (b) Evolution of the different order parameters as a function of tunneling $zJ/U_s$ for filling 1/2, corresponding to a cut through the CDW(1,0) lobe at $U_l/U_s = 0.3$ in panel (a). With increasing $zJ/U_s$, the system evolves from CDW over LSS to SF. Here, all order parameters change continuously across the phase transitions. Figure and caption adapted and reprinted with permission from Ref.~\cite{Dogra2016Phase}~\textcopyright~2016 by the American Physical Society.}
\label{fig:eBHModelPhaseDiagram}
\end{figure}

All phases of the system can be identified using the order parameters $\beta_{e/o}$ and $\Theta$. Zero superfluid order parameters $\beta_e=\beta_o=0$ indicate an incompressible insulating state $\partial\bar{n}/\partial\mu=0$, with $\bar{n}=\sum_j\langle \hat{n}_j\rangle/N_\Lambda=N/N_\Lambda$ being the average density. This incompressible state is a Mott insulator if $\Theta=0$, otherwise a charge-density-wave state. On the other hand $\beta_e=\beta_o\neq0$ and $\Theta=0$ corresponds to a superfluid sate, while $\beta_{e/o}\neq0$ with $\beta_e\neq\beta_o$ and $\Theta\neq0$ signals the existence of a so-called lattice supersolid. The phase diagram of the system is shown in Figure~\ref{fig:eBHModelPhaseDiagram}.  The system can be studied in a similar way for other integer ratios of $\lambda_c/\lambda_{\rm latt}$.

\begin{figure}[t!]
\centering
\includegraphics [width=\textwidth]{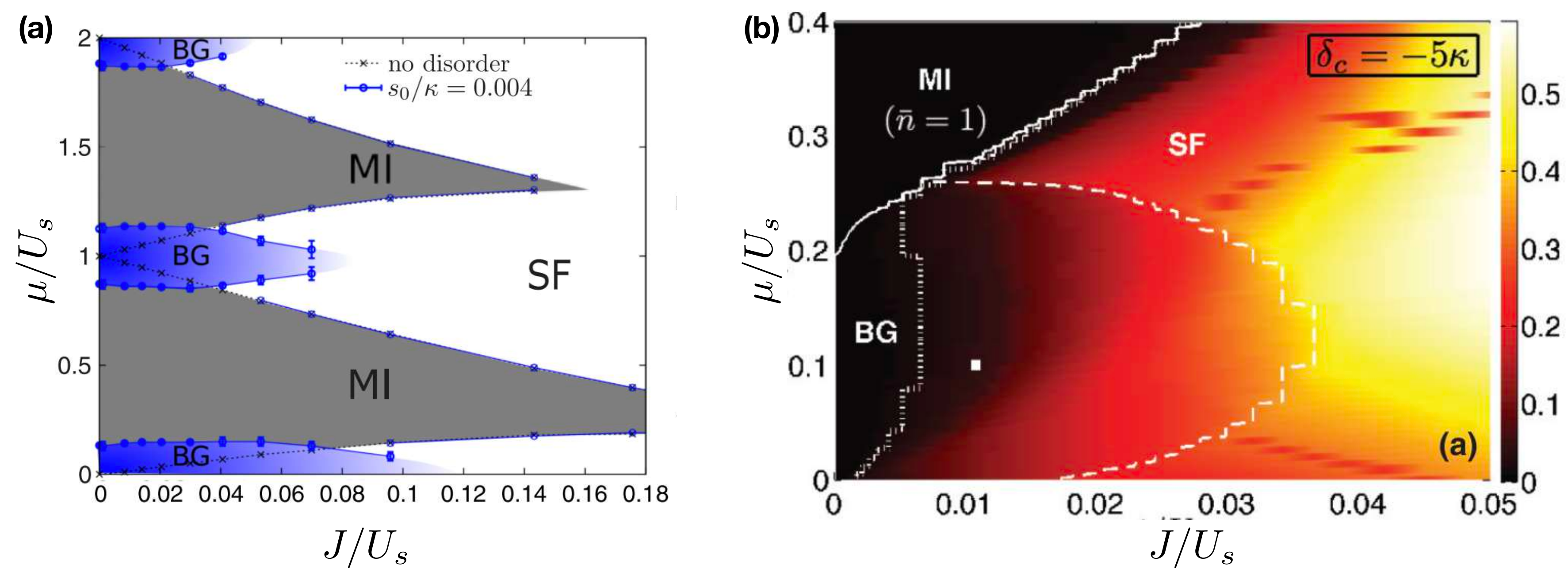}
\caption{Phase diagram of a cavity-induced incommensurate extended Bose-Hubbard model 
in the $\mu$-$J$ plane for the irrational ratio of $\lambda_c/\lambda_{\rm latt}=830/785$. 
The phase diagram of the system includes Mott-insulator (MI), superfluid (SF), and Bose-glass (BG) states, 
obtained in 1D via quantum Monte Carlo simulations (a) 
and in 2D using mean-field calculations (b). In panel (b) the dotted and solid curves delineate the regions with a small, arbitrarily 
chosen threshold for average superfluid order parameter and the density fluctuations, respectively, 
while inside the dashed curve the number of photons is at least 100 times larger than outside.
The region with finite compressibility and vanishing superfluid order parameter corresponds to the Bose-glass phase.
In the Bose-glass phase, the system forms clusters with checkerboard density distribution, where the atoms  
scatter photons in phase into the cavity mode. 
Panel (a) adapted and reprinted with permission from Ref.~\cite{Habibian2013Bose} and panel (b) from Ref.~\cite{Habibian2013Quantum}~\textcopyright~2013 by the American Physical Society.} 
\label{Fig:Habibian2013BoseFig2and3}
\end{figure}

\paragraph{Incommensurate: Bose-glass phase}
\label{sec:ieBH}
We now consider the case where the ratio $\lambda_c/\lambda_{\rm latt}$ is not a rational number~\cite{Habibian2013Bose, Habibian2013Quantum, Niederle2016Ultracold, Kubala2020Ergodicity}. Therefore, the external classical lattice is not commensurate with the pump and dynamical potentials, resulting in a non-periodic configuration. One may envisage the cavity-mediated interaction as a disorder in this case, since the amplitude of the interaction proportional to $A_\mathbf{j}$ oscillates at the cavity wavelength $\lambda_c$, which is incommensurate with respect to the external lattice. The phase diagrams of the system in 1D (along the cavity axis $x$) and 2D are shown in Figures~\ref{Fig:Habibian2013BoseFig2and3}(a) and (b), obtained from quantum Monte Carlo and mean-field simulations, respectively~\cite{Habibian2013Bose}. In addition to the superfluid and Mott-insulator states, the system exhibits another phase with interesting properties. It comprises a compressible state $\partial\bar{n}/\partial\mu\neq0$ with vanishing superfluid density. The atomic density in this phase exhibits a checkerboard density-wave pattern with an envelope at the beating wavenumber $|k_{\rm latt}-k_c|$, such that the Bragg density order parameter $\Theta=\sum_\mathbf{j}A_\mathbf{j}\langle\hat{n}_\mathbf{j}\rangle/N_\Lambda$ acquires a nonzero value. The envelope arises due to the fact that the system tries to maximize $\Theta$ by convoluting the (checkerboard) density as $A_\mathbf{j}$ oscillates at the cavity wavelength $\lambda_c$ and becomes out of phase with respect to the external lattice with the incommensurate wavelength $\lambda_{\rm latt}$. Therefore, the atoms can constructively scatter photons from the pump laser into the cavity. This state has the characteristic of a Bose-glass (BG) phase typical in disordered systems.

The effect of the cavity-mediated interactions is particularly evident in the 1D case, where the pump term is a constant shift in the chemical potential [cf.\ Equation~\eqref{eq:eBH}]: At small hopping amplitudes the sizes of the Mott lobes reduce and they give way for the Bose-glass states (compare it with the dashed curves for the phase diagram without cavity-mediated interactions $U_l=0$). By increasing the hopping amplitude, the Bose glass melts and gives way to the superfluid state. Note that in 1D the Bose-glass state corresponds exactly to the superradiant phase, where the cavity field is populated macroscopically. However, in 2D a Bose-glass state can also arise due to the pump potential which is also incommensurate with respect to the external optical lattice. The color map in Figure~\ref{Fig:Habibian2013BoseFig2and3}(b) represents the average superfluid order parameter $\sum_\mathbf{j} \beta_\mathbf{j}/N_\Lambda$ and the dashed line delineates the superradiant phase. In the single-particle case, the incommensurate cavity potential can lead to the Anderson-type localization of the atom and mobility edges---energy eigenstates that separate coexisting extended and localized states---even in the purely optomechanical regime~\cite{Zhou2011Cavity, Rojan2016Localization, Major2018Single}.

In Ref.~\cite{Yin2020Localization}, a closely related scenario to the schemes discussed in this section has been studied: it has been assumed that non-interacting bosonic atoms are already trapped in a 1D static, external quasiperiodic potential---composed of two incommensurate lattices and described by the single-band Aubry-Andr\'{e} model---along the cavity axis and pumped by a laser in the transverse direction. If the strength of disorder in the external Aubry-Andr\'{e} lattice exceeds a critical value, it leads to the Anderson localization of the atoms. As a consequence, the atoms scatter pump photons into the cavity mode completely in phase and constructively, driving the Dicke-superradiant instability even in vanishingly small pump strengths. Such superradiant instability with vanishing pump strengths has also been predicted for fermionic atoms, but due to a completely different mechanism---Fermi surface nesting~\cite{Piazza2014Umklapp}.

\subsubsection{Higher-order corrections to cavity-induced extended Bose-Hubbard models}
\label{sec:corrections-cavity-eBHm}

In deriving the lattice Hamiltonian~\eqref{eq:H-latt}, we have assumed that the external classic lattice $V_{\rm ext}(\mathbf{r})$ and its lowest-band Wannier functions are not modified by the transverse pump and dynamical potentials. However, this is not strictly true, specially in the strong pump and the deep superradiant regime. Therefore, one has to find \emph{self-consistently} the Wannier functions of the lowest Bloch band of the total lattice potential, 
$V_{\rm ext}(\mathbf{r})+\hbar V(\mathbf{r})+\hbar U(\mathbf{r})\hat{a}^\dag\hat{a}+\hbar \eta(\mathbf{r})(\hat{a}^\dag+\hat{a})$.
That is, the steady-state cavity field $\hat{a}_{\rm ss}$, Equation~\eqref{eq:ss-a-lattice}, depends on the matrix elements~\eqref{eq:A-B-coeff-cBHM} calculated using the Wannier functions, which in turn depend on the cavity field through the total lattice potential. Correspondingly, the hopping amplitude $J$ and the strength of the short-range contact interaction $U_s$ must be also determined self-consistently. This is a genuine many-body cavity-QED nonlinear effect, where the coupled matter-field dynamics depend crucially on the atomic density. Such a nonlinear effect is most prominent when the dynamic cavity potential is the sole lattice potential in the system, e.g., by longitudinally pumping a cavity~\cite{Maschler2005Cold} where the existence of overlapping, competing Mott-insulator states as well as bistable behavior has been predicted~\cite{Larson2008Mott, Larson2008Quantum}. 
 
As mentioned already, in deriving the lattice Hamiltonian~\eqref{eq:H-latt} only the on-site matrix elements of the spatial part of the pump and dynamic cavity potentials, Equation~\eqref{eq:A-B-coeff-cBHM}, have been kept. Although much smaller compared to the on-site matrix elements, retaining the off-site matrix elements of the potentials alters the lattice Hamiltonian~\eqref{eq:H-latt} crucially. Namely, after adiabatic elimination of the steady-state cavity field $\hat{a}_{\rm ss}$ these off-site contributions modify the bare hopping amplitude $J$ and result also in cavity-induced correlated two-particle hoppings proportional to $[\sum_{\langle \mathbf{j},\mathbf{j}'\rangle}(\hat{b}_\mathbf{j}^\dag\hat{b}_{\mathbf{j}'}+\text{H.c.})]^2$ in the extended Bose-Hubbard Hamiltonian. As a consequence, phase boundaries between different phases are modified~\cite{Maschler2005Cold,Maschler2008Ultracold}. In particular, in the vicinity of phase boundaries different photon-number states are associated with different atomic states, so that the ground state of the system in the proximity of phase boundaries contains atomic states of different nature correlated with different photon numbers due to photon-number fluctuations. A scenario for enhancing the off-site matrix elements while suppressing the on-site ones which can result in additional quantum phases was discussed in Refs.~\cite{Caballero-Benitez2015Quantum, Caballero-Benitez2016Quantum}.

\subsubsection{Experimental realizations of the extended Bose-Hubbard model}
\paragraph{Quantum phases from competing interactions}
The experimental realization of the above described commensurate extended Bose-Hubbard model with cavity-mediated global-range interactions (Section~\ref{sec:ceBH}) requires to create a static but tunable 3D optical lattice inside an optical cavity. The depth of this lattice controls the strength of the on-site short-range collisional interactions between the atoms, while the cavity is used to induce competing global-range interactions. For a sufficiently deep 3D optical lattice, the BEC will---for small enough global-range interactions---undergo a quantum phase transition from a superfluid phase to a Mott-insulating phase featuring an integer number of particles with equal occupations of even and odd sites~\cite{Greiner2002Quantum}. In contrast, cavity-mediated global-range interactions favor superradiant states with an occupation imbalance between even and odd sites. For small enough short-range interactions, the system will undergo a transition from the SF phase to a lattice supersolid phase with a modulated superfluid state. If both short-range and global-range interactions are strong, the system will enter a superradiant, insulating charge-density wave state with a periodically modulated integer site occupation.

\begin{figure}[t!]
\centering
\includegraphics[width=0.75\columnwidth]{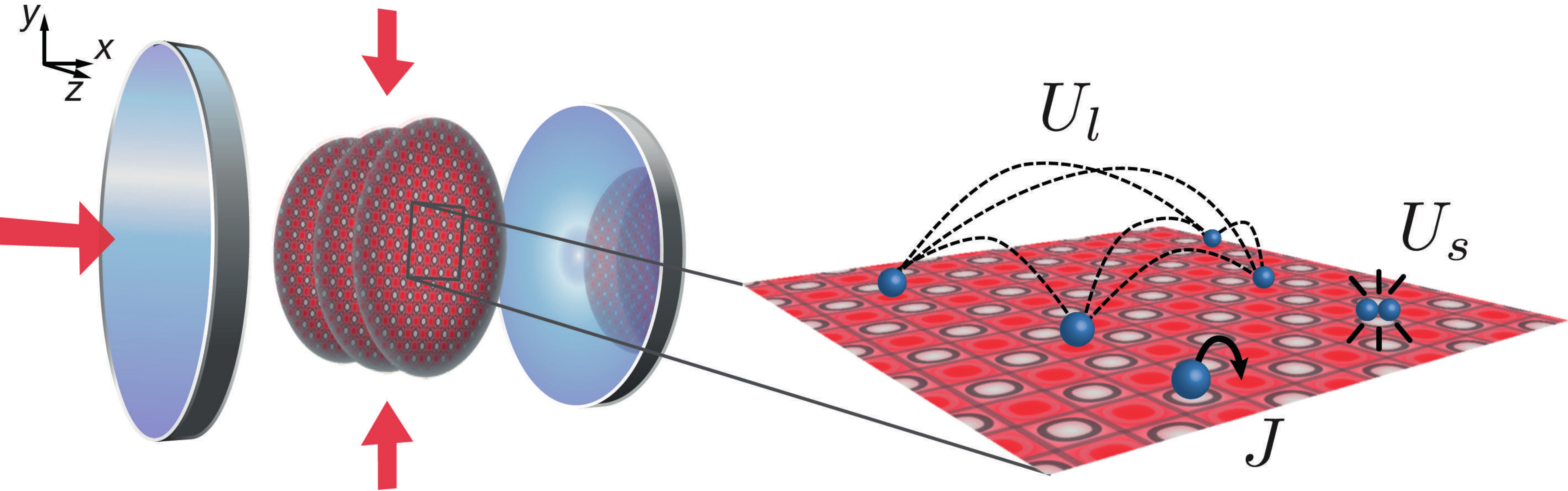}
\caption{Experimental setup to realize the commensurate extended Bose-Hubbard model. A BEC is trapped at the center of an optical cavity and loaded into a deep optical lattice along the $z$ axis, which slices the cloud into 2D layers in the $x$-$y$ plane. A TEM$_{00}$ mode of the cavity is pumped to create an intra-cavity lattice along the $x$ direction, which forms together with a free space lattice along the $y$ direction a square optical lattice of tunable depth $V_{\rm latt}^{(0)}$. The lattice along the $y$ direction simultaneously acts as a transverse pump with lattice depth $V_0=V_{\rm latt}^{(0)}$ for the orthogonally-polarized  TEM$_{00}$ cavity mode inducing cavity-mediated global-range interactions. The figure on the right displays the fundamental processes determining the physics within a 2D layer: atoms are subject to global interactions with strength $\propto U_l$,  favoring a density modulation, and to repulsive on-site interactions with strength $U_s$, favoring localization of the atomic wave function on individual lattice sites. Finally, tunneling between neighboring lattice sites with amplitude $J$ favors delocalization of the atomic wavefunction. Figure adapted and reprinted with permission from Ref.~\cite{Landig2016Quantum} published at 2016 by the Nature Publishing Group.}
\label{fig:eBHModel}
\end{figure}

Figure \ref{fig:eBHModel} displays the basic experimental setup together with the relevant atomic processes. A BEC trapped at the center of an optical cavity by external fields is loaded into a deep optical lattice along the $z$ axis, effectively freezing out tunneling in this direction. This slices the cloud into 2D pancakes in the $x$-$y$ plane. These 2D layers are further exposed to a classical optical lattice in the $x$-$y$ plane, formed by one free space lattice along the $y$ direction and one intra-cavity optical standing wave along the $x$ direction. By a  choice of perpendicular polarizations and a large enough relative detuning, these two fields do not interfere such that the resulting potential is a square lattice. This corresponds to the external optical lattice $V_\mathrm{ext}(\boldsymbol{r})$ introduced in Equation (\ref{eq:Vext-latt}) with adjustable lattice depth $V_{\rm latt}^{(0)}$, allowing to tune the strength of the short-range interactions as well as the hopping amplitude. At the same time, the standing-wave potential along the $y$ axis acts additionally as a transverse pump lattice with lattice depth $V_0=V_{\rm latt}^{(0)}$ [cf.\ Equation~\eqref{eq:H-latt}] for another cavity mode with the similar polarization, inducing the cavity-mediated global interactions. The resonance frequency of the cavity is chosen with an adjustable detuning $\Delta_c$ with respect to the frequency of this pump field. As a result, the strength of the cavity-mediated global-range interactions $\propto A^2U_l$ can be tuned by changing $\Delta_c$ or by varying $V_{\rm latt}^{(0)}$. Most importantly, this arrangement allows to change the ratio of short-range and global-range interactions which can thus be brought into competition. The physics of this system is captured in a wide range of parameters by a Hamiltonian of the form of Equation~\eqref{eq:eBH-commensurate}. It is, however, important to note that due to the twofold role of the transverse pump as the external lattice and the pump field simultaneously, the transverse pump is not a perturbative term anymore. This requires to calculate the Wannier functions, $J$, $U_s$, and other coefficients consistently using $V_{\rm latt}^{(0)}=V_0$. The fundamental processes underlying the different terms in the Hamiltonian are visualized in Figure~\ref{fig:eBHModel}.

In order to detect different quantum phases reached by tuning the parameters, two observables are necessary. First, the presence or absence of an intra-cavity light field is a measure whether the atomic system has an imbalanced occupation of the even and odd lattice sites or not [see Equations~\eqref{eq:ss-a-lattice} and~\eqref{eq:theta}], allowing to distinguish the superfluid and Mott-insulating phases from the lattice-supersolid and the charge-density-wave phases. Second, absorption images of the atomic cloud after ballistic expansion are used to probe the coherence of the gas. Observation of narrow momentum peaks and a high condensate fraction indicate the presence of coherence typical for the superfluid and lattice-supersolid phases. In contrast, broadened peaks and a vanishing condensate fraction indicate the loss of coherence and can be associated with the loss of superfluidity and the formation of an insulating phase (i.e., the Mott-insulator and charge-density-wave phases). Combining both observables---related to $\beta_{e/o}$ and $\Theta$ in the mean-field description of Section~\ref{sec:ceBH}---thus allows to distinguish all the possible phases of the system.

\begin{figure}[t!]
\centering
\includegraphics[width=1\columnwidth]{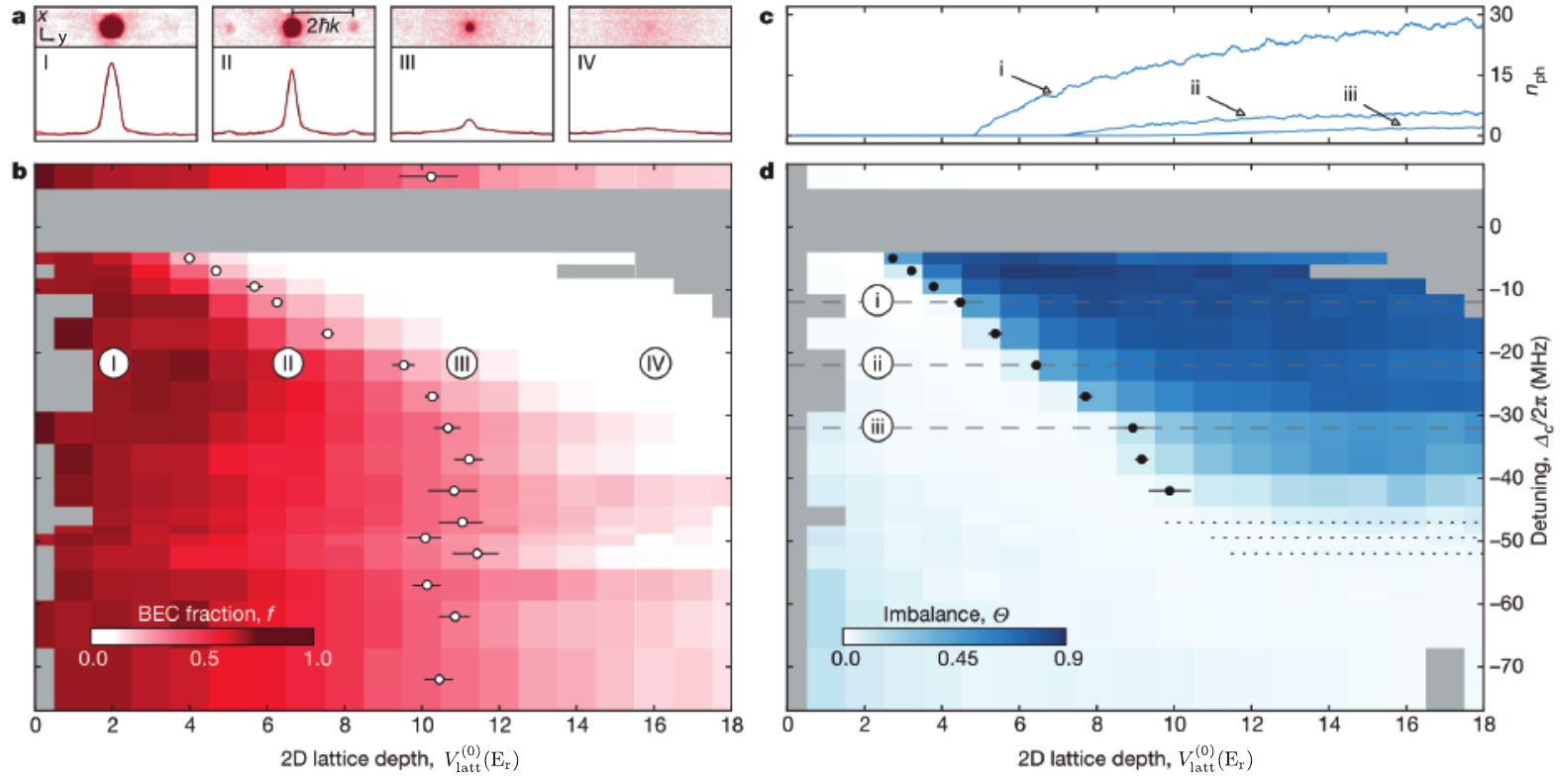}
\caption{Experimental observables of the cavity-induced extended Bose-Hubbard model as a function of the cavity-pump detuning $\Delta_c$ and the lattice depth $V_{\rm latt}^{(0)}$. Panel (b) shows the BEC fraction extracted from bimodal fits to absorption images after ballistic expansion of the atomic cloud [examples are given in panel (a)]. Panel (d) shows the imbalance $\Theta$ between even and odd lattice sites, calculated from the measured mean intra-cavity photon number $n_{\rm ph}$ [example traces of $n_{\rm ph}$ are shown in panel (c)]. The white and black data points indicate the threshold for transitions between superfluid and insulating phases [panel (b)] and between balanced and imbalanced site occupations [panel (d)], respectively. Figure adapted and reprinted with permission from Ref.~\cite{Landig2016Quantum} published at 2016 by the Nature Publishing Group.}
\label{fig:eBHModelETHRawData}
\end{figure}

This scheme was experimentally implemented in Ref.~\cite{Landig2016Quantum}, and Figure~\ref{fig:eBHModelETHRawData} displays the measurement of both observables as a function of the lattice depth $V_{\rm latt}^{(0)}$ and the relative cavity-pump detuning $\Delta_c$. The BEC fraction vanishes for increasing lattice depth $V_{\rm latt}^{(0)}$ and a threshold for the transition from the superfluid to insulating phases can be defined, as indicated by the white data points; see Figure~\ref{fig:eBHModelETHRawData}(a) and~(b). Figure~\ref{fig:eBHModelETHRawData}(c) and~(d) show respectively the simultaneously measured intra-cavity photon number and the imbalance $\Theta=\langle \hat \Theta \rangle$ between even and odd sites as defined in Equation~\eqref{eq:theta}, but assuming completely localized atoms on even or odd sites. This imbalance is directly calculated from the mean intra-cavity photon number using Equation (\ref{eq:ss-a-lattice}). Also for this observable, a threshold can be defined that indicates the transition from a normal phase with balanced occupation of even and odd sites to a superradiant state with an imbalanced occupation.

\begin{figure}[t!]
\centering
\includegraphics[width=1\columnwidth]{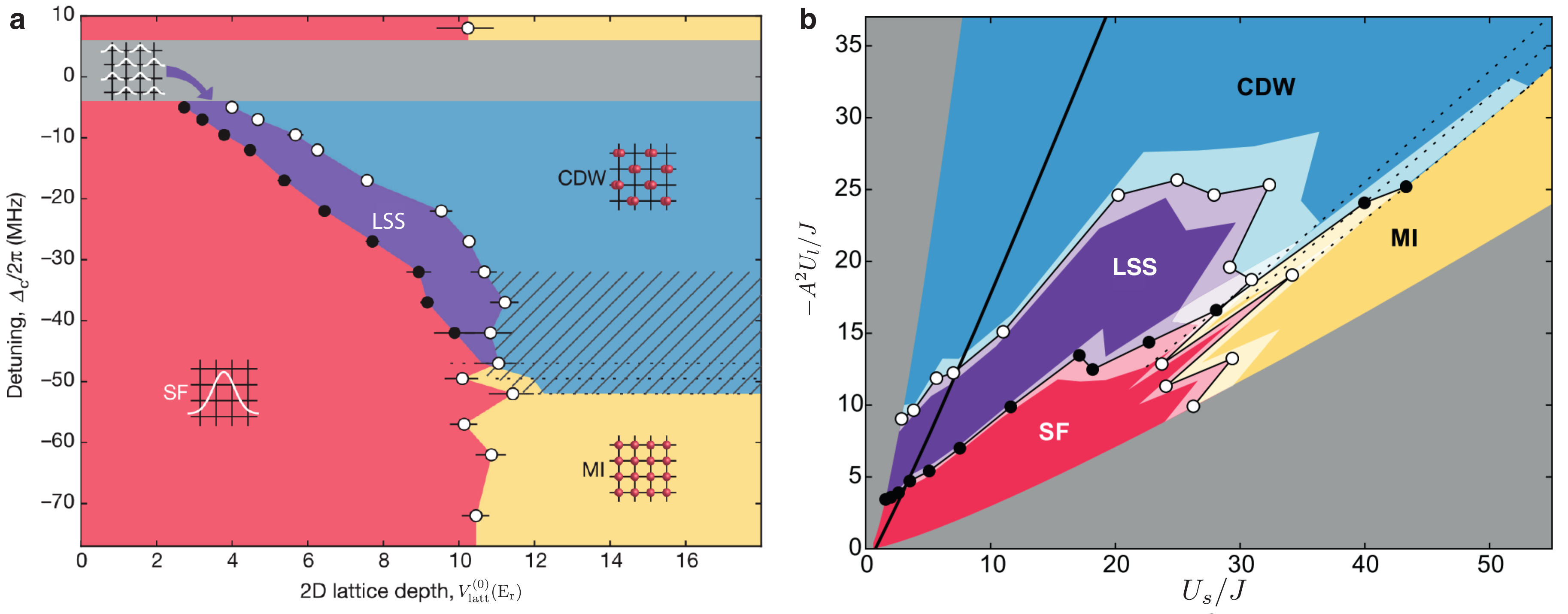}
\caption{Phase diagram of the cavity-induced extended Bose-Hubbard model constructed from the thresholds identified in Figure~\ref{fig:eBHModelETHRawData}. All four theoretically expected phases can be identified: superfluid (SF), lattice supersolid (LSS), Mott insulator (MI), and charge-density wave (CDW).  Panel (a) shows the phase diagram as a function of the experimental parameters $\Delta_c$ and $V_{\rm latt}^{(0)}$, while panel (b) shows the same phase diagram as a function of the Hamiltonian parameters $U_s/J$ and $-A^2U_l/J$, calculated from the experimental parameters. Figure adapted and reprinted with permission from Ref.~\cite{Landig2016Quantum} published at 2016 by the Nature Publishing Group.}
\label{fig:eBHModelETHPhaseDiagram}
\end{figure}

Combining the measurement results a phase diagram can be constructed, as shown in Figure~\ref{fig:eBHModelETHPhaseDiagram}. All four expected phases [i.e., superfluid (SF), lattice supersolid (LSS), Mott insulator (MI), and charge-density wave (CDW)] can be identified. The transition line between the superfluid and the lattice supersolid shifts to smaller values of $V_{\rm latt}^{(0)}$ when approaching cavity resonance, as expected from self-organization without the additional lattices (see Section~\ref{sec:SR_discrete}). The transition line between the lattice supersolid and the charge-density wave follows also this trend, indicating that the insulating phase is entered for smaller lattice depths if a density modulated state is favored. This has been attributed to a reduced nearest-neighbor tunneling due to an increasing energy offset between even and odd lattice sites in the superradiant phases. The transition between the charge-density wave and the Mott insulator phase is especially interesting, since the system has to transition between two different incompressible density patterns while tunneling is strongly suppressed. An experimental exploration of this first-order phase transition is discussed below in Section~\ref{sec:MI-CDW-trans}.

The experimental setup also included a shallow harmonic potential trapping the atomic cloud, leading to an inhomogeneous density profile. As a consequence, it is energetically more costly for the atoms to be away from the trap center and the insulating phases will form a wedding cake structure, also allowing for the coexistence of phases. Furthermore, since the global character of the cavity-mediated interactions leads to a break-down of the local-density approximation, a direct comparison of the experimental results with theoretically calculated phase diagrams is challenging. The role of the trapping potential in this system has, however, been theoretically discussed in Refs.~\cite{Li2013Lattice, Sundar2016Lattice}.

\begin{figure}[t!]
\centering
\includegraphics[width=1\columnwidth]{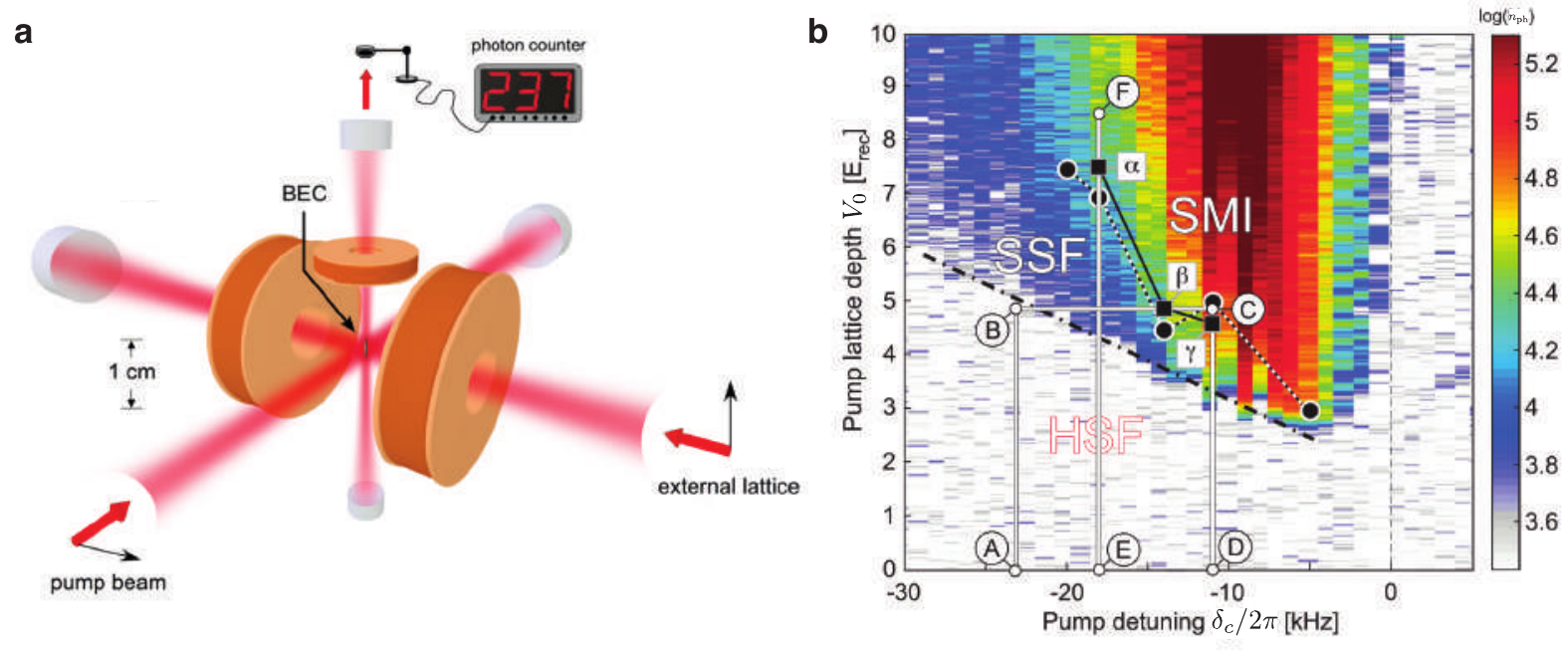}
\caption{Extended Bose-Hubbard model realized in the emergent checkerboard lattice of the superradiant phase. (a) Schematic representation of the experimental setup. A BEC is magnetically trapped at the center of an optical cavity (oriented vertically) and sliced into 2D layers by an external lattice. A second lattice, not interfering with the first one, is applied in the perpendicular direction and also acts simultaneously as a pump beam for the cavity mode inducing global-range interactions. Photons leaking from the cavity are recorded with a photon counter and measure the density modulation of the atomic system. (b)~The measured mean intra-cavity photon number $n_\mathrm{ph}$  is shown as a function of the dispersively shifted pump-cavity detuning $\delta_c$ and the pump lattice depth $V_0$. Combining this measurement with absorption images taken after time of flight expansion of the atomic cloud allows to identify the different phases of the system. The homogeneous superfluid (HSF) phase undergoes a phase transition with increasing pump lattice depth into a self-organized superfluid (SSF) phase and then finally into a self-organized Mott-insulating (SMI) state. Figure adapted and reprinted with permission from Ref.~\cite{Klinder2015Observation}~\textcopyright~2015 by the American Physical Society.}
\label{fig:eBHModelHamburg}
\end{figure}

As already mentioned, even without an external optical lattice important aspects of the competition between short-range and cavity-induced global-range interactions can be explored in the deep superradiant phase~\cite{Fernandez-Vidal2010Quantum, Lin2019Superfluid}. Indeed, such a setup was the first experimental scheme to explore the competition between short-range and global-range interactions~\cite{Klinder2015Observation}. Figure~\ref{fig:eBHModelHamburg}(a) shows the experimental scheme of the Hamburg group where the atomic cloud was sliced into 2D pancakes, however, no additional lattice was applied along the cavity axis. This scheme relies on the fact that the emergent superradiant checkerboard lattice can become sufficiently deep such that the self-organized superfluid (SSF) phase undergoes eventually a transition into a self-organized Mott-insulator (SMI) phase. Since there is no underlying 3D optical lattice, this system can not enter a Mott insulating phase without imbalanced occupations between even and odd sites.

The phase diagram can again be constructed by combining measurements of the intra-cavity field with the interpretation of absorption images after ballistic expansion of the atomic cloud. The resultant phases are shown in Figure~\ref{fig:eBHModelHamburg}(b) on top of the measured mean intra-cavity photon number as a function of pump-cavity detuning and pump lattice depth. Besides the homogeneous superfluid phase (HSF), the self-organized superfluid and the self-organized Mott insulator could be identified.

\paragraph{Metastability at the MI-CDW transition}
\label{sec:MI-CDW-trans}
The phase transition between the Mott-insulator state and the charge-density-wave phase described above is a first order phase transition with an according parameter region of metastability [see hatched region in Figure \ref{fig:eBHModelETHPhaseDiagram}(a)]. Metastability, hysteresis, and a self-induced switching between the Mott insulator and the charge-density wave have been observed experimentally using the real-time access typical to cavity-based quantum gas experiments~\cite{Hruby2018Metastability}. 

\begin{figure}[t!]
\centering
\includegraphics[width=1\columnwidth]{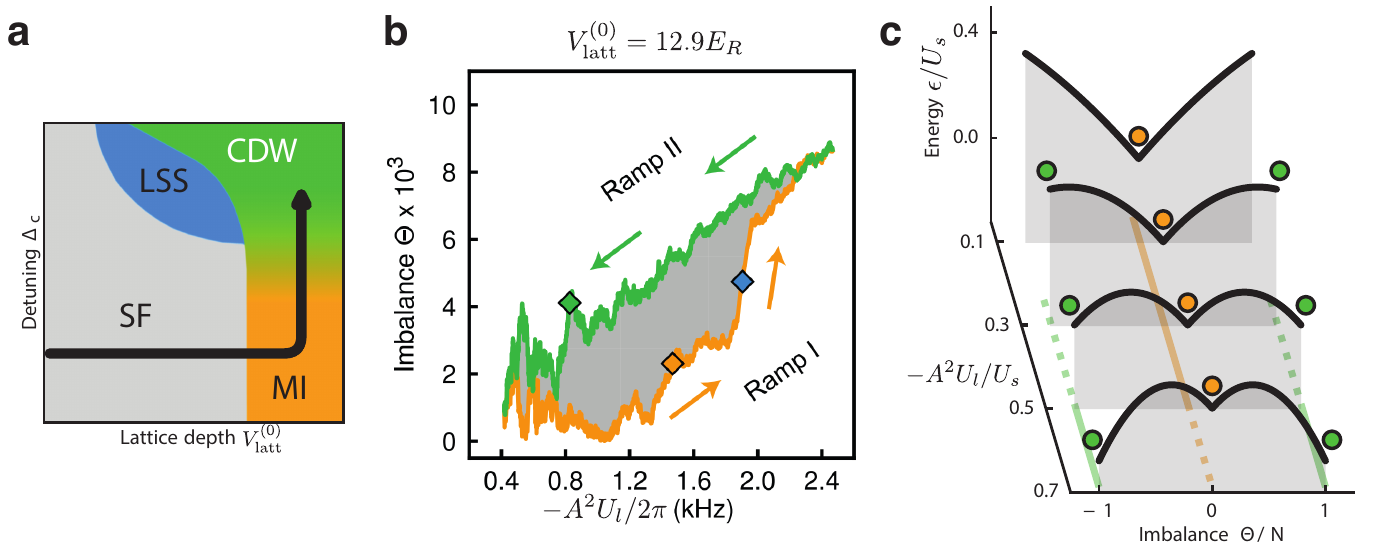}
\caption{(a) Schematic phase diagram of the system with a superfluid (SF, gray), a lattice supersolid (LSS, blue), a Mott insulator (MI, orange), and a charge-density wave (CDW, green) phase. The shaded region between the MI and CDW indicates a region of hysteresis between the phases. The black arrow illustrates the experimental sequence: The atoms are prepared in the SF phase and the 3D optical lattice is ramped up to increase $U_s$, which brings the system into the MI phase. Subsequently, a detuning ramp toward cavity resonance is carried out which increases the strength of the global-range interaction. (b) Non-normalized imbalance, as derived from the intra-cavity light field. The imbalance created during ramp I is shown in orange and the imbalance during ramp II is shown in green. Arrows indicate the ramp directions. (c) Mean-field results from the toy model assuming unit filling. In the presence of short- and global-range interactions, atoms placed in a lattice potential can show metastable behavior. States (indicated by circles) can be protected by an energy barrier and the present state of the system depends on its history, leading to hysteresis. The Mott insulator (orange line) and the charge-density wave (green lines) are stable (solid), metastable (dashed), or unstable. Figure and caption adapted and reprinted with permission from Ref.~\cite{Hruby2018Metastability} published in 2018 by the United States National Academy of Science.
}
\label{fig:eBHModelMetastability}
\end{figure}

Figure~\ref{fig:eBHModelMetastability}(a) shows the experimental sequence for this study. After the system is prepared in the superfluid phase at a detuning $\Delta_c$ corresponding to vanishing global-range interactions, the depth $V_{\rm latt}^{(0)}$ of the square lattice is ramped up such that the system undergoes a phase transition to the Mott-insulator state. This constitutes the starting point for the experiments exploring the first-order phase transition between the Mott-insulating and charge-density-wave states. In the next step, the absolute value of the detuning is first reduced (ramp I) and then increased again (ramp II) over time, which brings the global-range interactions $\propto U_l$ into competition with the short-range interactions $\propto U_s$ and back, driving the system from the Mott insulator to the charge-density-wave state and back. The light field leaking out of the cavity is a measure for the spatial ordering of the atoms, captured by the  imbalance $\Theta$. While $\Theta \approx 0$ in the Mott-insulator state, the charge-density-wave phase is characterized by a large imbalance $\Theta$. A typical single trajectory measuring the evolution of the imbalance during the two ramps is shown in Figure \ref{fig:eBHModelMetastability}(b). The trajectory encloses a hysteresis loop and encompasses a characteristic jump, indicated by the blue symbol.

A simple toy model can be used to explain the occurrence of a first order phase transition in this system. Neglecting tunneling for simplicity in $\hat{H}_{\rm ceBH}$~\eqref{eq:eBH-commensurate} in deep insulating phases, the system can be characterized by the competition of short-range and global-range interactions captured by the Hamiltonian $\hat{H}_{\rm insul}=(U_s/2)\sum_{\ell\in e,o}\hat{n}_\ell(\hat{n}_\ell-1)
+(A^2U_l/N_\Lambda)\left(\sum_{e}\hat{n}_e-\sum_{o}\hat{n}_o\right)^2$. Figure~\ref{fig:eBHModelMetastability}(c) shows the average ground-state energy per particle as a function of the imbalance $\Theta$ and the ratio of the interactions $U_l/U_s$. In the limit of small global-range interactions, a global energy minimum at zero imbalance stabilizes the system in the Mott-insulator phase. With increasing global-range interactions, $\mathbf{Z}_2$-symmetric local minima at finite imbalance emerge. By further increasing global-range interactions, the Mott-insulator state becomes a local minimum in the energy landscape, separated from the charge-density-wave states by an energy barrier resulting from competing short- and global-range interactions.

Not captured by this simple model is the observed jump in the imbalance [blue symbol in Figure~\ref{fig:eBHModelMetastability}(b)], where the atomic system is strongly re-organized via tunneling between neighboring sites in a dynamic process. Time-resolved studies of this regime indicate that the process is a collective tunneling of several thousand atoms within the expected tunneling time of a single double well. The microscopic description of the observed dynamics relies on the harmonic trapping potential experienced by the atoms. The atomic system accordingly forms a wedding-cake structure with a Mott-insuolator core and a superfluid surface. Particles in the surface possess a high mobility and can locally organize into a density-modulated state. Local self-organization leads to a global energy offset between even and odd sites. Depending on the position in the trap, this offset can bring the tunneling between neighboring sites into resonance, which leads to an avalanche dynamics and very fast collective tunneling.

\subsection{Two-component extended Bose-Hubbard model}
\label{sec:2comp-extended-BH}

As discussed in Section~\ref{sec:spinor-selfordering}, when the internal states of atoms become relevant~\cite{Zoubi2009Quantum}, the exchange of cavity photons by the atoms can result in global-range spin-spin and spin-density interactions among the atoms in addition to the density-density interactions~\cite{Mivehvar2019Cavity}. When the atoms are trapped in an external lattice, the cavity-induced long-range spin interactions can then compete directly with contact-interaction-induced short-range spin interactions in the deep insulating phases, opening the possibility for realizing intriguing lattice spin models including topological spin orders and frustrated spin states.   

Cavity-induced lattice spin models can be obtained by imposing an external optical lattice potential in the spinor models discussed in Section~\ref{sec:spinor-selfordering}. The simplest model of these kinds is the lattice version of a transversely-driven spinor BEC with standing-wave pump lasers $\Omega_{1,2}(y)=\Omega_{01,02}\cos(k_cy)$ inside a single-mode linear cavity. Namely, a state-independent external, static optical lattice potential $V_{\rm ext}(\mathbf{r})$, Equation~\eqref{eq:Vext-latt}, is added to the effective two-component Hamiltonian $\hat{\mathcal H}_{1,\rm eff}$, Equation~\eqref{eq:single-particle-H-den-spinor}. Following exactly the same procedure as discussed in Section~\ref{sec:1comp-extended-BH}, for a strong external optical lattice and after adiabatic elimination of the cavity field one obtains up to $1/\Delta_a^2$ a two-component extended Bose-Hubbard model,
\begin{align}
\label{eq:2comp-eBH}
\hat{H}_{\rm 2eBH}&=
-J\sum_{\tau,\langle \mathbf{j},\mathbf{j}'\rangle}\left(\hat{b}_{\tau\mathbf{j}}^\dag\hat{b}_{\tau\mathbf{j}'}+\text{H.c.}\right)
+\frac{1}{2}\sum_{\tau,\mathbf{j}} U_{s,\tau} \hat{n}_{\tau\mathbf{j}}(\hat{n}_{\tau\mathbf{j}}-1)
+ U_{s,\uparrow\downarrow} \sum_{\mathbf{j}} \hat{n}_{\uparrow\mathbf{j}}\hat{n}_{\downarrow\mathbf{j}}
\nonumber\\
&+\sum_{\tau,\mathbf{j}}\left(\hbar V_0B_\mathbf{j}^{(y)}-\mu\right)\hat{n}_{\tau\mathbf{j}}
+N_\Lambda U_l\Big(\frac{1}{N_\Lambda}\sum_\mathbf{j}A_\mathbf{j}\hat{s}_{x,\mathbf{j}}\Big)^2,
\end{align} 
where $\hat{b}_{\tau\mathbf{j}}$ is the bosonic annihilation operator destroying a boson at spin state $\tau$ and site $\mathbf{j}$, and satisfies the bosonic commutation relation $[\hat{b}_{\tau\mathbf{j}},\hat{b}_{\tau'\mathbf{j}'}^\dag]=\delta_{\tau,\tau'}\delta_{\mathbf{j},\mathbf{j}'}$. Here we have assumed the balanced Raman condition as well as $V_{\uparrow}(\mathbf{r})=V_{\downarrow}(\mathbf{r})=V_0\cos^2(k_cy)$. The last term in the Hamiltonian $\hat{H}_{\rm 2eBH}$ is the cavity-mediated global spin-spin interaction. It is indeed nothing but the lattice version of the continuum spin Hamiltonian $\hat{H}_{\rm spin}$, Equation~\eqref{eq:eff-spin-H}, in a special case where only the $x$ component of the Heisenberg term is present (recall that for standing-wave pump lasers other spin-spin interactions are absent; see Section~\ref{subsubsec:effective-spin-H_spin-slefordering}).  

For the case of the commensurate lattice with $\lambda_c=\lambda_{\rm latt}$, the global spin-spin interaction in Equation~\eqref{eq:2comp-eBH} reduces to the imbalance of the $x$ component of the spin on even and odd sites, $\propto U_l(\sum_{e}\hat{s}_{x,e}-\sum_{o}\hat{s}_{x,o})^2$. Therefore, for large enough (negative) cavity-mediated interaction strength $U_l$, a ground state with spin imbalance between even and odd sites $\sum_{e}\langle\hat{s}_{x,e}\rangle\neq\sum_{o}\langle\hat{s}_{x,o}\rangle$ is favored. The ground state in this regime is either a spin-density wave or a lattice supersolid depending the superfluid order parameter is zero or finite~\cite{Guan2019Two}. 

More interesting lattice spin models can be obtained by considering different pump-laser configurations as discussed for the continuum models in Section~\ref{sec:spinor-selfordering}. Therefore, a variety of emergent magnetic orders can be explored in the framework of lattice cavity QED with multi-component quantum gases; see also the next section for the fermionic variants.

\subsection{Extended Fermi-Hubbard models}
\label{sec:extended-FH}

Similar to the bosonic atoms discussed in the previous section, ultracold fermionic atoms can also be loaded into the external lattice~\eqref{eq:Vext-latt} inside the cavity, yielding extended Fermi-Hubbard-like Hamiltonians with global, all-to-all atomic interactions~\cite{Camacho-Guardian2017Quantum}. The only difference with respect to the extended Bose-Hubbard models, Equations~\eqref{eq:eBH} and \eqref{eq:2comp-eBH}, is that the only nonzero onsite atomic interaction would be the inter-species interaction $U_{s,\uparrow\downarrow} \sum_{\mathbf{j}} \hat{n}_{\uparrow\mathbf{j}}\hat{n}_{\downarrow\mathbf{j}}$ for two-component fermionic atoms due to the Pauli exclusion principle.

\subsubsection{Many-body localization in the presence of cavity-mediated global interactions}
\label{sec:MBL-vs-cavity-interactions}

\begin{figure}[t!]
\centering
\includegraphics [width=\textwidth]{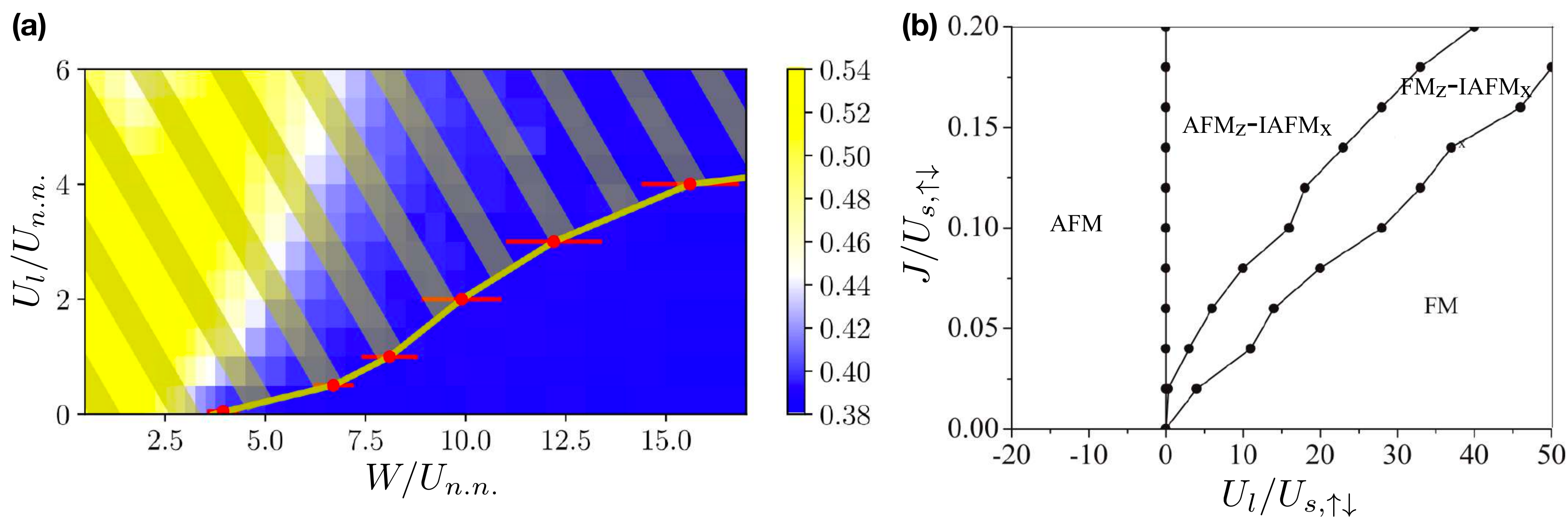}
\caption{Steady-state phase diagram of cavity-induced commensurate, 1D extended Fermi-Hubbard model 
for spin-polarized (a) and spin-1/2 (b) fermions. 
(a) In addition to the cavity-mediated global interactions, fermions experience nearest-neighbor
interactions with strength $U_{n.n.}$ and random onsite disorder characterized by $W$. 
While the disorder drives the system into the many-body localized phase (the blue region characterized
by the mean gap ratio shown as the background color map), the global interactions favor eigenvector-thermalization-hypothesis
phase (the yellow region). The simulation has been done at half-filling for a lattice of sixteen sites.
The solid yellow line indicates the phase boundary between the two phases, obtained via finite-size scaling analysis.
(b) The cavity-mediated global spin interaction with strength $U_l$ induces different magnetic orders: 
ferromagnetic (FM), antiferromagnetic (AFM), and incommensurate antiferromagnetic (IAFM) orders.
The subscripts in the figure indicate the direction of the magnetic order in the internal spin space. 
The magnetic orders in the repulsive cavity-mediated interaction regime, $U_l\propto\Delta_c>0$,
are induced by quantum fluctuations of the photonic field.
Panel (a) is adapted and reprinted with permission from Ref.~\cite{Sierant2019Many} published in 2019 by the SciPost and panel (b) from Ref.~\cite{Fan2018Magnetic}~\textcopyright~2018 by the American Physical Society.} 
\label{Fig:Sierant2019ManyFig2_Fan2018MagneticFig3}
\end{figure}

The effect of the cavity-mediated global, all-to-all commensurate interactions $\propto U_l [\sum_\ell (-1)^\ell \hat{n}_\ell]^2$ on the many-body localization of spin-polarized fermionic atoms in a 1D external, static optical lattice with nearest-neighbor atomic interactions $U_{n.n.}\sum_\ell\hat{n}_\ell\hat{n}_{\ell+1}$ and random onsite disorder $\sum_\ell \epsilon_\ell \hat{n}_\ell$, with $\epsilon_\ell$ being uncorrelated random variables uniformly distributed in $[-W,W]$, has been explored via an exact diagonalization for a small system size. It has been found that the many-body localization persists in the presence of the cavity-mediated global interactions, but occurs at larger disorder values~\cite{Sierant2019Many}; see Figure~\ref{Fig:Sierant2019ManyFig2_Fan2018MagneticFig3}(a). Similar results have also been predicted for one-component bosonic atoms inside cavities with on-site contact and random interactions~\cite{Sierant2019Many}, where the existence of Bose glass phases is also anticipated~\cite{Zhang2020Phase,Zhang2020The}.

\subsubsection{Cavity-fluctuation-induced lattice magnetic orders}
\label{sec:lattice-magnetic-orders-fermi}

The steady-state phase diagram of spin-1/2 (i.e., two-component) fermionic atoms subject to the cavity-mediated global, commensurate spin interaction $\propto U_l [\sum_\ell (-1)^\ell \hat{s}_{x,\ell}]^2$ in a 1D lattice at half filling has been mapped out using a numerical density-matrix-renormalization-group method~\cite{Fan2018Magnetic}; see Figure~\ref{Fig:Sierant2019ManyFig2_Fan2018MagneticFig3}(b). The phase diagram exhibits a rich structure containing anisotropic magnetic orders: FM (AFM) indicates a state where both $\hat{s}_x$ and $\hat{s}_z$ exhibit ferromagnetic (antiferromagnetic) correlations, while FM$_z$-IAFM$_x$ (AFM$_z$-IAFM$_x$) is a state with a ferromagnetic (antiferromagnetic) correlation for $\hat{s}_z$ and an incommensurate---with respect to the static optical lattice--- antiferromagnetic correlation for $\hat{s}_x$. In the absence of the cavity-mediated global interaction, $U_l=0$, a repulsive onsite interaction $U_{s,\uparrow\downarrow}>0$ already favors an AFM order at the half filling. The attractive cavity-mediated interaction $U_l<0$ further stabilizes the AFM order. In contrast, the repulsive cavity-mediated interaction $U_l>0$ competes with the repulsive onsite interaction, stabilizing the other magnetic orders in the phase diagram. Interestingly, the region of the positive cavity-mediated interaction $U_l\propto\Delta_c>0$ (i.e., the blue pump-cavity detuning region) corresponds to a non-superradiant regime where the coherent field amplitude is vanishingly small, $\alpha=\langle \hat{a} \rangle \approx0$. Nonetheless, cavity-field fluctuations  $\langle \delta\hat{a}^\dag \delta\hat{a} \rangle/\langle \hat{a}^\dag\hat{a} \rangle$ is considerably large (in the order of one) in this regime, signaling that the FM, FM$_z$-IAFM$_x$, and AFM$_z$-IAFM$_x$ magnetic orders are induced by quantum fluctuations of the photonic field.

\subsubsection{Cavity-induced superfluid pairing in a lattice}
\label{sec:cavity-induced-lattice-pairing}

\begin{figure}[t!]
  \begin{center}
    \includegraphics[width=1\columnwidth]{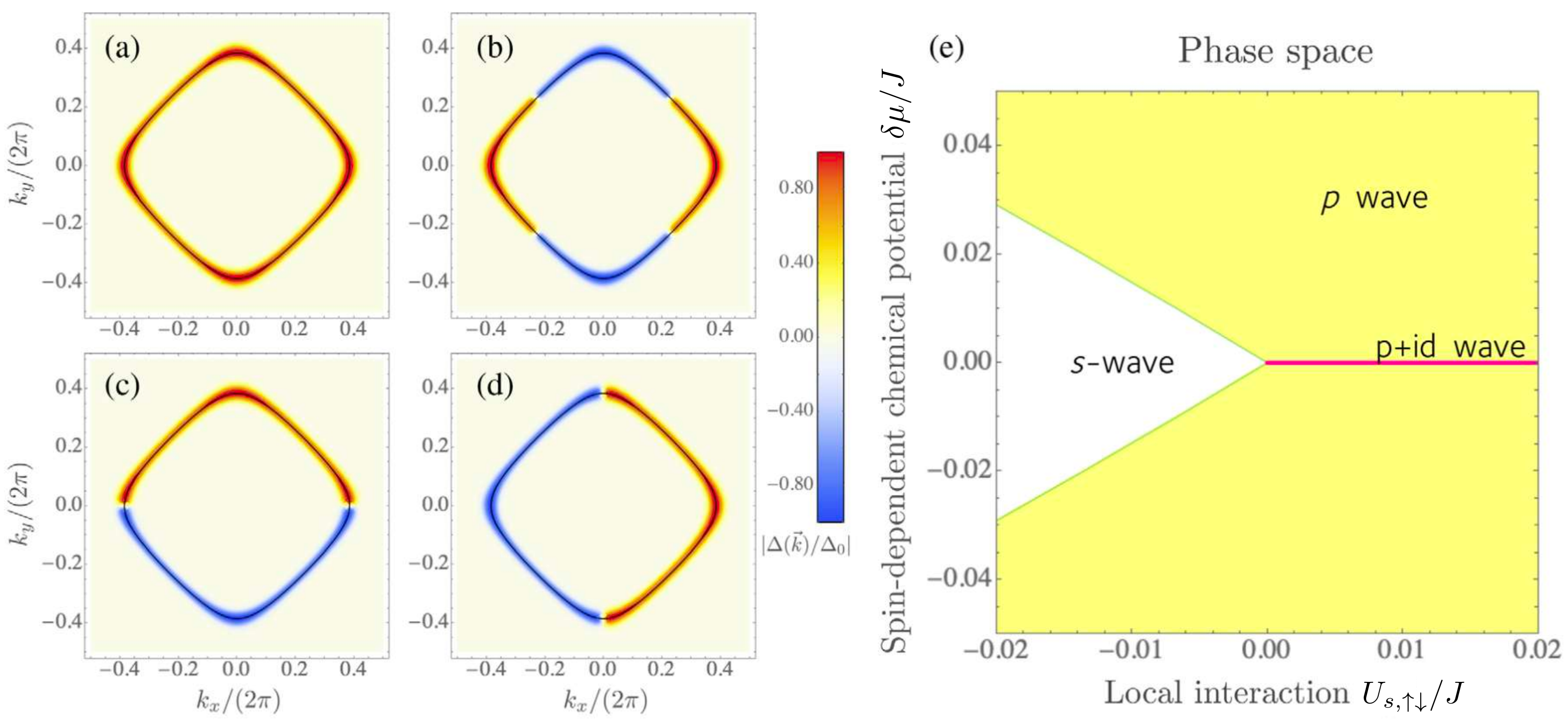}
    \caption{Cavity-induced pairing in a 2D Fermi gas trapped in an external, square optical lattice.
    (a)-(d)~Solutions of the gap equation for the superfluid order parameter induced via cavity-mediated interactions~\eqref{eq:paring-H-lattice}. The initial state for the iterative solution is chosen with (a) $s$-wave, (b) $d$-wave, and (c)-(d) $p$-wave symmetry, respectively. The cavity wave number is set to $k_c=0.02\pi$ (with lattice spacing $a_{\rm latt}=1$). The other parameters are $\mathcal{D}_0=-0.1J$, $U_{s,\uparrow\downarrow}=0$, and $\mu=-0.5J$. 
    (e) Dominant pairing symmetry as a function of the spin-dependent chemical potential $\delta\mu$ splitting the energy of the two spin states, and the local Feshbach interaction $U_{s,\uparrow\downarrow}$. The cavity-mediated interaction strength is $\mathcal{D}_0=-0.1J$, and the chemical potential is $\mu=-3J$. Figure adapted and reprinted with permission from Ref.~\cite{Schlawin2019Cavity}~\textcopyright~2019 by the American Physical Society.
  }
    \label{fig:cavity_unconventional_pairing}
  \end{center}
\end{figure}

In Section~\ref{subsec:cavity-induced pairing} we described a scenario for superfluid pairing mediated via cavity fluctuations in the blue-detuned regime, $\Delta_c>0$. Here we consider a lattice scenario for cavity-induced pairing in the red-detuned regime $\Delta_c<0$ corresponding to attractive cavity-mediated interactions $\mathcal{D}_0<0$. In particular, two ring cavities oriented along axes of a 2D external optical lattice~\eqref{eq:Vext-latt} mediate global interactions 
\begin{align} \label{eq:paring-H-lattice}
\hat{H}_{\rm int}=\mathcal{D}_0\sum_{\mathbf{k},\mathbf{k}',\mathbf{k}_c}\sum_{\tau,\tau'}
\hat{f}_{\tau,\mathbf{k}\pm\mathbf{k}_c}^\dag \hat{f}_{\tau,\mathbf{k}}
\hat{f}_{\tau',\mathbf{k}'\mp\mathbf{k}_c}^\dag \hat{f}_{\tau',\mathbf{k}'},
\end{align}
among spin-1/2 fermionic atoms in the external optical lattice, where $\mathbf{k}_c=k_c\hat{e}_i$ with $i=x,y$. In Ref.~\cite{Camacho-Guardian2017Quantum}, such cavity-mediated interactions have been considered in a regime where $|\mathbf{k}_c|\sim k_{\rm latt}$, which is characterized by the strong competition between a cavity-induced superfluid paired phase and other types of non-paired, charge- or spin-density phases. On the other hand, in the regime where $|\mathbf{k}_c| \ll k_{\rm latt}$ considered in Ref.~\cite{Schlawin2019Cavity}, the superfluid pairing instability dominates. Furthermore, different types of pairing compete with one another, and can give rise to interesting exotic topological superfluid states. This phenomenon is a consequence of an accidental degeneracy of competing paired phases, shown in Figure~\ref{fig:cavity_unconventional_pairing}, due to the global range of the cavity-mediated interaction, leaving the system in a frustrated state. The perturbative addition of local interactions splits this degeneracy and creates topological $p+id$ wave superfluid-paired state featuring Majorana fermions. The resulting phase diagram is shown in Figure~\ref{fig:cavity_unconventional_pairing}(e).

These results highlight the promise of cavities for inducing novel mechanisms for pairing as well as stabilizing more exotic superconducting states. This direction of investigation has recently received a lot of attention also in the context of solid-state materials~\cite{Schlawin2019CavityA,Curtis2019Cavity,Gao2020Photoinduced,Chakraborty2020Non}.

\section{Cavity-induced synthetic gauge potentials and topological states}
\label{sec:synthetic-gauge-fields-TI}

Gauge potentials and quantum gauge theories play central roles in our understanding of Nature. The simplest example is electromagnetism with its Abelian vector and scaler potentials describing the coupling between charged matter and electromagnetic fields. Quantum electrodynamics (QED), a relativistic gauge theory describing the coupling of a relativistic charged matter field to the Abelian electromagnetic gauge potentials, is among the most successful and most accurate theories of physics. In the Standard Model of elementary particles, gauge fields are the mediators of fundamental interactions among the elementary particles. Gauge potentials are also of great significance in condensed-matter physics~\cite{Frohlich1993Gauge}. In fact, they are the essential ingredients of topological states of matter~\cite{Hasan2010Colloquium,Qi2011Topological}. In addition to their fundamental significance, topological insulators are also of great importance from a technological point of view: they possess conducting edge or surface states which are topologically protected and robust against external perturbations. These topological edge states can have exotic properties. In the case of quantum Hall insulators, such edge states carry precisely one quantum of conductance. They can also host anyonic excitations, quasiparticles that do not obey Bose or Fermi statistics. 

Since ultracold atoms are charge neutral, they do not couple to gauge potentials the same way charged particles (such as electrons) do. However, a variety of gauge potentials and minimal couplings can be engineered via the light-matter interaction for neutral atoms, to mimic phenomena encountered not only in condensed-matter physics but also in high-energy physics~\cite{Dalibard2011Artificial, ZHAI2012SPIN, Goldman2014Light, Zhai2015Degenerate, Zhang2018Topological, Cooper2019Topological}. The induction of such synthetic gauge potentials generally relies on geometric phases acquired through adiabatic motion of quantum particles with internal structures. That is, the light-matter interaction couples parametrically the atomic internal dynamics to its external degree of freedom (i.e., position), resulting in the appearance of a geometric vector potential under special conditions. The laser-induced gauge potentials are normally \emph{static}, in the sense that the back-action of the quantum matter on these gauge fields is negligible. They are solely background gauge potentials, unaffected by the matter dynamics (though they can still have an externally imposed time dependence) and described by extra fixed terms in the Hamiltonian of the system. For instance, the synthetic electric and magnetic fields induced by the light-matter interaction for quantum gases need not to obey the Maxwell's equations. However, gauge potentials encountered in quantum gauge theories such as QED, quantum chromodynamics, etc., and strongly correlated many-body condensed-matter systems are normally \emph{dynamic}. That is, the dynamics of these gauge potentials are governed by their own Hamiltonians and affected by the back-action of the matter. Motivated by these dynamical gauge potentials, some proposals have been put forward to simulate dynamical gauge potentials for quantum gases~\cite{Wiese2013Ultracold}. A simplest interacting gauge potential toward full dynamical gauge potentials is perhaps a density-dependent gauge potential~\cite{Clark2018Observation, Goerg2019Realization}.

An alternative, natural route to implement dynamic gauge potentials for neutral atoms is the framework of cavity QED. This hinges on the non-linear, dynamical nature of cavity fields which can induce gauge potentials. Unlike lasers interacting with atoms in free space, cavity fields are dynamical and affected by the atomic back-action. Therefore, cavity-induced gauge potentials inherit the dynamical nature of the cavity fields and respond naturally to the atomic back-action. Furthermore, as radiation fields decay out of a cavity, quantum-gas--cavity systems with induced gauge potentials are naturally out of equilibrium. This leads to nonequilibrium topological phases in the framework of cavity QED, a scenario beyond common condensed-matter and free-space quantum-gas systems. Guided by this nonlinear coupled dynamics of matter and radiation in cavity QED, proposals have been put forward to induce dynamical gauge fields~\cite{Deng2014Bose, Mivehvar2014Synthetic, Dong2014Cavity, Mivehvar2015Enhanced, Dong2015Photon, Kollath2016Ultracold, Sheikhan2016Cavity-chiral, Zheng2016Superradiance, Ballantine2017Meissner, Halati2017Cavity, Mivehvar2019Cavity, Halati2019Cavity, Colella2019Hofstadter} and topological phases~\cite{Pan2015Topological, Sheikhan2016Cavity-topological, Mivehvar2017Superradiant} for quantum gases in cavities, with the experimental realization of a dynamical spin-orbit coupling~\cite{Kroeze2019Dynamical}. Another peculiarity of cavity-induced gauge potentials and topological states, in addition to their nonequilibrium dynamical nature, is the appearance of them on the onset of the superradiance phase transition. Since the threshold of the superradiant quantum phase transition depends crucially on atomic properties, the onset of a cavity-induced gauge potential and topological phase is also sensitive to atomic structures and is self-consistent.

Here we will review some of the cavity-based schemes for inducing gauge potentials for quantum gases. We start with the cavity-QED implementation of a synthetic magnetic field corresponding to a position-dependent vector potential, the simplest gauge potential with the $U(1)$ gauge symmetry. Our focus shall be mainly on lattice models where the synthetic magnetic field is induced via a cavity-assisted Peierls phase. We then study rich physics arising from this dynamical synthetic magnetic field including topological effects, chiral currents, fractal energy bands, and Meissner-like states. We then consider cavity-assisted two-photon Raman schemes to engineer dynamical spin-orbit coupling for neutral atoms, where the interplay between the dynamical spin-orbit coupling, contact interactions, and cavity-mediated long-range interactions results in rich physics. Finally, we discuss nonequilibrium topological phases induced in quantum gases via the cavity field dynamics.

\subsection{Cavity-induced synthetic gauge potentials}
\label{sec:synthetic-gauge-potentials}

\subsubsection{Dynamical synthetic magnetic fields}
\label{subsec:synthetic-B}

\begin{figure}[t!]
\centering
\includegraphics [width=\textwidth]{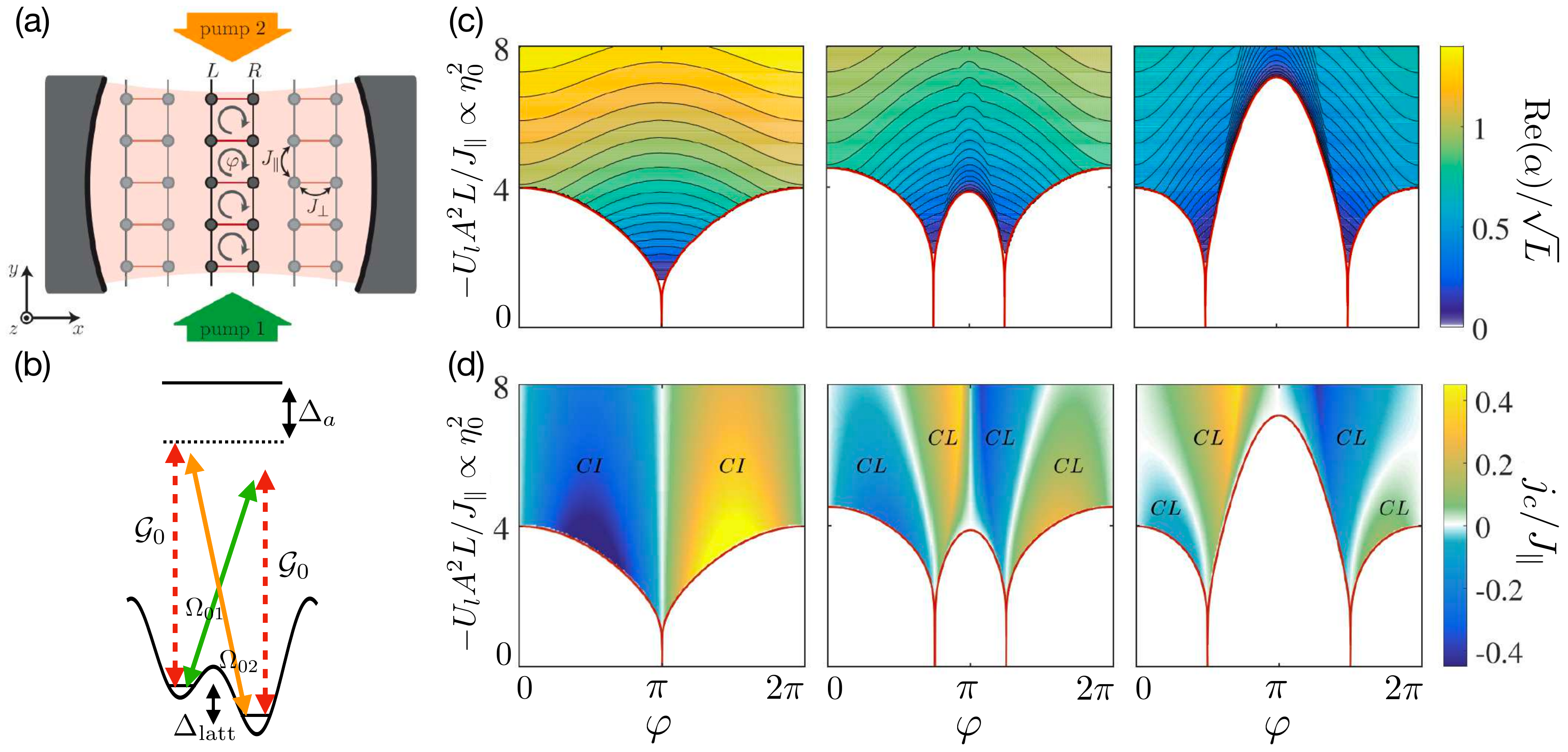}
\caption{Cavity-induced synthetic magnetic field for fermionic atoms. 
(a)~The schematic sketch of the system. The external lattice provides 
disjoint ladders. (b)~A single rung in one ladder,
showing all the transitions induced by the pump lasers and the cavity field.
The real part of the cavity-field amplitude $\alpha$~(c) and the chiral current $j_c$~(d) are shown
as a function of the flux $\varphi$ and $-U_lA^2L/J_\|\propto\eta_0^2$ for three representative filling $n=1/2,3/8,1/4$,
from left to right. The solid red curves mark the onset of the superradiant phase transition, above which a
synthetic magnetic field appears for the fermionic atoms. As a consequence of the latter,
a chiral current is induced in the system. Depending on the filling $n$, the atomic ground state corresponds 
to either a chiral insulator (CI) or a chiral liquid (CL).
Figure adapted and reprinted with permission from Refs.~\cite{Kollath2016Ultracold, Sheikhan2016Cavity-chiral}~\textcopyright~2016 by the American Physical Society.} 
\label{Fig:Sheikhan2016CavityFig8-9}
\end{figure}

Following Ref.~\cite{Kollath2016Ultracold}, consider spin-polarized fermionic atoms trapped in a strong external classical 2D lattice in the $x$-$y$ plane with the superlattice structure along the $x$ direction. The optical lattice is located inside an initially empty linear cavity and created using standing-wave lasers with wavelength $\lambda_{\rm latt}^{(y)}$ along the $y$ direction, and with wavelengths $\lambda_{\rm latt}^{(x)}$ and $2\lambda_{\rm latt}^{(x)}$ along the $x$ direction (which is the cavity axis). The phase difference of the two lasers along the $x$ direction is chosen such that to create potential offsets $\pm\hbar\Delta_{\rm latt}$ between adjacent sites of the lattice, resulting in an imbalanced superlattice structure along the cavity (i.e., $x$) axis. Therefore, the external lattice is separated into a set of decoupled ladders as depicted schematically in Figure~\ref{Fig:Sheikhan2016CavityFig8-9}(a). The atoms can tunnel along the legs of the ladders (i.e., the $y$ direction) due to their kinetic energy, while the tunneling along the rungs (i.e., the $x$ direction) is strongly suppressed due to the potential offset $\hbar\Delta_{\rm latt}$. The atoms are dispersively pumped in the transverse direction by two running-wave pump lasers with space-dependent Rabi couplings $\Omega_{1,2}(y)=\Omega_{01,02}e^{\pm ik_cy}$ and frequencies $\omega_{p1,p2}$ such that $\omega_{p2}-\omega_{p1}=2\Delta_{\rm latt}$. The atoms are strongly coupled to a standing-wave mode of the cavity with the bare frequency $\omega_c$ such that two-photon Raman transitions involving the pump lasers and the cavity mode are close to resonant, $\omega_{p2}-\omega_{c}\approx\omega_{c}-\omega_{p1}\approx\Delta_{\rm latt}$; see Figure~\ref{Fig:Sheikhan2016CavityFig8-9}(b). The double $\Lambda$ scheme ensures that the system is not optically pumped into one leg due to the cavity photon losses. The cavity mode is initially in the vacuum state; however, it can be populated due to the constructive, collective Raman scattering from the pump lasers by the atoms. This in turn restores the hopping along the rungs (i.e., the cavity axis along $x$), imprinting a phase into the wavefunction of the atoms due to the running-wave pump lasers.

In the tight-binding limit, one ladder of the system is described in the rotating frame of the lasers by the Hamiltonian,
\begin{align}
\label{eq:H-lattice-ladder}
\hat{H}_{\rm latt}&=-J_\|\sum_{\ell=1,2}\sum_m\left(\hat{f}_{\ell,m+1}^\dag\hat{f}_{\ell,m}+\text{H.c.}\right)
+\hbar\eta_0(\hat{a}^\dag+\hat{a})\sum_m\left(Ae^{-im\varphi}\hat{f}_{2,m}^\dag\hat{f}_{1,m}+\text{H.c.}\right)\nonumber\\
&-\hbar\delta_c\hat{a}^\dag\hat{a},
\end{align}
where $\hat{f}_{\ell,m}$ is the fermionic operator annihilating a particle at leg $\ell$ and rung $m$, $J_\|$ the kinetic-energy hopping amplitude along the legs, $\delta_c=(\omega_{p1}+\omega_{p2})/2-\omega_c-U_0\sum_{\ell,m}B_{\ell,m}^{(x)}\hat{n}_{\ell,m}$ the dispersively shifted cavity detuning with respect to the pumps with $U_0=\mathcal{G}_0^2/\Delta_a$ being the dispersive shift per atom [where $\Delta_a=(\omega_{p1}+\omega_{p2})/2-\omega_a$ is the atomic detuning with respect to the pumps], and $\eta_0=\Omega_{01}\mathcal{G}_0/\Delta_a=\Omega_{02}\mathcal{G}_0/\Delta_a$ the balanced two-photon Raman coupling. The coefficient $B_{\ell,m}^{(x)}$ is the same as the one defined in Equation~\eqref{eq:A-B-coeff-cBHM}, while the coefficient $A$ has now the form 
\begin{align}
A=\int W_{2,m}^*(\mathbf{r})\cos(k_cx)
\exp{\left[-ik_c\left(y-\frac{1}{2}m\lambda_{\rm latt}^{(y)}\right)\right]}W_{1,m}(\mathbf{r})d\mathbf{r}.
\end{align}
The phase $\varphi=k_c\lambda_{\rm latt}^{(y)}/2=\pi\lambda_{\rm latt}^{(y)}/\lambda_c$ arises due to the photon-assisted tunneling. In particular, the hopping amplitude $\propto\eta_0(\hat{a}^\dag+\hat{a})$ along the rungs is an operator depending on the cavity field operator $\hat{a}$, which in turn depends on atomic properties highlighting the highly nonlinear nature of the system. In the absence of the cavity photon, the system consists of decoupled 1D legs along the $y$ direction. The nonzero cavity field restores the hopping along the rungs, where the atoms collect the total phase $\varphi$ while traversing a plaquette counterclockwise: $-m\varphi+(m+1)\varphi=\varphi$. This phase simulates an electromagnetic vector potential through the Peierls substitution, where the vector potential corresponds to a synthetic transverse magnetic field perpendicular to the $x$-$y$ plane. The phase and correspondingly the synthetic magnetic field can be tuned by changing the ratio between the wavelengths $\lambda_{\rm latt}^{(y)}/\lambda_c$ of the external lattice along the $y$ axis and the cavity.  

The steady state of the system can be obtained in the thermodynamic limit using a mean-field approach; see also Section~\ref{subsubsec:MF-fermi}. To this end, the cavity field operator $\hat{a}$ is replaced with its steady-state expectation value $\alpha=\eta_0\langle A\hat{K}_\bot+A^*\hat{K}_\bot^\dag\rangle/(\delta_c+i\kappa)$, where $\hat{K}_\bot=\sum_m e^{-im\varphi}\hat{f}_{2,m}^\dag\hat{f}_{1,m}$ is the directed atomic tunneling in all rungs. Substituting the steady-state field amplitude $\alpha$ in Equation~\eqref{eq:H-lattice-ladder} results in an effective atom-only Hamiltonian for a single ladder
\begin{align}
\label{eq:H-ladder}
\hat{H}_{\rm ladd}=-J_\|\sum_{\ell=1,2}\sum_m\left(\hat{f}_{\ell,m+1}^\dag\hat{f}_{\ell,m}+\text{H.c.}\right)
-\sum_m\left(J_\bot e^{-im\varphi}\hat{f}_{2,m}^\dag\hat{f}_{1,m}+\text{H.c.}\right),
\end{align}
with the self-consistent hopping amplitude 
\begin{align}
\label{eq:self-consistent-Jbot}
J_\bot=-U_lA\langle A\hat{K}_\bot+A^*\hat{K}_\bot^\dag\rangle/2, 
\end{align}
along the cavity axis, where $U_l=4\hbar\eta_0^2\delta_c/(\delta_c^2+\kappa^2)$ [cf.\ definition of $U_l$ in Equation~\eqref{eq:eBH}]. The self-consistency condition~\eqref{eq:self-consistent-Jbot} can be solved, for instance, graphically and has a nontrivial solution only for $\delta_c<0$. Furthermore, the existence of a nontrivial solution depends also crucially on the system's parameters including the filling $n=N/2L$ with $N$ being the total number of the atoms and $L$ the size of the system (i.e., the number of the rungs), the flux $\varphi$, and the pump strength $\eta_{0}$ ($\propto \sqrt{U_{l}}$), which in turn they determine the energy bands and the Fermi surface (i.e., Fermi points). Therefore, for a given parameter set including a fixed $\eta_0$, the transverse hopping $J_\bot$ is varied continuously and correspondingly $\langle A\hat{K}_\bot+A^*\hat{K}_\bot^\dag\rangle$ is calculated. The system has a nontrivial self-consistent solution if the left- and right-hand sides of Equation~\eqref{eq:self-consistent-Jbot} intersect in a nonzero value.

Recall that a nontrivial solution of the self-consistency equation implies a nonzero cavity field amplitude $\alpha=-J_\bot(\delta_c+i\kappa)/2\hbar\eta_0A\delta_c$. Thanks to the restored hopping $J_\bot$ along the rungs of the ladder for a nontrivial solution of Equation~\eqref{eq:self-consistent-Jbot}, fermionic atoms become subject to a synthetic magnetic field which induces a chiral current $j_c=\sum_{m}\langle\hat{j}_{1,m}-\hat{j}_{2,m}\rangle/(L-1)$ with $\hat{j}_{\ell,m}=-iJ_\|(\hat{f}_{\ell,m+1}^\dag\hat{f}_{\ell,m}-\text{H.c.})$ being the local current operator along the leg $\ell$.

Figures~\ref{Fig:Sheikhan2016CavityFig8-9}(c) and~(d) show, respectively, the real part of the cavity-field amplitude $\alpha$ and the chiral current $j_c$ as a function of the flux $\varphi$ and $-U_lA^2L/J_\|\propto\eta_0^2$ for three representative filling $n=1/2,3/8,1/4$~\cite{Sheikhan2016Cavity-chiral}. The cavity field and the chiral current build up only above a pump-strength threshold $\eta_{0c}(\varphi)$, indication of the appearance of the cavity-induced synthetic magnetic field beyond the nonlinear pump-strength threshold $\eta_{0c}(\varphi)$. This is in sharp contrast to the free-space implementation of this scheme (in square lattices)~\cite{Aidelsburger2013Realization, Miyake2013Realizing}. For a given filling $n$, however, the pump-strength threshold $\eta_{0c}$ is strongly suppressed at the critical flux $\varphi_c=2\pi n$ (and $2\pi-\varphi_c$) and the hopping is restored even for an infinitesimal pump strength. This behavior is intimately related to the band structure and the Fermi points. 
For critical fluxes $\varphi_c$, the cavity field amplitude increases slowly above the zero pump strength. On the other hand, for $\varphi<\varphi_c\leq\pi$ and $\varphi>2\pi-\varphi_c$ ($\varphi_c<\varphi<2\pi-\varphi_c$) the cavity-field amplitude acquires a nonzero value only beyond a nonzero pump-strength threshold $\eta_{0c}$, exhibiting a first (second) order phase transition into the self-ordered superradiant phase. Depending on the filling $n$, the atomic ground state corresponds to a band insulator or a liquid, carrying a chiral edge current [indicated respectively by CI and CL in Figures~\ref{Fig:Sheikhan2016CavityFig8-9}(c) and~(d)]. In some parameter regimes, there exists a second solution. This second solution vanishes at hight pump strengths, signaling the unstable nature of it. Furthermore, the exact diagonalization of the corresponding master equation for a small-size system reproduces solely the first stable solution (above the pump-strength threshold $\eta_{0c}$). The field amplitude and the chiral current exhibit similar behavior as functions of the filling $n$ and $U_lA^2L/J_\|\propto\eta_0^2$ for representative fluxes. In particular, the cavity field and the chiral current appear again only above a pump-strength threshold $\eta_{0c}(n)$, where the threshold approaches zero at the critical filling $n=\varphi_c/2\pi$.

This cavity-based scheme to implement a synthetic magnetic field in a ladder can be straightforwardly generalized into a square/rectangle optical lattice~\cite{Sheikhan2016Cavity-topological, Colella2019Hofstadter}, corresponding to the quantum Hall effect for electrons. This allows one to study the Harper-Hofstadter model with fractal energy bands---known as the Hofstadter's butterfly---arising from a dynamical synthetic magnetic field. Indeed, the cavity-induced synthetic magnetic field in a finite-size system leads to topological edge states~\cite{Sheikhan2016Cavity-topological}. In an infinite system, the dynamic nature of the magnetic field induces a nontrivial deformation of the Hofstadter butterfly in the superradiant phase due to the delicate interplay between the collective superradiant scattering inducing the synthetic magnetic field and the emerging underlying fractal band structure~\cite{Colella2019Hofstadter}. This is in sharp contrast to the free-space implementation of this scheme as first proposed in Ref.~\cite{Jaksch2003Creation}, where the energy bands show the conventional Hofstadter-butterfly structure.

\begin{figure}[t!]
\centering
\includegraphics [width=.7\textwidth]{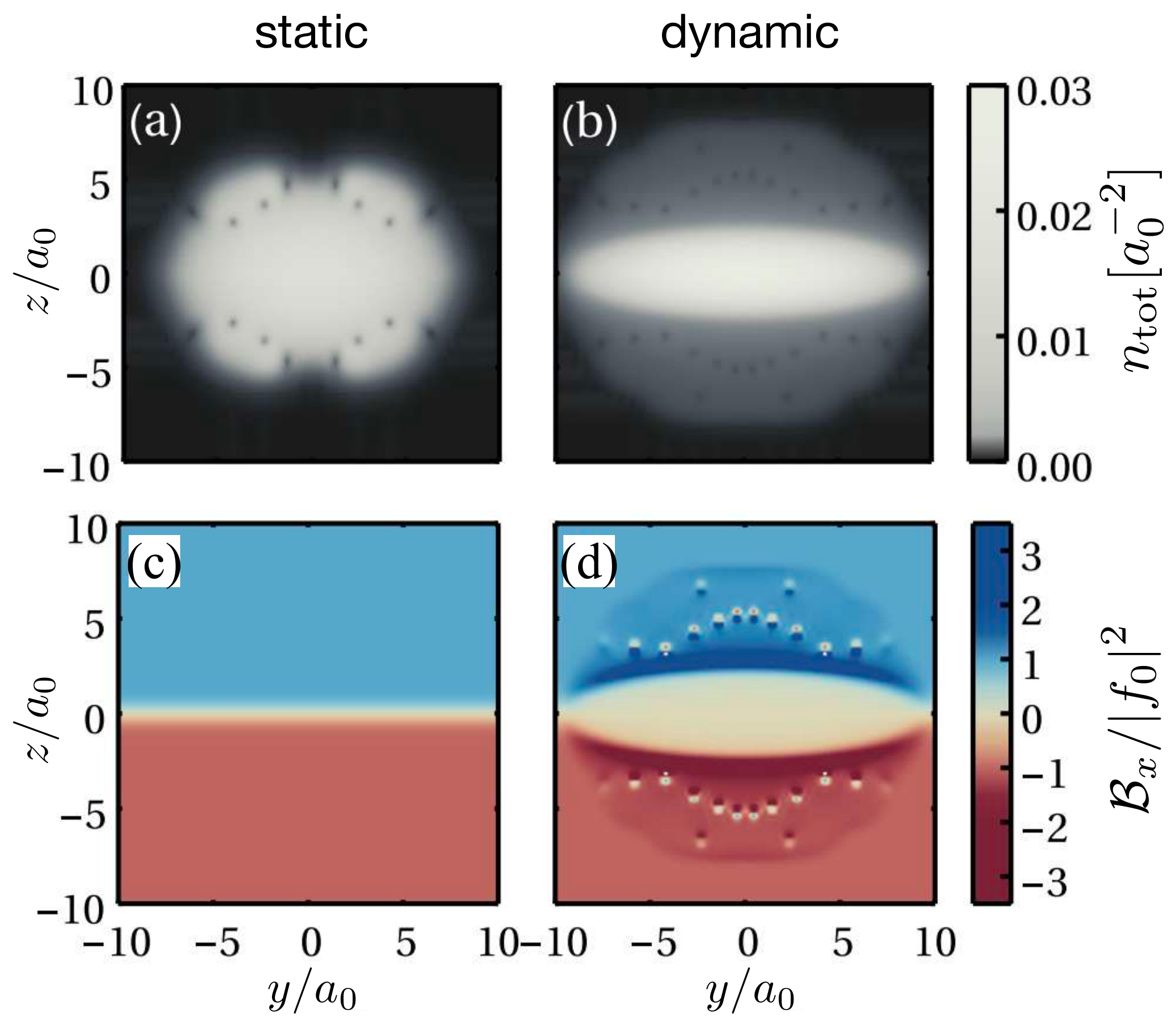}
\caption{A BEC in a synthetic magnetic field. 
(b)~The density profile of the BEC in a dynamic, synthetic magnetic field induced by a near-confocal multimode cavity. 
There is no vortex in the bulk of the BEC (i.e., the high density region) due the repulsion of the synthetic magnetic field from the inside of the BEC. 
Vortices are moved to the edge of the BEC. 
(d)~The spatial profile of the cavity-induced dynamic magnetic field, affected
by the atomic back-action. The synthetic magnetic field is expelled from the bulk of the condensate, similar to
the Meissner effect in superconductors.
The BEC density (a) in a static, synthetic magnetic field (c), included for the sake of comparison.
Here $a_0$ is the harmonic oscillator length of the trapping potential and
$f_0$ is a constant specifying the overall amplitude of the pump.
Figure adapted and reprinted with permission from Ref.~\cite{Ballantine2017Meissner}~\textcopyright~2017 by the American Physical Society.} 
\label{Fig:Ballantine2017Meissner_Fig2}
\end{figure}

In the above discussed schemes~\cite{Kollath2016Ultracold, Sheikhan2016Cavity-chiral, Sheikhan2016Cavity-topological, Colella2019Hofstadter}, although the cavity-induced synthetic magnetic field emerges dynamically owing to the backaction of the atoms, the spatial profile of the synthetic magnetic field itself is fixed by the cavity wave number $k_c$. The latter can also be made dynamical as proposed in Refs.~\cite{Zheng2016Superradiance, Ballantine2017Meissner}, providing a local dynamical coupling between the matter and the gauge potential as in the Ginzburg-Landau theory. A variant of the scheme discussed in detail above was considered in Ref.~\cite{Zheng2016Superradiance}, where an external 1D lattice with an energy offset $\hbar\Delta_{\rm latt}$ between adjacent sites is located perpendicular (i.e., along the $y$ direction) to the axis of a single-mode linear cavity. The suppressed hopping is restored by a cavity-assisted ``directional'' hopping containing a Peierls phase, which is proportional to the phase of the cavity field and is dynamic and must be determined self-consistently. In an infinite lattice under certain initial conditions, a dynamical vector gauge potential appears in the system, inducing a net, steady-state directed atomic current. In contrast, for a finite system the atomic current shows a transient pulse behavior due to a superradiant pulse (similar to those observed by illuminating degenerate quantum gases in free space) and a nonequilibrium gauge potential. A generalization of this scheme to a two-leg ladder coupled to two cavity modes revealed even more interesting features, including the emergence of an electromotive force due to dynamical instabilities according to simulated Faraday's induction law for the neutral atoms inside the cavity~\cite{Colella2021Open}. In a different approach, coupling a BEC to many nearly degenerate transverse modes of a near-confocal cavity in a particular configuration induces a synthetic magnetic field $\mathcal{B}_x$, which its spatial profile, proportional to the spatial derivative of the transverse intensity profile of the cavity light field, can evolve freely in response to the atomic state~\cite{Ballantine2017Meissner}. In particular, the cavity-induced synthetic magnetic field is expelled from the bulk of the condensate due to the diamagnetic response of the system, reminiscent of the Meissner effect in superconductors; see Figure~\ref{Fig:Ballantine2017Meissner_Fig2}(d). Correspondingly, vortices in the BEC density also disappear as shown in Figure~\ref{Fig:Ballantine2017Meissner_Fig2}(b).  
 
\subsubsection{Dynamical synthetic spin-orbit coupling}
\label{subsec:synthetic-SOC}

As the name suggest, the spin-orbit (SO) coupling refers to the coupling between the spin and the (linear or angular) momentum of a particle and it is a relativistic effect. A charged particle moving with velocity ${\mathbf v}$ in an electric field $\mathbf{E}$ experiences a magnetic field $\mathbf{B}=\mathbf{E}\times{\mathbf v}/c^2$ in its own rest frame according to the Lorentz transformation up to the first order in $(v/c)^2$ (with $c$ being the speed of the light). Consequently, the magnetic moment $\hat{\boldsymbol \mu}$ of the particle proportional to its charge $q$ and its spin $\hat{\mathbf s}$ couples to this magnetic field $\hat{\mathcal H}_{\rm SO}=-\hat{\boldsymbol \mu}\cdot\mathbf{B}\propto q\hat{\mathbf s}\cdot(\mathbf{E}\times\hat{\mathbf p})$, resulting in the SO coupling. Depending on the spatial form of the electric field, different forms of SO coupling can arise.

Such a coupling between the spin and the external degree of freedom is absent for an atom, as atoms are charge neutral. The SO coupling can be simulated for neutral atoms via different methods such as dressing the atoms by laser lights, where in the simplest case two atomic (ground) pseudospin states are needed to encode a spin-1/2 degree of freedom. As for the simulation of synthetic dynamical magnetic fields for the atoms using cavity fields described in Section~\ref{subsec:synthetic-B}, dynamical SO coupling can also be engineered for neutral atoms based on cavity fields. We will review here a few concrete schemes based on cavity-assisted two-photon Raman processes. The Raman technique using (non-copropagating) plane-wave lasers has so far been the only method to experimentally engineer an equal contribution of the Rashba and Dresselhaus dynamical SO couplings for quantum gases inside cavities~\cite{Kroeze2019Dynamical}---it was also the first method to implement static SO coupling for quantum gases in free space~\cite{Lin2011Spin,Wang2012Spin}. In Raman-based schemes for inducing SO coupling, it is required to employ at least one running-wave light field, in order to impart spin-dependent momentum recoil to the atoms.

\begin{figure}[t!]
\centering
\includegraphics [width=\textwidth]{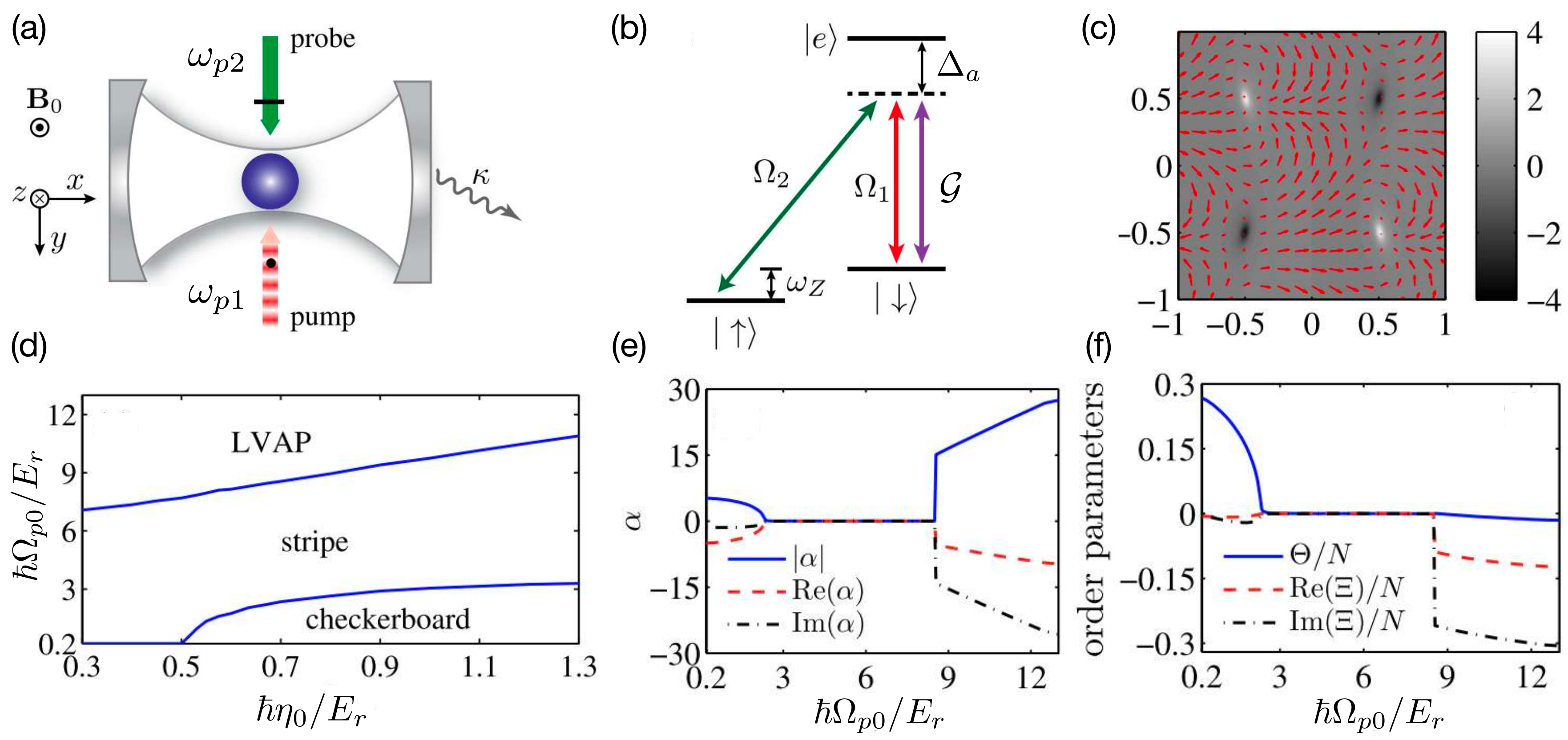}
\caption{Cavity-induced dynamical spin-orbit coupling. 
The schematic sketch of the system (a) and the atom-photon coupling (b). 
(c) The transverse spin $\mathbf{s}_\bot(\mathbf{r})=(s_x,s_y)$ in the 
lattice of vortex-antivortex pairs (LVAP) phase shows texture with opposite 
windings ($\pm1$) at even and odd sites. (d) The
mean-field phase diagram of the system in the parameter plane of $\eta_0$-$\Omega_{p0}$,
exhibiting checkerboard, stripe, and LVAP phases. These phases are characterized by
the cavity field amplitude $\alpha$ and the order parameters $\{\Theta,\Xi\}$,
shown respectively in panels~(e) and (f) for a vertical cut through the phase diagram at $\hbar\eta_0=0.7E_r$. 
The LVAP phase arises due to the competition between the static and dynamic SO coupling.
Figure adapted and reprinted with permission from Ref.~\cite{Deng2014Bose}~\textcopyright~2014 by the American Physical Society.} 
\label{Fig:Deng2014BoseFig1and2}
\end{figure}

\paragraph{Linear-cavity-based schemes} Consider $\Lambda$-type three level bosonic atoms, with two ground spin states $\{\ket{\downarrow},\ket{\uparrow}\}$ and an electronic excited state $\ket{e}$, inside a linear cavity as in Ref.~\cite{Deng2014Bose}. The magnetic quantum number $m_\tau$ of the atomic internal states satisfy $m_\downarrow-m_e=0$ and $m_\uparrow-m_e=1$.  A bias magnetic field applied along the $z$ direction defines the quantization axis and produces a Zeeman shift $\hbar\omega_Z$ between the spin states. The atoms are driven in the transverse direction by a standing- and a running-wave laser with frequencies $\omega_{p1}=\omega_p+\Delta\omega_p$ and $\omega_{p2}=\omega_p$, respectively. The first (second) laser is linearly polarized along the $z$ ($x$) directions, therefore, inducing the atomic transition $\ket{\downarrow}\leftrightarrow\ket{e}$ ($\ket{\uparrow}\leftrightarrow\ket{e}$) with the position-dependent Rabi frequency $\Omega_1(\mathbf{r})=\Omega_{01}\cos(k_cy)$ [$\Omega_2(\mathbf{r})=\Omega_{02}e^{ik_cy}$]. Furthermore, the transition $\ket{\downarrow}\leftrightarrow\ket{e}$ is also strongly coupled into an initially empty standing-wave mode of the cavity linearly polarized along $z$, with the coupling strength $\mathcal{G}(\mathbf{r})=\mathcal{G}_0\cos(k_cx)$. 
The system is depicted schematically in Figures~\ref{Fig:Deng2014BoseFig1and2}(a) and (b). Note that the atom-pump detuning $\Delta_a=\omega_p-\omega_a>0$ is chosen to be positive, i.e., blue detuned. 

As in the case of double-$\Lambda$ system discussed in Section~\ref{subsec:multi-com-atoms-cavity}, in the large atom-pump detuning limit, $|\Omega_{01,02}/\Delta_a|\ll1$ and $|\mathcal{G}_0/\Delta_a|\ll1$, the atomic excited state can be adiabatically eliminated. This results in an effective Hamiltonian for the ground spin states, 
\begin{align}
\label{eq:H-SOC}
\hat{H}_{\rm SO}=\int \hat\Psi^\dag(\mathbf{r})\hat{\mathcal H}_{1,\rm eff}\hat\Psi(\mathbf{r})d\mathbf{r}
+\frac{1}{2}\sum_{\tau,\tau'=\downarrow,\uparrow}g_{\tau\tau'}\int\hat\psi_\tau^\dag(\mathbf{r})\hat\psi_{\tau'}^\dag(\mathbf{r})
\hat\psi_{\tau'}(\mathbf{r})\hat\psi_{\tau}(\mathbf{r})d\mathbf{r}
-\hbar\Delta_c\hat{a}^\dag\hat{a},
\end{align}
where $\Delta_c\equiv\omega_{p1}-\omega_c$, and 
$\hat{a}$ and $\hat\Psi=(\hat\psi_\uparrow,\hat\psi_\downarrow)^\top$ are
the photonic and two-component atomic bosonic annihilation field operators, respectively.
The single-particle atomic Hamiltonian density reads,
\begin{align} \label{eq:single-particle-H-den-SOC}
\hat{\mathcal H}_{1,\rm eff}=
-\frac{\hbar^2}{2M}\boldsymbol{\nabla}^2+V_{\rm ext}(\mathbf{r})
+\hbar
\begin{bmatrix}
\delta & 
\Omega_{p}(\mathbf{r})+\Omega_{c}(\mathbf{r})\hat{a}
\\
\Omega_{p}^*(\mathbf{r})+\Omega_{c}^*(\mathbf{r})\hat{a}^\dag & 
V(\mathbf{r})+U(\mathbf{r})\hat{a}^\dag\hat{a}+\eta(\mathbf{r})(\hat{a}^\dag+\hat{a})
\end{bmatrix},
\end{align}
with the state-independent harmonic trap $V_{\rm ext}(\mathbf{r})$, and the classical potential $V(\mathbf{r})=\Omega_1^2(\mathbf{r})/\Delta_a=V_0\cos^2(k_cy)$, the cavity potential $U(\mathbf{r})=\mathcal{G}^2(\mathbf{r})/\Delta_a=U_0\cos^2(k_cx)$ per photon, and the interference potential $\eta(\mathbf{r})=\mathcal{G}(\mathbf{r})\Omega_1(\mathbf{r})/\Delta_a=\eta_0\cos(k_cx)\cos(k_cy)$ for the spin-$\downarrow$ component. The two spin states are coupled via the classical $\Omega_{p}(\mathbf{r})=\Omega_1(\mathbf{r})\Omega_2^*(\mathbf{r})/\Delta_a=\Omega_{p0}\cos(k_cy)e^{-ik_cy}$ and cavity-induced $\Omega_{c}(\mathbf{r})=\mathcal{G}(\mathbf{r})\Omega_2^*(\mathbf{r})/\Delta_a=\Omega_{c0}\cos(k_cx)e^{-ik_cy}$ two-photon Raman-Rabi processes. Here it has been assumed without loss of generality $\{\Omega_{01},\Omega_{02},\mathcal{G}_0\}\in \mathbb{R}$. The maximum strength of the potentials and Raman-Rabi couplings are defined as $V_0=\Omega_{01}^2/\Delta_a$, $U_0=\mathcal{G}_0^2/\Delta_a$, $\eta_0=\mathcal{G}_0\Omega_{01}/\Delta_a$, $\Omega_{p0}=\Omega_{01}\Omega_{02}/\Delta_a$, $\Omega_{c0}=\mathcal{G}_0\Omega_{02}/\Delta_a$. Here, $\delta\equiv\omega_Z+\Delta\omega_p+\Omega_{02}^2/\Delta_a$ is the effective two-photon detuning and can be modified independently via the external bias magnetic field tuning $\omega_Z$. Both classical and quantum Raman-Rabi interactions $\Omega_{p}(\mathbf{r})$ and $\Omega_{c}(\mathbf{r})\hat{a}$ induce coupling between internal and external degrees of freedom of the atom, resulting in a dynamical SO coupling. The interplay of this dynamical SO coupling with the classical and quantum optical potentials leads to interesting phenomena as discussed in the following.  

The mean-field approach $\hat\psi_\tau(\mathbf{r})\to\psi_\tau(\mathbf{r})=\langle\hat\psi_\tau(\mathbf{r})\rangle$ and $\hat{a}\to\alpha=\langle\hat{a}\rangle$ can be used to describe the system; see Section~\ref{subsubsec:MF-bose}. The corresponding coupled Gross-Pitaevskii equations for the condensate wavefunctions are solved self-consistently with the steady-state field amplitude
\begin{align}
\alpha_{\rm ss}=\frac{\eta_0\Theta+\Omega_{c0}\Xi}{\Delta_c-U_0B+i\kappa},
\end{align}  
where $B=\int |\psi_\downarrow(\mathbf{r})|^2\cos^2(k_cx)d\mathbf{r}$, $\Theta=\int |\psi_\downarrow(\mathbf{r})|^2\cos(k_cx)\cos(k_cy)d\mathbf{r}$ [cf.\ Equation~\eqref{eq:order-paramter-op}], and $\Xi=\int \psi_\downarrow^*(\mathbf{r})\psi_\uparrow(\mathbf{r})\cos(k_cx)e^{ik_cy}d\mathbf{r}$. Here it has been assumed the cavity decay $\kappa$ is large so that the cavity field follows the atomic dynamics instantly. The cavity field has two contributions: i) from collective photon scattering from the first pump with spin-down atoms without changing their internal state characterized by $\Theta$, and ii) from collective photon scattering from the second pump via the spin-flipping process quantified by $\Xi$. The first scattering process favors a checkerboard density pattern, while the second process results in the cavity-induced SO coupling favoring a lattice of vortex-antivortex pairs (LVAP). Therefore, $\Theta$ and $\Xi$ can be identified as order parameters, characterizing different phases of the system. 

The mean-field phase diagram of the system in the parameter plane of $\eta_0$-$\Omega_{p0}$ for a fixed $\delta$ is presented in Figure~\ref{Fig:Deng2014BoseFig1and2}(d). In order to reduce the number of the free parameters, the ratio $\Omega_{01}/\mathcal{G}_0$ has been fixed at 10, implying $\Omega_{c0}=0.1\Omega_{p0}$. In relatively small $\Omega_{p0}$ the checkerboard phase arises beyond a critical threshold $\eta_{0c}$, similar to the one-component checkerboard self-ordering discussed in Section~\ref{sec:SR_discrete}. For chosen immiscible two-body contact interaction strengths $g_{\downarrow\uparrow}^2>g_{\downarrow\downarrow}g_{\uparrow\uparrow}$, in this phase the density structure of the spin-$\downarrow$ component  is determined by the potentials $V(\mathbf{r})$, $U(\mathbf{r})$, and $\eta(\mathbf{r})$, while the density structure of the spin-$\uparrow$ component is mainly fixed by the contact interactions. In moderate $\Omega_{p0}$, the system undergoes a phase transition into the density-modulated stripe phase due to the classical potential $V(\mathbf{r})$ and the ``static'' SO coupling induced by the classical pump Raman-Rabi coupling $\Omega_{p}(\mathbf{r})$. The cavity field amplitude is zero in this phase, signaling that it is a normal (i.e., non-superradiant) state. Note that the stripe phase here does not possess supersolid characteristics as the translational symmetry along the transverse ($y$) direction is broken explicitly by the classical Raman-Rabi coupling $\Omega_{p}(\mathbf{r})$ and the classical potential $V(\mathbf{r})$. The relative phase $\Delta\phi(\mathbf{r})$ of the two condensate wave functions is periodically modulated along the $y$ direction to minimize the Raman-coupling energy; see also Section~\ref{subsubsec:disorder-induced-slefordering} for more details. The LVAP phase is, however, favored in the strong $\Omega_{p0}$ limit where the cavity-induced ``dynamical'' SO coupling by the Raman-Rabi coupling $\alpha\Omega_{c}(\mathbf{r})$ is comparable with the static SO coupling owing to the large number of photons in the system. In this phase, the relative phase of the two condensate wave functions exhibits a lattice of vortex-antivortex pairs, hence the name LVAP of this phase. Consequently, the transverse spin $\mathbf{s}_\bot(\mathbf{r})=(s_x,s_y)$ forms a hedgehog (quadrupole) texture with the winding number $\pm1$ at the even/odd sites breaking the $\mathbf{Z}_2$ symmetry of the system; see Figure~\ref{Fig:Deng2014BoseFig1and2}(c). The cavity field amplitude $\alpha$ and the order parameters $\{\Theta,\Xi\}$ are shown in Figures~\ref{Fig:Deng2014BoseFig1and2}(e) and~(f), respectively, for a vertical cut through the phase diagram shown in Figure~\ref{Fig:Deng2014BoseFig1and2}(d). Note that $|\Theta/\Xi|\gg1$ ($|\Xi/\Theta|\gg1$) in the checkerboard (LVAP) phase, while both order parameters as well as the cavity field are zero in the stripe phase.

We now revisit the model described in Sections~\ref{subsec:spinor-BEC-self-ordering} and~\ref{subsec:ExperimentSpinSO} for realizing density and spin self-ordering. As it was shown, the Hamiltonian of Equation~\eqref{eq:single-particle-H-den-spinor} for two pure running-wave pump lasers leads to a transverse conical, chiral spin-spiral texture. In fact, this emergent spin order is due to the cavity-induced dynamical SO coupling in the system. This has been studied in detail in Ref.~\cite{Ostermann2021Many}, by also including the two-body contact interactions between the atom in the model. This is the first and only cavity-based scheme where a dynamical SO coupling has been realized for ultracold atoms so far~\cite{Kroeze2019Dynamical}, as will be discussed in more details in Section~\ref{subsubsec:synthetic-SOC-experiment}. This cavity-induced SO-coupled BEC system has also been studied in 1D along the direction of the pumps using a finite-size density matrix renormalization group (DMRG) algorithm in the matrix product state form~\cite{Halati2019Cavity}. By adiabatically eliminating the cavity field and discretizing the space, the system amounts to an interacting Bose-Hubbard ladder pierced with a cavity-induced synthetic magnetic field (similar to the ladder model for the fermionic atoms with the cavity-induced synthetic magnetic field discussed in Section~\ref{subsec:synthetic-B}). In the continuum limit, a Meissner superfluid with a net chiral (i.e., spin) current is dynamically stabilized in some parameter regimes. 

\paragraph{Ring-cavity-based schemes} Another approach for inducing a dynamical SO coupling in the framework of cavity QED is to utilize ring cavities~\cite{Mivehvar2014Synthetic, Dong2014Cavity, Mivehvar2015Enhanced, Dong2015Photon}. Since the natural modes of ring cavities are plane waves, this provides the important ingredient of a SO coupling based on the Raman scheme. Reference~\cite{Dong2014Cavity} has considered a scenario where non-interacting $\Lambda$-type bosonic atoms are placed inside one arm of a ring cavity. The atoms are Raman-dressed by one running-wave laser and one counterpropagating plane-wave mode of the ring cavity pumped longitudinally through a cavity mirror. The single-particle energy dispersion of the system has two branches, similar to the free-space SO-coupled atoms. However, the energy dispersion develops a loop structure in some parameter regimes due to the nonlinearity of the system. The nonlinearity also leads to dynamical instabilities, which are more pronounced in the limit of low cavity-photon number. An interacting $\Lambda$-type BEC has also been considered inside a ring cavity, where the atoms are Raman-dressed instead by two distinct counterpropagating plane-wave modes of the cavity pumped longitudinally through a cavity mirror~\cite{Mivehvar2015Enhanced}. The cavity fields induce a synthetic SO coupling as well as long-range interactions between the atoms, where the latter stems from the dynamical nature of the photonic fields and their backaction on the atomic fields (see Section~\ref{subsubsec:cavity-induced-int}). The strength of the cavity-mediated interactions can be tuned readily by changing the pump amplitudes and frequencies as well as the decay rate of the cavity modes. The interplay between the single-particle energy dispersion on one hand, and the two-body contact and cavity-induced global interactions on the other hand determines the many-body ground state of the system. In a wide range of parameters, the stripe phase is favored over the plane-wave phase owing to the cavity-induced interactions. Notably, the stripe phase is energetically stabilized even for condensates with attractive intraspecies and interspecies contact interactions for sufficiently large cavity-induced interactions. Furthermore, for a symmetric double-well energy dispersion it is also possible by tuning the cavity-mediated interactions to obtain a stripe phase with an arbitrary superposition of atoms in the left and right wells, in sharp contrast to a stripe phase in a free-space symmetric double-well energy dispersion consisting of an equal number of atoms in the left and right wells. Both Refs.~\cite{Dong2014Cavity, Mivehvar2015Enhanced} have ignored the unpumped degenerate counterpropagating mode(s) corresponding to the pumped mode(s) of the ring cavities. This is a good first approximation for strongly pumped cavities in short time scales, as the probably of scattering a photon into a strongly pumped mode is much higher than the probability of scattering a photon into the degenerate empty mode due to the Bose-enhancement factor. By building on Ref.~\cite{Mivehvar2015Enhanced}, the effect of the unpumped degenerate counterpropagating modes in this model has been included and studied in Ref.~\cite{Ostermann2019Cavity}; though the two-body interactions were omitted. These unpumped modes are populated solely by the coherent backscattering of photons from the pumped modes via the atoms and play an important role in the dynamics of the system in long times. The system exhibits three fundamentally different steady-state phases, as the pump strengths and pump frequencies are varied, characterized by distinct spatial density distributions and spin textures: a combined density and spin wave, a continuous spin-spiral texture with a homogeneous atomic density, and a spin spiral with a modulated density. The spin-spiral states are topologically nontrivial, as discussed in Section~\ref{subsec:spinor-BEC-self-ordering}, and are intimately related to cavity-induced SO coupling appearing beyond a critical pump power. The topologically trivial density-wave--spin-wave state spontaneously breaks two continuous symmetries, the $U(1)$ freedom of the total phase of the two condensate wave functions and a $U(1)$ screw-like symmetry corresponding to a continuous spatial translation accompanied by phase rotations of the unpumped cavity-field amplitudes and condensate wave functions, a characteristic of a supersolid state as discussed in Section~\ref{sec:supersolid}. The transitions between different phases are either topological and first order, or non-topological and second order. Remarkably, all the phase transitions can be detected nondestructively at the cavity output via monitoring the unpumped cavity modes.

\subsubsection{Experimental realization of dynamical spin-orbit coupling}
\label{subsubsec:synthetic-SOC-experiment}

\begin{figure}[t!]
\centering
\includegraphics[width=1\columnwidth]{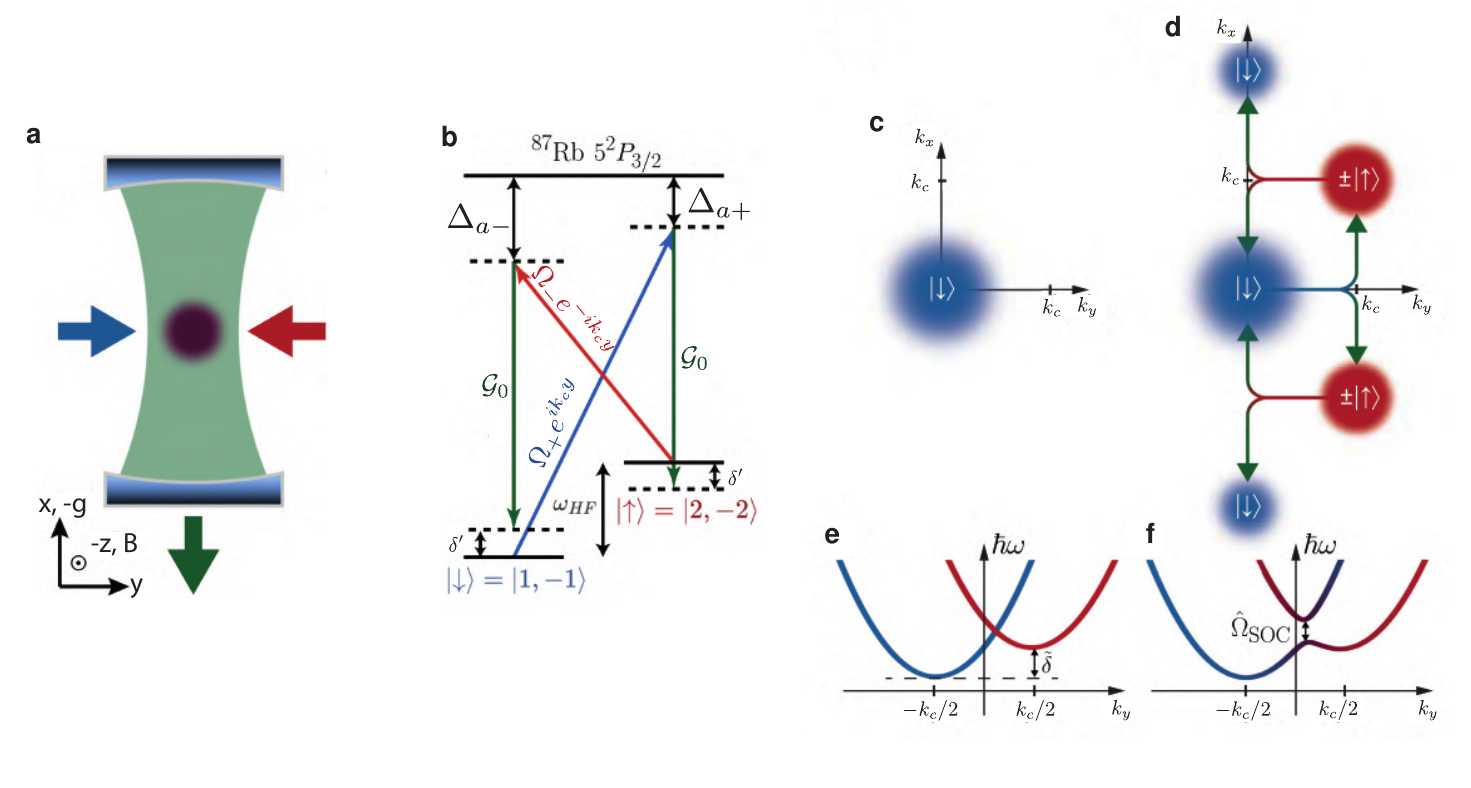}
\caption{Schematic of the cavity-induced dynamical spin-orbit coupling. Two Raman pump beams (red and blue arrows), polarized along the cavity axis, counterpropagate through a BEC of Rb atoms (purple) inside a TEM$_{00}$ cavity mode. The cavity emission (green arrow) is detected by a single-photon counter, and the atoms are imaged after ballistic expansion. (b) Level diagram illustrating the cavity-assisted Raman coupling between two hyperfine levels of the $^{87}$Rb atoms acting as the spin states. $\delta'$ is the Raman detuning. (c),(d) Momentum-space depiction of the emergence of SO coupling. (c) Initially atoms are in a spin-polarized state $\ket{\downarrow}$. (d) If the transverse pump strength is sufficiently strong, SO coupling emerges and the spin components are in different momentum states. The $\pm$ sign of the $\ket{\uparrow}$ spin component indicates the $\mathbf{Z}_2$ symmetry-broken phase freedom. (e),(f) Energy-momentum dispersion relation of each spin state, transitioning from free (e) to SO-coupled (f) dispersion bands. The coupling strength $\hat{\Omega}_\mathrm{SOC}$ is proportional to $\hat{a}$ and $\hat{a}^\dag$ and therefore arises dynamically as the atoms scatter pump photons into the cavity. The zero of the momentum has been shifted with respect to the lab frame by $-k_c/2$ in the plot. Figure and caption adapted and reprinted with permission from Ref.~\cite{Kroeze2019Dynamical}~\textcopyright~2019 by the American Physical Society.}
\label{fig:expSOCscheme}
\end{figure}

The first experiment demonstrating cavity-assisted dynamical spin-orbit coupling made use of the experimental scheme shown in Figure~\ref{fig:expSOCscheme}(a) and (b)~\cite{Kroeze2019Dynamical}. It is a variant of the scheme presented in Section~\ref{subsec:ExperimentSpinSO}, however, employing running-wave fields for the pumps instead of the standing-wave fields. More precisely, it is a special case of the Hamiltonian of Equation~\eqref{eq:single-particle-H-den-spinor} for two pure running-wave pump lasers corresponding to $R=0$ in Section~\ref{subsec:spinor-BEC-self-ordering}. A BEC is loaded into a single-mode cavity and transversally illuminated with two balanced counterpropagating running-wave laser fields. The frequencies of these laser fields are chosen such that they, in combination with vacuum fluctuations of the cavity mode, each induce a Raman transition between two atomic hyperfine states, labeled as $\ket{\downarrow}$ and $\ket{\uparrow}$. For pump rates below threshold, the atomic spins and positions are disordered, and the Raman scattering rates are low. Above a critical pump strength, spin and density of the gas self-order and the cavity mode is populated with a coherent field due to superradiance. The frequency of the emerging cavity field is the center frequency between the two coherent pump-field frequencies.

Above threshold, SO coupling emerges in this scheme. In contrast to a situation with standing-wave pump fields where no net momentum is transferred to a cloud (cf.\ Section~\ref{subsec:ExperimentSpinSO}), during the self-organization with the running-wave pumps momentum is imparted to the system while the atomic spins are flipped. This can also be understood from the momentum-space representation shown in Figure~\ref{fig:expSOCscheme}(c) and (d). While the cloud is spin polarized in $\ket{\downarrow}$ below the critical pump strength and only occupies the zero momentum state, above threshold additional momentum states are occupied. Since the Raman scattering process is spin selective, the spin state $\ket{\uparrow}$ is only populated if a net momentum along the positive $y$ axis is transferred to the atoms. In contrast along the $x$ axis, the accompanying momentum change averages to zero due to the standing-wave mode of the cavity. Going into an according co-moving reference frame, opposite spin states thus move in opposite directions, which thus realizes a SO coupling. Since this process depends self-consistently on the evolution of the atomic cloud and the cavity field in order to maximize the superradiant collective photon scattering, the SO coupling has a dynamical character.

The relevant physics is captured---neglecting dispersive shifts---by a Hamiltonian obtained from Equation~\eqref{eq:single-particle-H-den-spinor} for two pure counterpropagating running-wave pump lasers $\Omega_{1,2}(y)=\Omega_\pm e^{\pm ik_cy}$ (corresponding to $R=0$) after transferring to the co-moving frame,
\begin{align}
  \hat{H} = -\hbar\Delta_c \hat{a}^\dagger\hat{a} + \int \hat{\Psi}^\dagger(\mathbf{r})
  \begin{pmatrix} 
  \frac{1}{2M}\left(\hat{\mathbf{p}} + \frac{\hbar k_c}{2}\mathbf{e}_y\right)^2 - \hbar\tilde{\delta} & 
  \hbar\hat{\Omega}_\mathrm{SOC} \cos{k_c x} \\ 
  \hbar\hat{\Omega}_\mathrm{SOC}^\dag \cos{k_c x} & 
  \frac{1}{2M}\left(\hat{\mathbf{p}} - \frac{\hbar k_c}{2}\mathbf{e}_y\right)^2\\ \end{pmatrix}
    \hat{\Psi}(\mathbf{r}) d\mathbf{r}\,,
\end{align}
where $\hbar\tilde{\delta}$ is the effective two-level spin splitting and $\hat\Psi=(\hat\psi_\uparrow,\hat\psi_\downarrow)^\top$ is a spinor containing the atomic annihilation operators $\hat{\psi}_{\uparrow,\downarrow}$. The strength of the two-photon Raman-Rabi rate giving rise to the SO coupling is an operator, given by
\begin{align}
   \hat{\Omega}_\mathrm{SOC} = \frac{\mathcal{G}(y,z) \Omega_+}{8\sqrt{2}\Delta_{a+}} \hat{a}^\dagger + \frac{\mathcal{G}(y,z) \Omega_-}{8\sqrt{2}\Delta_{a-}} \hat{a},
\end{align}
where $\Delta_{a\pm}$ are the atomic detunings with respect to the two pump fields. Considering the dispersion relation of both spin states in presence of the SO coupling, one finds that the energy bands are displaced from one another due to the opposite momentum recoils and coupled via the Raman-Rabi term. This leads to a double-minima lowest energy-band structure and a band gap given by $\hat{\Omega}_\mathrm{SOC}$ proportional to $\hat{a}^\dagger$ and $\hat{a}$, shown schematically in Figure~\ref{fig:expSOCscheme}(e) and (f).

\begin{figure}[t!]
\centering
\includegraphics[width=1\columnwidth]{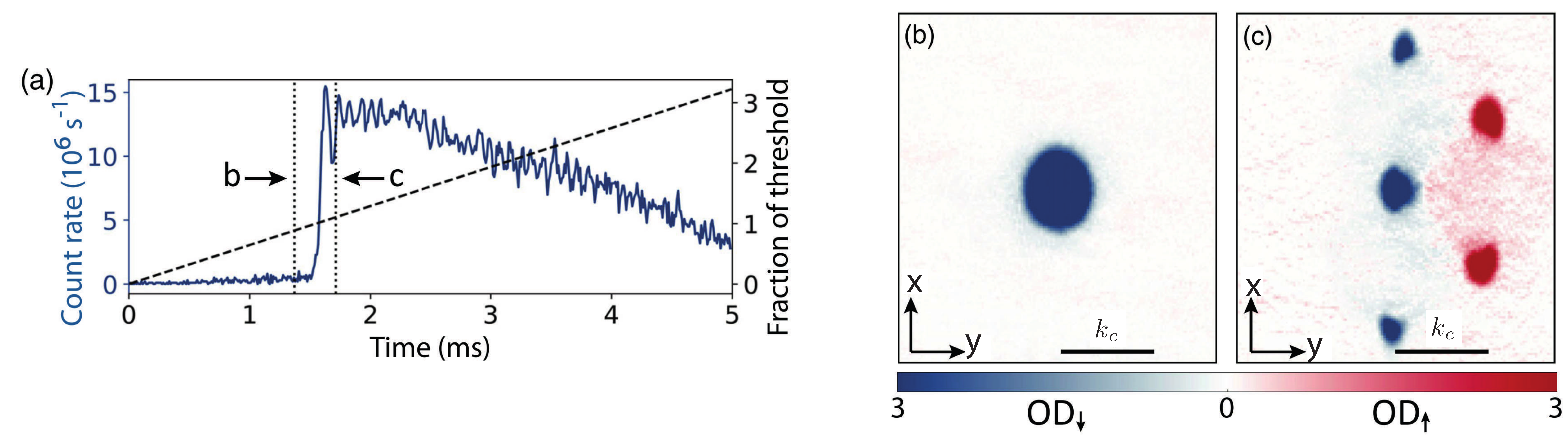}
\caption{Experimental realization of dynamical spin-orbit coupling. (a) Cavity emission detected by single-photon counters (solid blue curve) and optical power in the Raman beams (dashed black line), both as a function of time. Steady-state SO coupling persists up to a few milliseconds. (b),(c) Spin-resolved momentum distribution in time of flight, taken at the points labeled in (a). All atoms are in the $\ket{\downarrow}$ state (blue) just below threshold (b). Above threshold~(c), spin-up atoms (red) have acquired a net momentum in the $y$ direction, as shown by the spin-colored Bragg peaks at nonzero momenta. Also shown are second-order diffraction peaks along the cavity direction due to the reverse Raman process. Figure adapted and reprinted with permission from Ref.~\cite{Kroeze2019Dynamical}~\textcopyright~2019 by the American Physical Society.}
\label{fig:expSOCdata}
\end{figure}

The superradiant state displaying SO coupling corresponds in real space to an atomic spinor state that exhibits a helical pattern (i.e., spin spiral) along the pump axis according to $\ket{\downarrow} \pm e^{ik_c y} \cos(k_c x) \ket{\uparrow}$. This state corresponds to the chiral spin-spiral phase in Figure~\ref{Fig:Mivehvar2019Cavity_Fig1}(b) and Figure~\ref{Fig:Mivehvar2019Cavity_Fig3}(a). The experimental observables of this state are the intra-cavity light field that rises to a finite value above threshold and the spin-selective occupation of momentum states according to the expectations described above; see Figure~\ref{fig:expSOCdata}. The lifetime of the superradiant state is only on the order of a few milliseconds, most likely limited by a dephasing of the two pump beams.

\subsection{Cavity-induced topological states}
\label{sec:TI}

\subsubsection{Su-Schrieffer-Heeger model and topological superradiant Peierls insulator}
\label{subsubsec:SSH}

The landscape of the superradiant optical lattice
\begin{align}
\label{eq:Vsr1d}
\hbar V_{\rm SR}(x)=\hbar U_0|\alpha|^2\cos^2(k_cx)+2\hbar\eta_0(\alpha+\alpha^*)\cos(k_cx),
\end{align} 
depends crucially on the sign of $U_0\propto\Delta_a^{-1}=(\omega_p-\omega_a)^{-1}$ as illustrated in Figure~\ref{fig:potentials}. The lattice potential $\hbar V_{\rm SR}(x)$ is $\lambda_c$ periodic due to the interference term $\propto\cos(k_cx)$. For a conventional red-detuned pump laser (and cavity field) with respect to the relevant atomic transition $U_0\propto\Delta_a^{-1}<0$, the superradiant lattice has a single global minimum in every unit cell of length $\lambda_c$. More precisely, the unit cell contains two lattice sites with an energy offset between them. For the blue-detuned repulsive case $U_0\propto\Delta_a^{-1}>0$, in stark contrast $\hbar V_{\rm SR}(x)$ is homopolar and its unit cell possesses two lattice sites with no energy offset between them. In the deep lattice limit, this latter case maps to the Su-Schrieffer-Heeger (SSH) model, a non-interacting 1D tight-binding fermionic model with staggered (i.e., different intracell and intercell) hopping amplitudes~\cite{Asboth2016A}. The SSH model has two distinct dimerizations: i) the intracell hopping amplitude is larger than the intercell hopping amplitude, ii) the intracell hopping amplitude is smaller than the intercell hopping amplitude. In the former case, the system is a trivial insulator at half filling. Whereas in the latter case, the Bloch bands have a nontrivial topological structure, characterized by a non-zero bulk topological invariant---the winding number---and topologically protected edge states. 

The self-ordering of spin-polarized, low-field-seeking (i.e., blue detuned $\Delta_a^{-1}>0$) fermionic atoms inside a linear cavity in 1D along the cavity axis has been studied theoretically in Ref.~\cite{Mivehvar2017Superradiant}. The system is described by the 1D limit of the effective Hamiltonian~\eqref{eq:H_eff_1comp} with $U_0\propto\Delta_a^{-1}>0$; see also Section~\ref{subsubsec:MF-fermi}. Note that the two-body contact interaction is identically zero owing to the Pauli exclusion principle and $V_{\rm ext}(x)=\hbar V(x)=0$.

At low pump strengths, the system is in the normal Fermi-gas state with no photon inside the cavity. By increasing the pump strength, the system becomes unstable toward the superradiant phase $\alpha\neq0$, where the Fermi gas crystallizes under the emergent superradiant potential $\hbar V_{\rm SR}(x)$, Equation~\eqref{eq:Vsr1d}, in turn enhancing constructive photon scattering from the pump laser into the cavity mode. Across the superradiant phase transition, the relative phase of the cavity field with respect to the pump laser is locked at either 0 or $\pi$ (under the assumption $\Delta_c\gg\kappa$), spontaneously breaking the $\mathbf{Z}_2$ symmetry (i.e., the invariance under the transformation $\alpha\to-\alpha$ and $x\to x+\lambda_c/2$) of the system. That is, the system chooses spontaneously one of the two possible superradiant lattice configurations shifted by $\lambda_c/2$ with respect to one another, corresponding to the two distinct dimerizations of the SSH model.     

\begin{figure}[t!]
\centering
\includegraphics[width=1\columnwidth]{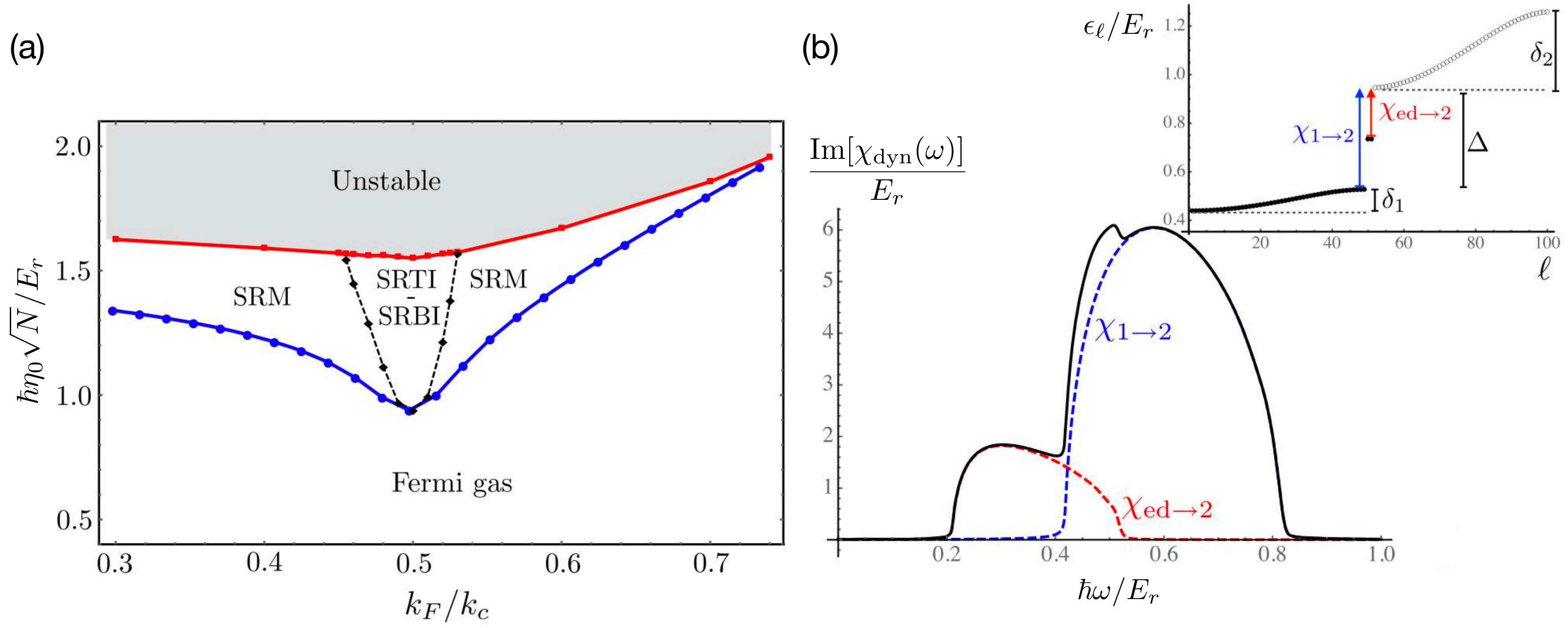}
\caption{Topological superradiant Peierls insulator in 1D spin-polarized fermionic atoms inside a linear cavity
in the blue-detuned repulsive regime $U_0\propto\Delta_a^{-1}>0$.
(a) The phase diagram of the system in the $\eta_0$-$k_F$ parameter space at the finite temperature $k_BT=0.01E_r$.
Above the critical pump strength (the blue curve), the Fermi gas crystallizes as a consequence of the superradiant phase transition.
The superradiant crystalline phase is either a metal (SRM) or an insulator. Depending on the spontaneously chosen 
dimerization (i.e., the cavity field phase) via the $\mathbf{Z}_2$ symmetry breaking process, 
the superradiant insulating state is either a band insulator (SRBI) or 
a topological insulator (SRTI), characterized by their Zak phases. 
(b) The imaginary part of the optical polarizability $\chi_{\rm dyn}(\omega)$ for a finite-size system, corresponding to the atomic absorption
of the cavity photons. The edge states within the energy gap open an additional absorption channel 
in a well-defined frequency range (the red dashed curve). The inset shows the energy spectrum of a finite-size system,
with a pair of edge states within the superradiant-opned energy gap.
Figure adapted and reprinted with permission from Ref.~\cite{Mivehvar2017Superradiant}~\textcopyright~2017 by the American Physical Society.}
\label{fig:Mivehvar2017Superradiant_Fig3and4}
\end{figure}

The phase diagram of the system in the parameter space of the pump strength $\eta_0$ versus the Fermi momentum $k_F$ at a finite temperature is shown in Figure~\ref{fig:Mivehvar2017Superradiant_Fig3and4}(a)~\cite{Mivehvar2017Superradiant}. The emergent lattice $\hbar V_{\rm SR}(x)$ opens up gaps in the quadratic energy dispersion of the free Fermi gas across the superradiant phase transition. For atomic densities around half filling $k_F=k_c/2$, the superradiant state is an insulator, while for filling sufficiently away from half filling the system corresponds to a superradiant metal (SRM). The transition to the insulating state with a dimerized structure at half filling occurs through the same mechanism underlying the Peierls instability in 1D electron-phonon models~\cite{Rylands2020Photon}. Depending on the spontaneously chosen dimerization via the superradiant $\mathbf{Z}_2$ symmetry breaking, the superradiant insulating state is either a trivial band insulator (SRBI) or a topological insulator (SRTI), characterized by the Zak phase. For this system the Zak phase is $\mathbf{Z}_2$ quantized and serves as a topological invariant, with the value 0 ($\pi$) corresponding to the SRBI (SRTI). Remarkably, the value of the Zak phase coincides exactly with the relative phase of the cavity field with respect to the pump laser, providing a nondestructive way to read out the topological Zak phase via the leaked cavity field. Note that the superradiant phase terminates at large pump strengths, indicated by the gray area in the phase diagram of Figure~\ref{fig:Mivehvar2017Superradiant_Fig3and4}(a); see also Figure~\ref{fig:PBandOrganization} for a similar behavior in self-ordering of low-field-seeking bosonic atoms. We also note that the critical pump strength at half filling becomes infinitesimally small at zero temperature due to the Fermi nesting~\cite{Piazza2014Umklapp}.

The SRTI state possesses a pair of edge states within the superradiant-opened energy band gap in a finite size system, another indication of the nontrivial topology of the state, represented in the inset of Figure~\ref{fig:Mivehvar2017Superradiant_Fig3and4}(b). These edge states can be detected indirectly via the spectral properties of the cavity output. Figure~\ref{fig:Mivehvar2017Superradiant_Fig3and4}(b)
shows for a finite-size system the imaginary part of the optical polarizability $\chi_{\rm dyn}(\omega)$ [see Equation~\eqref{eq:dyn_polarization}], corresponding to the atomic absorption with respect to the propagation of cavity photons. The edge states modify the polarization function significantly and open an additional absorption channel in a well-defined frequency range as shown in Figure~\ref{fig:Mivehvar2017Superradiant_Fig3and4}(b).

\subsubsection{Spin-orbit-coupled-induced topological superradiant phase}
\label{subsubsec:SOC-top-SR}

\begin{figure}[t!]
\centering
\includegraphics[width=1\columnwidth]{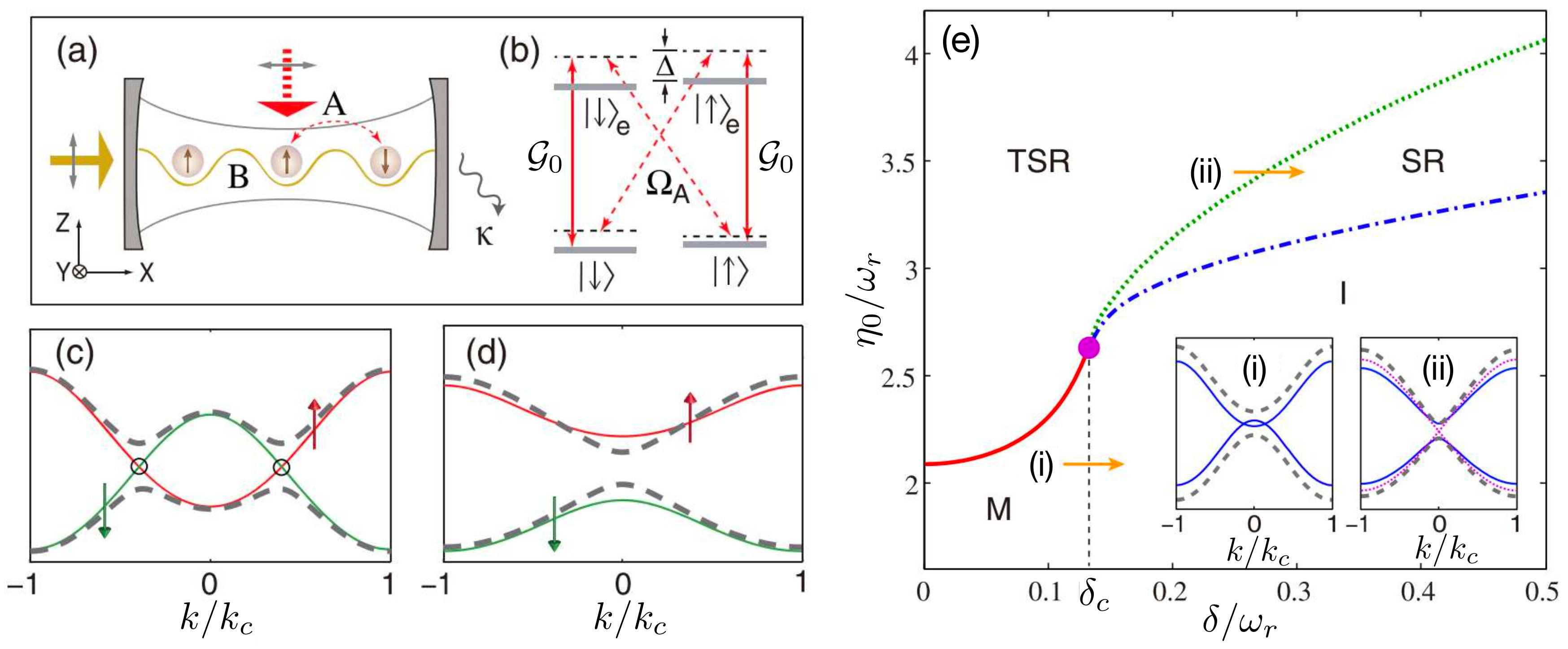}
\caption{Topological superradiant state in 1D spin-1/2 fermionic atoms inside a linear cavity in the blue-detuned repulsive regime, $\Delta_a>0$.
The schematic representation of the setup~(a) and the atom-photon coupling scheme~(b). 
The Bloch energy bands before (solid) and after (dashed) the superradiant phase transition
for $\delta<\delta_c$ (c) and $\delta>\delta_c$ (d). For $\delta<\delta_c$, the 
superradiance-induced gap opening in finite momenta marked by circles results 
in the band inversion. (e)  The phase diagram of the system at half filling
and finite temperature $k_BT=0.005E_r$ is represented  
in the $\eta_0$-$\delta$ parameter plane, exhibiting metallic (M), insulating (I), superradiant (SR),
and topological superradiant (TSR) states. The various boundaries merge at a tetracritical point (the purple dot).
The insets show the evolution of the energy bands (solid-dotted-dashed) 
across the phase boundaries indicated by arrows.
Note, in particular, that the band gap closes and re-opens across the TSR-SR phase boundary.
Figure adapted and reprinted with permission from Ref.~\cite{Pan2015Topological}~\textcopyright~2015 by the American Physical Society.}
\label{fig:Pan2015Topological_Fig1and4}
\end{figure}

A closely related scheme to the spinor bosonic self-ordering discussed in Section~\ref{sec:spinor-selfordering} has also been considered for 1D spin-1/2 fermionic atoms in the repulsive regime, i.e., where the pump laser and the cavity field are blue detuned with respect to relevant atomic transitions~\cite{Pan2015Topological}; see Figure~\ref{fig:Pan2015Topological_Fig1and4}(a) and (b). The transverse pump laser and the cavity mode A couple the two  ground spin states $\{\ket{\downarrow},\ket{\uparrow}\}$ via a double-$\Lambda$ scheme (i.e., two independent Raman processes), while the longitudinally pumped cavity mode B provides a background spin-independent ``classical'' optical lattice potential $V_{\rm ext}(x)=V_{\rm latt}^{(0)}\cos^2(k_cx)$ for the atoms. The system is described by a special case of the Hamiltonian density~\eqref{eq:single-particle-H-den-spinor}, 
\begin{align} \label{eq:single-particle-H-den-spinor-fermion}
\hat{\mathcal H}_{1,\rm eff}=
\begin{pmatrix}
\frac{p_x^2}{2M}+V_{\rm ext}(x)+\hbar U(x)\hat{a}^\dag\hat{a}+\hbar\delta & 
\hbar\eta(x)(\hat{a}+\hat{a}^\dag)
\\
\hbar\eta(x)(\hat{a}+\hat{a}^\dag)& 
\frac{p_x^2}{2M}+V_{\rm ext}(x)+\hbar U(x)\hat{a}^\dag\hat{a}
\end{pmatrix},
\end{align}
where $\hat{a}$ is the annihilation operator of the mode A, $U(x)=U_0\cos^2(k_cx)$, and $\eta(x)=\eta_0\cos(k_cx)$, with $U_0$ and $\eta_0$ defined as before. However, note that here $U_0\propto\Delta_a^{-1}>0$ as the pump laser and the cavity field are assumed to be blue detuned with respect to the relevant atomic transitions. The energy difference (i.e., the effective two-photon detuning) $\hbar\delta$ between the two spin states can be tuned via an external bias magnetic field, where the latter also fixes the quantization axis along the $z$ direction.

At small pump strengths $\eta_0$, there is no constructive photon scattering from the pump laser into the cavity mode, $\langle \hat{a}\rangle=0$. Therefore, there is no (Raman) coupling between the two spin states and the atoms only experience the background optical lattice potential $V_{\rm ext}(x)$ and the Zeeman field $\hbar\delta$, with Bloch energy bands corresponding to different spins being independent from each other. The normal Fermi gas at half filling is either a gapless metal (M) or a gapped insulator (I) depending on $\delta$; see Figures~\ref{fig:Pan2015Topological_Fig1and4}(c) and (d).

By increasing the pump strength above the critical value $\eta_{0c}(\delta)$, the system enters the superradiant phase with $\langle \hat{a}\rangle\neq0$. As a consequence, the cavity-assisted space-dependent Raman processes (with periodicity twice as the background lattice potential) induce a spin-orbit coupling which mixes different spin Bloch bands and may open gaps in the bulk energy spectrum within the first Brillouin zone; see Figures~\ref{fig:Pan2015Topological_Fig1and4}(c) and (d). Depending on $\delta$, this interband mixing in the superradiant phase can result in a band inversion, leading to topologically nontrivial states. 

The steady-state phase diagram of the system at half filling in the $\eta_0$-$\delta$ parameter plane is illustrated in Figure~\ref{fig:Pan2015Topological_Fig1and4}(e). In particular, for $\delta<\delta_c$ the system is in the metallic state below the superradiant threshold and a gap opens in the bulk energy spectrum across the superradiant phase transition. Intriguingly, a pair of topological edge modes with localized wave functions appears within the superradiant-opened energy gap in a finite-size system. Furthermore, the topologically nontrivial nature of this superradiant state (TSR) is confirmed by a nonzero bulk winding number, a topological invariant counting the number of times the momentum-space spin texture sweeps the circle $S^1$ (i.e., does a full $2\pi$ rotation) across the first Brillouin zone. In contrast, for $\delta>\delta_c$ the system is in the insulating state below the superradiant threshold and the trivial insulating gap persists across the superradiant phase transition, resulting in a topologically trivial superradiant (SR) state. However, by further increasing the pump strength the trivial superradiant state crosses into the topological superradiant phase. This is due to the increased interband mixing, which closes and then re-opens the lowest band gap, resulting in a band inversion; see the inset of Figure~\ref{fig:Pan2015Topological_Fig1and4}(e-ii).  

The topological phase transition across the SR-TSR boundary can be detected in the abrupt change of the momentum-space density distribution of the spin-up and spin-down components via a spin-selective time-of-flight imaging technique. However, since the cavity field here is driven by the self-ordering of the atomic spin, this provides a nondestructive tool to monitor the atomic spin texture and the SR-TSR phase transition. In particular, the derivative of the cavity photon number exhibits a pronounced peak across the SR-TSR phase boundary.

\section{Superradiant self-ordering with emergent discrete rotational symmetry: quasicrystals}
\label{sec:qc}

The preceding sections, especially Sections~\ref{sec:SR_discrete}, \ref{sec:supersolid}, and \ref{sec:multimode}, were focused on cavity-QED systems with the discrete ($\mathbf{Z}_2$) and continuous [$U(1)$ and $SO(3)$] symmetries, and how the self-ordering of the atoms into ``crystalline'' structures in the superradiant phase spontaneously breaks these symmetries. We noted that these hybrid atom-cavity systems exhibit fundamentally different steady-state phase diagrams, depending on underlying symmetries and their \emph{spontaneous breaking}. This is the common paradigm of condensed-matter physics, described by Landau theory. That said, in rare exotic situations in some condensed-matter models, global~\cite{ZAMOLODCHIKOV1989INTEGRALS, Coldea2010Quantum, Chen2015Quantum, Sagi2016Emergent} and local gauge~\cite{Baskaran1988Gauge, Senthil2004Deconfined, tHooft2007Emergent, Barcelo2016From, Witten2018Symmetry} symmetries \emph{emerge} in the proximity of certain quantum phase transitions and/or in some quantum phases. 

Recently, a novel many-body cavity-QED setup has been proposed in Ref.~\cite{Mivehvar2019Emergent}, where a discrete eightfold rotational symmetry $C_8$ (i.e., a rotation by 45\textdegree) \emph{emerges} in the low-energy states across the superradiance quantum phase transition. Since the eightfold rotational symmetry $C_8$ is not consistent with any translational symmetry, this results in an emergent ``quasicrystalline'' superradiant potential for the atoms in analogy to natural quasicrystalline materials. Quasicrystals are quasiordered (i.e., orientationally ordered) materials with no translational symmetry, rather with crystallographically forbidden rotational symmetries such as five-, seven-, eightfold rotational symmetries, as discovered from their diffraction patterns first by Shechtman \textit{et al.}\ in 1984~\cite{Shechtman1984Metallic}.  

%========================================================================
\subsection{Four crossed linear cavities}
\label{sec:4cavities}

%----------------------------------------------------------------------------------------------------------------------------
\subsubsection{Superradiant phase transition}
\label{sec:SR4cavities}

\begin{figure}[t!]
\centering
\includegraphics [width=\textwidth]{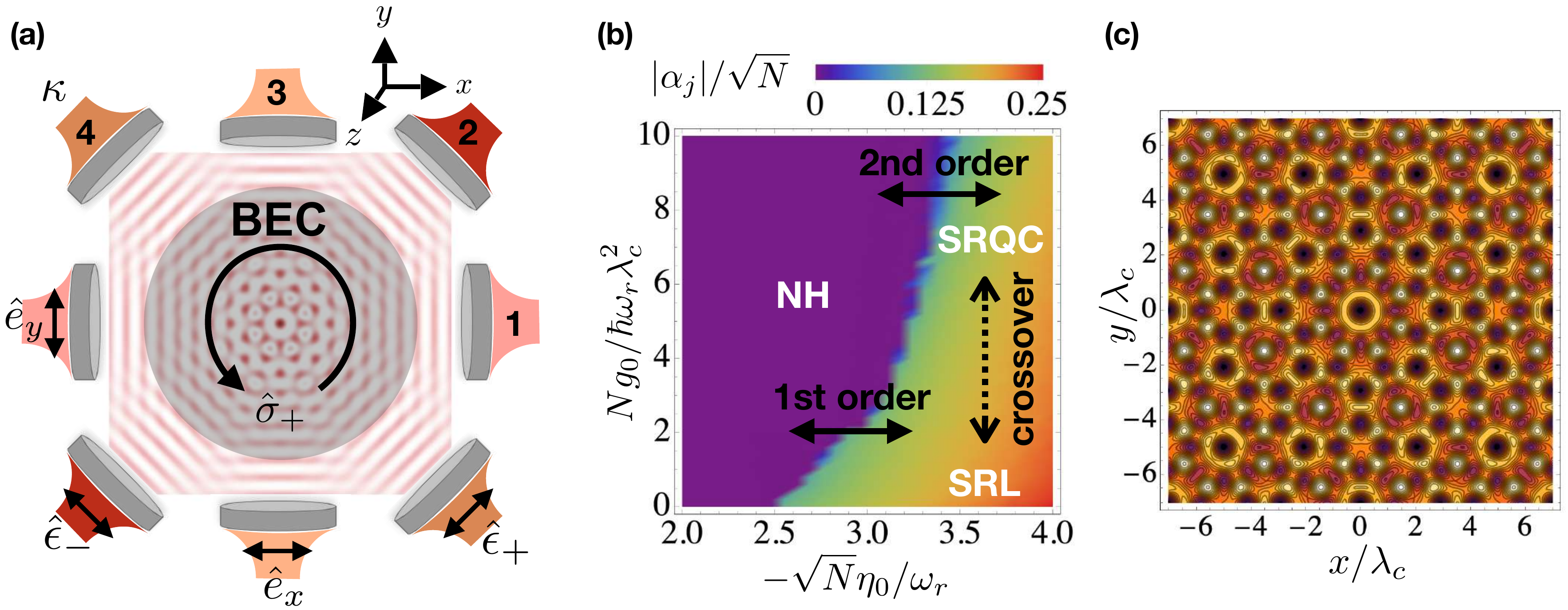}
\caption{Self-ordering of a transversely driven BEC inside four crossed linear cavities, leading to the \emph{emergence} of an eightfold ($C_8$) quasicrystalline symmetry. (a) Schematic sketch of the setup. A 2D BEC is transversely driven (in the $z$ direction) by a right circularly-polarized laser (denoted by $\hat\sigma_+$) and is coupled dispersively to an in-plane polarized mode of each cavity (denoted by $\hat{e}_{x,y}$ and $\hat{\epsilon}_\pm$). (b) Phase diagram of the system as a function of the strength of the two-body repulsive interaction $g_0$ and the pumping strength $\eta_0$. Beyond the pump-strength threshold $\eta_{0c}(g_0)$, the system exhibits a (first- or second-order) quantum phase transition from the normal homogeneous (NH) state into the self-ordered superradiant state, where a quasicrystalline optical potential emerges for the atoms. Depending on the strength of the two-body interaction, in the superradiant phase either the condensate localizes in one or few of deepest minima of the quasicrystal potential (denoted as SRL), or a quasicrystalline density order is stabilized in the system (denoted as SRQC). (c) A typical emergent quasicrystalline potential. The center of the quasicrystal is fixed via a process of the spontaneous breaking of an approximate $\otimes_{j=1}^4\mathbf{Z}_2$ symmetry. 
Figure adapted and reprinted with permission from Ref.~\cite{Mivehvar2019Emergent}~\textcopyright~2019 by the American Physical Society.} 
\label{Fig:Mivehvar2019Emergent_Fig1andFig2}
\end{figure}

In the setup proposed in Ref.~\cite{Mivehvar2019Emergent}, four identical linear cavities are arranged symmetrically in a plane with a common center such that they make a 45\textdegree\ angle with one another. A  BEC is tightly confined in the direction perpendicular to the plane at the intersection of these initially empty four cavities, and is strongly coupled to an in-plane polarized mode of each cavity $\hat{a}_j$. The BEC is also driven with the Rabi frequency $\Omega_0$ by a spatially uniform, right circularly polarized pump laser propagating perpendicular to the cavity-BEC plane as depicted in Figure~\ref{Fig:Mivehvar2019Emergent_Fig1andFig2}(a). The involved atomic internal states satisfy the magnetic selection rule $\Delta m=m_e-m_g=1$. Generalizing the procedure of Section~\ref{sec:basic-model} (see also Section~\ref{subsec:degenerate_cavity_modes}), in the dispersive regime of a large atom-pump detuning $\Delta_a$ the system is described by the effective many-body Hamiltonian $\hat{H}_{\rm eff}=\int \hat{\psi}^\dag(\mathbf{r}) \hat{\mathcal H}_{1,\rm eff}\hat{\psi}(\mathbf{r})d\mathbf{r}+\hat{H}_{\rm int}-\hbar\Delta_c\sum_j\hat{a}^\dag_j\hat{a}_j$ with the effective single-particle atomic Hamiltonian density in the rotating-frame of the laser,
\begin{align} \label{eq:q-1-H}
\hat{\mathcal H}_{1,\rm eff}=-\frac{\hbar^2}{2M}\boldsymbol{\nabla}^2+V_{\rm ext}(\mathbf{r})
+\frac{\hbar}{\Delta_a}\Big |\Omega_0+\sum_{j=1}^4\hat{a}_j\mathcal{G}_j(\mathbf{r})\Big |^2,
\end{align}
and the two-body contact interaction Hamiltonian $\hat{H}_{\rm int}=g_0\int \hat{\psi}^\dag(\mathbf{r})\hat{\psi}^\dag(\mathbf{r})\hat{\psi}(\mathbf{r})\hat{\psi}(\mathbf{r})d\mathbf{r}$. The atom-cavity couplings are given by $\mathcal{G}_j(\mathbf{r})=e^{i\theta_j}\mathcal{G}_{0}\cos(\mathbf{k}_{j}\cdot\mathbf{r})$, where $\mathbf{k}_1=k_c\hat{e}_x$, $\mathbf{k}_3=k_c\hat{e}_y$, and $\mathbf{k}_{2,4}=k_c(\hat{e}_x\pm\hat{e}_y)/\sqrt{2}$ with $k_c=|\mathbf{k}_j|$ being the cavity (and pump laser) wave number and $\hat{e}_{x}$ ($\hat{e}_{y}$) the unit vector along the $x$ ($y$) direction. The phase factors $\theta_1=\pi/2$, $\theta_2=5\pi/4$, $\theta_3=\pi$, and $\theta_4=3\pi/4$ arise due to the projection of the in-plane linear polarizations of the cavity fields onto the right circular polarization~\cite{Mivehvar2019Emergent}. 

The mean-field phase diagram of the system in the $\{Ng_0/\hbar\omega_r\lambda_c^2,\sqrt{N}\eta_0/\omega_r\}$ parameter space is presented in Figure~\ref{Fig:Mivehvar2019Emergent_Fig1andFig2}(b), where $\eta_0=\Omega_0\mathcal{G}_0/\Delta_a$ as before. Beyond the pump-strength threshold $\eta_{0c}(g_0)$, there is a quantum phase transition from the normal homogeneous (NH) state into the superradiant phase $\alpha_j=|\alpha_j|e^{i\gamma_j}\neq0$, where not only are the absolute values of all the field amplitudes equal $|\alpha|\equiv|\alpha_1|=\cdots=|\alpha_4|\neq0$, but also the phase of each field amplitude is locked at $\gamma_j=\gamma_0-\theta_j$ or $\gamma_0-\theta_j+\pi$, where $\gamma_0$ is a common phase introduced due to the nonzero cavity-field decay rate $\kappa$. The nature of the phase transition (i.e., first or second order) depends on the strength of the two-body interaction $g_0$ and is indicated in the figure.    

%--------Figure------------ 

%----------------------------------------------------------------------------------------------------------------------------
\subsubsection{Emergent eightfold quasicrystalline rotational symmetry}
\label{sec:emergent-c8-symmetry}

The single-particle Hamiltonian density $\hat{\mathcal H}_{1,\rm eff}$, Equation~\eqref{eq:q-1-H}, possesses a symmetry denoted by $\tilde{C}_8$: an eightfold rotation $C_8$ around the $z$ axis with the transformation $x\to x'=(x+y)/\sqrt{2}$ and $y\to y'=(y-x)/\sqrt{2}$ followed by the field transformation $\hat{a}_1\to\hat{a}_2e^{-i(\theta_1-\theta_2)}$, $\hat{a}_2\to\hat{a}_3e^{-i(\theta_2-\theta_3)}$, $\hat{a}_3\to\hat{a}_4e^{-i(\theta_3-\theta_4)}$, and $\hat{a}_4\to\hat{a}_1e^{-i(\theta_4-\theta_1)}$. Although $\hat{\mathcal H}_{1,\rm eff}$ is \textit{not} invariant under the sole action of the $C_8$ symmetry, an eightfold rotational symmetry $C_8$ \emph{emerges} in the low-lying energy states of the superradiant phase as a consequence of the amplitude and phase locking of the cavity fields: $|\alpha|\equiv|\alpha_1|=\cdots=|\alpha_4|\neq0$, and  $\gamma_j-\gamma_0+\theta_j=0$ or $\pi$~\cite{Mivehvar2019Emergent}. This leads to the formation of a quasicrystalline superradiant optical potential with a $C_8$ rotational symmetry for the atoms [see Figure~\ref{Fig:Mivehvar2019Emergent_Fig1andFig2}(c)], where the location of the $C_8$ rotational axis (i.e., the center of the quasicrystal) is determined spontaneously via the phases $\gamma_j$ of the cavity electromagnetic fields. The two possible choices of phase, $\gamma_0-\theta_j$ or $\gamma_0-\theta_j+\pi$, for each field amplitude correspond to an \emph{approximate} $\mathbf{Z}_2$ symmetry for each cavity. The approximate nature of the $\mathbf{Z}_2$ symmetry is due to the lack of a translational symmetry in the system. That is, a parity transformation of a cavity field $\hat{a}_j\to-\hat{a}_j$ cannot be compensated ``exactly'' by any translation because of incommensurate cavity wavevectors along a given cavity axis. However, the parity transformation of each cavity field can be ``approximately'' compensated by a translation, leading to an approximate $\mathbf{Z}_2$ symmetry for each cavity. Therefore, at the onset of the superradiant phase transition, the center of the emergent quasicrystal is fixed through the spontaneous breaking of an approximate $\otimes_{j=1}^4\mathbf{Z}_2$ symmetry. The sixteen $\otimes_{j=1}^4\mathbf{Z}_2$-symmetry-broken states form the low-lying energy sector of the system's Hilbert space.

%----------------------------------------------------------------------------------------------------------------------------
\subsubsection{Localized state versus extended quasicrystalline state}
\label{sec:localized-qc-states}

In the weakly interacting regime, the quasicrystal potential dominates over the two-body interactions. Consequently, the condensate localizes in one or few of deepest minima of the quasicrystal potential since the quasicrystal potential lacks a translation symmetry and can act as a disordered potential. In this superradiant localized (SRL) phase the atoms scatter more photons from the pump laser into the cavity modes [see Figure~\ref{Fig:Mivehvar2019Emergent_Fig1andFig2}(b)] due to constructive interference: the more atoms are condensed in the same potential minimum, the more photons are scattered with the same phase. 
As the two-body contact interaction is increased, it is energetically more favorable for the atoms to occupy different global and local minima of the quasicrystal optical potential. For sufficiently strong two-body contact interactions, the system crosses over into an extended phase, where a quasicrystalline density order is stabilized in the system. The momentum distribution in this superradiant quasicrystalline (SRQC) phase exhibits a dense, fractal structure with an eightfold rotational symmetry, a characteristic of quasicrystalline order indicating that the two-dimensional momentum space is spanned by four incommensurate cavity wavevectors $\{\mathbf{k}_1,\cdots,\mathbf{k}_4\}$~\cite{Mivehvar2019Emergent}. We stress that the SRQC phase in this setup is a genuine many-body state as the two-body repulsive interaction between the atoms is necessary for its stabilization.

%========================================================================
\subsection{Different cavity setups and other quasicrystals}
\label{sec:x-cavities}

In the preceding section, we discussed a scenario where a quasicrystalline superradiant optical potential with an eightfold rotational symmetry emerges for ultracold atoms coupled to four-crossed linear cavities, where the atoms in turn self-order in this emergent quasicrystalline potential. This proposal is an extension of the ETH Zurich group's two-crossed cavity setup (see Section~\ref{sec:supersolidity-two-cavities}). Other superradiant quasicrystalline optical potentials with different rotational symmetries can also be implemented by employing different cavity setups, although they might become progressively more challenging experimentally. However, they can provide a platform to study many fundamental open questions concerning the formation and nature of quasicrystals. For instance, it is still not completely clear whether quasicrystals are only entropy-stabilized high-temperature states or can also be thermodynamically stable at low temperatures. The conditions and nature of quasicrystal growth are also under debate with a lack of a generally accepted model. Further enriching the physics in cavity-based composite atom-light quasicrystals compared to solid-state quasicrystalline materials is the interplay between two-body contact and cavity-mediated long-range interactions, both tunable in principle.

\section{Superradiant self-organization without a steady state}
\label{sec:dynamics}
So far we mostly considered situations, where atomic and photonic degrees of freedom reach a steady state. In the present section we will instead discuss some interesting quantum-gas--cavity scenarios with non-steady states. In order to put the results presented in the following into the general context of quantum many-body dynamics, some remarks are in order. First and foremost, a hybrid atom-cavity setup is intrinsically an open quantum system due to the photon leakage out of the cavity. This means that the dynamics always leads to an attractor (which needs not to be unique), defined by the eigenvalue of the master equation~\eqref{eq:master-eq} with vanishing real part, i.e., with only damping. This scenario has to be distinguished from the case of isolated systems, where the dynamics are classified in terms of ergodic versus non-ergodic behavior. In the context of quantum many-body systems, this characterization is still an active field of research, with the phenomenon of many-body localization being a prominent example~\cite{Polkovnikov2011Nonequilibrium}.

Besides being dissipative, the atom-cavity systems we consider are also driven, as external lasers in conjunction with cavity fields are used to generate the dispersive light-matter coupling discussed throughout this review. In this case, the long-time attractor does not necessarily need to be a steady state, even though the Liouville operator governing the dynamics has no explicit time dependence (in the frame rotating with the driving laser). This situation will be considered in  Section~\ref{sec:non-stationary-CW}, while the case of explicitly time-dependent Liouvillians will be discussed in Section~\ref{sec:non-stationary-modulated}.

Some of the non-steady states discussed below can be classified, in a broader sense, as so-called time crystals. Such phases, which spontaneously break some time-translation symmetry and show robust periodic oscillations in time, have recently received considerable amount of attention, with examples from different physical platforms~\cite{Khemani2019Brief}. We will also encounter in the following non-steady phases which do not oscillate periodically in time but rather exhibit irregular behavior, similar to the ergodic dynamics of isolated systems. In the context of open quantum systems, the definition and characterization of ergodic or chaotic behavior is still an open issue, with very recent interesting developments \cite{Can2019Spectral,Can2019Random,Xu2019Extreme,Luitz2019Exceptional,Shirai2020Thermalization,Sa2020Complex,Sa2020Spectral,Wang2020Hierarchy}. In this context, atom-cavity setups seem a very promising platform for further investigation.

Before proceeding, let us note that in a driven-dissipative setting like the ones considered here, even a steady attractor can be very nontrival, since it is not in general of the thermal Gibbs type~\cite{Piazza2014Quantum}. We refer here to the discussion given in Section~\ref{subsec:thermal_nonthermal}.

\subsection{Non-steady-state phases with time-independent driving}
\label{sec:non-stationary-CW}
In this section, we discuss two examples, where a stable, non-steady behavior emerges without an explicit time-dependent external driving. This type of phenomena shares strong similarities with limit cycles, quantum synchronization, and some types of dissipative time crystals.

\subsubsection{Atomic blue-detuned laser driving}

A possible scenario through which stable non-steady attractors can emerge is via the competition between different steady attractors. This can be achieved in the simplest single-mode superradiant self-ordering scenario by driving the atoms with a laser blue-detuned with respect to the atomic resonance, $\Delta_a>0$, as first pointed out in Ref.~\cite{Piazza2015Self}. Blue-detuned laser fields lead to repulsive optical potentials but still allow for superradiant self-ordering in some limited parameter regimes, as discussed in Section~\ref{subsubsec:blue_detuned_SR}. To better understand the phenomenon, instead of considering the two-dimensional configuration of that section, we re-examine here the simpler one-dimensional blue-detuned model introduced in Section~\ref{subsubsec:SSH}, this time for bosonic atoms instead of fermionic atoms.

\begin{figure}[t!]
  \begin{center}
    \includegraphics[width=\columnwidth]{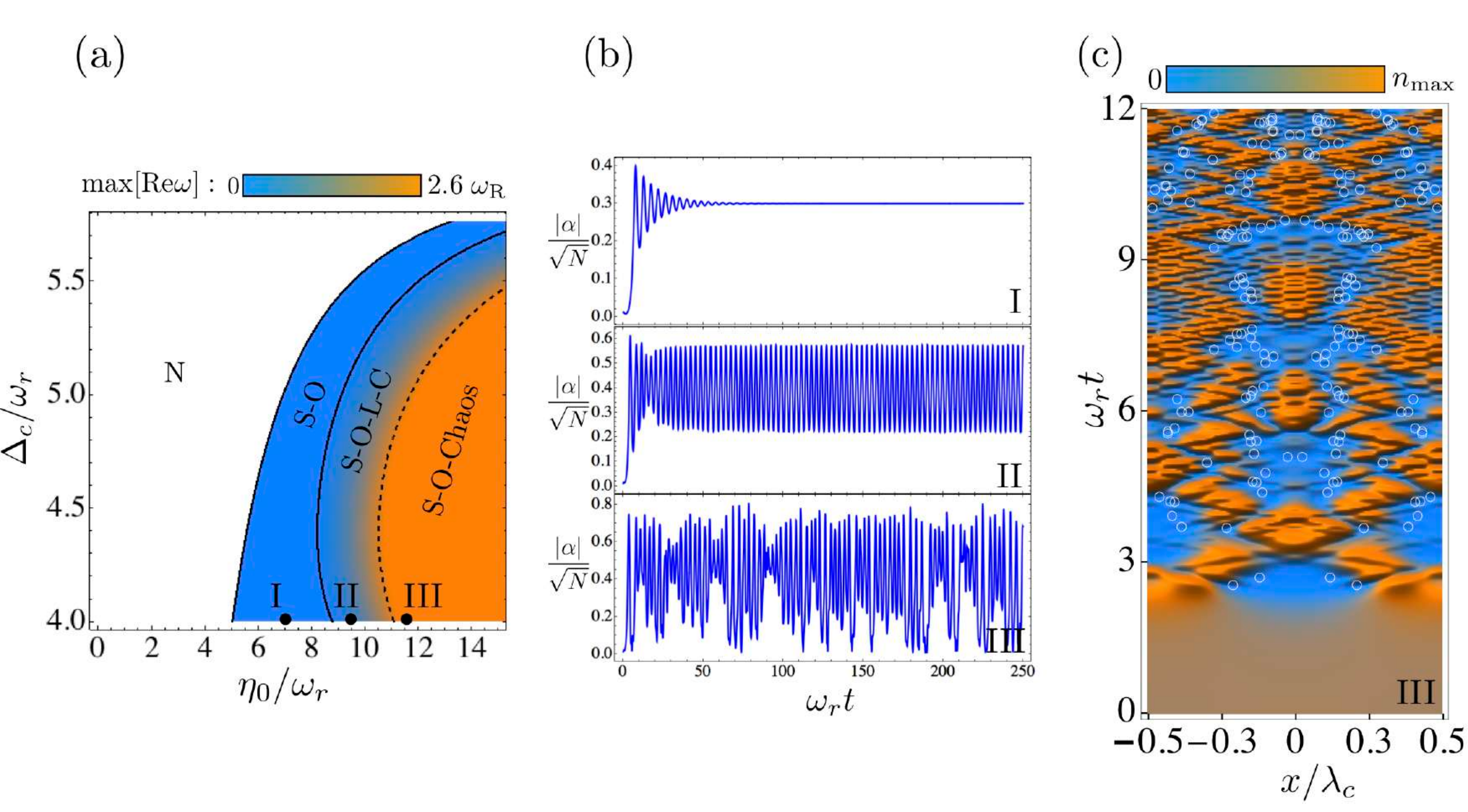}
    \caption{Emergence of stable, non-steady-state behavior for a BEC inside a cavity, driven transversely by a blue-detuned laser. A quasi-one-dimensional BEC is trapped along the axis of a linear cavity and coupled to a single standing-wave mode of it. The transverse, driving laser is blue detuned from the atomic transition, $\Delta_a>0$.  
    (a) Phase diagram of the system. While the bare cavity detuning $\Delta_c$ is positive in the graph, the dispersively-shifted detuning is still negative: $\delta_c<0$.
    The phase diagram features the usual steady-state normal (N) and superradiant self-organized (S-O) states, as well as non-steady-state superradiant self-organized limit-cycle (S-O-L-C) and chaotic (S-O-Chaos) phases. (b)~Time evolution of the coherent cavity field in the three different self-organized phases: S-O, S-O-L-C, and S-O-Chaos in order.  (c)~Time evolution of the BEC density in the chaotic-like phase, where the white circles indicate the space-time coordinates where solitons are nucleated. The parameters are $U_0N=12.1\omega_r$ and $\kappa=10\omega_r$. Panel (c) also features a finite short-range atom-atom repulsion $ng_0=\hbar\omega_r$. Figure adapted and reprinted with permission from Ref.~\cite{Piazza2015Self}~\textcopyright~2015 by the American Physical Society.}
    \label{fig:LC_blue_detuned_piazza}
  \end{center}
\end{figure}

The phase diagram of this 1D system is shown in Figure~\ref{fig:LC_blue_detuned_piazza}(a). The competition which destabilizes the steady states can be immediately understood from Equation~(\ref{eq:Vsr1d}), expressing the optical potential felt by the atoms in the superradiant phase. For a laser which is blue detuned from the atomic transition, one has $U_0\propto\Delta_a^{-1}>0$. This implies that the $\cos^2(k_c x)$ part of the potential pushes the atoms out of the self-ordered pattern induced by the $\cos(k_c x)$ part, thereby favoring a homogenous density and thus the non-superradiant (normal) phase. At very small photon numbers, the $\cos(k_c x)$ term dominates so that a steady-state superradiant phase is stabilized. However, for sufficiently strong pumps the $\cos^2(k_c x)$ term becomes equally important, so that the system cannot decide between the normal and the superradiant phases. This leads to self-sustained oscillations between these two states which are stabilized by the cavity loss. These limit-cycle oscillations are shown in the middle panel of Figure~\ref{fig:LC_blue_detuned_piazza}(b). By further increasing the pump strength the number of harmonics increases by the known mechanism of period doubling, eventually crossing over into irregular behavior exemplified in the lower panel of Figure~\ref{fig:LC_blue_detuned_piazza}(b). Within a (semi-)classical description of the system based on coupled nonlinear equations for the cavity coherent field and the BEC wavefunction (see Section~\ref{subsubsec:MF-bose}) this irregular behavior can be associated with a chaotic attractor~\cite{Lichtenberg2013Regular}. Indeed, the dissipative, i.e., non-Hamiltonian, nature of the equations of motion leads to the contraction of the phase-space volume to within a finite region. Still, within the attractor region the dynamics is chaotic, that is, it features at least one positive Lyapunov exponent.

What this classically known scenario implies for the full quantum state of the system is an open question. Differently from the closed-system case, the study of quantum dissipative chaos is still in its beginning and matter of active research~\cite{Can2019Random,Xu2019Extreme,Denisov2019Universal,Sa2020Spectral,Wang2020Hierarchy}. It is thus also an open question whether this type of classical dynamics underlie some kind of ergodic or thermal-like behavior of our quantum many-body system. This is a matter of future research, for which the atom-cavity systems considered in this review article seem an ideal platform. From the point of view of the atoms, the chaotic attractor corresponds to a fast increase of the kinetic energy~\cite{Piazza2015Self,Lin2020Pathway}, which reaches a value much larger than that in the limit-cycle attractor. The chaotic attractor is thus characterized by a more effective redistribution, i.e., a behavior which is ``more ergodic'' than for the limit-cycle attractor. Still, the increase in the kinetic energy is counteracted by the cavity loss via the mechanism of cavity cooling, which eventually stops the heating~\cite{Piazza2015Self}.

A further interesting aspect connected with the chaotic attractor is the phenomenon of phase slippage and related soliton nucleation in the BEC~\cite{Piazza2015Self}, as shown in Figure~\ref{fig:LC_blue_detuned_piazza}(c). This behavior is peculiar to superfluids excited out of equilibrium, whereby spatial gradients in the condensate phase are converted into localized excitations carrying a phase slip. In the present case the phase gradients are caused by the modulation of the cavity field and have a characteristic length scale fixed by $\lambda_c$. On the other hand, the characteristic size of a soliton is set by the healing length, whose value $(2Mg_0n)^{-1/2}$ depends on the interatomic repulsion and is set to $0.1\lambda_c$ in Figure~\ref{fig:LC_blue_detuned_piazza}(c). The main contribution to the BEC kinetic energy can be attributed to soliton proliferation. The latter is counteracted by cavity cooling, in a situation which resembles turbulence. This interesting analogy should be further investigated in the future.

\begin{figure}[t!]
  \begin{center}
    \includegraphics[width=0.9\columnwidth]{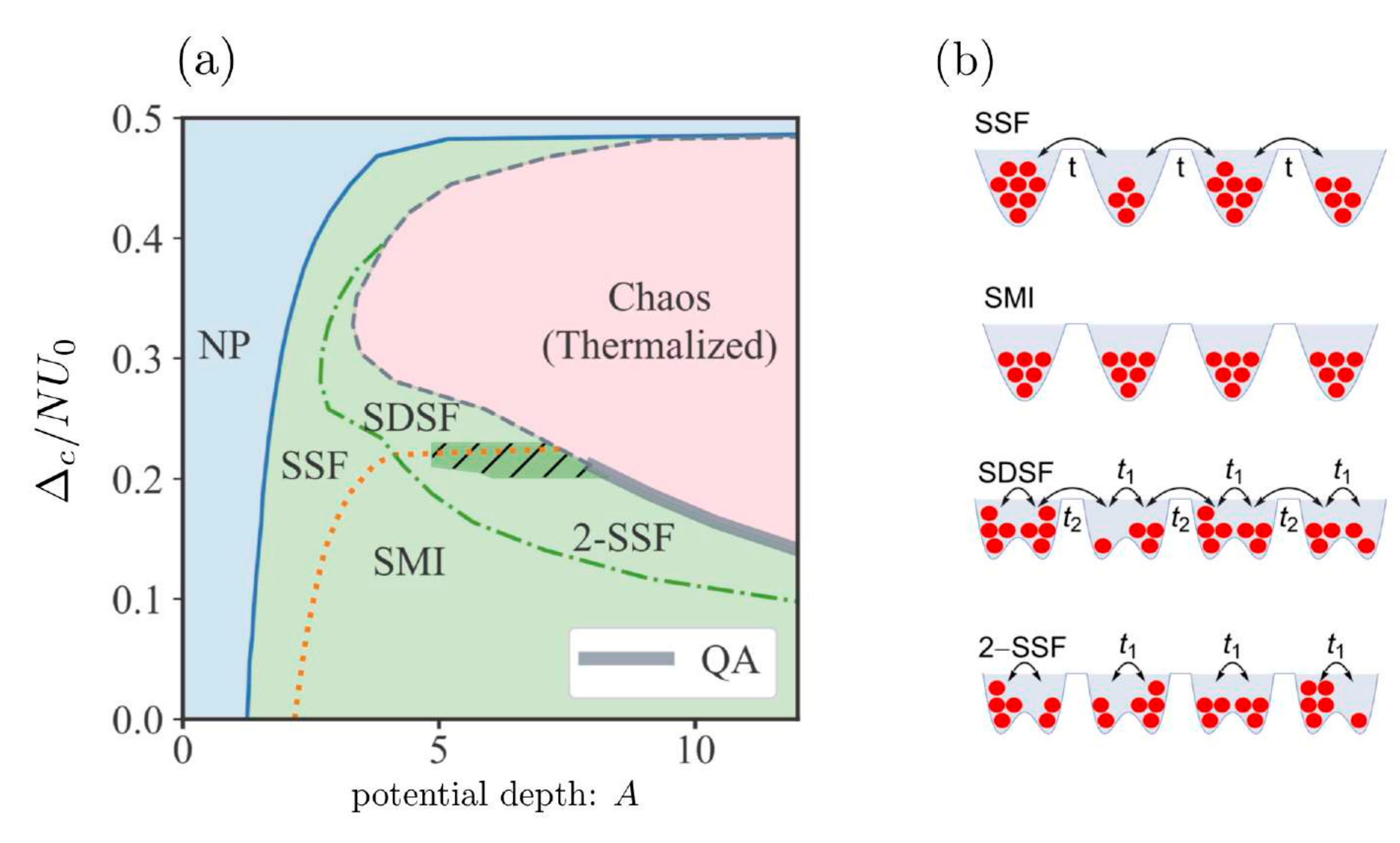}
    \caption{(a) Extension of the mean-field phase diagram of Figure~\ref{fig:LC_blue_detuned_piazza}(a) by including non-mean-field atomic correlations as well as an harmonic trap for the atoms. The dimensionless potential depth is defined as $A=N\eta_0\sqrt{U_0/\omega_r}/\sqrt{(\Delta_c-NU_0B)^2+\kappa^2}$, with the overlap $B=\int dx n(x)\cos^2(k_c x)$. Besides the normal (NP) and self-organized superfluid (SSF) phases present in Figure~\ref{fig:LC_blue_detuned_piazza}(a), a further mean-field-type phase appears as a self-organized dimerized superfluid (SDSF). Moreover, beyond-mean-field phases are also stabilized: a self-organized Mott insulator (SMI) phase [also present in the phase diagram of Figure~\ref{fig:eBHModelHamburg}(b)] and self-organized second-order superfluid (2-SSF) phase. Additionally, non-steady-state phases are also present: A quasi-periodic attractor (QA) related to the limit-cycle phase of Figure~\ref{fig:LC_blue_detuned_piazza}(a) and a chaotic attractor corresponding to the chaotic phase of Figure~\ref{fig:LC_blue_detuned_piazza}(a). (b) Pictorial representation of the different steady-state phases. Figure adapted and reprinted with permission from Ref.~\cite{Lin2020Pathway}~\textcopyright~2020 by the American Physical Society.}
    \label{fig:LC_blue_detuned_lin}
  \end{center}
\end{figure}

The Gross-Pitaevskii study of Ref.~\cite{Piazza2015Self} has been extended to include non-mean-field correlations between the bosonic atoms via a multiconfigurational time-dependent Hartree method for indistinguishable particles (MCTDH-X)~\cite{Lin2020Pathway}, as well as to include the effect of harmonic trapping of the atomic cloud. The phase diagram for the one-dimensional case is shown in Figure~\ref{fig:LC_blue_detuned_lin}(a). A non-mean-field, steady-state phase peculiar to this blue-detuned case is found. In this so-called self-organized second-order superfluid phase, the phase coherence is limited to within each double well, i.e., the unit cell of the dimerized superradiant lattice. Within this beyond-mean-field description a limit-cycle phase is also identified, with the only difference with respect to the mean-field version discussed above being the quasi-periodic nature of the oscillations, which has been attributed to the combination of the external trap and the short-range atomic repulsion~\cite{Lin2020Pathway}. The chaotic attractor is also present within this beyond-mean-field description, indicating that this phenomenon is not an artifact of the (semi-)classical description. This opens the interesting perspective for further investigating of this type of chaotic phases in a fully quantum setting.

\begin{figure}[t!]
  \begin{center}
    \includegraphics[width=0.9\columnwidth]{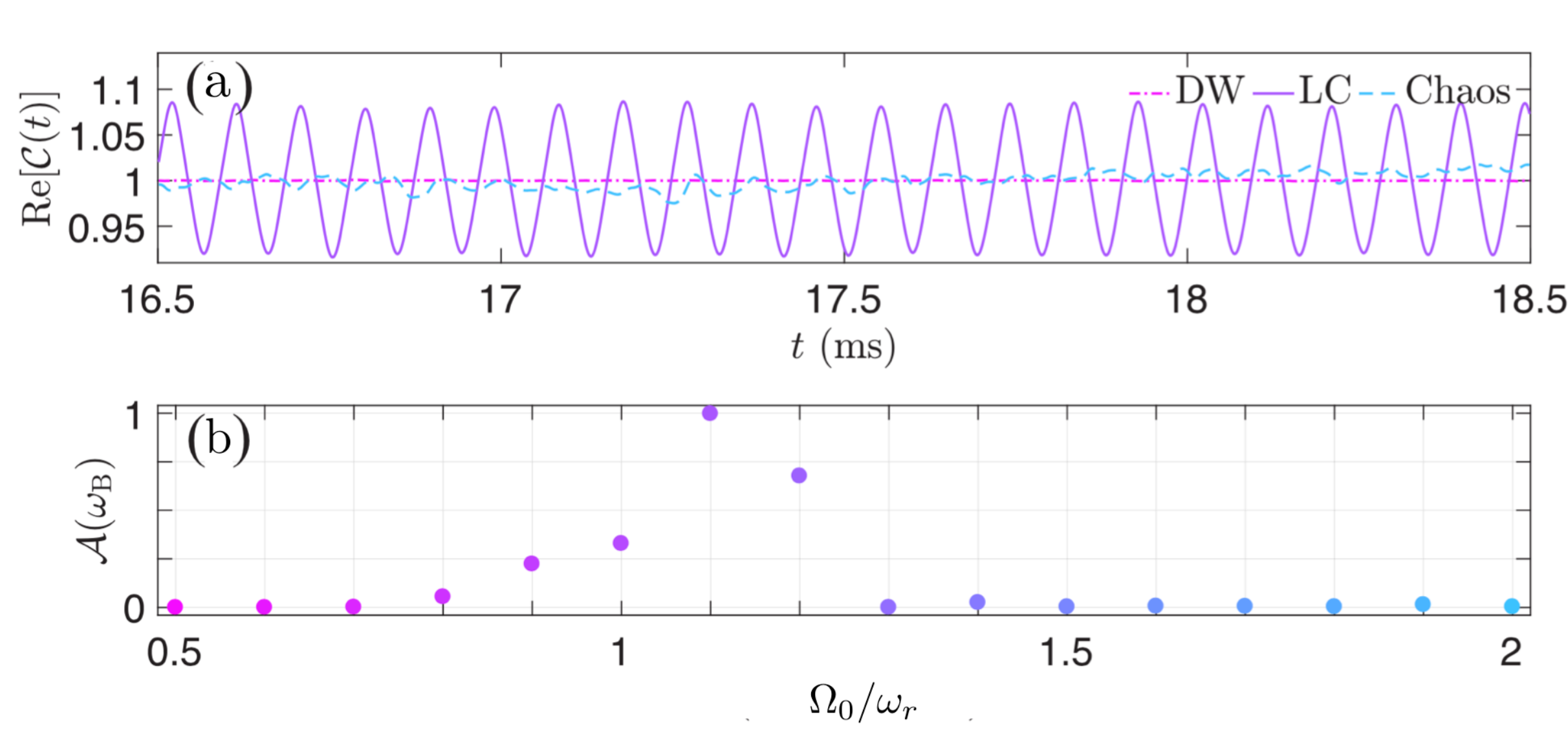}
    \caption{Time crystals in non-steady-state self-organized phases in the repulsive, blue-detuned regime.
    (a) Temporal coherence $\mathcal{C}(t)$ (defined in the text) for the self-organized phases shown in the phase diagram of Figure~\ref{fig:LC_blue_detuned_piazza}(a) [here the DW corresponds to the S-O phase in Figure~\ref{fig:LC_blue_detuned_piazza}(a)].
    The emergence of temporal coherence in the limit-cycle (LC) phase is a characteristic of time crystals.
     (b)~The amplitude of the oscillations of $\mathcal{C}(t)$ as a function of the pump strength across the three phases (the LC phase extends roughly between 0.7 and 1.5). Here the simulations are done for a two-dimensional BEC inside a standing-wave cavity pumped transversally with a standing-wave laser. Adapted and reprinted with permission from Ref.~\cite{Kessler2019Emergent}~\textcopyright~2019 by the American Physical Society.}
    \label{fig:LC_blue_detuned_cosme}
  \end{center}
\end{figure}

Further insight into the limit-cycle phase and its relation to time crystals is provided by the analysis of the temporal coherence presented in Figure~\ref{fig:LC_blue_detuned_cosme}. These results~\cite{Kessler2019Emergent} are obtained using the truncated Wigner approximation, which goes beyond the mean-field approximation by including the noise induced by cavity loss [see Equation~\eqref{eq:Heisberg-eq-a-loss}], as well as the quantum noise in the initial state. In Ref.~\cite{Kessler2019Emergent}, the latter has been included by occupying low-lying BEC and cavity modes with vacuum fluctuations. Observables are then computed by averaging over a set of stochastically-generated trajectories. Even in the presence of these forms of noise, the temporal coherence of the limit-cycle phase is clearly visible in Figure~\ref{fig:LC_blue_detuned_cosme}(a) as finite-amplitude harmonic oscillations in the quantity $\mathcal{C}(t)=\langle\alpha^*(t)\alpha(t_1)\rangle$, where the averaging is performed over an ensemble of trajectories. The emergence of temporal coherence in the limit-cycle phase is a characteristic of time crystals. On the contrary, the chaotic phase does not show appreciable temporal coherence. The amplitude of the oscillations of temporal coherence is shown in Figure~\ref{fig:LC_blue_detuned_cosme}(b) as a function of the ramping of the pump strength from the self-organized steady-state phase, across the limit-cycle phase over to the chaotic phase.

We conclude this section by noting that the above described limit-cycle phase has not been yet experimentally demonstrated, though experiments at the ETH Zurich group have been carried out in the blue-detuned regime, as discussed in Section~\ref{subsubsec:blue_detuned_SR}. One possible explanation is the two-dimensional geometry of the ETH Zurich experiment, which is different from the two-dimensional geometry studied in Ref.~\cite{Kessler2019Emergent} and presented in Figure~\ref{fig:LC_blue_detuned_cosme}, where the pump laser is orthogonal to the cavity axis. Another possible explanation is the large value of the cavity loss rate in the experiment, reducing the size of the limit-cycle region in the phase diagram. Addressing this issue requires further investigation.

\subsubsection{Dissipation-induced instability in spinor BECs}
\label{sec:dissipation-induced-instability-spinor-BEC}
Non-steady-state long-time dynamic behavior can also be induced by a competition of coherent and dissipative couplings, as has been experimentally demonstrated in Ref.~\cite{Dogra2019Dissipation} and theoretically discussed in Refs.~\cite{Chiacchio2019Dissipation, Buca2019Dissipation}. We consider the model already introduced in Section~\ref{par:SpinTextureFormation}, where a spinor BEC with equal populations $N$ of the Zeeman states $m_F=+1$ and $m_F=-1$ is coupled to a single standing-wave cavity mode. The linear polarization of a standing-wave transverse pump field with lattice depth $V_0$ can be adjusted with an angle $\varphi$ with respect to the $y$ axis [see Figure~\ref{fig:FormationSpinTexture}(a)], which changes the ratio of scalar and vector polarizability of the illuminated atoms. Generalizing the effective Dicke Hamiltonian of Equation~(\ref{eq:DickeHamiltonianZeeman}) for a single-component BEC to both Zeeman states, the many-body Hamiltonian of the system can be expressed as
\begin{align}
  \begin{split}
  \hat{H} =& -\hbar \delta_c \hat{a}^\dagger\hat{a} 
  + 2\hbar\omega_r\left(\hat{J}_{z,+}+\hat{J}_{z,-}\right) \\ 
  &+ \frac{\hbar}{2}\left[ \eta_D (\hat{a}^\dagger + \hat{a}) \left(\hat{J}_{x,+}+\hat{J}_{x,-}\right) 
  -i\eta_S  (\hat{a}^\dagger - \hat{a}) \left(\hat{J}_{x,+}-\hat{J}_{x,-}\right) \right].
  \end{split}
\end{align}
The subscripts $\pm$ of the effective angular momentum operators $\hat{J}_{x,\pm}$, $\hat{J}_{y,\pm}$, and $\hat{J}_{z,\pm}$ indicate the Zeeman states $m_F=\pm1$. If the two spin components occupy the same checkerboard pattern, the expectation value for the density modulation $x_D=(\langle\hat{J}_{x,+}\rangle+\langle\hat{J}_{x,-}\rangle)/N$ becomes non-zero and the system self-organizes its density. If on the other hand the two spin-components occupy the opposite checkerboard patterns, the expectation value for the spin modulation $x_S=(\langle\hat{J}_{x,+}\rangle-\langle\hat{J}_{x,-}\rangle)/N$ is non-zero. The parameters $\eta_D=\eta_{s0} \cos\varphi$ and $\eta_S=\eta_{v0} \sin\varphi$ [cf.\ Equation~\eqref{eq:DickeHamiltonianZeeman}] describe the coherent coupling strengths to the density and spin modulated states, respectively. The coupling with the density modulated state is mediated via the real quadrature $(\hat{a}^\dagger + \hat{a})$ of the cavity field, which is thus simultaneously occupied with a finite value of $x_D$. While the imaginary quadrature $i(\hat{a}^\dagger - \hat{a})$ is occupied with the emergence of the spin modulated state, $x_S\neq0$. Without dissipation, this system shows stable self-organization with either a density or a spin modulation above a critical coupling strength, when either the eigenfrequency $\omega_D$ of the density modulated state or $\omega_S$ of the spin modulated state have softened to zero. Density and spin modulated states are distinguishable via the phase of the cavity field; see Section~\ref{par:SpinTextureFormation}. 

\begin{figure}[t!]
\centering
\includegraphics[width=0.75\columnwidth]{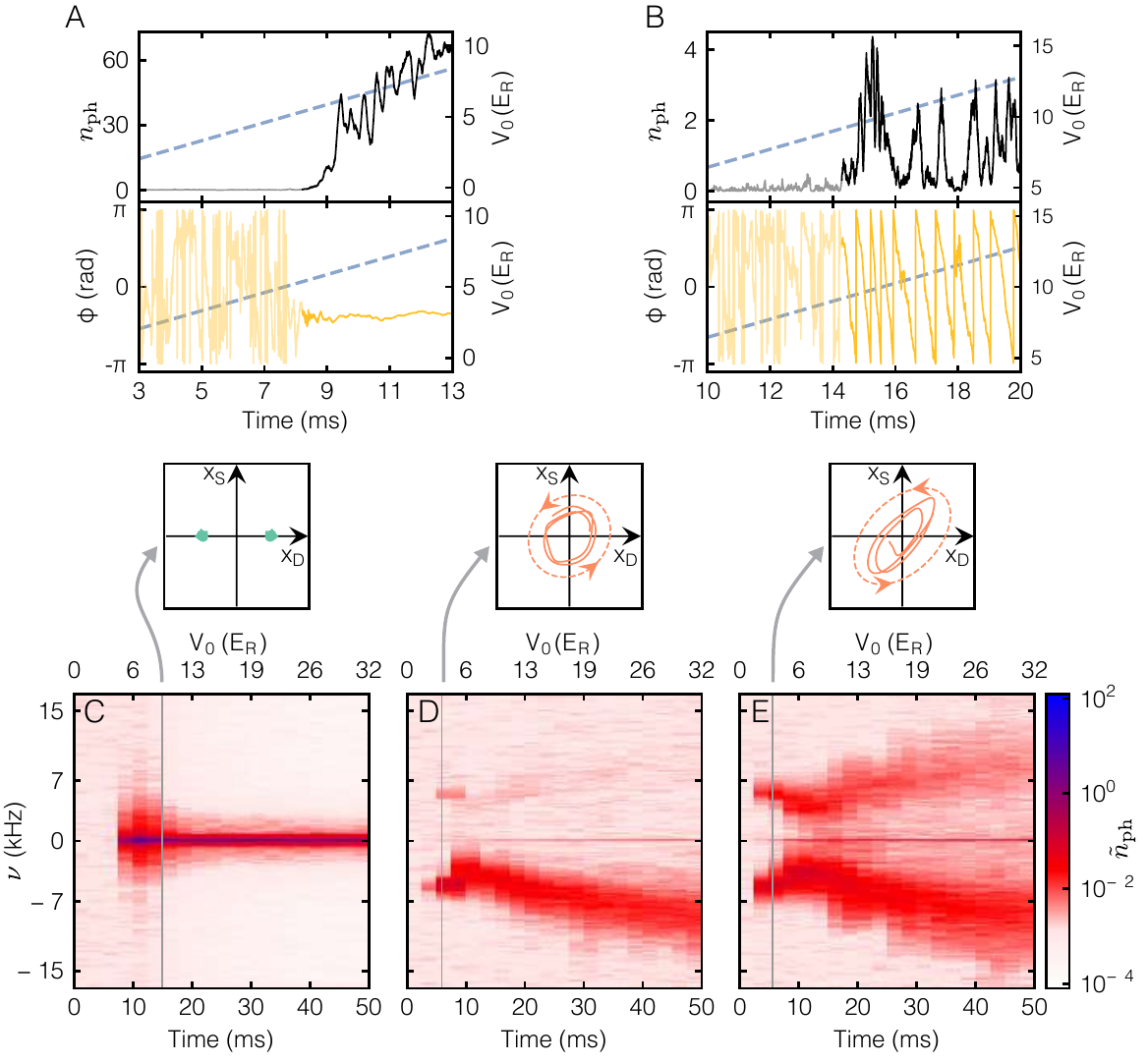}
\caption{Dissipation-induced dynamical behavior in a two-component BEC coupled to a single mode of a linear cavity via both scalar and vector polarizabilities. (A and B) Time evolution of the mean number of photons (black) in the cavity and the corresponding phase of the light field modulo $2\pi$ (yellow), while the transverse pump-lattice depth $V_0$ is ramped up over time (dashed line). Panel (A) displays data for a polarization angle and detuning where stable self-organization takes place, while panel (B) shows data for parameters where a transition to a dissipation-induced instability happens. Panels (C) to (E) show spectrograms of the mean intra-cavity photon number as a function of frequency while the pump-lattice depth is increased over time. The corresponding insets show the evolution of the state of the system in the $x_D$-$x_S$ plane. Panel (C) displays the situation for stable self-organization, and panel (D) for the dissipation-induced instability with degenerate mode frequencies $\omega_S \approx \omega_D$. The observation of a single red sideband corresponds to a linearly running phase of the light field, or a circular motion in the $x_D$-$x_S$ plane. Panel (E) shows the situation for non-degenerate modes, which leads to the appearance of also a blue sideband and a corresponding elliptic motion in the $x_D$-$x_S$ plane. Figure adapted and reprinted with permission from Ref.~\cite{Dogra2019Dissipation} published in 2019 by the American Association for the Advancement of Science.}
\label{fig:DissipationInducedInstabilityExp}
\end{figure}

A non-zero cavity-field dissipation rate $\kappa$ leads to a phase shift $\phi_\kappa = \tan^{-1}(-\kappa/\delta_c)$ of the light field scattered from the pump field into the cavity mode. As a consequence, a pure density modulation, for instance, thus leads to an occupation of both the real and the imaginary quadratures of the cavity field. The light field in the imaginary quadrature then drives self-organization of the spin mode. This process exerts an effective force onto the system and triggers an instability constantly driving it between density and spin modulation already far below the threshold for self-organization. In the experiment, this was observed via a constantly evolving phase of the intra-cavity light field and the associated emergence of sidebands in the spectrum of the cavity field; see Figure~\ref{fig:DissipationInducedInstabilityExp}.

Adiabatically eliminating the cavity field, linearized equations of motion for the amplitudes $x_D$ and $x_S$ can be derived~\cite{Dogra2019Dissipation}:
\begin{align}\label{eq:EquationsOfMotionDissipative}
  \frac{d^2}{dt^2}\begin{pmatrix}x_D \\ x_S\end{pmatrix} = \begin{pmatrix} -\omega_D^2 & -K^2 \\ K^2 & -\omega_S^2\end{pmatrix} \begin{pmatrix}x_D \\ x_S\end{pmatrix},
\end{align}
where $(\omega_D^2-\omega_S^2)\propto \sin\delta\phi$ with $\delta\phi=2\tan^{-1}(-\eta_S/\eta_D)-\pi/2$ captures the relative coupling strength of the BEC to the density and the spin modulated state. The important term is the dissipative coupling strength $K^2\propto V_0 \sin^2\phi_{\kappa}\cos\delta\phi$ which can be increased by increasing the cavity-induced phase shift (via the detuning $\delta_c$) or by making the two modes more degenerate (via the polarization angle $\varphi$ affecting the ratio $\eta_D/\eta_S$). Expressing the state of the system by a vector in a plane spanned by the two amplitudes $x_D$, $x_S$, this dissipative coupling induces a force which is always orthogonal to the instantaneous position vector of the system. Accordingly, solutions to the equations of motion~(\ref{eq:EquationsOfMotionDissipative}) show that the system undergoes amplified rotations  with fixed chirality in the $x_D$-$x_S$ plane. This mean-field description does not fully cover the experimental observations. Specifically, the theoretical model predicts an amplified motion, while in the experiment an evolution similar to limit-cycles without increasing amplitude was observed. This behavior has been attributed to collisional interactions of the atoms.

\begin{figure}[t!]
\centering
\includegraphics[width=.9\columnwidth]{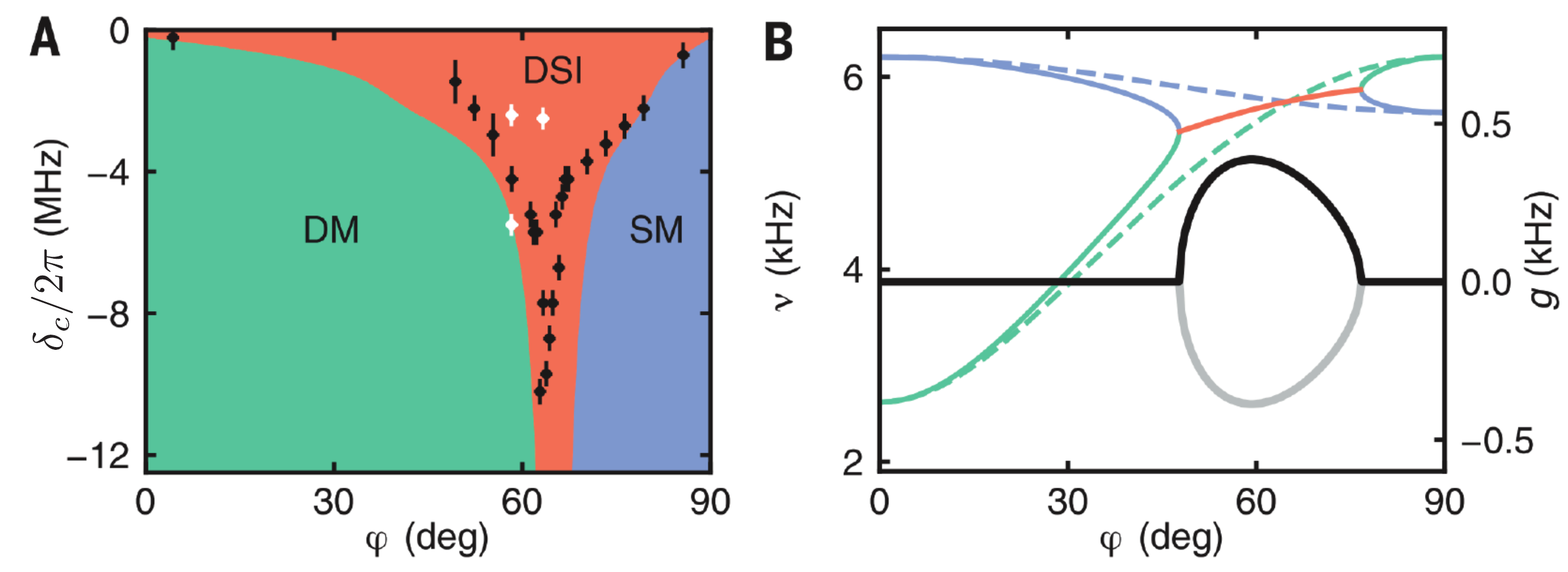}
\caption{Dissipation-induced dynamical behavior in a two-component BEC coupled to a single mode of a linear cavity via both scalar and vector polarizabilities. Panel (A) shows the boundary of the dissipation-induced structural instability (DSI) regime   as a function of dispersively shifted cavity detuning $\delta_c$  and angle of the pump field polarization $\varphi$. Data points indicate where the sidebands from Figure~\ref{fig:DissipationInducedInstabilityExp} become dominant, the colored background is the result of a mean-field calculation. Panel (B) shows the frequencies of the two eigenmodes in presence (solid lines) and absence (dashed lines) of dissipation. Without dissipation the two modes are the spin modulation (SM, blue) and the density modulation (DM, green). Dissipation leads to level attraction and finally synchronization in the instability dominated region (orange). The corresponding gains of the amplified and damped modes are shown in black and grey, respectively. Around the critical angle ($\varphi\approx 65\degree$), the eigenmodes become degenerate, and the DSI extends to large absolute values of $\delta_c$. Figure adapted and reprinted with permission from Ref.~\cite{Dogra2019Dissipation} published in 2019 by the American Association for the Advancement of Science.}
\label{fig:DissipationInducedInstabilityEigen}
\end{figure}

An analysis of the eigenspectrum reveals that the emergence of this dynamical instability is connected to the presence of exceptional points with the according level attraction as shown in Figure~\ref{fig:DissipationInducedInstabilityEigen}(b). This behavior is well known for systems described by non-Hermitian matrices such as Equation~(\ref{eq:EquationsOfMotionDissipative}).  After the instability has been induced, the two eigenmodes of the system synchronize and oscillate at a mean frequency. It was shown that the dissipative processes in the system generate a nonreciprocal coupling between the two collective spins that eventually induces this instability in a certain parameter regime, shown in Figure~\ref{fig:DissipationInducedInstabilityEigen}(a). Going beyond adiabatic elimination of the cavity field, it was though shown that the instability should occur in the long-time limit for basically all parameters due to additional anti-damping from the cavity field fluctuations. In the bad-cavity limit that applies to the experiment, the instability is, however, confined to a finite parameter region~\cite{Chiacchio2019Dissipation}.

An analysis studying the full quantum model of this system confirmed that the inclusion of cavity field fluctuations renders the system unstable indeed also in regions where the mean-field model predicts stability~\cite{Buca2019Dissipation}. Using the framework of Lindblad master equation, the eigenspectrum of the system was studied. The interpretation was that the presence of equally spaced imaginary parts of the eigenvalues prevents a dephasing of the dynamics and thus leads to a long-time non-steady-state dynamics. This can be contrasted to steady-state situations, where the eigenfrequencies are dense and incommensurate and thus lead to dephasing as it would be the case for a closed system. An analysis of the correlation functions of the system revealed  indications of beyond mean-field behavior such as entanglement induced by dissipation. The described behavior can be connected to dissipative time crystals breaking a continuous symmetry, where the description in a rotating frame is given by a time-independent master equation~\cite{Buca2019Dissipation}.

\subsection{Non-steady-state phases with time-dependent driving}
\label{sec:non-stationary-modulated}

Non-trivial dynamics without reaching a steady state can also be induced by applying a time-dependent transverse pump lattice, modulated either in amplitude or in phase; see Figure~\ref{fig:ParametricDrive}(a). Restricting the discussion to the self-organization of the atomic density in the low energy sector, the modulation results in a Dicke model with an explicitly time-dependent coupling rate, even in the frame rotating with the natural pump frequency.

\subsubsection{Time-modulated pump fields}
\begin{figure}[t!]
\centering
\includegraphics[width=0.8\columnwidth]{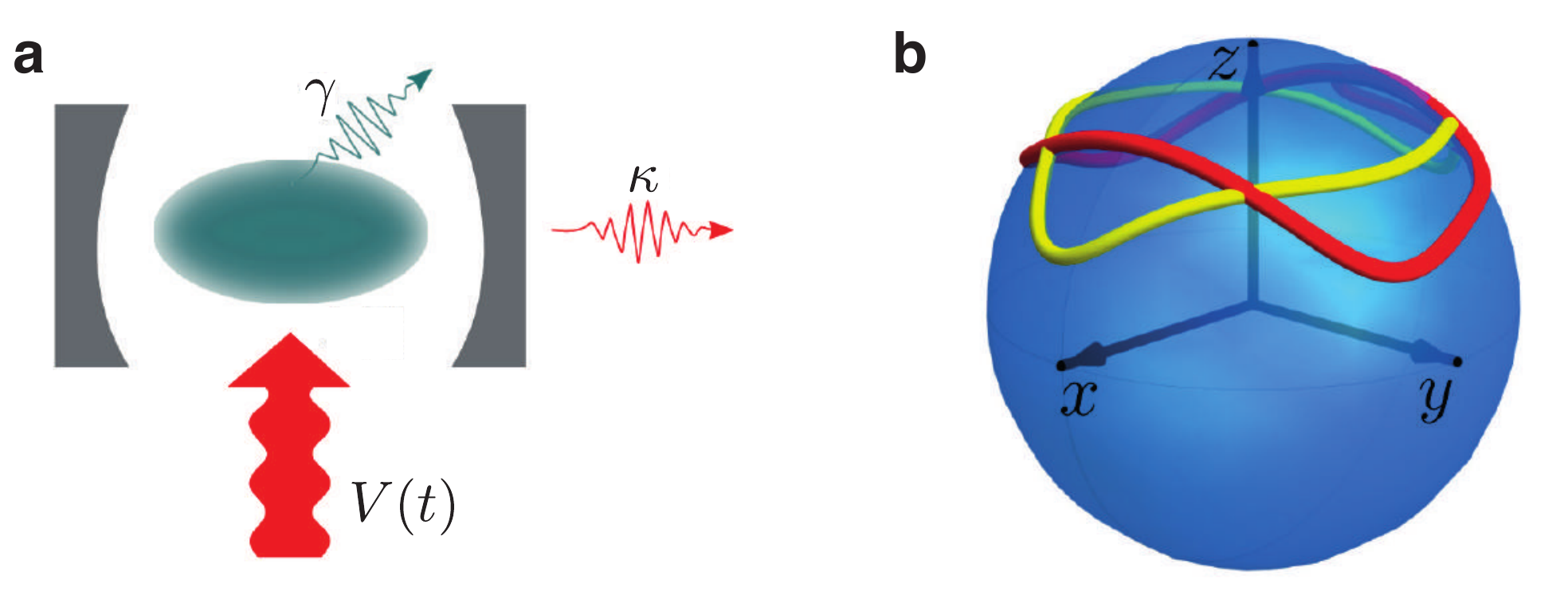}
\caption{Parametrically driven Dicke model. (a) Sketch of the setup. A BEC inside a dissipative single-mode cavity is driven transversely by a pump laser with an oscillating intensity $V(t)$. The atoms are coupled to the environment with a dissipation rate $\gamma$, and photons with a rate $\kappa$. (b) Exemplary trajectories of the atomic order parameter on a Bloch sphere as a function of time in the steady state for different driving parameters. Figure adapted and reprinted with permission from Ref.~\cite{Chitra2015Dynamical}~\textcopyright~2015 by the American Physical Society.}
\label{fig:ParametricDrive}
\end{figure}

First investigations of a parametrically driven Dicke model showed that additional minima in the energy landscape of the system appear already for infinitesimally small modulation if the modulation frequency is close to an eigenfrequency of the undriven system~\cite{Bastidas2012Nonequilibrium}. For strong enough modulation, the predicted phase diagram features metastable phases and according nonequilibrium first-order phase transitions. Including dissipation both via the cavity and via the atoms stabilizes the system against the instability seen on resonance for the closed system, such that a finite modulation strength is required to enter the non-steady-state phases~\cite{Chitra2015Dynamical}. This work also investigated possible trajectories of the Bloch vector in the non-steady-state phases. Examples of such trajectories are shown in Figure~\ref{fig:ParametricDrive}(b). For non-steady states in the so-called ``dynamical normal phase'', the phase of the cavity field rotates in more or less complicated trajectories and the field amplitude averages to zero. For the atomic degree of freedom this corresponds to a quasi-resonant switching between the two ordered density patterns allowed by the \textbf{Z}$_2$ symmetry. Likewise, the superradiant phase displays oscillations, however, around a mean non-zero photon number. The simulations also showed the existence of chaotic behavior: While the normal and the self-organized phases seem to be robust against parametric heating also in presence of the modulated drive, the above introduced dynamical normal phase shows ergodic-like behavior similar to the one characterizing the chaotic phases discussed in Section~\ref{sec:non-stationary-CW}. Numerical simulations also taking into account atomic collisional interactions and a trapping potential confirm the existence of the parametrically driven phases~\cite{Molignini2018Superlattice}.

The response of a BEC to a modulated field in the case of a cavity in the sideband-resolved limit has been numerically studied using an open system truncated Wigner approximation~\cite{Cosme2018Dynamical}. Also in this work, the pump field amplitude was assumed to be modulated periodically. In presence of this modulation, which effectively imprints two sidebands at the modulation frequency onto the pump field, the system was found to either relax to a steady state or enter dynamical states. Depending on the modulation frequency, density waves involving different higher atomic momentum modes with multiple recoils being transferred along the pump or the cavity direction can be excited. 

Another theoretical proposal for a periodically driven atom-cavity system is based on a phase modulated rather than an amplitude modulated pump field~\cite{Cosme2019Time}. Such a phase modulation results in a periodically shaken optical pump lattice which the atoms are exposed to. This configuration is expected to lead to self-organization in specific density waves, depending on the modulation frequency. Different to the case of an amplitude modulated pump field, here the atoms can also order in a state with density maxima at the nodes of the emergent superradiant optical lattice, dynamically switching between the symmetry broken states. The induced oscillations can have a very low frequency, set by the beating between the mode eigenfrequency and the modulation frequency. This beating leads to a subharmonic response also at incommensurate frequency ratios, and has been interpreted as incommensurate time crystal.

Periodic phase modulation of the cavity-induced optical lattice can also modify the properties of the superradiant phase transition. For instance,  periodic phase modulation of a cavity field has been predicted to yield a first-order superradiant phase transition with a hysteresis loop, provided the modulation frequency is chosen appropriately and the bosonic gas is sufficiently-weakly interacting~\cite{Luo2018Self}. The periodic phase modulation of a cavity field can also result in cavity-assisted long-range hoppings for fermionic atoms in an external lattice inside the cavity, leading to topological superradiant states~\cite{Feng2019Topological}.

\subsubsection{Experimental realizations of time-modulated pump fields}
The first experimental realization of a modulated atom-cavity system employed the beating between two different pump-field frequencies~\cite{Georges2018Light}. One field is red detuned with respect to the cavity resonance and drives the self-organization phase transition in the system for sufficiently strong amplitudes. The amplitude of the second pump field is increased from zero only after self-organization has set in. The frequency of this second field is closer to, and blue detuned with respect to the cavity resonance, such that it counteracts atomic self-organization. Indeed, a suppression of the self-organized density-wave state was observed with increasing amplitude of the second pump field. This behavior was present for a wide range of modulation frequencies. It was argued~\cite{Cosme2018Dynamical, Georges2018Light} that this mechanism of suppressing the self-organized density wave and enhancing atomic coherence in this setup bears analogy with light-induced restoration of superconductivity in hight-$T_c$ cuprates. From dynamical point of view, this suppression of the self-organization can be understood from (reduced) rescaling of the atom-field coupling $\eta_0$ in the corresponding  time-independent effective Magnus Hamiltonian~\cite{Cosme2018Dynamical}. Accordingly, the self-organization threshold is pushed to stronger atom-field couplings. Resonant features as reported in Ref.~\cite{Chitra2015Dynamical} were not observed employing this scheme, except for rapid excitations if the modulation frequency hit multiples of the vibrational frequency of the pump lattice.

\begin{figure}[t!]
\centering
\includegraphics[width=1\columnwidth]{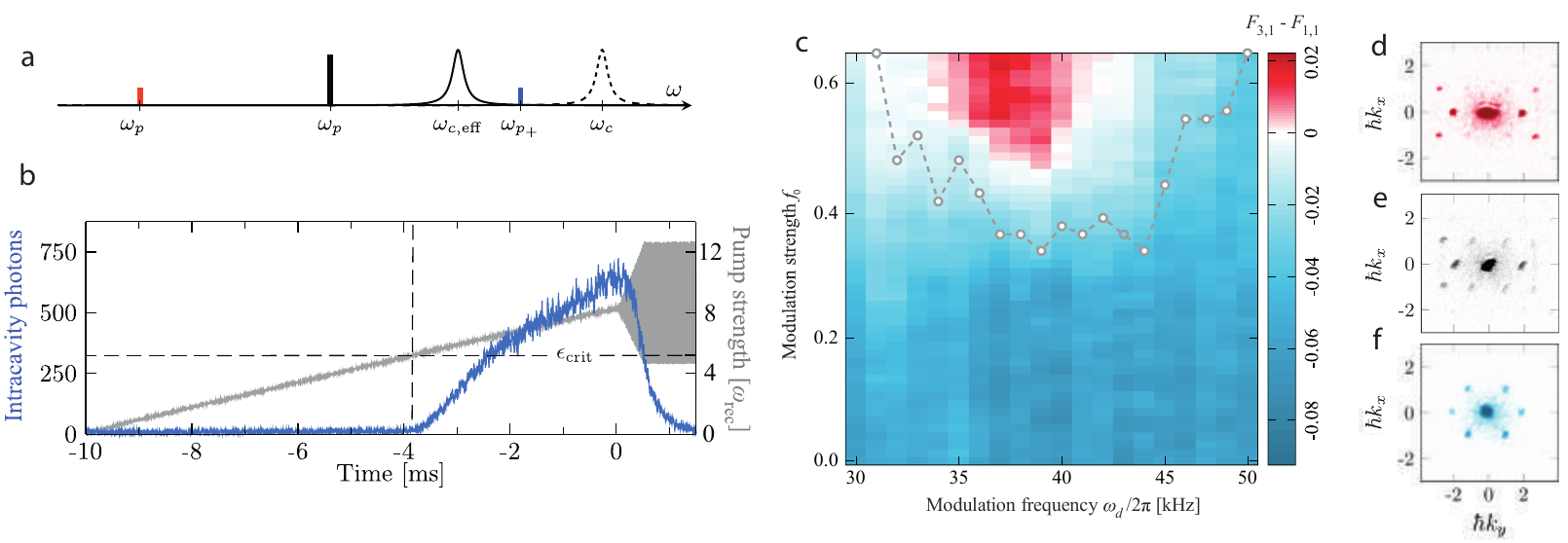}
\caption{Dynamical self-organization of a BEC in a standing-wave cavity driven transversely by an amplitude-modulated pump lattice. (a) The carrier of the pump field $\omega_p$ is red detuned with respect to the dispersively shifted cavity resonance at $\omega_{c, {\rm eff}}$. The amplitude modulation leads to sidebands at frequencies $\omega_{p\pm}$. (b) The driving protocol is shown in grey, where the pump lattice depth is first increased linearly such that the atomic system self-organizes. Deep in the self-organized phase the amplitude modulation is ramped up. The blue line indicates the mean intra-cavity photon number. (c) Dynamical phase diagram showing the difference of the occupations in a state induced by the modulation $F_{3,1}$ (a density wave with three recoil momenta along the pump direction and one recoil momentum along the cavity direction) and in the state induced by standard self-organization $F_{1,1}$ (one recoil momentum along both pump and cavity directions) as a function of modulation strength and modulation frequency. (d-f) Absorption images of the atomic cloud after ballistic expansion for a modulation frequency of $2\pi \times$\SI{40}{\kilo \hertz} and modulation strength (d) $f_0=0.60$, (e) $f_0=0.47$, and (f) $f_0=0.10$. Figure adapted and reprinted with permission from  Ref.~\cite{Georges2020Nonequilibrium}.}
\label{fig:ParametricDriveExp}
\end{figure}

When an amplitude modulation was implemented experimentally by imprinting instead two sidebands onto the pump field as suggested in Ref.~\cite{Cosme2018Dynamical}, a resonant feature was observed in addition to the suppression of superradiance described above~\cite{Georges2020Nonequilibrium}. The experimental scheme and observations are displayed in Figure~\ref{fig:ParametricDriveExp}. The pump intensity is modulated at a frequency $\omega_d$ according to $V(t) = V_0 [1+f_0 \cos(\omega_d t)]$, which imprints frequency sidebands at $\omega_{p\pm}=\omega_p\pm\omega_d$. The higher frequency sideband is close to and blue detuned with respect to the dispersively-shifted cavity resonance. Operating in the sideband-resolved regime is important to select specific atomic excitations~\cite{Kessler2014Steering}: In a certain range of modulation frequencies $\omega_d$, a different set of momentum states is occupied than for usual self-organization induced by an unmodulated pump field. For the given parameters, this state corresponds to a transfer of three recoil momenta along the pump axis and one recoil momentum along the cavity axis. Such resonant  occupation of specific momentum states, corresponding to a nonequilibrium density wave, has also been theoretically predicted (see above) if the modulation frequency is close to multiples of the atomic momentum-excitation energy~\cite{Cosme2018Dynamical,Georges2020Nonequilibrium}. Interestingly, the observed density wave is a subradiant non-steady state where the cavity occupation is suppressed while higher order momentum states are occupied. As shown in Figure~\ref{fig:ParametricDriveExp}(b), the cavity field decays rapidly once the amplitude modulation is switched on. At the end of the sequence absorption images are taken after ballistic expansion; examples are displayed in Figures~\ref{fig:ParametricDriveExp}(d)-(f). Away from the resonance, momentum peaks signaling standard self-organization are observed, while on resonance the occupation of higher momentum states is visible.

Theoretical analysis provided together with the experiment shows that the self-organization in this nonequilibrium density wave takes place at unstable positions, and that the atomic pattern is expected to periodically switch between symmetry broken states, similar to the discussion in Ref.~\cite{Chitra2015Dynamical}. The theoretically predicted transition to chaotic behavior for increasing modulation depth could not be observed experimentally due to driving-induced heating and accordingly shorter lifetimes.

A recent experimental work~\cite{Kessler2020Observation} on time-modulated pump fields employed a heterodyne detection scheme to gain access to the phase of the light field leaking from the cavity. The observed period doubling of the atomic response to the modulation was interpreted as a dissipative time crystal.

\section{Cavity-enhanced quantum measurement in quantum gases}
\label{sec:quantmeasure}

In the previous sections we demonstrated that ultracold quantum gases trapped together with light within high-Q optical cavities constitute an incredibly rich field of modern AMO physics with growing connections and application prospects towards solid-state physics. In many respects the specific quantum properties or quantum fluctuations of driving laser light fields and cavity modes can be neglected and a mean-field description in terms of coherent fields yields a valid and reliable description (see Section~\ref{subsec:MF} for more details). However, from a wider perspective quantum-gas cavity QED, i.e., the combination of cavity quantum electrodynamics with ultracold quantum gas physics, goes well beyond the mean-field picture of the coherent cavity-field mode evolution. In its full blossom the field can address and study novel phenomena, where the quantum statistical nature of both light and ultracold matter play equally important roles. While the light fields create forces and dynamical potentials with quantum properties, the atoms constitute a dynamical refractive index with genuine quantum properties as well. This wider view has several intriguing consequences as we will discuss in the following sections.

%==========================================================
\subsection{Projective measurements}
Beyond the mean-field approximation, a quantum description of the electromagnetic modes allows for dynamic field-matter entanglement beyond classical correlations. Hence, a quantized cavity mode can still induce forces and long-range interaction via photons even for zero average field-mode amplitude. Consequently, this opens a completely new venue for projective measurement-based preparation of special many-body  states. In particular, a wide class of emerging atomic states can be probabilistically prepared via the choice of optical geometry. As generic basic examples, atom number-squeezed as well as Schr{\"o}dinger-cat states can be prepared in close analogy to cavity-based spin squeezing~\cite{Leroux2010Implementation}.

\begin{figure}[t!]
  \begin{center}
    \includegraphics[width=0.75\columnwidth]{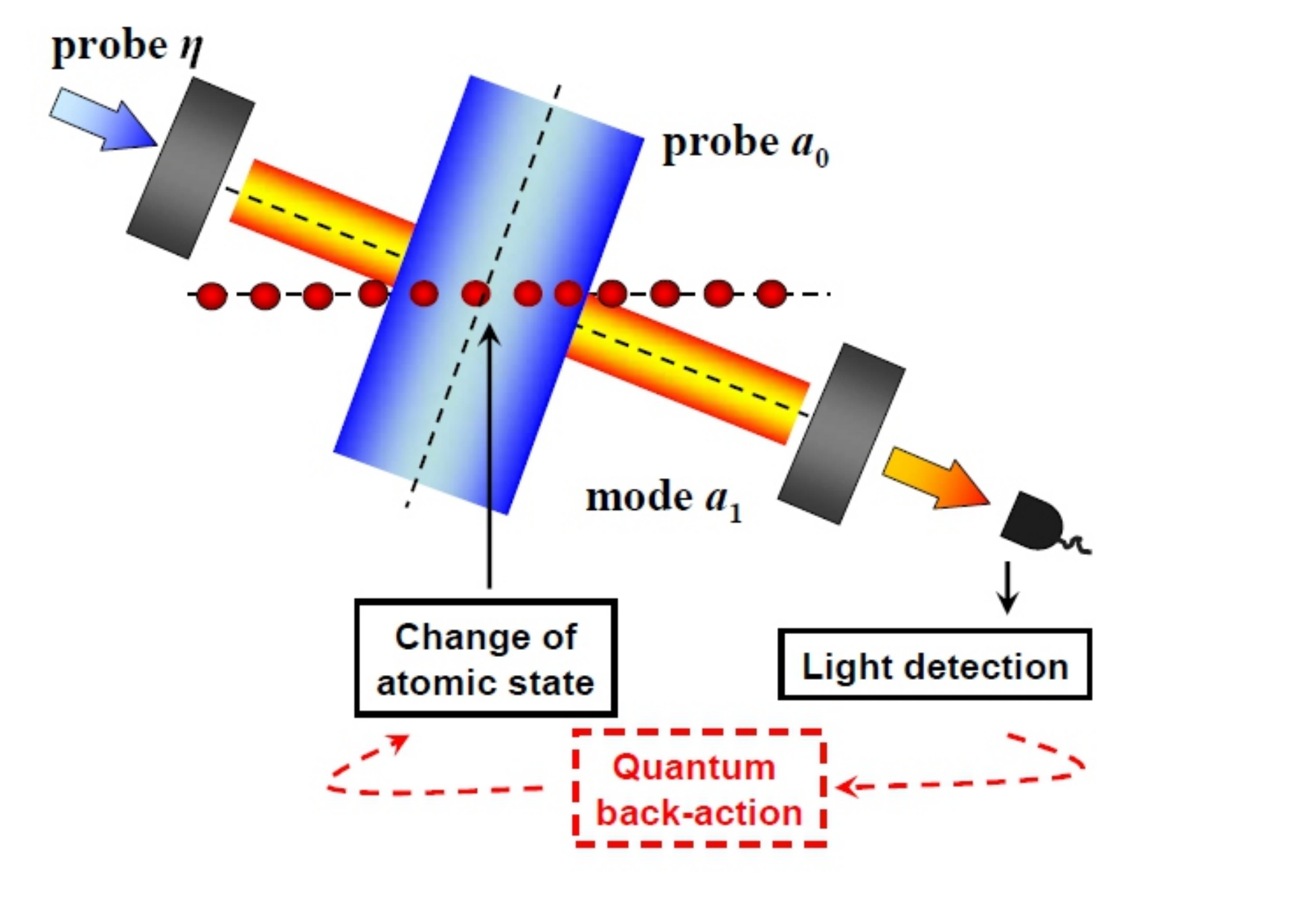}
    \caption{Cavity-based non-destructive measurement scheme: Light scattered by the quantum gas from a pump laser is collected and builds up in a cavity mode. The field leaking through the cavity mirrors can be analyzed by various detection schemes and lead to quantum measurement back-action on the atomic dynamics. Figure adapted and reprinted with permission from Ref.~\cite{Mekhov2009Quantum}~\textcopyright~2009 by the American Physical Society.}
    \label{fig:quantmeasure_scheme}
  \end{center}
\end{figure}

\begin{figure}[t!]
  \begin{center}
    \includegraphics[width=0.75\columnwidth]{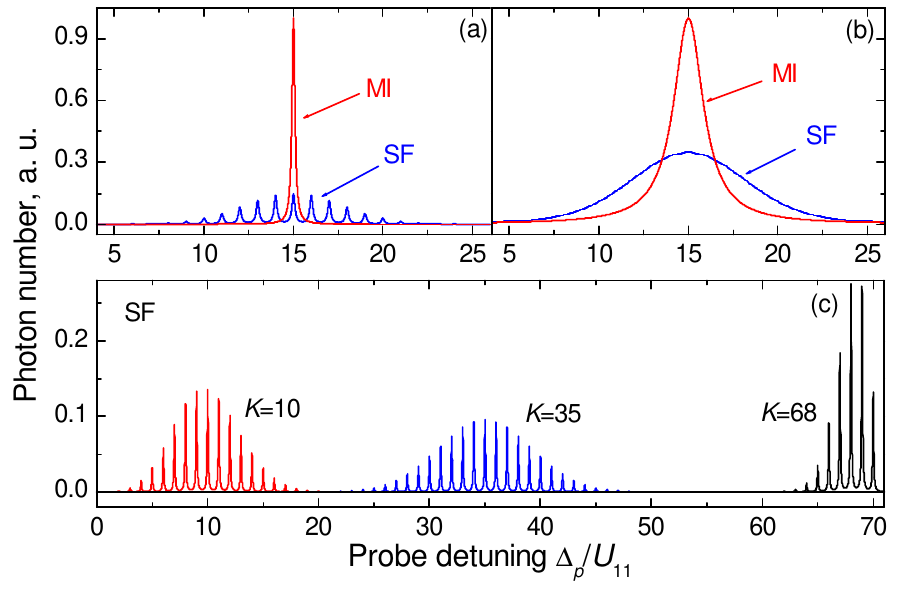}
    \caption{Transmission spectra of a cavity filled with a BEC in an external lattice. The transmission peaks directly map out the full atom-number distribution of the lattice sites within the cavity mode as shown in Figure~\ref{fig:quantmeasure_scheme}. (a) A single Lorentzian for a Mott-insulating (MI) phase reflects the non-fluctuating atom number. Many Lorentzians for a superfluid (SF) phase indicate atom-number fluctuations, which are imprinted on the positions of narrow resonances in the spectrum. For a very narrow bandwidth cavity, i.e. when the cavity decay rate $\kappa$ is smaller than the cavity light shift per atom $U_0$,  all cavity transmission are resolved. Here we see the example of a superfluid of $N=30$ atoms in a lattive with $M=30$ sites, where $K=15$ sites within the cavity are illuminated.  (b) The same as in panel~(a) but for a worse cavity with $\kappa = U_0$, which gives a smooth broadened contour for SF. Although the individual peaks are not resolved, the spectra for SF and MI states are still very different. (c) Same spectra as above for SF with $N=M=70$ ($\kappa=0.05U_0$) and different number of sites illuminated $K=10,35,68$. The transmission spectra have different forms, since different atom distribution functions correspond to different $K$. Adapted and reprinted with permission from Ref.~\cite{Mekhov2007Probing} published in 2006 by the Nature Publishing Group.}
    \label{fig:quantmeasure_example}
  \end{center}
\end{figure}

As we generally deal with open systems with respect to the cavity-field dynamics, the output light continuously provides information about the intrinsic dynamics with minimal perturbation~\cite{Mekhov2009Quantum}; see  Figure~\ref{fig:quantmeasure_scheme}. Ultimately it can serve as a quantum non-demolition (QND) probe of the spatial distribution, phase diagram, or the phase-transition dynamics of the ultracold quantum gas. As an example, a Mott-insulator state in an optical lattice inside a cavity exhibits clearly distinct light scattering properties compared to a superfluid state. In this way the corresponding phase transition between the two states can be directly monitored from angle resolved scattering~\cite{Mekhov2007Light,Mekhov2009Quantum} or cavity transmission measurements~\cite{Mekhov2007Probing} as shown in Figure~\ref{fig:quantmeasure_example}. The QND detection of the Bose-glass phase~\cite{Habibian2013Bose,Kozlowski2015Probing} and few-body complexes of polar molecules~\cite{Mekhov2013Quantum} were suggested as well. In addition to being able to monitor non-destructively the orbital state (i.e., Mott insulator or superfluid) of one-component atoms in an optical lattice inside a cavity through cavity output fields, it is also feasible to probe non-destructively the spinorial state of multi-component atoms through cavity transmission spectrum. This provides a novel approach for non-destructive detection of various magnetic orders in quantum gases~\cite{Padhi2014Spin}. The QND property also allows for intriguing and promising applications for quantum enhanced sensing as in, e.g., a supersolid-based accelerometer in a ring cavity~\cite{Gietka2019Supersolid}; see Section~\ref{sec:quantum-sensing} for more details.

From a fundamental physics point of view such systems allow detailed studies of quantum measurement theory for many-body quantum systems ranging from the quantum Zeno effect~\cite{Murch2008Observation,Kozlowski2016Non} to quantum measurement-induced ordering~\cite{Mazzucchi2016Quantum}. Implementing weak measurement schemes on the field modes even allows controlled steering of the atom-field quantum evolution~\cite{Mazzucchi2016Collective}, and generation and observation of multi-partite entanglement via nonlocal cavity-enhanced measurements~\cite{Elliott2015Multipartite}. The generation of squeezed light has also been suggested~\cite{Caballero-Benitez2015QuantumProperties}.
As one might expect the genuine quantum nature of measurement back-action also allows to generate dynamics beyond the mean-field level. For fermions theoretical models show the appearance of measurement-induced antiferromagnetic order~\cite{Mazzucchi2016Quantum} as well as the appearance of many-body localization effects~\cite{Sierant2019Many}.
By trapping atoms inside an optical cavity, one creates optical potentials and forces, which are not externally prescribed but are quantum dynamical variables themselves. Ultimately,  cavity QED with quantum gases requires a self-consistent solution for light and particles, which enriches the picture of quantum many-body states of atoms trapped in quantum potentials. A proper quantum treatment of interaction and back-action turns out to be particularly important to implement quantum simulations or to exploit quantum-enhanced optimization using cavity-mediated tailored and externally controlled interactions~\cite{Torggler2017Quantum}. Interestingly, a mean-field treatment of the fields here leads to a significantly reduced success probability hinting for a real quantum advantage~\cite{Torggler2019A}.  

\subsection{Measurement-induced dynamics and quantum feedback}

\begin{figure}[t!]
 \begin{center}
    \includegraphics[width=\columnwidth]{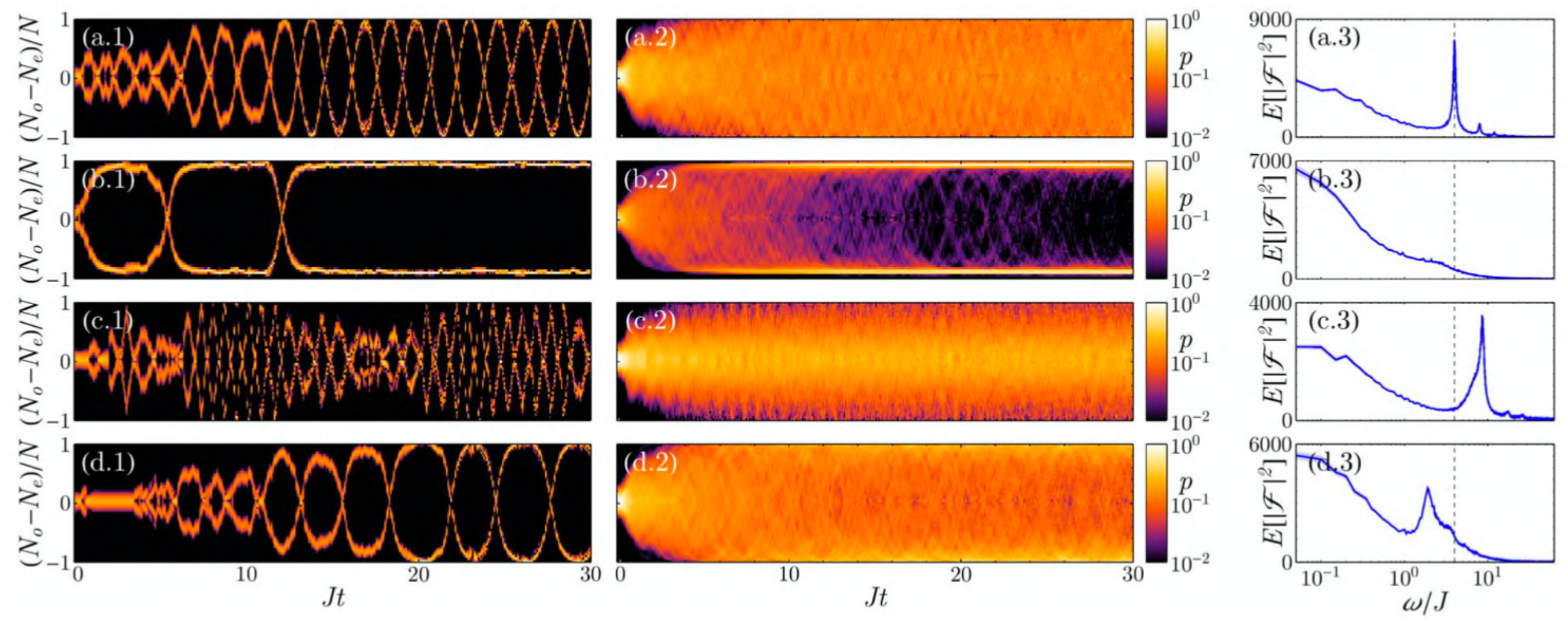}
    \caption{Probability distribution of the population imbalance between odd and even sites for single quantum trajectories (column1) and averages over 200 trajectories (column 2) for different values of the feedback gain. Column 3 shows the power spectrum averaged over 200 trajectories. In the absence of feedback [panel (a)] the oscillations of the population between even and odd sites are visible only in a single trajectory. For an increased feedback strength $z$ beyond a creitical value $z >  z_c$ [panel (b)] the imbalance between odd and even sites soon becomes frozen for each individual quantum trajectory . Below the critical feedback strength  $z < z_c$ the frequency of the oscillations can be tuned above [panel (c)] or below [panel (d)] the frequency defined by the tunneling amplitude $J$. Again, the phase varies from run to run and the oscillatory dynamics is visible only in a single quantum trajectory. Adapted and reprinted with permission from Ref.~\cite{Mazzucchi2016QuantumC} published in 2016 by the Optical Society.}
    \label{fig:quantmeasure_feedback} 
  \end{center}
\end{figure}

One can make use of the continuously measured output-field properties for a feedback on the coupled atom-field system. Applying weak measurements and feedback to a quantum system induces phase transitions driven by fundamental quantum fluctuations due to measurements. Non-Markovianity and nonlinearity of feedback enables simulating special spin-bath problems and Floquet time crystals with long-range and memory time interactions~\cite{Mazzucchi2016QuantumB,Ivanov2020Feedback}. As an example we show in Figure~\ref{fig:quantmeasure_feedback} the simulation of quantum trajectories for a continuous measurement on an ultracold gas in a lattice with continuous monitoring of the cavity outputs~\cite{Mazzucchi2016QuantumC}. The interplay between measurement projection and nonlinear dynamics creates an intriguing dynamical pattern for the atom number distribution on even or odd sites along the cavity axis.  

Classical feedback derived from the output of a cavity field has been experimentally demonstrated~\cite{Kroeger2020Continuous}. A BEC trapped inside a linear cavity was brought to self-organization by applying a transverse pump field. Using the field leaking out from the cavity as a signal, the intensity of the transverse pump laser was regulated by an electronic circuit such that the mean intra-cavity photon number was stabilized to a given value. While this photon number decreases without feedback due to atom losses, applying the feedback scheme allowed to stabilize the system over multiple seconds.

\subsection{Quantum sensing with ultracold gases in cavities }
\label{sec:quantum-sensing}

\begin{figure}[t!]
 \begin{center}
    \includegraphics[width=\columnwidth]{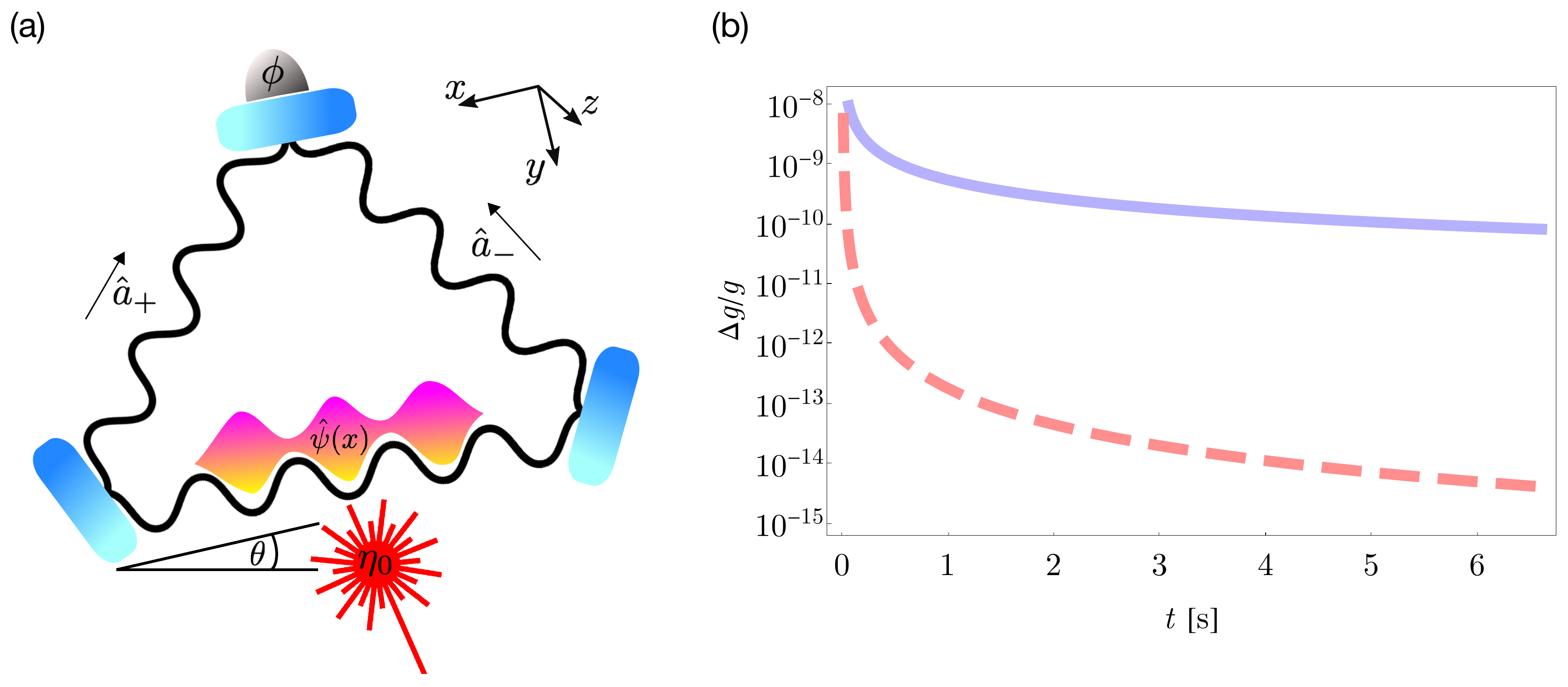}
    \caption{Supersolid-based gravimeter in a ring cavity.
    (a)~Schematic sketch of the setup. A transversely driven 1D BEC is coupled to two degenerate counterpropagating modes $\hat{a}_\pm$ of a ring cavity. The cavity is tilted with respect to the horizontal direction by an angle $\theta$ such that the BEC experiences the gravitational potential $V_{\rm ext}(x)=Mgx\sin\theta$ (cf.\ Section~\ref{sec:supersolid-ring-cavity-transverse-pump} and Figure~\ref{Fig:Mivehvar2018Driven_Fig1andFig3}). 
    The relative phase $\phi$ of the two cavity modes is measured in the cavity output, in order to monitor the motion of the BEC along the cavity axis under the gravitational force.
    (b)~Relative sensitivity for state-of-the-art experimental parameters as a function of time. The red dashed curve presents the sensitivity for $\kappa = 0$ (for the sake of comparison) while the blue solid curve takes into account the effective friction arising from photon loss $\kappa\neq0$. Figure reprinted with permission from Ref.~\cite{Gietka2019Supersolid}~\textcopyright~2019 by the American Physical Society.}
    \label{fig:Gietka2019Supersolid_Fig1_3} 
  \end{center}
\end{figure}

Measurement-induced back-action and feedback are intriguing tools to study genuine quantum physics in action and the measurement outcomes provide in-depth information on the quantum gas dynamics~\cite{Sandner2015Self}.  Furthermore, as quantum-gas--cavity systems are extremely sensitive to external perturbations, these hybrid systems constitute a new class of quantum sensors~\cite{Goldwin2014Backaction,Gietka2017Quantum,Bariani2015Atom,Gietka2020Magnetometry}.  A remarkable example is a recently proposed supersolid-based gravimeter in a ring cavity, which exhibits Heisenberg-like-scaling sensitivity with respect to the atom number~\cite{Gietka2019Supersolid}. For state-of-the-art experimental parameters, the relative sensitivity $\Delta g/g$ of such a gravimeter has been predicted to be of the order of $10^{-10}$--$10^{-8}$ for a condensate of a half a million atoms within a few seconds; see Figure~\ref{fig:Gietka2019Supersolid_Fig1_3}. Detailed numerical studies on the dynamical behavior of such ring-cavity systems were carried out in Ref.~\cite{Qin2020Self}. Recently, strong evidence for the creation of a supersolid in a ring-cavity geometry as a prerequisite for such a sensor was observed in the experiment~\cite{Schuster2020Supersolid} as discussed in detail in Section~\ref{sec:supersolid-ring-cavity-longitudinal-pump}.   

Also the real-time monitoring of Bloch-Zener oscillations via the light field leaking from a cavity has been theoretically discussed~\cite{Venkatesh2009Atomic,Peden2009Nondestructive,Venkatesh2013Bloch} and experimentally explored~\cite{Georges2017Bloch, Kessler2016In}. Such a method holds the promise to help to increase the data acquisition speed in precision measurements of forces. Recently, more refined modeling of the minimally invasive detection of quantum atomic currents revealed the great potential of cavities to study the presence and consequences of synthetic gauge fields as well as dynamical quantum phases in particular for atomic models of quantum transport~\cite{Laflamme2017Continuous,Uchino2018Universal}. First results on real-time studies of $g^{(2)}$ correlations in the BEC-BCS crossover regime revealed the great potential of this approach~\cite{Roux2021Cavity}. 

\section{Conclusions, outlook, and open challenges}
\label{sec:conclusions}

Starting from a short historical overview and an introduction of the basic underlying physical models and mechanisms, the central goal of this review has been to present the state of the art in the field of quantum-gas--cavity QED (QG-CQED). As some comprehensive and nice reviews on the classical gas aspects~\cite{Domokos2003Mechanical}, simple cavity-lattice models~\cite{Mekhov2012Quantum}, and the early quantum-gas developments~\cite{Ritsch2013Cold} already exist, for the sake of compactness here we concentrated on more recent important developments and outstanding achievements in the field. Nevertheless, the overwhelming recent growth in theory and experiment has rendered the goal of being comprehensive, detailed enough, and complete rather futile. Even well exceeding the initially planned length of about 100 pages did not allow us to include everything, which would deserve to be mentioned here in sufficient detail. Even while writing this review article intriguing new developments continuously surfaced, we finally had to strongly restrict ourselves and leave some of the most recent developments for future reviews. Hence, we here have to apologize for some important omissions. Nevertheless, we hope that on the one hand this review gives a good and solid starting point to new researchers interested in entering the field, and on the  other hand fosters new and closer collaborations among the existing diverse community.   

Let us wrap things up shortly here: starting from early theoretical ideas and abstract proposals somewhat in the style of \textit{Gedankenexperiments} at the turn of the century~\cite{Horak2000Coherent,Gardiner2001Cavity,Horak2001Manipulating},  the field of QG-CQED has enormously grown and expanded in various directions. First aims were driven by theoretical visions of cavity-induced ground-state cooling towards implementing a continuous BEC and a continuous-wave atom lasers~\cite{Salzburger2007Atom,Jaksch2001Uniting}, or creating macroscopic superposition states of larger particle numbers~\cite{Maschler2007Entanglement}. Nevertheless, it soon became clear that using a simpler implementation starting from a zero-temperature degenerate quantum gas prepared by conventional means within an optical cavity, one already realizes a very versatile and well controllable test ground for many-body physics in an extreme quantum limit. Here cavity-induced thermalization limits the available measurement time, but this process is in many cases slow enough such that it does not hinder typical experimental sequences.  As a particular example, it proved straightforward to emulate a wide class of optomechanical Hamiltonians at zero temperature~\cite{Brennecke2008Cavity,Gupta2007Cavity,Purdy2010Tunable}, a goal far out of reach of other optomechanical systems at that time~\cite{Aspelmeyer2014Cavity}.

Almost a decade after theoretical suggestions, a series of cutting edge foundational experiments trapping a BEC within a cavity mode has been implemented in Z\"{u}rich~\cite{Brennecke2007Cavity}, T\"{u}bingen~\cite{Slama2007Superradiant}, Berkeley~\cite{Gupta2007Cavity}, and Paris~\cite{Colombe2007Strong}. These allowed transforming \textit{Gedankenexperiments} into real setups in laboratories and thus sparked a great deal of interest well beyond the ultracold-atom community, in particular, in the solid-state and quantum-information communities. The subsequent multitude of theoretical models, predictions, and suggested applications initiated another wave of experimental efforts, encompassing BECs in recoil-resolved linear and ring cavities~\cite{Wolke2012Cavity, Wolf2018Observation},  in multi-mode cavities~\cite{Kollar2015An}, in crossed linear cavities~\cite{Leonard2017Supersolid}, and in folded ring cavities~\cite{Naik2018Bose}, as well as most recently also of degenerate Fermi gases coupled to a linear cavity~\cite{Roux2020Strongly}.

A second decisive boost to the field came from the discovery that cavity-mediated global interactions can induce self-ordering of trapped atomic clouds~\cite{Domokos2002Collective,Black2003Observation}. Most interestingly, in the zero-temperature regime this corresponds to a reversible quantum phase transition. In a seminal experiment, this self-ordering phase transition was explored in real time~\cite{Baumann2010Dicke}. Even more interestingly, a direct mapping of the corresponding Hamiltonian to the Dicke model allowed to confirm an almost 50-year-old prediction of the Dicke superradiant phase transition~\cite{Nagy2010Dicke,Baumann2010Dicke}, a prediction that was controversially disputed in literature for decades with its actual direct realization being still disputed and sought for in solid-state platforms~\cite{Frisk2019Ultrastrong,Andolina2019Cavity}. Nevertheless, this proof-of-principle demonstration highlighted the remarkable possibilities for QG-CQED as a versatile platform for quantum simulation of many-body Hamiltonians with all-to-all interactions. With these possibilities at hand a new wave of theoretical work was triggered.

With the subsequent extension and refinement of experimental apparatus, experiments went well beyond the existing theoretical capabilities. Soon it became clear that existing 1D mean-field models, which allow quite general understanding of the emerging physical phenomena, could not adequately account for all observations. This became particularly obvious for the case of 3D intra-cavity lattices or setups with several cavity modes dynamically involved~\cite{Klinder2015Observation, Landig2016Quantum}. As a notable example, the clear experimental observation of signatures of a lattice-supersolid phase required new theoretical approaches such as dynamical mean-field theories~\cite{Li2013Lattice}, Keldysh-type approaches~\cite{Piazza2013Bose}, or multi-configuration time-dependent Hartree methods~\cite{Lode2017Fragmented}. As experiments become more complex but also more precise, these theoretical efforts strongly grow in various directions such as, e.g., a thrust to make use of powerful DMRG methods~\cite{Halati2020Numerical}, which turns out to be rather delicate due to the competition of local and cavity-mediated global interactions. 

In a next phase theory and experiments pushed the field ahead further by exploiting Zeeman sub-levels of atomic ground-state manifolds which added an additional spin degree of freedom to the dynamics and paved the way for exploring emergent magnetic orders in real time~\cite{Mivehvar2017Disorder,Landini2018Formation,Kroeze2018Spinor,Mivehvar2019Cavity}. Here the dynamical nature of the cavity fields allowed to extend models for synthetic gauge fields to the dynamical regime~\cite{Kroeze2019Dynamical} previously inaccessible in free-space configuration. New realms of physics also became accessible by introducing repulsive optical potentials in the blue-detuned regime allowing to emulate a nonequilibrium  form of the Su-Schrieffer-Heeger model~\cite{Mivehvar2017Superradiant} as well as entering the domain of nonlinear dynamical instabilities, deterministic chaos, and a new arising area of time-crystal formation~\cite{Piazza2015Self,Griesser2011Nonlinear}. Using complex resonator geometries even allowed to study quasicrystal formation beyond naturally existing crystal symmetries~\cite{Mivehvar2019Emergent}.     

Beyond studying fundamental aspects of quantum physics, the application perspective of QG-CQED has been theoretically discussed already since the early stage of the field as the extreme sensitivity of the systems to external perturbations and the real-time observation capabilities hint for high measurement precision and accuracy. In this context several new proposals for acceleration, force, and magnetic-field sensors have recently been put forward~\cite{Gietka2019Supersolid, Gietka2020Magnetometry} and are currently refined. In a parallel development the large versatility of cavity-induced interactions, i.e., in principle almost arbitrary forms of all-to-all couplings, allowed to envisage novel types of quantum simulators~\cite{Gopalakrishnan2012Exploring,Gopalakrishnan2011Frustration} or annealers~\cite{Torggler2017Quantum,Marsh2020Enhancing}. In some special cases it could be shown that these implementations, which typically require a number of laser frequencies, only require a minimal q-bit overhead so that implementations beyond the capabilities of classical computers are within reach of current experimental technology~\cite{Torggler2019A}. Recently, in an another direction QG-CQED-based setups for active atomic clocks in the form of so-called superradiant lasers were theoretically proposed~\cite{Bohnet2012A,Hotter2019Superradiant} and many efforts started for their experimental implementation and technical refinement~\cite{Norcia2016Superradiance,Chen2019Continuous}. Here the cavity-induced all-to-all coupling among clock atoms is centrally responsible for the synchronization of the atomic dipoles at very low photon numbers in the spirit of the successful micro-maser-based clock devices.

Let us also mention that cavity QED with quantum matter recently became possible also in solid-state platforms~\cite{Thomas2019Exploring,Paravicini-Bagliani2019Magneto,Tan2020Interacting,Sentef2018Cavity,Mazza2019Superradiant,Kiffner2019Manipulating,Schlawin2019CavityA,Curtis2019Cavity,Allocca2019CavityA,Chakraborty2020Non,Ashida2020Quantum,Sentef2020Quantum}, potentially attracting the interest of an even larger community. Those platforms partially differ from their ultracold-atomic counterparts both in terms of models and parameter regimes. Nevertheless, a large number of phenomenological features are still common, so that we believe that this review article should be useful also for readers interested in solid-state many-body cavity QED.

Although it is very difficult to make predictions about future directions, we finally point out some recent promising developments and comment on a possible road map. As a recent central advancement, first cavity-QED experiments with fermionic atoms have recently been presented~\cite{Roux2020Strongly}. This will bring the field even closer to emulate genuine solid-state Hamiltonians and condensed-matter phenomena, such as momentum-space pairing in high-$T_c$ superconductors~\cite{Colella2018Quantum,Schlawin2019Cavity,Colella2019Antiferromagnetic}. With more and more quantum-gas species trapped within resonators, one can also envisage multi-species configurations soon~\cite{Griesser2012Cooperative}, which will pave the way to molecule formation via dark-state-induced collective photoassociation~\cite{Winkler2005Atom} and subsequently superradiant quantum chemistry~\cite{Herrera2016Cavity}. Important new possibilities will arise from more complex resonator geometries covering two and three dimensions, as already demonstrated in 2D~\cite{Leonard2017Supersolid}, and by the help of higher-order transverse modes~\cite{Kollar2015An}. Similarly, the whole manifold of longitudinal modes available in any optical resonator opens the possibility to emulate inter-particle forces of virtually arbitrary form. Here the natural frequency-periodic identical nature of cavity and frequency-comb spectra can certainly help to reduce the complexity of corresponding setups~\cite{Torggler2020Self}. In addition, the combination of high-resolution optical access with QG-CQED experiments may make new schemes possible for the shaping of cavity-induced interactions, but also for the simultaneous preparation of multiple atomic systems, or for the {\sl in-situ} imaging of self-ordered atomic structures. Finally, the open system physics of non-steady states and in particular dynamic periodic solutions of many-body systems, often called ``time crystals,'' has drawn strongly increased recent attention. Here the nonlinear nature of the coupled atom-field dynamics and real-time observation capabilities of the cavity fields together with the excellent control over the system make QG-CQED setups a prime candidate for corresponding experimental studies and tests~\cite{Kessler2019Emergent,Kessler2020Observation}.

\section*{Acknowledgements}
We are grateful to Nishant Dogra, Peter Domokos, Francesco Ferri, J\"urg Fr\"ohlich, Yudan Gao, Sarang Gopalakrishnan, Andreas Hemmerich, Jonathan Keeling, Ronen Kroeze, Benjamin Lev, Brendan Marsh, Igor Mekhov, Heiko Rieger, and Claus Zimmermann, for critical reading and valuable feedback on our manuscript. We thank Jean-Philippe Brantut, R. Chitra, Tilman Esslinger, Petr Karpov, Corinna Kollath, Giovanna Morigi, and Oded Zilberberg for stimulating discussions. We also thank Elvia Colella for the assistance to create Figure~\ref{fig:coupling_scheme}.

\section*{Funding}
F.\,M.\ is supported by the Lise-Meitner Fellowship M2438-NBL of the Austrian Science Fund FWF. F.\,M.\ and H.\,R.\ acknowledge the International Joint Project No.~I3964-N27 of the the Austrian Science Fund FWF and the National Agency for Research of France ANR.
T.\,D.\ acknowledges funding from the Swiss National Science Foundation SNF: NCCR QSIT and the project ``Cavity-assisted pattern recognition'' (Project No.\ IZBRZ2\_186312). T.\,D.\ and H.\,R.\ acknowledge funding from EU Horizon 2020: ITN grant ColOpt (Project No.\ 721465).

\setcounter{section}{0}
\setcounter{equation}{0}
\setcounter{figure}{0}
\setcounter{table}{0}
\renewcommand{\thesection}{A\arabic{section}}
\renewcommand{\theequation}{A\arabic{equation}}
\renewcommand{\thefigure}{A\arabic{figure}}
\renewcommand{\thetable}{A\arabic{table}}
\newpage
\section{Appendix: The most important and commonly-used symbols throughout the paper}
\label{sec_app:symbols}

\begin{center}
\begin{longtable}{p{1.5cm}p{12cm}}  

    \label{tab:symbols}\\
\hline
             \textbf{Symbol} & \textbf{Definition} \\
\hline
\endfirsthead
\multicolumn{2}{c}%
{\tablename\ \thetable\ -- \textit{Continued from previous page}} \\
\hline
             \textbf{Symbol} & \textbf{Definition} \\
\hline             
\endhead
\hline \multicolumn{2}{r}{\textit{Continued on next page}} \\
\endfoot
\hline
\endlastfoot
      \multicolumn{2}{l}{\textit{Fundamental constants}}\\
      $c$ & speed of light \\
      $\epsilon_0$  & vacuum permittivity \\
      $h$ & Planck's constant, $\hbar=h/2\pi$\\
\hline   
      \multicolumn{2}{l}{\textit{Properties concerning the atoms}}\\
      $M$ & atomic mass\\
      $\mathbf{d}_{ge}$ & matrix element of the induced atomic electric dipole moment for the transition $\ket{g}\leftrightarrow\ket{e}$\\
      $\omega_a$ &  atomic transition frequency \\
      $\Delta_a$ & atomic detuning, relative frequency between the pump field and the atomic resonance: $\Delta_a=\omega_p-\omega_a$\\
       $\gamma$ & decay rate of the atomic dipole\\ 
       $\Gamma$ & natural linewidth of the atomic transition: $\Gamma=2\gamma$\\
       $a^s_{\tau \tau'}$ & atomic s-wave scattering length between internal atomic states $\tau, \tau'$ \\
       $g_{\tau\tau'}$ & strength of the two-body contact interactions: $g_{\tau\tau'}=4\pi a^{s}_{\tau\tau'}\hbar^2/m$\\
       $N$ & number of atoms \\
       $T$ & temperature of the atomic cloud\\
       $V_{\rm ext}(\mathbf{r})$ & external trapping potential \\
       $\hat{\psi}_\tau(\mathbf{r},t)$ & atomic annihilation field operator for internal state $\tau$\\
       $\hat{b}_j,\hat{c}_j(\hat{f}_j)$ & bosonic (fermionic) atomic annihilation operator for momentum state/lattice site $j$\\
       $\psi_\tau(\mathbf{r},t)$ & condensate wave function for internal state $\tau$\\

\hline
      \multicolumn{2}{l}{\textit{Properties concerning the pump field}}\\
      $\omega_p$ & frequency of the pump field \\
      $\lambda_p$ & wave length of the pump field: $\lambda_p \simeq \lambda_c$\\
      $k_p$ & wave number of the pump field: $k_p=2\pi/\lambda_p \simeq k_c$\\
      $\boldsymbol{\epsilon}_p$ & polarization vector of the pump field\\
      $\omega_r$ & recoil frequency: $\omega_r = \hbar k_c^2 / 2M$\\
      $P$ & optical power of the pump field\\
      $\Omega_0$ & maximum Rabi frequency of the pump field \\
      $\Omega(\mathbf{r})$ & position-dependent Rabi frequency of the pump field \\ 
      $\hbar V_0$ & maximum pump-lattice depth: $\hbar V_0=\hbar\Omega_0^2/\Delta_a$\\
      $\hbar V(\mathbf{r})$ & position-dependent pump lattice: $\hbar V(\mathbf{r})=\hbar\Omega^2(\mathbf{r})/\Delta_a$\\
      $\eta_0$ & maximum two-photon Rabi frequency: $\eta_0 = \Omega_0\mathcal{G}_0/\Delta_a$\\
      $\eta(\mathbf{r})$ & position-dependent two-photon Rabi frequency: $\eta(\mathbf{r}) = \Omega(\mathbf{r})\mathcal{G}(\mathbf{r})/\Delta_a$\\
\hline
      \multicolumn{2}{l}{\textit{Properties concerning the cavity}}\\
      $R^c_i$ & radius of curvature of mirror $i$\\
      $l_\mathrm{res}$ & resonator length \\
      $g_i$ & $g$-parameter of the cavity: $g_i=1-l_\mathrm{res}/R^c_i$\\
      $w_0$ & beam waist radius of the cavity mode:  $w_0^2=(\lambda l_{\rm res}/\pi)\sqrt{\frac{g_1g_2(1-g_1g_2)}{(g_1+g_2-2g_1g_2)^2}}$\\
      $\mathcal{V}$ & mode volume of the cavity: $\mathcal{V}=\pi w_0^2 l_\mathrm{res}/4$\\
      $\mathcal{R}_i, \mathcal{T}_i, \mathcal{L}_i$ & reflectivity, transmissivity, and losses of mirror $i$: $\mathcal{R}_i + \mathcal{T}_i + \mathcal{L}_i = 1$\\
      $\nu_\mathrm{FSR}$ & free spectral range of the cavity: $\nu_\mathrm{FSR} = c/2l_\mathrm{res}$\\
      $\nu_{lmn}$ & resonance frequency of the cavity mode with longitudinal index $l$ and transverse mode indices $(mn)$\\
      $\omega_c$ & cavity resonance frequency: $\omega_c=2\pi\nu_{lmn}$ \\
      $\Delta_c$ & cavity detuning, relative frequency between the pump field and the cavity resonance: $\Delta_c=\omega_p-\omega_c$\\
      $\tilde{\Delta}_c$ & complex cavity detuning including cavity dissipation: $\tilde{\Delta}_c\equiv \Delta_c + i\kappa$ \\
      $\delta_c$ & dispersively-shift cavity detuning: $\delta_c\equiv \Delta_c - \int U(\mathbf{r})\hat{n}(\mathbf{r})d\mathbf{r}$ \\
      $\tilde{\delta}_c$ & dispersively-shift complex cavity detuning: $\tilde{\delta}_c\equiv \Delta_c - \int U(\mathbf{r})\hat{n}(\mathbf{r})d\mathbf{r} + i\kappa=\delta_c+i\kappa=\tilde{\Delta}_c- \int U(\mathbf{r})\hat{n}(\mathbf{r})d\mathbf{r}$ \\
      $\lambda_c$ & wave length of the cavity field\\
      $k_c$ & wave number of the cavity field: $k_c=2\pi / \lambda_c$\\
      $\boldsymbol{\epsilon}_c$ & polarization vector of the cavity field\\
      $\mathcal{E}_{mn}(\mathbf{r})$ & Hermite-Gaussian wavefunction of $\text{TEM}_{mn}$ mode\\
      $\mathcal{E}_0$ & maximum electric field strength of a single photon in the TEM$_{00}$ mode with volume $\mathcal{V}$: $\mathcal{E}_0 = \sqrt{\hbar \omega_c / 2\epsilon_0 \mathcal{V}}$\\
      $\varphi^{\rm Gouy}_{mn}(x)$ & Gouy phase shift associated with $\text{TEM}_{mn}$ mode: $\varphi^{\rm Gouy}_{mn}(x)=(1+m+n) \arctan(\lambda_c x / \pi w_0^2)$\\
      $\varphi^{\rm offset}_{mn}$ & Phase offset due to boundary conditions of the cavity:  $\varphi^{\rm offset}_{mn}= (m+n) \arctan(\lambda_c l_{\rm res} / 2\pi w_0^2)$ \\
      $\hat{a}_\nu$ & photonic annihilation operator for mode $\nu$ with mode indices $\nu = (lmn)$\\
      $\alpha_\nu$ & coherent field amplitude for mode $\nu$  with mode indices $\nu = (lmn)$: $\alpha_\nu=\langle\hat{a}_\nu\rangle$  \\
      $\mathcal{G}_0$ & maximum single-photon vacuum Rabi frequency: $\mathcal{G}_0=\mathbf{d}_{ge}\cdot \boldsymbol{\epsilon}_c \mathcal{E}_0/\hbar$\\
      $\mathcal{G}(\mathbf{r})$ & position-dependent vacuum Rabi frequency: $\mathcal{G}(\mathbf{r})=\mathcal{G}_0\mathcal{E}(\mathbf{r})/\mathcal{E}_0$\\
      $\hbar U_0$ &  maximum cavity-lattice depth per photon (or maximum intra-cavity light frequency shift per atom): $\hbar U_0= \hbar \mathcal{G}_0^2/\Delta_a$\\
      $\hbar U(\mathbf{r})$ & position-dependent cavity lattice per photon: $\hbar U(\mathbf{r})=\hbar\mathcal{G}^2(\mathbf{r})/\Delta_a$\\
       $\kappa$ & decay rate of the cavity field through cavity mirrors \\
      $\Delta\nu$ & full width at half maximum of cavity resonance peak: $\Delta\nu = 2\kappa / 2\pi$\\
      $\mathcal{F}$ & Finesse of the cavity: $\mathcal{F}=\pi\nu_\mathrm{FSR}/\kappa$\\
      $C$ & cooperativity parameter: $C=\mathcal{G}_0^2/\kappa \gamma$\\
\end{longtable}
\end{center}

\setcounter{section}{0}
\renewcommand{\thesection}{}

\clearpage
\phantomsection
\addcontentsline{toc}{section}{References}

%\bibliography{../../bibtex_files/Review_MB_cavity_QED}
%\bibliographystyle{tfq}

\end{document}